\documentclass[
11pt, 
english, 
singlespacing, 
headsepline, 
]{MastersDoctoralThesis} 

\usepackage[utf8]{inputenc} 
\usepackage[T1]{fontenc} 
\usepackage{multirow}
\usepackage{mathpazo} 

\usepackage[style=numeric-comp,sorting=none,backend=biber]{biblatex}
\usepackage[colorlinks=true, pdfencoding=auto,psdextra]{hyperref}

\DeclareFieldFormat[article, inbook, incollection, inproceedings, misc, thesis, unpublished]{title}{#1}

\renewbibmacro{in:}{}
\DeclareSourcemap{
    \maps[datatype=bibtex]{
        \map{
            \step[fieldsource=doi, final] 
            \step[fieldset=archiveprefix, null] 
            \step[fieldset=arxivid, null] 
            \step[fieldset=eprint, null] 
        }
    }   
}

\usepackage[autostyle=true]{csquotes} 

\makeatletter
\def\blfootnote{\gdef\@thefnmark{}\@footnotetext}
\makeatother

\usepackage{mathtools}
\usepackage{lettrine}
\usepackage{amsmath,amssymb}
\usepackage{dsfont}
\usepackage{subcaption}
\usepackage{ stmaryrd }
\usepackage{pdfpages}
\usepackage{enumerate}

\usepackage{soul}
\usepackage{physics}
\usepackage{accents}
\usepackage{xcolor}
\usepackage{bbm}
\usepackage{graphicx}
\usepackage{diagbox}
\usepackage{cleveref}
\usepackage{tikz-cd}
\usepackage{subcaption}
\usepackage{float}
\newenvironment{Align}
  {\begin{equation}\begin{aligned}}
  {\end{aligned}\end{equation}\noindent\ignorespacesafterend}
\newenvironment{Align*}{\begin{equation*}
\begin{aligned}}
{\end{aligned}
\end{equation*}\noindent}

\newcommand{\ldv}[1]{\mathcal{L}_{#1}}

\newcommand{\If}{\mathbbm{i}}
\newcommand{\Lf}{\mathbbm{L}}
\newcommand{\R}{\mathbb{R}}
\newcommand{\C}{\mathbb{C}}
\newcommand{\D}{{\mathbbm{D}}}

\newcommand{\V}{{\mathbbm{V}}}
\newcommand{\W}{{\mathbbm{W}}}

\newcommand\sbullet[1][.5]{\mathbin{\vcenter{\hbox{\scalebox{#1}{$\bullet$}}}}}

\newcommand{\dbm}{\mathbbm{d}}
\newcommand{\defeq}{\vcentcolon=}
\newcommand{\os}{\,\hspace{.045cm}\hat{=}\,\hspace{.045cm}}

\newcommand{\spl}[1]{\mathrm{SL}(#1,\mathbb{R})}
\newcommand{\gl}[1]{\mathrm{GL}(#1,\mathbb{R})}

\newcommand{\ops}[1]{#1^{(S)}}

\newcommand{\pairing}[2]{\langle #1\, , \, #2 \rangle}
\newcommand{\coad}[1]{\mathrm{ad}^*_{#1}}
\newcommand{\ad}[1]{\mathrm{ad}_{#1}}

\newcommand{\updown}[2]{^{#1}_{\phantom{#1}#2}}
\newcommand{\downup}[2]{_{#1}^{\phantom{#1}#2}}
\newcommand{\da}{{a^\dag}}

\newcommand{\h}{\mathfrak{h}}
\newcommand{\Fock}{\mathfrak F}
\newcommand{\QCSop}{\hat{\boldsymbol{\chi}}}

\newcommand{\stintbt}[2]{\int_{#1}\dd \Sigma_{#2}\,}
\newcommand{\hg}{\hat{g}}

\newcommand{\intpull}[2]{\int_{#1} \phi^*}
\newcommand{\lem}{L_{\text{EM}}}
\newcommand{\DSint}[2]{\int_{\partial\Sigma} \dd \sigma_{#1 #2}\,}
\newcommand{\Sint}[1]{\int_\Sigma \dd \Sigma_{#1}\,}

\newcommand{\X}{\mathtt{x}}
\newcommand{\intpE}[2]{\int \dd\Omega_{#1,#2}\,}

\makeatletter
\DeclareRobustCommand{\loplus}{\mathbin{\mathpalette\dog@lsemi{+}}}
\DeclareRobustCommand{\lotimes}{\mathbin{\mathpalette\dog@lsemi{\times}}}
\DeclareRobustCommand{\roplus}{\mathbin{\mathpalette\dog@rsemi{+}}}
\DeclareRobustCommand{\rotimes}{\mathbin{\mathpalette\dog@rsemi{\times}}}

\newcommand{\dog@rsemi}[2]{\dog@semi{#1}{#2}{-90,90}}
\newcommand{\dog@lsemi}[2]{\dog@semi{#1}{#2}{270,90}}
\newcommand{\dog@semi}[3]{%
  \begingroup
  \sbox\z@{$\m@th#1#2$}%
  \setlength{\unitlength}{\dimexpr\ht\z@+\dp\z@\relax}%
  \makebox[\wd\z@]{\raisebox{-\dp\z@}{%
    \begin{picture}(1,1)
    \linethickness{\variable@rule{#1}}
    \roundcap
    \put(0.5,0.5){\makebox(0,0){\raisebox{\dp\z@}{$\m@th#1#2$}}}
    \put(0.5,0.5){\arc[#3]{0.5}}
    \end{picture}%
  }}%
  \endgroup
}
\newcommand{\variable@rule}[1]{%
  \fontdimen8  
  \ifx#1\displaystyle\textfont3\else
    \ifx#1\textstyle\textfont3\else
      \ifx#1\scriptstyle\scriptfont3\else
        \scriptscriptfont3\relax
  \fi\fi\fi
}
\makeatother

\thesistitle{At the Corner of Quantum and Gravity} 
\supervisor{Jerzy \textsc{Kowalski-Glikman}} 
\examiner{} 
\degree{Doctor of Philosophy} 
\author{Ludovic \textsc{Varrin}} 
\addresses{} 

\subject{Physical Sciences} 
\university{\href{http://www.ncbj.gov.pl}{National Centre for Nuclear Research}} 
\department{\href{https://www.ncbj.gov.pl/en/theoretical-physics-division}{Department BP2}} 
\group{\href{}} 
\faculty{\href{}{Factulty of Physics}} 

\AtBeginDocument{
\hypersetup{pdftitle=\ttitle} 
\hypersetup{pdfauthor=\authorname} 
\hypersetup{pdfkeywords=\keywordnames} 
}

\usepackage{tocloft}

\begin{document}

\frontmatter 

\pagestyle{plain} 


\begin{titlepage}
\begin{center}

{\scshape\LARGE \univname\par}\vspace{1.5cm} 
\textsc{\Large Doctoral Thesis}\\[0.5cm] 

\HRule \\[0.4cm] 
{\huge \bfseries \ttitle\par}\vspace{0.4cm} 
\HRule \\[1.5cm] 
 
\begin{minipage}[t]{0.4\textwidth}
\begin{flushleft} \large
\emph{Author:}\\
\href{https://google.com}{\authorname} 
\end{flushleft}
\end{minipage}
\begin{minipage}[t]{0.4\textwidth}
\begin{flushright} \large
\emph{Supervisor:} \\
\href{https://www.ncbj.gov.pl/}{\supname}\\ 
\end{flushright}
\end{minipage}\\[2cm]
 
\vspace{-1 cm}
\large \textit{A thesis submitted in fulfillment of the requirements\\ for the degree of \degreename}\\[0.3cm] 
\textit{in the}\\[0.4cm]
\groupname\deptname\\[-2cm] 
\includegraphics[width=0.5\textwidth]{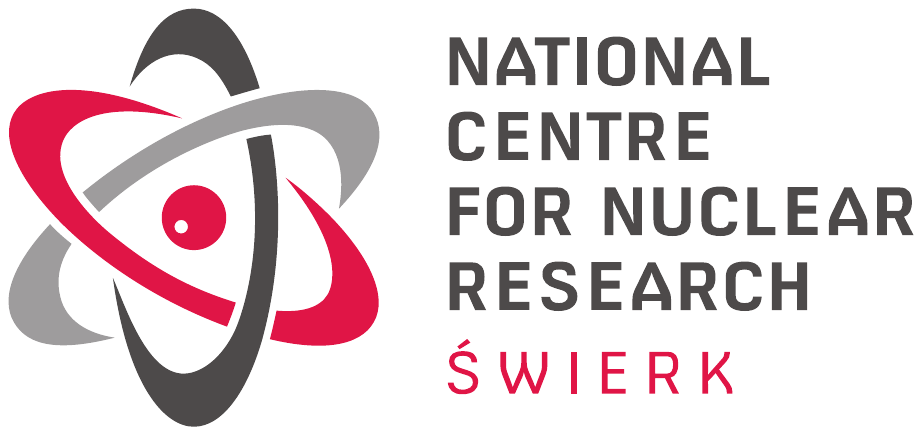}\\[-3cm]
\vfill
{\large March 23, 2026} 
 
\end{center}
\end{titlepage}


\begin{declaration}
\addchaptertocentry{\authorshipname} 
\noindent I, \authorname, declare that this thesis titled, \enquote{\ttitle} and the work presented in it are my own. I confirm that:

\begin{itemize} 
\item This work was done wholly or mainly while in candidature for a research degree at the National Centre for Nuclear Research.
\item Where any part of this thesis has previously been submitted for a degree or any other qualification at the National Centre for Nuclear Research or any other institution, this has been clearly stated.
\item Where I have consulted the published work of others, this is always clearly attributed.
\item Where I have quoted from the work of others, the source is always given. With the exception of such quotations, this thesis is entirely my own work.
\item I have acknowledged all main sources of help.
\item Where the thesis is based on work done by myself jointly with others, I have made clear exactly what was done by others and what I have contributed myself.\\
\end{itemize}
 
\noindent Signed:\\
\rule[0.5em]{25em}{0.5pt} 
 
\noindent Date:\\
\rule[0.5em]{25em}{0.5pt} 
\end{declaration}

\cleardoublepage





\begin{abstract}
\addchaptertocentry{\abstractname} 
In the presence of spacetime boundaries, diffeomorphisms in gravitational theories can become physical and acquire non-vanishing Noether charges.
These charges obey an algebra which, within the extended phase-space formalism, faithfully realizes diffeomorphism algebra.
The corner proposal takes this algebra of physical corner symmetries as a fundamental ingredient of quantum gravity, in close analogy with the role of the
Poincar\'e group in quantum field theory.
In this thesis we develop the quantum corner framework in the two-dimensional setting. We give the full representation theory of the two-dimensional extended corner
symmetry group, which may be interpreted either as the symmetry group of two-dimensional gravity or as the corner symmetry group relevant for four-dimensional spherically
symmetric gravity. Within the corner proposal, the resulting representation spaces are then interpreted as candidate Hilbert spaces for quantum gravity.
This representation-theoretic structure naturally enables a description of local subsystems. In particular, we present a gluing procedure that constructs
quantum states associated
with an entangling corner between two spacetime subregions, and use it to compute the entanglement entropy between the two regions.
To connect the quantum observables to the classical corner charges, we construct the coadjoint orbits of the quantum corner symmetry group and relate
them to the classical structure through twisted moment maps and to the quantum structure through generalized Perelomov coherent states.
This provides a notion of semiclassical limit within the corner framework.
Finally, in the context of static, spherically symmetric spacetimes, we show that a distinguished family of coherent states reproduces the horizon area law for entropy in the semiclassical limit, yielding a quantum, symmetry-based explanation of the Bekenstein--Hawking formula.

\end{abstract}

\begin{extraAbstract}
\addchaptertocentry{\extraAbstractname}\begin{otherlanguage}{polish}
W teoriach grawitacji w obecności brzegów czasoprzestrzeni, dyfeomorfizmy mogą stać się symetriami fizycznymi i mogą posiadać niezerowe ładunki Noether.
Ładunki te tworzą algebrę, która w ramach formalizmu rozszerzonej przestrzeni fazowej wiernie realizuje algebrę dyfeomorfizmów.
Hipoteza narożnikowa (\textit{corner proposal}) traktuje tę algebrę fizycznych symetrii narożnikowych jako fundament kwantowej grawitacji,
w ścisłej analogii do roli grupy Poincar\'ego w kwantowej teorii pola.
W niniejszej pracy rozwijamy formalizm kwantowych narożników w przestrzeni dwuwymiarowym.
Przedstawiamy pełną teorię reprezentacji dwuwymiarowej rozszerzonej grupy symetrii narożnikowych, którą można interpretować albo jako grupę symetrii
dwuwymiarowej grawitacji, albo jako grupę symetrii narożnikowych konfiguracji o symetrii sferycznej w czterowymiarowej grawitacji. W ramach hipotezy narożnikowej
otrzymane przestrzenie reprezentacji algebry narożnikowej interpretuje się jako przestrzenie Hilberta dla kwantowej grawitacji. Taka struktura reprezentacyjna naturalnie
umożliwia opisanie lokalnych podukładów: przedstawiamy procedurę „sklejania”, która konstruuje stany kwantowe związane z narożnikiem jako splątanie pomiędzy dwoma podobszarami czasoprzestrzeni, oraz wykorzystujemy ją do obliczenia entropii splątania między tymi regionami.
Aby powiązać obserwable kwantowe z klasycznymi ładunkami narożnikowymi, konstruujemy orbity reprezentacji sprzężonej kwantowej grupy symetrii narożnikowych i wiążemy
je ze strukturą klasyczną poprzez skręcone mapy momentu oraz ze strukturą kwantową poprzez uogólnione stany koherentne Perelomova. 
W ten sposób definiujemy pojęcie granicy semiklasycznej w ramach formalizmu narożnikowego. Na koniec, w kontekście statycznych, 
sferycznie symetrycznych czasoprzestrzeni, pokazujemy, że wyróżniona rodzina stanów koherentnych odtwarza w granicy semiklasycznej 
regułę pola powierzchni dla entropii horyzontu, dostarczając kwantowego, opartego na symetriach wyjaśnienia wzoru Bekensteina--Hawkinga.
\end{otherlanguage}
\end{extraAbstract}

\begin{acknowledgements}
\addchaptertocentry{\acknowledgementname} 

This thesis is the result of years of intense work. It would not have been possible without the support of collaborators and friends, whom I would like to thank wholeheartedly.\par
First of all, to my PhD advisor, Jerzy. I feel incredibly grateful to have had the chance to be supervised by such a knowledgeable and curious physicist.
Your deep understanding of physics and the patience you demonstrated sitting through my ramblings about formal mathematics were invaluable to me.
Although you might not have realized it at the time, the topic you gave me could hardly have been better suited to my interests, and I will always be grateful for that.
I hope we can continue working together for a long time.\par
To my spiritual co-advisor, Luca. Your kindness and guidance got me through the hardest times of these past few years. You taught me so much, both in physics and in the ins and outs of the academic world, and I could not imagine what my PhD would have looked like without your presence.
I appreciate that you tried to save me from self-desctruction by means of representation theory, even though I went back to it the moment you looked away.
I am eternally grateful to have met you. You are, in your own words, \textit{un illuminato}.\par
I also want to thank my friends and colleagues who were exiled to the cyclotron building with me.
We may have lost the battle against the astrophysicists, but the war is not over.
We will obtain offices in the main building, as is our fundamental right.
Daniele, I could not have gone through the jungle of administrative duties without your help.
Andrés, Esau, Jonas, Joseph, René, thank you for making lunches something to look forward to. Giuliano, thank you for sharing my frustrations with the man.
Margherita and Yashwanth, thank you for the incredible nights out.\par
Outside of work, I would like to thank my friends back in Switzerland for their unconditional love and support through the hardest times I faced while living abroad.
Sage, for always being there to listen to my stories even when we lose touch for a while. Baptiste, Colas, Nathan, Sam, for the amazing musical
projects that allowed me to disconnect when physics was not playing nice. Lucie, for always being a positive presence in my life and visiting me in Warsaw. Anto, for your regular ``kefa frère'' and for giving me access to your paid accounts.\par
Emi, thank you for your support, presence, and the wonderful times we had together in the past \textbf{couple} of years.\par
Last but not least, I would like to thank my parents for everything they did for me, for their emotional and financial support and for their
unwavering belief in my decision to follow my dreams. Merci du fond du c\oe ur, vous avez toujours cru en moi, même quand moi je n'y croyais pas.

\end{acknowledgements}


\tableofcontents 

\mainmatter 

\pagestyle{thesis} 

\chapter{Introduction}
\vspace{-1.5pt}
The modern structure of theoretical physics is built on a beautiful story of successive unifications.
As a student, one usually first learns about classical Newtonian mechanics and thermodynamics as separate theories based on distinct fundamental physical laws.
As one progresses through their studies, they find themselves in the unfortunate position of having to study statistical physics.
\footnote{Goodstein's book on statistical physics \cite{Goodstein1975StatesOfMatter}
infamously begins with "\textit{Ludwig Boltzmann, who spent much of his life studying statistical mechanics, died in 1906, by his own hand. Paul Ehrenfest,
carrying on the work, died similarly in 1933. Now it is our turn to study statistical mechanics.}"}
The frustrating difficulties of the field put aside, one quickly realizes that
thermodynamics is nothing but Newton's theory applied to a very large number of particles, and that thermodynamics is simply an emergent behavior of complex systems
subject to the same fundamental laws that govern the trajectories of macroscopic objects. A similar story befalls Maxwell's theory of electricity and magnetism.
At first, the electrical and magnetic fields were described as two distinct physical objects obeying separate equations.
Einstein's theory of special relativity \cite{Einstein1905Electrodynamics} then showed
that they were one and the same, and that the apparent distinction only appears because of the relative motion between the observer and the field.
Often, these unifications originate from generalizing a theory and understanding it as a special-case limit of the new, more general setting. This is the nature of the relationship
between classical mechanics and relativistic mechanics, or between Newton's theory and quantum mechanics.
Eventually, the student is confronted with perhaps the greatest unificatory achievement of modern physics: quantum field theory.
Not only can they now describe relativistic quantum particles in a unified framework, but the pesky distinction between forces
and particles has all but disappeared. Every physical phenomenon that the student has ever heard of is described in terms of fluctuating quantum fields.
Fortunately, the student has not heard about the small caveat that these particles and forces
constitute only about five percent of the energy content of the universe, and as such, they rejoice in this uncontested victory of reductionism. In the back of their mind however,
there still lies one small worry.
In parallel with the quantum field theory lecture, they also learned Einstein's theory of general relativity \cite{Einstein:1916vd}, another successful unification of the description of physical phenomena
from the perspective of different observers, which also happens to describe gravity as the geometry of spacetime. How this fits into the quantum picture of the universe was not discussed by the professor.
``I'll make sure to enroll in the lecture that explains this last piece of the puzzle next semester'' they think to themselves. Of course, that lecture never came.\par

The question of quantum gravity has been unresolved for more than a century. General relativity emerged in 1915, and quantum mechanics in 1926.
Although it is usually believed that quantum gravity was born in the second half of the 20th century,
some early remarks were made by Einstein, Klein, Pauli, and others
\cite{Einstein1916Naeherungsweise,Klein1928Funfdimensionalen,Rosenfeld1930Quantelung,Rosenfeld1930Gravitationswirkungen,BlokhintsevGalperin1934,Heisenberg1938Grenzen,
Fierz1939Kraeftefreier,PauliFierz1939Elektromagnetischen}. Later, Dirac, Bergmann, Arnowitt, Deser, and Misner worked out the canonical structure of general relativity, leading 
to the famous ADM formalism \cite{Misner1957Feynman,Dirac1958Hamiltonian,Dirac1959Fixation,ArnowittDeserMisner1962Original,Bergmann1989Canonical}.
This, in turn, led to the famous Wheeler--DeWitt equation \cite{dewitt1967quantum,DeWitt1967QuantumII,DeWitt1967QuantumIII,Wheeler1968Superspace}.
Together with the canonical formalism, this is called canonical quantum gravity and constitutes one of the earliest systematic nonperturbative approaches
to quantum gravity. While remarkable, this formalism encounters a number of issues. Some of which are the problem of time \cite{Kuchar1991ProblemOfTime,Isham1993CanonicalQuantumGravity}, the regularizing and ordering of the kinetic term, and the anomaly problem.\par
Later, in the seventies, as the understanding of quantum field theory was deepened by Wilson's theory of renormalization
\cite{Wilson1971StrongInteractions,WilsonKogut1974EpsilonExpansion}, t'Hooft and his collaborators proved that general relativity is not renormalizable \cite{tHooft1973Algorithm,tHooftVeltman1974Gravitation,DeserVanNieuwenhuizen1974EinsteinMaxwell,DeserVanNieuwenhuizen1974DiracEinstein}.\footnote{In what might be the first instance of the tradition of starting with Yang-Mills theory as a warmup for the gravitational case, t'Hooft and Veltman won the Nobel prize by showing that Yang-Mills theory is renormalizable \cite{tHooft1971MassiveYangMills,tHooftVeltman1972GaugeFields}}
This is an important step, as it suggests that general relativity may not admit a satisfactory quantization, but should instead be viewed as a low-energy effective theory of gravity, much like Fermi’s four-fermion theory in relation to the full electroweak theory.
Another very important result from this decade is the discovery that black holes radiate
\cite{Bekenstein1973,Bekenstein1974,Hawking:1975vcx} (see also \cite{Zurek1982,Page1983Comment} for latter comments). This implies that black holes are thermodynamic systems.
In particular, they possess an entropy which, remarkably, scales with the area of the horizon rather than with its volume. This “area law” also appears as the leading behavior of entanglement entropy in local quantum field theory \cite{Bombelli:1986rw,Srednicki:1993im}. Explaining its microscopic origin has therefore become
a standard benchmark for candidate theories of quantum gravity.
For brevity, let us now jump ahead to the most famous candidates for quantum gravity: string theory and Loop Quantum Gravity (LQG).
We refer the interested reader to \cite{Stachel1999EarlyHistoryQG,Rovelli2002BriefHistoryQG} for a more detailed historical account.\par
String theory originated as a model for hadrons and the strong interaction \cite{Veneziano1968CrossingSymmetric,Nambu1970FactorizationVeneziano,Nielsen1970AlmostPhysical,Susskind1970DualSymmetricI}. It was later realized that the string spectrum contains the graviton---a massless spin-2 particle characteristic of gravity---which
led to the proposal that string theory could provide a unified description of all interactions
\cite{Yoneya1973ZeroSlope,ScherkSchwarz1974Nonhadrons,Yoneya1974Gravidynamics}. Although this perspective was initially largely ignored by the community, it gained
major traction following the famous work of Green and Schwarz on the anomaly cancellation \cite{GreenSchwarz1984AnomalyCancellation}, which sparked the
first superstring revolution. Throughout the 1980s and 1990s, substantial progress was made in the development of superstring theory and supergravity
\cite{Polyakov1981BosonicString,Polyakov1981FermionicString,BelavinPolyakovZamolodchikov1984,GrossHarveyMartinecRohm1985,CandelasHorowitzStromingerWitten1985,
HorowitzLykkenRohmStrominger1986,GreenSchwarzWitten1987Vol1,GreenSchwarzWitten1987Vol2}. Later, the second superstring revolution, led by Witten,
brought nonperturbative aspects of string theory and M-theory to the forefront
\cite{HullTownsend1995Unity,Polchinski1995DirichletRR,Duff1996MTheory,BanksFischlerShenkerSusskind1997}.
String theory is an exceptionally rich theoretical framework,
but this very richness is often regarded as one of its main weaknesses.
In particular, the existence of an enormous number of consistent vacua---often estimated to be of order $10^{500}$ \cite{BoussoPolchinski2000Fluxes}---appears to obstruct the extraction of unique low-energy predictions from the theory.
While some authors invoke the anthropic principle to the rescue, others question the scientific status of a framework with such a restricted predictive power \cite{Susskind2003AnthropicLandscape,Woit2006NotEvenWrong,Smolin2006TroubleWithPhysics}.
Other commonly discussed issues include the absence of manifest background independence in most formulations of the theory, as well as the lack of experimental evidence for supersymmetric particles, whose simplest low-energy realizations are increasingly constrained by collider searches \cite{ArkaniHamed2012Future}.\par
Loop quantum gravity, on the other hand, is a nonperturbative approach to quantum gravity with no ambition of providing a unified description of all interactions.
It is based on Ashtekar's connection formulation of general relativity \cite{Ashtekar1987NewHamiltonian} and on the loop-like solutions to the Wheeler--DeWitt
equation found by Jacobson and Smolin \cite{JacobsonSmolin1988NonPerturbative}. The loop representation was first introduced in
\cite{RovelliSmolin1988KnotTheory,RovelliSmolin1990LoopSpace} and subsequently developed in \cite{AshtekarIsham1992Holonomy,AshtekarRovelliSmolin1992Weaving,AshtekarLewandowski1995Projective,Loll1995VolumeOperator,RovelliSmolin1995Discreteness,Rovelli:1996dv,Rovelli1996LoopBlackHolePhysics,Thiemann1996AnomalyFree,AshtekarLewandowski1997AreaOperators,Krasnov:1996tb,ReisenbergerRovelli1997SumOverSurfaces,Lewandowski1995Operators,Ashtekar2000QuantumMechanicsGeometry}.
LQG provides an extremely interesting framework in which to address some of the standard problems of quantum gravity.
However, it also faces a number of important challenges. Not least of all is the connection to classical general relativity and the emergence of a smooth spacetime structure in the classical limit \cite{NicolaiPeetersZamaklar2005OutsideView,AlexandrovRoche2011CriticalOverview}.\par
Quantum gravity is therefore in a peculiar position: its main candidate frameworks are now more than thirty years old, and although substantial progress has been made in developing their internal consistency, their connection to observable physics remains limited and, in some respects, increasingly elusive.
This naturally raises an important question. Since a complete theory of quantum gravity appears to be out of reach for the foreseeable future,
is there an alternative approach that allows one to make qualitative statements about its structure without requiring a fully detailed, top-to-bottom description?
Were one to undertake this task, what would be a natural starting point?
In this thesis we put forward a proposal inspired by the historical development of particle physics. A bottom-up approach
to quantum gravity, built around the concept lying at the heart of every unification story mentioned earlier: \textit{symmetries}.\par
The word symmetry comes from the Greek $\sigma\upsilon\mu\mu\epsilon\tau\rho\iota\alpha$ (\textit{symmetria}), whose implied meaning is ``measured together''.
Its use in descriptions of the natural world can
be traced back to the Platonic solids. Kepler's failed theory of planetary configurations, for instance, was based on these five regular polyhedra.
A more successful use of symmetries emerged from the study of permutations, developed through the work of Lagrange, Abel, Galois, and Cauchy in the eighteenth
and nineteenth centuries. Building on this body of ideas, Cayley introduced the modern abstract notion of a group, soon followed by Lie's seminal work on continuous
groups \cite{Lie1874UeberGruppen,Lie1880TheorieI,LieEngel1888Transformationsgruppen}. Together with the foundations of analytical mechanics by Lagrange, Hamilton and Jacobi
\cite{Lagrange1811MecaniqueAnalytiqueI,Lagrange1815MecaniqueAnalytiqueII,Hamilton1834GeneralMethod,Hamilton1835SecondEssay,Jacobi1866Vorlesungen},
this culminated with the Noether's famous work on the connection
between continuous symmetries and conserved quantities \cite{Noether:1918zz}. Nowadays, symmetries occupy a central place in
theoretical physics: gauge symmetries describe interactions in the Standard Model \cite{YangMills1954,Weinberg1967ModelLeptons,Salam1968WeakElectromagnetic}, spontaneous symmetry breaking accounts
for the origin of mass \cite{EnglertBrout1964BrokenSymmetry,Higgs1964GaugeBosons,GuralnikHagenKibble1964},
and Wigner's classification of unitary irreducible representations of the Poincar\'e group \cite{Wigner:1939cj} underpins quantum field theory.
The latter will be of particular interest in this thesis, as it provides the main analogy guiding our considerations of quantum gravity.\par
Although it is undeniable that symmetries play an essential role in modern theoretical physics, their physical status has long been a subject of debate. Which symmetries are physical, and which are gauge symmetries characterizing redundancies of the description? This turns out to be a subtle and important question, especially in the presence of boundaries.
The idea that boundaries can play an essential role in the physics of gauge theories was already appreciated in the condensed-matter community with the discovery of
the \textit{quantum Hall effect} in the 1980s \cite{vonKlitzingDordaPepper1980,Laughlin1981QuantizedHall}. For a system of electrons confined to a finite two-dimensional surface,
one can show that chiral excitations appear at the edge (boundary), and these are responsible for the quantized conductance characteristic of the quantum Hall effect.
It is now well understood that this phenomenon can be described by a $(2+1)$-dimensional $U(1)$ Chern--Simons theory in which, to maintain gauge invariance of the
action in the presence of a boundary, an additional term must be included. This modification induces non-trivial boundary dynamics, which in turn accounts for the chiral
edge excitations \cite{ZhangHanssonKivelson1989,FrohlichKerler1991Universality,FrohlichZee1991LargeScale,Balachandran:1991dw,Wen:1992vi}. The dynamical fields localized at the spatial boundary are referred to as
edge modes, edge states, or boundary modes.
Even though it was not fully understood at the time, this provided an early indication that gauge theories exhibit qualitatively new features in the presence
of spatial boundaries. In the quantum Hall setting, the gauge symmetry itself is responsible for the emergence of edge modes, with directly observable consequences.
This brings us back to the gauge-versus-physical debate: in what sense, if at all, are gauge symmetries physical? It is often asserted that gauge symmetries are pure
redundancy and therefore carry no physical information. This viewpoint is reinforced by the standard link between symmetries and observables provided by Noether's
theorem: the Noether charge of a gauge symmetry vanishes, seemingly confirming
that it conveys no physical content.
How, then, can this be reconciled with the manifest physical role of gauge symmetry in Chern--Simons theory? The key point is that boundaries can give rise
to non-vanishing charges associated with gauge symmetries, and these charges are localized entirely at the edge. In the quantum Hall effect,
the $U(1)$ gauge symmetry acquires boundary charges that correspond precisely to the electric charges carried by the edge modes.\par
On the gravitational side, the first appearance of gravitational boundary charges traces back to the work of Regge and Teitelboim \cite{Regge:1974zd} and to the birth
of asymptotic symmetries in the seminal work of Bondi, Sachs and Metzner \cite{BONDI1960,Sachs1961,Bondi1962,Sachs1962,Sachs1962b}.
As we discuss in more detail later in this thesis, there has been a vast amount of work on asymptotic symmetries, as well as on boundary symmetries more generally.
\footnote{For a modern review and further references, we refer the reader to the introduction of \cite{Ciambelli:2022vot}.}
A striking recent development is the so-called \emph{infrared triangle} \cite{Strominger:2017zoo}, whose modern formulation is largely due to Strominger and collaborators.
The basic observation is that three structures which had long been studied separately in gauge theory and gravity are in fact deeply intertwined: asymptotic symmetries,
soft theorems, and memory effects. On the one hand, the Bondi--Metzner--Sachs (BMS) group provides the natural asymptotic symmetry group of asymptotically
flat spacetimes at null infinity. On the other hand, Weinberg's soft graviton theorem states that scattering amplitudes
exhibit a universal factorization property in the limit where an external graviton becomes soft \cite{Weinberg1964PhotonsGravitons}. The modern insight is that these two facts
are not independent. The leading soft graviton theorem can be reinterpreted as the Ward identity associated with BMS supertranslations
\cite{Strominger2014BMS,HeLysovMitraStrominger2015}. In parallel, gravitational memory---the permanent relative displacement of detectors after the passage of
gravitational radiation---was shown to be the classical observable encoding the same infrared physics \cite{StromingerZhiboedov2016Memory,PasterskiStromingerZhiboedov2016Memories}.
In this way, asymptotic symmetries, soft theorems, and memory effects form the three corners of a triangle of dualities.
This perspective was subsequently extended in several directions. In particular, subleading soft graviton theorems were related to generalized asymptotic symmetries,
often described in terms of superrotations or Virasoro-like enhancements of the BMS group
\cite{CachazoStrominger2014SoftGraviton,KapecLysovPasterskiStrominger2014,Campiglia:2014yka,CampigliaLaddha2015NewSymmetries}.\par
\begin{figure}[H]
    \centering
    \includegraphics[width=\linewidth]{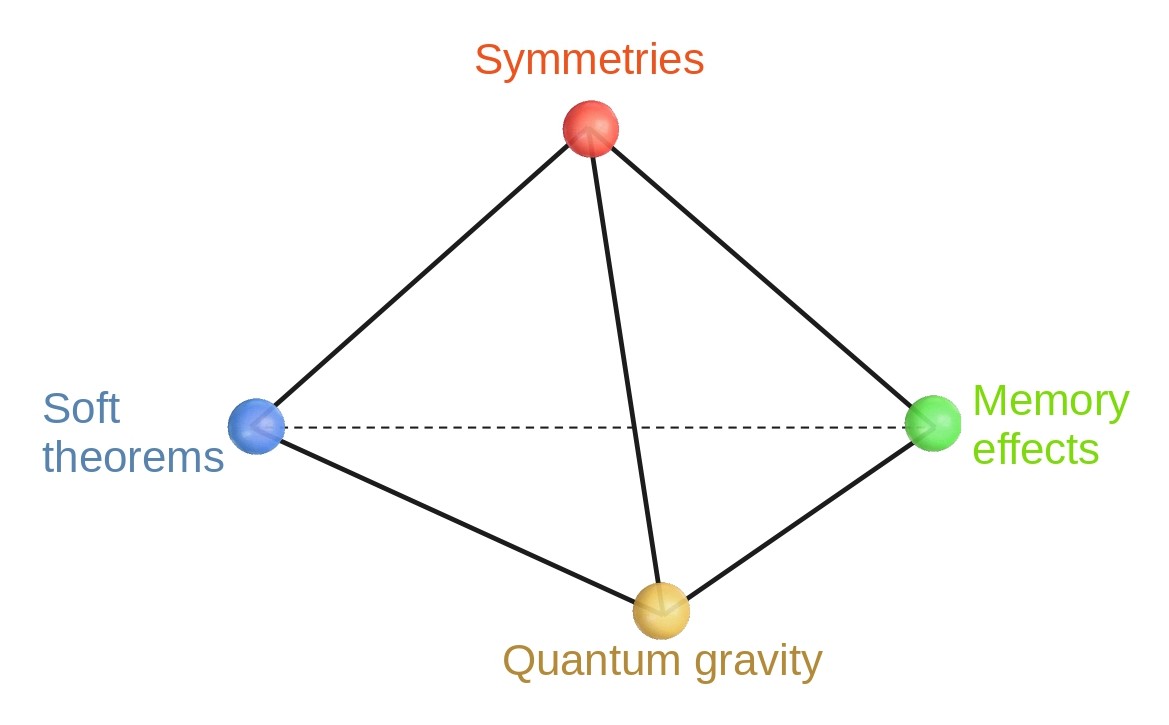}
  \caption{Reproduced from \cite{Ciambelli:2023bmn}, the symmetries pyramid represents the perspective of the corner proposal as a generalization of Strominger's infrared triangle. In this picture, symmetries are the fundamental structure from which memory effects, soft theorems, and quantum gravity can be derived.}
    \label{fig:symmetrypyramid}
\end{figure}

The central proposal of this thesis is that symmetries are at the core of quantum gravity. From this viewpoint, the infrared triangle
is naturally promoted to a \emph{symmetries pyramid}, a generalization due to Ciambelli (see the last section of \cite{Ciambelli:2023bmn}), reproduced
here in Figure~\ref{fig:symmetrypyramid}. In this picture, symmetries occupy the apex of the pyramid, from which memory effects, soft theorems, and quantum
gravity itself may be understood to descend.
This brings us to the main topic of the present work, whose origins lie in the seminal paper of Donnelly and Freidel \cite{Donnelly:2016auv}, who showed that four-dimensional Einstein--Hilbert gravity
admits non-vanishing Noether charges on codimension-2 boundaries---called \textit{corners}---which obey the \textit{extended corner symmetry} algebra.
The associated group is
\begin{equation}\label{eq1}
    \mathrm{ECS} = \mathrm{Diff}(S)\ltimes \qty(\spl{2} \ltimes \R^2)^S,
\end{equation}
where $S$ denotes the corner, and the exponent indicates that there is a copy of the algebra in parentheses at each point of $S$.
It was later shown by Ciambelli and Leigh \cite{Ciambelli:2021vnn} that the ECS is a subgroup of the \textit{universal corner symmetry} (UCS) group,
\begin{equation}\label{eq2}
    \mathrm{UCS} = \mathrm{Diff}(S)\ltimes \qty(\mathrm{GL}\qty(2,\R)\ltimes \R^2)^S,
\end{equation}
which captures the maximal set of symmetries that can admit non-vanishing Noether charges in any gravitational theories.
While the meaning of this statement will be clarified later, we now explain how these symmetry groups are expected to play a
fundamental role in quantum gravity through the lens of the \textit{corner proposal}.\par
Motivated by the universality of the group \eqref{eq2}, the corner proposal asserts that the UCS should be taken as the fundamental ingredient of gravity, much as the Poincar\'e
group is the fundamental ingredient of quantum field theory.
While the groups \eqref{eq1} and \eqref{eq2} were originally obtained either from topological considerations and the existence
of a spacetime manifold, or directly from a specific gravitational theory, the idea behind the corner program is to reverse this line of reasoning
and instead start from the symmetry groups themselves, remaining agnostic about their origin.
This framework is particularly relevant for quantum gravity, since symmetries provide one of the most robust handles for relating a classical theory to its quantum realization.
In particular, in direct analogy with Wigner's classification, quantum-gravitational states should fall into unitary irreducible representations of the corner
symmetry group. This provides a handle on the space of possible states without requiring a complete dynamical theory, which is why the approach is often
described as \textit{bottom-up}. As mentioned earlier, one of the first benchmark tests for quantum-gravity proposals is to account for the statistical origin of the
entropy area law. The present work is devoted to explaining the emergence of the area law within the corner framework, thereby providing a quantum, symmetry-based derivation
of the Bekenstein--Hawking entropy formula.
\par
This thesis is organized as follows. Chapter~\ref{chapter2} begins by presenting the general mathematical framework needed to treat corner symmetries
in gravitational theories. The second part of the chapter is devoted to the universal corner symmetries discussed above, and to an explicit example of their
emergence in spherically symmetric Einstein--Hilbert gravity. Chapter~\ref{chapter3} develops the mathematical formalism underlying the quantum description of corner
charges and algebras. The resulting objects are referred to as \emph{quantum corners}. The chapter starts with a general discussion of the necessary background and
then applies it to the particular case of interest in this thesis. Finally, Chapter~\ref{chapter4} addresses the physics of quantum corners, from the description
of local subsystems to the emergence of the area law.
\par
This thesis is based on the author's contributions to the review articles and original works \cite{Ciambelli:2023bmn,Assanioussi:2023jyq,Ciambelli:2024qgi,Varrin:2024sxe,Neri:2025fsh,Ciambelli:2025ztm,Kowalski-Glikman:2025arealaw,Varrin:2025okc}.
Whenever material is reproduced from---or closely based on---these references, this is stated explicitly. Additionally, the author's specific contribution is indicated.
As an example, parts of the introduction were taken from the author's contribution to \cite{Assanioussi:2023jyq}.

\chapter{Gravitational Charges}\label{chapter2}
Noether's famous theorem \cite{Noether1918} states
that, for each continuous symmetry of the system, there exists an associated conserved quantity.
In the context of gravity, the symmetries are simply the freedom in the choice of coordinates, i.e. diffeomorphisms.
So what are the associated conserved quantities? The naive answer is that diffeomorphisms are gauge symmetries encoding redundancies in the description and, as such, carry vanishing Noether charges. This is the hallmark of a non-physical symmetry. The modern perspective however, is
more subtle and requires a more precise definition of gauge and physical symmetries. Since the latter plays such an important role in the corner proposal and in this thesis,
we spend some time here to give a precise definition. We start with the famous definition of Weyl \cite{Weyl1952Symmetry}
\begin{center}
    ``\textit{A symmetry is a transformation of the state of a physical system that leaves its essential properties unchanged.}"
\end{center}
In particular, a symmetry transformation acts on the dynamical degrees of freedom of a theory. Now what is a physical symmetry? In physics, symmetries are usually
separated into two categories: $(i)$ global symmetries, which are parametrized by constant transformation parameters and:
$(ii)$ local symmetries, for which the parameters can be different at each point of spacetime.
The latter are usually interpreted as gauge redundancies of the description, and hence as non-physical. This is precisely the point that requires refinement.
As mentioned in the introduction, local symmetries can have physical consequences in the presence of spacetime boundaries.
Since these effects are captured by the appearance of a non-vanishing Noether charge, we will use it to classify symmetries: physical symmetries are those that admit
a non-vanishing Noether charge. With this criterion, some local symmetries become physical in the presence of a boundary, while others do not.\par
A widely studied example of this phenomenon is that of asymptotic symmetries, where the relevant boundary is taken to be asymptotic infinity. In this thesis, however, our main focus will be on finite spacetime subregions.
In that setting, the charges arise because one restricts attention to a finite region of spacetime. One of the main difficulties in the presence of finite-distance boundaries is the appearance of fluxes,
which spoil the integrability of the charges for diffeomorphisms that move the boundary.
In the context of this thesis, this is a consequential obstacle, since we would like to treat any diffeomorphism admitting a non-vanishing Noether charge as the
Hamiltonian generator of a canonical transformation.
As we will see, this issue is tied to how the boundary is embedded in spacetime, and it can be resolved by treating that embedding dynamically.\par
Let us now outline the organization of this chapter and summarize its main results. Section~\ref{sec:covariantphasespace} is devoted to the covariant phase-space
formalism, the modern framework used to analyze symmetries and Noether charges.
We begin with a general introduction to Cartan calculus on spacetime and to its field-space generalization, which lies at the core of the formalism.
We then present the construction for a general Lagrangian theory and formulate Noether's theorems in this language, thereby providing a systematic procedure
to obtain the symplectic structure and the Noether charges associated with the symmetries of the theory. Finally, we introduce the extended phase-space formalism,
which treats the boundary embedding as a dynamical field and resolves the issues of fluxes and integrability.
To illustrate this less familiar aspect of the covariant phase-space formalism, we apply it to global Poincar\'e symmetries in electromagnetism.\par
Next, in Section~\ref{sec:cornersymmetriesingravity} we discuss the concrete emergence of corner charges in gravity.
We begin by explaining how the universal corner symmetries arise from topological considerations of embedded surfaces. This
is where the UCS group, and its important subgroups such as the ECS, first appear. We then apply the covariant phase-space formalism to spherically symmetric
four-dimensional Einstein--Hilbert gravity. We begin by performing the dimensional reduction,
recasting the theory as two-dimensional dilaton gravity, and then compute the diffeomorphism Noether charges.
This shows that the group of physical symmetries is indeed the two-dimensional version of the ECS, thereby justifying the considerations that follow in the next chapters.\par
We conclude this introduction with a few remarks about the notation in this chapter:
\begin{itemize}
    \item Greek indices $\mu,\nu,\ldots$ denote spacetime indices in four dimensions.
    \item Latin indices $a,b,\ldots$ label coordinates tangential to the corner, while $i,j,\ldots$ label coordinates normal to the corner.
    \item Cartan calculus on spacetime is denoted by $\dd$, $\iota$, and $\mathcal{L}$, standing respectively for the exterior derivative, interior product, and Lie derivative. Spacetime vector fields are denoted by capital letters $V,W,\ldots$.
    \item The corresponding field-space operators are denoted by $\dbm$, $\If$, and $\Lf$, and field-space vectors by $\V,\W,\ldots$.
\end{itemize}
\par
Subsections~\ref{sec:spacetimecartancalculus} and \ref{sec2.2}, the example in Section~\ref{sec:extendedphasespaceandambiguities}, and parts of Subsection~\ref{sec:Lagrangeformalism}
are based on the author's contributions to the review \cite{Assanioussi:2023jyq}. Parts of Subsection~\ref{subsec:universalcornersymmetries} and of Subsection~\ref{sec:extendedphasespaceandambiguities}
are based on the author's contributions to the review \cite{Ciambelli:2023bmn}. Subsection~\ref{subsec:cornersymmetriessss}
is based on the original work \cite{Kowalski-Glikman:2025arealaw}.
All original calculations underlying the new results in that subsection were first performed by the author of this thesis.

\section{Covariant Phase Space}\label{sec:covariantphasespace}

The Covariant Phase-Space (CPS) formalism is a geometric framework in which the symplectic
structure is defined on the space of solutions of the theory, rather than on a space of initial data used to parametrize those solution.
This allows to define a phase space without having to choose an initial Cauchy slice which breaks manifest covariance.
The formalism's mathematical construction hinges on the variational bicomplex developed independently by
Vinogradov \cite{Vinogradov1977AlgebroGeometric,Vinogradov1978SpectralSequence,Vinogradov1984CSpecI,Vinogradov1984CSpecII} and Tulczyjew \cite{Tulczyjew1980EulerLagrange}
and later refined by Anderson \cite{Anderson1991,AndersonTorre1996}.
In this thesis, however, we will focus on presenting the covariant phase space in an accessible way that does not require advanced prior knowledge of differential
geometry. While the origin of the CPS formalism can be traced back to Lagrange himself, its most recent incarnation originated in the seventies
\cite{Gawedzki1972Geometrization,Kijowski1973FiniteDimensional,KijowskiSzczyrba1976CanonicalStructure} and was further developed in the eighties
\cite{Witten1986OpenSuperstrings,Crnkovic:1986ex,Crnkovic:1987tz,AshtekarBombelliReula1991,Hayward:1993my}.
It was finally put into its modern form by Iyer, Lee, Wald and Zoupas \cite{Lee:1990nz,Wald:1993nt,Iyer_1994,Iyer_1995,Wald:1999wa}.
For modern and comprehensive reviews we refer the reader to \cite{Margalef-Bentabol:2020teu,Harlow:2019yfa}.\par
The formalism hinges on the generalization of Cartan calculus to the space of field configurations. We therefore start in Subsection \ref{sec:spacetimecartancalculus} by
giving a refresher on the standard calculus on spacetime before moving on to its generalization to field space in Subsection \ref{sec2.2}. Next, 
in Subsection~\ref{sec:Lagrangeformalism}, we describe the formalism and the surprisingly simple form that Noether's theorems take within it. Finally, in Subsection \ref{sec:extendedphasespaceandambiguities}, 
we describe the extended phase space formalism which solves the question of integrability of the Noether charges. This is illustrated with the example of global
Poincaré symmetries in electromagnetism.

\subsection{Cartan calculus on spacetime}\label{sec:spacetimecartancalculus}

Given a $4$-dimensional spacetime $M$, let us denote the space of $p$-forms by $\Lambda^{p}(M)$ and  the space of all differential forms on the manifold by $\Lambda^{\sbullet}(M)$. The Cartan calculus consists of a set of three \textit{differentiations}: the exterior derivative $\dd$, the interior product $\iota_V$ and the Lie derivative $\ldv{V}$, where $V \in \mathrm{T}M$ is a spacetime vector field. The exterior derivative is a derivation of degree 1, which means that it raises the rank of the form by one
\begin{equation}
	\dd : \Lambda^{p}(M) \longrightarrow \Lambda^{p+1}(M).
\end{equation}
It is the antisymmetrisation of the usual differentiation operation on tensors and is defined as follows. Let $V_i$ be a collection of spacetime vector fields and $\alpha$ a $p$-form. Then
\begin{Align}
	\dd\alpha(V_1,\ldots,V_{p+1}) &= \sum_{i=1}^p (-1)^{i+1}\ V_i\left(\alpha(V_1,\ldots,V_{i-1},V_{i+1},\ldots, V_{p+1})\right)\nonumber\\
	&+ \sum_{i<j} (-1)^{i+j}\ \alpha([V_i,V_j],V_1\ldots,V_{i-1},V_{i+1},\ldots,V_{j-1},V_{j+1}, V_{p+1})
\end{Align}
where $[V_i,V_j]$ denotes the commutator of two vector fields.

On the other hand, the interior product is defined as the contraction of a differential form with a vector field $V$. It is a differentiation map of degree $-1$, which means that it lowers the rank of the form
\begin{equation}
	\iota_V: \Lambda^{p}(M) \longrightarrow \Lambda^{p-1}(M). 
\end{equation}
The interior product is sometimes called a \textit{contraction}, and its action on a $p$-form $\alpha$ is given by
\begin{equation}
	(\iota_V \alpha)(V_1,V_2,...,V_{p-1}) = \alpha(V,V_1,V_2,...,V_{p-1}).
\end{equation}
Finally, the Lie derivative can be given by a combination of the two former differentiations, called \textit{Cartan's magic formula}
\begin{Align}
	\ldv{V} \alpha = \dd(\iota_V \alpha) +  \iota_V(\dd \alpha)  
\end{Align}

The name ``differentiation" comes from the fact that these operators obey a graded Leibniz rule with respect to the exterior product:
\begin{align}
	\dd (\alpha \wedge \beta) &= (\dd \alpha) \wedge \beta + (-1)^p \alpha \wedge (\dd \beta),  \label{spacetimeexteriorcalculus1} \\
	\iota_V (\alpha \wedge \beta) &= (\iota_V \alpha) \wedge \beta + (-1)^p \alpha \wedge (\iota_V \beta), \label{spacetimeexteriorcalculus2} \\
	\ldv{V} (\alpha \wedge \beta) &= (\ldv{V} \alpha) \wedge \beta + \alpha \wedge (\ldv{V} \beta), \label{spacetimeexteriorcalculus3}
\end{align}
for all $\alpha \in \Lambda^{p}(M)$ and $\beta \in \Lambda^{q}(M)$. Note that by definition $\dd^2 = \iota_V^2 = 0$, and
\begin{align}
	\left[ \ldv{V},\ldv{w} \right] &= \ldv{\left[ v,w \right]}, \\
	\iota_V \iota_w + \iota_w \iota_V &= 0 .
\end{align}
for all $V,W \in \mathrm{T}M$. One can then show that the following relations between the three operators hold
\begin{align}
	\left[\ldv{V},\iota_W\right] &= \iota_{\left[V,W\right]}, \label{differentationsrelation1} \\
	\left[\ldv{V},\dd \right] &= 0. \label{differentationsrelation2}
\end{align}
Lastly, for a (pseudo-)Riemannian manifold, the metric tensor allows for the definition of the Hodge dual which takes $p$-forms to $(d-p)$-forms
\begin{equation}
    \star: \Lambda^p \longrightarrow \Lambda^{d-p},
\end{equation}
While the Hodge dual admits a coordinate-free definition, it is more convenient to introduce it in the coordinate language. For this reason, we take this opportunity to recast the Cartan calculus introduced earlier in a coordinate formulation, after which the definition of the Hodge star will follow naturally. 

A basis of the vector space $\Lambda^p(M)$ can be obtained by taking wedge products of the basis one-form $\dd x^\mu$. Any $\alpha \in \Lambda^p(M)$ can thus be written
\begin{equation}\label{eq:pformdefinition}
    \alpha = \frac{1}{p!}\alpha_{\mu_1\mu_2...\mu_p}\dd x^{\mu_1...\mu_p}
\end{equation}
where $\alpha_{\mu_1\mu_2...\mu_p}$ is totally antisymmetric and we used the notation
\begin{equation}
 	\dd x^{\mu_1...\mu_p} \defeq \dd x^{\mu_1}\wedge...\wedge \dd x^{\mu_p}.
\end{equation} 
The wedge product of two forms $\alpha \in \Lambda^p(M)$ and $\beta\in \Lambda^q(M)$ can then be written
\begin{equation}
    \alpha \wedge \beta = \frac{1}{p!q!}\alpha_{\mu_1...\mu_p}\beta_{\nu_1...\nu_q} \dd x^{\mu_1...\mu_p \nu_1...\nu_q}.
\end{equation}
Using the conventions of equation \eqref{eq:pformdefinition}, one can use the antisymmetrization operation
\begin{equation}
    T_{[\mu_1...\mu_p]} = \frac{1}{p!}\sum_{\sigma\in S_p} \mathrm{sgn}(\sigma)T_{\mu_{\sigma(1)}...\mu_{\sigma(p)}},
\end{equation}
where $S_p$ denotes the group of permutation of order $p$, to write the components of the $(p+q)$ form $\alpha\wedge \beta$ as
\begin{equation}
    \qty(\alpha\wedge \beta)_{\mu_1...\mu_{p+q}} = \frac{(p+q)!}{p!q!}\alpha_{[\mu_1...\mu_p}\beta_{\mu_{p+1}...\mu_{p+q}]}.
\end{equation}
Similarly, the exterior derivative and contraction operation of a $p-$form have the following components
\begin{Align}
    \qty(\dd \alpha)_{\nu\mu_1...\mu_p} &= (p+1)\partial_{[\nu}{\alpha_{\mu_1...\mu_p]}},\\
    \qty(\iota_V\alpha)_{\mu_2...\mu_p} &= V^{\mu_1} \alpha_{\mu_1 \mu_2...\mu_p},\label{eq:contractioncoordinate}
\end{Align}
where $V^{\mu}$ are the components of the vector in the dual basis to $\dd x^\mu$
\begin{equation}
    V = V^\mu \pdv{x^\mu}.
\end{equation}
Lastly, we can now define the Hodge dual operation by writing down the components of the dual form
\begin{equation}
    (\star \alpha)_{\mu_{1}...\mu_{d-p}} = \frac{1}{p!} \varepsilon_{\mu_1...\mu_{d-p}\nu_1...\nu_p}\alpha^{\nu_1...\nu_p},
\end{equation}
where $\epsilon_{\mu_1...\mu_d}$ is the Levi-Civita \textbf{symbol} with $\epsilon_{01...d} = 1$ and 
\begin{equation}\label{eq:levicivitatensordensity}
    \varepsilon_{\mu_1...\mu_d} = \sqrt{\abs{g}} \epsilon_{\mu_1...\mu_d},
\end{equation}
is the Levi-Civita \textbf{tensor density}. 

\subsection{Cartan calculus on field space}\label{sec2.2}

We now move to generalize the definitions and results of the previous section to the space $\Gamma$ of all possible field configurations $\{\varphi^I(x)\}$ on the spacetime of a given theory, where $x$ represents a spacetime point and $I$ stands for all the indices (spacetime, algebra, etc.) of the field. From now on the space $\Gamma$ will be referred to as the field space.
The field space $\Gamma$ can be viewed as an infinite dimensional manifold where each point corresponds to a field configuration $\varphi^I$ on spacetime.
In other words, it is a manifold with coordinates $\varphi^I$ which consists of the collection of all field components at all spacetime points (see for example \cite{KrigleMichor1997} for more details). 
For simplicity, we now restrict ourselves to a four-dimensional spacetime. Lastly before we begin, in order to avoid confusion, we give here the definition of the language we use when talking about symmetries. 
A symmetry is called internal if it acts ont the internal indices of the family of fields. It is called a spacetime symmetry if instead it acts on the fields via their spacetime indices and their spacetime argument.
In this section, we will denote internal symmetry generators by $\lambda$ and spacetime symmetry generators by $\xi$.
Furthermore, a symmetry is called global if the symmetry parameter does not depend on the spacetime point. A symmetry
is called local, if the symmetry parameter does depend on the spacetime point. Finally as mentioned in the introduction to this chapter, a symmetry is called physical if it
admits a non-vanishing Noether charge and gauge if it does not.

Vector fields $\V$ on $\Gamma$, i.e.~$\V \in T\Gamma$, can be expressed in a basis generated by the functional derivatives with respect to the field configurations $\varphi^I$
\begin{equation}\label{fieldspacevector}
	\V = \int \dd^4 x \, \V^I(x) \frac{\dbm}{\dbm \varphi^I(x)},
\end{equation}
Internal symmetries associated with a symmetry parameter $\lambda$ will be generated on field space by the vector field 
\begin{equation}
	\V_{\lambda} = \int \dd^4 x \, \delta_{\lambda}\varphi^I \frac{\dbm}{\dbm \varphi^I(x)},
\end{equation}
Spacetime symmetries associated with a symmetry parameter $\xi$ will be generated on field space by the vector field
\begin{equation}\label{fieldspacediffeovectorfield}
\V_\xi = \int \dd^4 x \, \delta_\xi\varphi^I \frac{\dbm}{\dbm \varphi^I(x)}.
\end{equation}
In particular, if $\xi\in \mathrm{T}M$ represents a diffeomorphism, we have
\begin{equation}
	\D_{\xi} = \int \dd^4 x \, \ldv{\xi} \varphi^I \frac{\dbm}{\dbm \varphi^I(x)},
\end{equation}
where, in this section, we denote by $\D_\xi$ the field-space vector field generating diffeomorphisms, to emphasize that it is the generator of this particular local spacetime symmetry.
We now want to define differential forms on $\Gamma$ in order to introduce the Cartan calculus. The starting point of the covariant phase space formalism is to interpret the variation of a field as the field space exterior derivative $\dbm$ acting on the field $\varphi$, and consequently producing a field space one-form\footnote{Another common notation for the field space exterior derivative is $\delta$ owning to the fact that it corresponds to the variation in field space.}:
\begin{equation}
	\mathbbm{d} : \Gamma \longrightarrow \Lambda^{(1)}(\Gamma).    
\end{equation}
This interpretation connects nicely to the usual meaning of the variation in a way that will soon become clear. The defining relation of the one-form $\dbm \varphi^I$ is given by its action on the basis vector fields on $\Gamma$
\begin{equation}
	\dbm\varphi^I\left(\frac{\dbm}{\dbm \varphi^b}\right) = \delta^a_b.
\end{equation}
It follows that the action of $\dbm \varphi^I$ on the vector $\V$ produces the vector components as in the usual spacetime setup
\begin{equation}
	\dbm \varphi^I (\V) =  \V^a.
\end{equation}

We can further introduce the interior product $\If_\V$ on the space of one-forms on $\Gamma$
. It maps the basis\footnote{Note that it suffices to define its action on $\dbm \varphi^I$, as these one-forms constitute a basis of the space of all one-forms.} $\dbm \varphi^I$ as

	\begin{Align}
		\If_\V: \Lambda^{(1)}(\Gamma) &\longrightarrow \Gamma\\
		\dbm \varphi^I &\mapsto \If_\V \dbm \varphi^I = \dbm \varphi^I(\V).
		\label{fieldspaceinteriorproduct}
	\end{Align}

This allows one to connect the field space exterior derivative to the usual interpretation of the variation of a field. That is, the variation of the field with respect to the symmetry generator $\lambda/\xi$ is the field space
contraction of the one form $\dbm \varphi^I$ with the field space vector constructed form the symmetry generator:
\begin{equation}\label{eq:variationascontraction}
	\delta_{\lambda/\xi}\varphi^I = \If_{\V_{\lambda/\xi}} \dbm \varphi^I.
\end{equation}

The last ingredient needed for Cartan calculus is a field space Lie derivative that we will denote $\Lf_\V$ and introduce through Cartan's magic formula:
\begin{equation}
	\Lf_\V = \dbm \If_\V + \If_\V \dbm.    
	\label{cartanfieldspace}
\end{equation}
Finally, it is straightforward to introduce an exterior product, and generalize $\dbm$ and $\If_\V$ to $p$-forms in such a way that the field space analogue of relations \eqref{spacetimeexteriorcalculus1}--\eqref{spacetimeexteriorcalculus3} holds. We then have a field space Cartan calculus with the standard relations
\begin{Align}
	\left[\Lf_\V,\If_\W \right] &= \If_{[\V,\W]},\label{eq:standardrelation}\\
	\left[\Lf_\V,\dbm\right] &= 0,
\end{Align}
where, $\V$ and $\W$ are field space vectors.

The introduction of Cartan calculus on the field space implies that, in the description of a theory, we now have two types of forms, those on spacetime and those on the field space. As we will see, the dynamical objects defining a theory will in general be $(p,q)$-forms, meaning a $p$-form on spacetime and a $q$-form on field space. We denote the space of $(p,q)$-forms by $\Lambda^{(p,q)}(M,\Gamma)$.
To complete Anderson's variational bicomplex~\cite{Anderson1991}, one introduces an exterior derivative on the total bicomplex. A standard choice is to define the total differential as the sum of the horizontal (spacetime) and vertical (field space) differentials:
$
d_{\mathrm{tot}} := \dd + \dbm.
$
In the conventional bicomplex setting this construction requires the two differentials to anticommute
so that \(d_{\mathrm{tot}}^{\,2}=0\). Since we will not rely on this abstract structure here and will focus instead on explicit computations, we adopt the more convenient convention
\begin{equation}
	\dbm \dd = \dd \dbm.
\end{equation}
To conclude this section note that equations \eqref{eq:variationascontraction} and \eqref{cartanfieldspace}, in the case of diffeomorphisms, implies 
\begin{equation}\label{eq:fieldspacelieequalspacetimelie}
	\Lf_{\D_\xi} \varphi^I(x) = \If_{\D_\xi} \dbm \varphi^I(x) = \ldv{\xi}\varphi^I(x),
\end{equation}
where we have used the fact that $\varphi^I(x)$ is a field space 0-form. More generally, the infinitesimal transformation of a field space tensor $\mathbb{T}$ can always be expressed in terms of the field space Lie derivative,
\begin{equation}
	\delta_\xi \mathbb{T} = \Lf_{\D_\xi} \mathbb{T}.
\end{equation}
This motivates the following key definition. The field space tensor $\mathbb{T}$ is called \textit{covariant} if its field space Lie derivative
coincides with its spacetime Lie derivative,
\begin{equation}
	\delta_\xi \mathbb{T} = \ldv{\xi}\mathbb{T}.
\end{equation}
This is a non-trivial condition. Indeed, since field space consists solely of dynamical fields, $\Lf_{\D_\xi}$ implements the diffeomorphism transformation
only on dynamical fields, whereas
$\ldv{\xi}$ also acts on background fields. Because symmetries, by definition, act only on dynamical fields, only covariant objects transform correctly under
symmetries.
The simplest situation is when there are no background structures, in which case field space tensors such as the Lagrangian and the symplectic potential are
covariant. In this thesis,
we restrict our attention to this case. For a more nuanced discussion, we refer the reader to \cite{Harlow:2019yfa}.

\subsection{Lagrangian formalism and Noether's theorems}\label{sec:Lagrangeformalism}

In this section, we use the Cartan calculus that we just presented to analyze a Lagrangian system, and we formulate Noether's second theorem for local symmetries.
For more details on the subject, we refer the reader to \cite{Lee:1990nz,Wald:1993nt,Iyer_1994,Iyer_1995,Wald:1999wa,Harlow:2019yfa,Ciambelli:2022vot}.
\par
In order to define the action of a theory, we integrate the Lagrangian density over the spacetime manifold $M$, which from here on, is assumed to be a globally
hyperbolic spacetime. The Lagrangian density is a spacetime $4$-form, also called a top-form. Moreover, it is a functional of the fields and their derivatives, and is therefore a function on field space i.e.~$0$-form \footnote{The dependence of the Lagrangian density on derivatives of the field does not change this fact. A proper treatment of the first and higher order derivatives of the fields in a geometrical language requires the notion of \textit{jet bundles}. This goes beyond what is needed for the material of this thesis. For a definition and application of the jet bundles in the covariant phase space formalism, we refer the reader to \cite{Anderson1991,KrigleMichor1997}}. The action of the theory is thus defined as
\begin{equation}
	S = \int_M L \left[\varphi^I\right],
\end{equation}
where $L \in \Lambda^{(4,0)}(M,\Gamma)$. By using the usual Leibniz rule, the field space exterior derivative of the Lagrangian density can always be written in the following form
\begin{equation}\label{VariationLagrangian}
	\dbm L[\varphi^I] = E_I \dbm \varphi^I + \dd \theta[\varphi^I,\dbm \varphi^I],
\end{equation}
where $E_I$ are the equations of motion and $\theta$ is a $(3,1)$-form, that is $\theta \in \Lambda^{(3,1)}(M,\Gamma)$, called the \textit{symplectic potential current}. The form $\theta$ contains all of the necessary information to provide a symplectic structure \cite{Arnold, SympGeo} for the field space of the theory. By integrating $\theta$ on a submanifold $\Sigma$ of a Cauchy surface we obtain
\begin{equation}\label{SympPot}
	\Theta = \int_\Sigma \theta,
\end{equation}
which is a $(0,1)$-form called the \textit{symplectic potential (or tautological form)} \cite{SympGeo}. We can now take the field space exterior derivative of the presymplectic potential to obtain the (pre)symplectic current,
\begin{equation}\label{PresympCurrent}
  \omega \defeq \dbm \theta \in \Lambda^{(d-1,2)}(M,\Gamma).
\end{equation}
Integrating $\omega$ over a Cauchy slice $\Sigma$ defines a closed $(0,2)$-form on field space,
\begin{equation}\label{PresympForm}
  \Omega \defeq \dbm \Theta = \int_\Sigma \dbm \theta = \int_\Sigma \omega,
\end{equation}
which we call the \textit{presymplectic form}.

Two comments are in order.
First, for the presymplectic form to be independent of the choice of Cauchy slice, the fields must satisfy appropriate boundary conditions. This issue is closely related to ambiguities in the definition of the symplectic structure and will be discussed later in this section.

Second, the term presymplectic is used to indicate that the form~\eqref{PresympForm} may be degenerate. Famously, this occurs in gauge theories, where the presymplectic form is degenerate along gauge orbits in $\Gamma$%
.\footnote{This is a direct consequence of Noether’s second theorem, as we will see shortly.}
As a result, the symplectic form cannot, in general, be inverted to define a Poisson bracket. One may nevertheless treat $\Gamma$ as a phase space by identifying the variables and their conjugate momenta from the expression for $\Omega$, and by introducing Poisson brackets via the canonical commutation relations. The resulting phase space is then a Poisson manifold, but not necessarily a symplectic one.
The standard resolution is to consider the gauge-reduced phase space obtained by quotienting the field space $\Gamma$ by the action of the gauge group~$G$,
\begin{equation}
	\tilde{\Gamma} \defeq \Gamma/G .
\end{equation}
A non-degenerate symplectic form $\tilde{\Omega}$ can then be defined on $\tilde{\Gamma}$ in such a way that $\Omega$ is the pullback of $\tilde{\Omega}$ from $\tilde{\Gamma}$ to the pre-phase space $\Gamma$. It turns out that the degeneracy of the presymplectic form is a necessary condition for $\tilde{\Gamma}$ to be a symplectic manifold whose symplectic form induces the presymplectic form $\Omega$ on $\Gamma$~\cite{Crnkovic:1986ex}.
When boundaries are present, this necessary condition is spoiled by the appearance of boundary gauge charges as we will shortly show. This issue can be remedied by extending the phase space with edge modes, as in the seminal work of Donnelly and Freidel~\cite{Donnelly:2016auv}. In this thesis, however, we will not focus on the classical phase-space analysis of gauge systems, but rather on the values of the charges and their algebras. This requires that diffeomorphisms generate non-degenerate directions in field space when boundaries are present.
We will therefore forgo these subtleties and make a slight abuse of language by referring to $\Omega$ as the symplectic form.\par
The three field space forms allow a natural derivation of the Noether current and the charges associated to the considered symmetry transformation.
To illustrate this, consider a vector field $\V \in T\Gamma$ that preserves $\Omega$, i.e.\ $\If_{\V} \omega = 0$.
In other words, $\V$ generates a symplectomorphism. Using the fact that $\omega$ is an exact field space form we get
\begin{equation}
	0 = \Lf_\V \omega = \dbm \If_\V \omega.
\end{equation}
Assuming trivial cohomology on the space of field space 1-forms, this implies that
\begin{equation}
	\If_\V \omega = -\dbm J_\V,
\end{equation}
for some $J_\V \in \Lambda^{(3,0)}$ called the Noether current. The reason for this name becomes clear once we integrate the above relation on a Cauchy slice
\begin{equation}\label{hamiltonequation}
	\If_\V \Omega = - \dbm H_{\V},
\end{equation}  
where
\begin{equation}\label{eq:Noethercharge}
	H_{\V} = \int_\Sigma J_{\V},
\end{equation}
is the Noether charge. Equation \eqref{hamiltonequation} is the field space equivalent of Hamilton's equation in symplectic geometry.

The particular form of the Noether current depends on the nature of the symmetry. For internal
symmetries we can compute
\begin{equation}
	-\dbm J_\lambda = \If_{\V_\lambda} \dbm \theta = \Lf_{\V_\lambda} \theta - \dbm \If_{\V_\lambda}\theta.
\end{equation}
If we further assume that $\Lf_{\V_\lambda} \theta = 0$, \footnote{Since the vector field $\V_\lambda$ generates an internal symmetry of the theory we have
\begin{equation*}
	0=\Lf_{\V_\lambda}\delta L \os \dd \qty(L_{\V_\lambda}\theta).
\end{equation*}	
This means that, at least on-shell, the term $L_{\V_\lambda} \theta$ can at most be an exact spacetime term. Since the symplectic potential current is only defined
up to spacetime exact terms anyways, we go forward with the assumption $\Lf_{\V_{\lambda}}\theta =0$.}
we get
\begin{equation}\label{eq:internalnoethercurrent}
	J_\lambda = \If_{\V_\lambda}\theta.
\end{equation}
up to field space exact terms.
For spacetime symmetries however, the story is slightly more subtle. As for the internal symmetries, we can compute
\begin{equation}
	-\dbm J_\xi = \If_{\V_\xi} \dbm \theta = \Lf_{\V_\xi}\theta - \dbm \If_{\V_\xi}\theta.
\end{equation}
From the background independence discussion below equation \eqref{eq:fieldspacelieequalspacetimelie}, we further know that $\Lf_{V_\xi} \theta = \ldv{\xi}\theta$. We then get
\begin{Align}\label{eq:diffnoethercurrentwithflux}
	 \dbm J_\xi &= \dbm \If_{\V_\xi}\theta -\iota_\xi \dd \theta - \dd \iota_\xi\theta\\
	&\os \dbm \qty(\If_{\V_\xi} \theta - \iota_\xi L) - \dd \iota_\xi \theta,
\end{Align}
where in the second line we have used the definition of the symplectic potential current and the fact that spacetime and field space Cartan calculus commutes. The hat denotes that the equality is only valid on-shell of the equations of motions
\begin{equation}
	E_I \os 0.
\end{equation}
The term $\dd \iota_\xi \theta$ therefore appears as an obstruction to the integration of the above equation. This is due to \textit{fluxes}, a known issue that spoils the integrability of Hamilton's equation.
For the time being we will assume that the flux term vanishes, and we will treat the general case in Subsection \ref{sec:extendedphasespaceandambiguities}. We define the diffeomorphism Noether current
\begin{equation}\label{eq:diffeomorphismnoethercurrent}
	J_{\xi} = \If_{\V_\xi} \theta - \iota_\xi L.
\end{equation}
\paragraph{Noether's theorem}
Noether's theorems are the first instance where the beauty of the CPS formalism becomes apparent. The vector field $\V_{\lambda} \in \mathrm{T}\Gamma$ is a symmetry of the action $S$ if its field space Lie derivative vanishes.
We can then write
\begin{equation}\label{eq:fieldspaceliederivativeofaction}
	\Lf_{\V_{\lambda}} S = \int_M \If_{\V_\lambda} \dbm L \os \int_{M} \dd(J_\lambda).
\end{equation}
If $\V$ is a symmetry of the action, the local Noether current is thus conserved on-shell
\begin{equation}
	\dd J_\lambda = 0.
\end{equation} 
This implies that the global Noether charge \eqref{eq:Noethercharge} is conserved on-shell for internal symmetries, which is the statement of Noether's first theorem. If the symmetry is local, Noether's second theorem \cite{Noether1918} states that the current is exact on-shell
\begin{equation}\label{eq:exactcurrentinternal}
	J_{\V_{\lambda}} = \dd Q_{\lambda}.
\end{equation}
The above equation is the generalization of Gauss's law to any local internal symmetry. Since diffeomorphisms are also local symmetries, Noether's second theorem applies to them as well. This can be shown by first computing the field space Lie derivative of the action
\begin{equation}
	\Lf_{\D_\xi} S = \int_M \If_{\D_\xi} \dbm L \os \int_{M} \dd(\If_{\D_\xi} \theta) = \int_{\partial M} \If_{\D_\xi} \theta,
\end{equation}
where we have used Stokes theorem to write the last line as an integral over the boundary of $M$. Next we compute the spacetime Lie derivative of the action
\begin{equation}
 \ldv{\xi}S = \int_M \dd \iota_\xi L = \int_{\partial M} \iota_\xi L.
\end{equation}
If the action is covariant, we can equate the two and find that the integral of the diffeomorphism Noether current over the boundary of $M$ vanishes. This implies that the current can 
at most be an exact spacetime form as in \eqref{eq:exactcurrentinternal}. When integrating the current over a Cauchy slice, we obtain
\begin{equation}\label{eq:cornerdiffcharges}
	H_\xi = \int_\Sigma J_\xi = \int_{\partial \Sigma} Q_{\xi}.
\end{equation}
We therefore see that the Noether charges associated with diffeomorphisms can only have support on a codimension-2 surface $\partial \Sigma$. This codimension-2 surface is what we call the \textit{corner}, and equation \eqref{eq:cornerdiffcharges} is the origin of the emergence of corner charges in gravitational theories.\\
\paragraph{Poisson bracket}
One of the most important aspects of symplectic geometry is that the symplectic two-form can be used to define a Poisson bracket on the phase space
of the theory. More importantly for us in the context of the corner proposal, we need the non-vanishing diffeomorphism charges to furnish a representation of the diffeomorphism algebra.
Given two symplectomorphisms $\V, \W\in \mathrm{T}\Gamma$ we can define the Poisson bracket of their Hamiltonian charge as
\begin{equation}\label{Poissonbracket}
	\qty{H_{\V},H_{\W}} = \Lf_\W H_\V.
\end{equation}
The antisymmetricity of this bracket comes from the antisymmetric interior product
\begin{equation}
	\qty{H_\V,H_\W} = \If_\W \dbm H_\V = -\If_\W \If_\V \Omega = \If_\V \If_\W \Omega = - \If_\V \dbm H_\W = -\qty{H_\V,H_\W}.
\end{equation}
The above equation also shows that, for symplectomorphisms, the Poisson bracket arises from the symplectic form. Let us now take a look at diffeomorphisms. For two vector fields $\xi,\zeta \in \mathrm{T}M$, the Poisson bracket \eqref{Poissonbracket} reads
\begin{equation}
	\qty{H_\xi,H_\zeta} = \Lf_{\D_\zeta} H_\xi = \If_{\D_\zeta}\dbm H_{\xi}.
\end{equation}
On the other hand, let us compute
\begin{Align}
	I_{\D_{\qty[\xi,\zeta]}} \Omega &= I_{\qty[\D_\xi,\D_\zeta]} \Omega \\
	&= \qty[\Lf_{\D_\xi}, \If_{\D_\zeta}]\Omega\\
	&= \Lf_{\D_\xi} \If_{\D_\zeta}\Omega\\
	&= \dbm \If_{\D_\xi} \If_{\D_\zeta} \Omega + \If_{\D_\xi}\dbm \If_{\D_\zeta} \Omega.
\end{Align}
In the first line we used that the field space vectors \eqref{fieldspacediffeovectorfield} represent the diffeomorphism algebra, in the second, \eqref{eq:standardrelation},
in the third, that $\D_\xi$ is a symplectomorphism, and in the fourth, Cartan's magic formula \eqref{cartanfieldspace}.
We can use Cartan's formula again to write
\begin{equation}
	\If_{\D_\xi} \dbm \If_{\D_\xi} = \If_{\D_\xi} \Lf_{\D_\xi}-\If_{\D_\xi} \If_{\D_\xi} \dbm.
\end{equation}
Using the fact that $\Omega$ is closed and that $\D_\xi$ is a symplectomorphism we finally get
\begin{equation}
	-\dbm H_{\qty[\xi,\zeta]} = I_{\D_{\qty[\xi,\zeta]}} \Omega =  - \dbm \If_{\D_\xi} \dbm H_\zeta = \dbm \qty{H_{\xi},H_\zeta}.
\end{equation}
From which we deduce
\begin{equation}\label{eq:projectiverepofdiffeomorphism}
	\qty{H_\xi,H_{\zeta}} = - H_{\qty[\xi,\zeta]}+ \kappa_{\xi,\zeta},
\end{equation}
where $\kappa_{\xi,\zeta}$ is a constant in field space $\dbm \kappa_{\xi,\zeta} = 0$. Thus, the algebra of charges represents the diffeomorphism algebra projectively. 
The field space constant $\kappa$ plays the role of a central extension. As we will see, the presence of a central extension is crucial in the quantum theory.
There is another caveat to the result \eqref{eq:projectiverepofdiffeomorphism}. In the derivations, we have used the fact that $\D_\xi$ is a symplectomorphism and an Hamiltonian
vector field. As mentioned earlier, this is not always the case, as the presence of a flux in equation \eqref{eq:diffnoethercurrentwithflux} spoils the former.
In the next section, we will present a formalism that allows for every non-vanishing Noether charge to be integrable: the extended phase space formalism. Coincidentally,
the formalism will also take care of the central term in \eqref{eq:projectiverepofdiffeomorphism}.  
\paragraph{Coordinate language}
For concrete computations, it is often convenient to work in coordinates rather than in the language of differential forms. One of the issues of expressing the covariant phase space formalism in coordinates is that the symplectic potential current and the Noether current's degrees depend on the spacetime dimension.
They are however always of the same ``corank": $\theta$ and $J_{\V}$ are always $(d-1)$-forms, where $d$ is the dimension of the spacetime. Because of this fact, the Hodge dual introduced in Section \ref{sec:spacetimecartancalculus} is
particularly well suited to the general coordinate description of the covariant phase space. The main observation is the following. Define the $p$-dimensional volume element as
\footnote{Note that we have used the Levi-Civita tensor density \eqref{eq:levicivitatensordensity} which means that the volume element knows about the metric.}
\begin{equation}\label{eq:volumeelement}
	\dd \Sigma_{\mu_1....\mu_{d-p}} = -\frac{1}{p!}\varepsilon_{\mu_1...\mu_{d-p}\rho_{1}...\rho_p} \dd x^{\rho_1...\rho_p},
\end{equation}
where the minus sign comes from the Lorentzian signature.
Then any differential $p$-form $\alpha$ can be written using the surface element and the component of the Hodge dual
\begin{equation}
	\alpha =  \frac{1}{(d-p)!}(\star \alpha)^{\mu_1...\mu_{d-p}}\dd \Sigma_{\mu_1...\mu_{d-p}},
\end{equation}
When integrating a $p$-form over a codimension $(d-p)$-hypersurface $\Sigma$, the object defined in equation \eqref{eq:volumeelement} serves as the natural volume element of the hypersurface
\begin{equation}
	\int_\Sigma \alpha = \int_{\sigma}\dd \Sigma_{\mu_1...\mu_{d-p}}  \frac{1}{(d-p)!}(\star \alpha)^{\mu_1...\mu_{d-p}}.
\end{equation} 
For example, the symplectic potential current is a $(d-1)$-form on space time, which means it can be written as
\begin{equation}\label{eq:symplecticpotentialcurrentdualdecomposition}
\theta  = \qty(\star \theta)^\mu \dd \Sigma_\mu.
\end{equation}
In the following, whenever a form and its Hodge dual have different degrees (equivalently, carry a different number of free indices),
we will suppress the explicit Hodge star to lighten notation. Concretely, we identify the components of the dual by the index structure and write
\[
(\star \theta)^\mu \equiv \theta^\mu,
\]
with the understanding that $\theta^\mu$ always denotes the Hodge-dual components of the original $(d-1)$-form $\theta$.
 To see how natural this decomposition is in the covariant phase space formalism, we write the Lagrangian $d$-form as
\begin{equation}
	L = \sqrt{-g} \dd^d x \mathcal{L}
\end{equation}
where $\mathcal{L}$ is the Lagrangian density---a 0-form on spacetime.
The field space derivative can then be written on-shell
\begin{equation}
	\dbm \qty(\sqrt{-g}\mathcal{L}) \os \sqrt{-g}\nabla_\mu \theta^\mu = \partial_\mu \qty(\sqrt{-g}\theta^\mu).
\end{equation} 
Therefore, in concrete calculations done in the coordinate language, the object that appears in the variation of the Lagrangian after integrating by parts is actually the Hodge dual of the symplectic potential current.
The Noether current can be written in a similar way through the use of the Hodge dual. For internal symmetries, the only difference between the Noether current and the symplectic potential current is a field space operation \eqref{eq:internalnoethercurrent}.
We can thus directly write the Noether current dual 
\begin{equation}
	J_\lambda^\mu = \If_{\V_\lambda} \theta^\mu,
\end{equation}
such that the associated Noether charge is given by
\begin{equation}
	H_\lambda = \int_\Sigma \dd \Sigma_\mu J^\mu_{\lambda}.
\end{equation}
For diffeomorphisms, one first notes that the $(d-1)$-form $\iota_\xi L$ can be written \eqref{eq:contractioncoordinate}
\begin{equation}
	\iota_\xi L = \frac{\mathcal{L}}{(d-1)!}\xi^\nu\epsilon_{\nu \mu_1...\mu_{d-1}}  \dd x^{\mu_1...\mu_{d-1}}.
\end{equation}
from which it follows that the diffeomorphism charge can be written as
\begin{equation}
	H_\xi = \int_\Sigma \dd \Sigma_\mu J_\xi^\mu,
\end{equation}
with
\begin{equation}
	J_{\xi}^\mu = \If_{\V_\xi} \theta^\mu - \xi^\mu \mathcal{L}.
\end{equation}
Consider the $(d-2)$-form $Q$. Similarly to symplectic potential current \eqref{eq:symplecticpotentialcurrentdualdecomposition}, it can be written as 
\begin{equation}
	Q = \frac{1}{2}Q^{\mu\nu}\dd \Sigma_{\mu\nu},
\end{equation}
where
\begin{equation}
	Q^{\mu\nu} = \frac{1}{(d-2)!} \varepsilon^{\mu\nu\rho_1...\rho_{d-2}} Q_{\rho_1...\rho_{d-2}}.
\end{equation}
If this is the Noether charge aspect of the Noether current $J$, we have
\begin{equation}
	J^\mu = \nabla_\nu Q^{\mu\nu},
\end{equation}
And Stokes theorem is written
\begin{equation}
	H_{\xi} = \int_{\Sigma}\dd \Sigma_\mu J^\mu_{\xi} = \frac12 \int_{\partial\Sigma}\dd \Sigma_{\mu\nu}Q^{\mu\nu}_{\xi}.
\end{equation}

\clearpage
\subsection{Extended phase space}\label{sec:extendedphasespaceandambiguities}
Let us now discuss in more detail the flux term in equation \eqref{eq:diffnoethercurrentwithflux}. As mentioned at the time, this term
spoils the integrability of Hamilton's equation.
\begin{equation}
	\If_{\D_\xi}\Omega = -\delta H_{\xi} - \dd \iota_\xi \theta.
\end{equation}
This is an expression of the fact that the charges are not conserved, or that the system is open. In order to gather intuition, let us compute the exterior derivative of the diffeomorphism
Noether charge 
\begin{equation}\label{eq:281}
	\dbm H_{\xi} = \int_\Sigma \dbm J_\xi = \int_\Sigma \dbm (\If_{\D_\xi}\theta - \iota_\xi L) - \int_{\partial\Sigma}\iota_\xi \theta.
\end{equation}
We note that, if the diffeomorphism is tangent to the boundary $\partial \Sigma$, the term arising from the flux drops out of the equation and Hamilton's equation is satisfied.
This implies that the fluxes are related to diffeomorphisms that move the position of the boundary. The extended phase space formalism takes this into account by adding
the embedding of the subregion under consideration to the field space of the theory. We will now describe it in more details.\par
In the local holographic context of the corner proposal, we always consider an embedded subregion $\phi: U \longrightarrow M$. The action functional for this subregion is then given by
\begin{equation}\label{eq:actionU}
	S_U = \int_U \phi^*(L).
\end{equation}
The appearance of fluxes can be understood as a consequence of assuming that the field space variations commutes with the pullback in the above expression.
 More precisely, we assumed something of the following form in a number of computations:
\begin{equation}\label{variationandintegrationcommute}
    \Lf_{\D_{\xi}} \int_U \mathcal{A} = \int_U \Lf_{\D_{\xi}} \mathcal{A},
\end{equation}
for a general functional $\mathcal{A}$.
However, the commutation of the variation and the integral is only true in the case where the variation does not move the surface we are integrating on.
If we now explicitly write the pullback by the embedding and do not assume that it is a field space constant, we find
\begin{equation}
\Lf_{\D_{\xi}} \int_S \phi^*\qty(\mathcal{A}) = \int_S \phi^* \qty(\Lf_{\D_{\xi}} \mathcal{A}) + \int_S \qty(\Lf_{\D_{\xi}} \phi^*) \qty(\mathcal{A}),
\end{equation}
The question now becomes:
``\textit{How can we keep track of the variation of the embedding at the phase space level?}" This is done by the introduction of a new field $\chi$ which is a vector
in spacetime and a one-form on field space:
\begin{equation}
    \chi \in TM \otimes T^*\Gamma,
\end{equation}
defined by the relation
\begin{equation}\label{chicondition}
    \If_{\D_{\xi}} \chi = -\xi,
\end{equation}
This field keeps track of the variation of the embedding in the following sense. For any functional $\mathcal{A}$ on field space, we have
\begin{equation} \label{embeddingvariation}
    \delta \phi^* (\mathcal{A}) = \phi^* (\ldv{\chi} \mathcal{A}).
\end{equation}
Note that this relation can be thought off as the equivalence between the passive and the active interpretation of diffeomorphisms. The $\chi$ field, which is sometimes
called an \textit{edge mode} because it lives on the corner, is then added to the phase space of the theory.\par
We will now calculate the symplectic form of the theory on the extended phase space that includes this new field. In the derivation we make use of the
relations
\begin{align}
    \dbm \chi &= -\frac{1}{2} [\chi,\chi], \label{equationa}\\
    \ldv{\chi} \iota_\chi L &= \frac{1}{2} \iota_{\qty[\chi,\chi]}L + \frac{1}{2} \dd (\iota_\chi i _\chi L), \label{equationb}\\
    \If_{\D_{\xi}} \iota_{\delta \chi} L &= \iota_{\qty[\xi,\chi]} L,\label{equationc}
\end{align}
which are proved at the end of the section.
We take the variation of the action \eqref{eq:actionU} 
    \begin{Align}
        \dbm S_U &= \int_U \qty[\dbm \phi^* (L) + \phi^* (\dbm L)]\\
        &= \intpull{U}{d}(\ldv{\chi} L +\dbm L)\\
        &\os \intpull{U}{d}(\dd \iota_\chi L + \dd \theta) \\
        &= \intpull{U}{d} (\dd \theta^{\mathrm{ext}})\;,
    \end{Align}
where we defined the extended symplectic potential current $\theta^{\mathrm{ext}}[\varphi, \dbm \varphi, \chi]$.
Integrating it over a codimension-1 hypersurface gives the extended symplectic potential
\begin{equation}
    \Theta^{\mathrm{ext}}= \intpull{\Sigma}{d-1} \theta^{\mathrm{ext}} = \intpull{\Sigma}{d-1} (\iota_\chi L + \theta).
\end{equation}
Taking another variation of the above yields the extended symplectic form
    \begin{Align}
        \Omega^{\mathrm{ext}}&= \dbm \Theta^{\mathrm{ext}}\\
        &= \int_\Sigma [\dbm \phi^*(\iota_\chi L + \theta) + \phi^*(\dbm(\iota_\chi L+ \theta))]\\
        &= \intpull{\Sigma}{d-1} (\ldv{\chi} \iota_\chi L + \ldv{\chi} \theta + \iota_{\dbm \chi} L - \iota_\chi \dbm L + \dbm \theta)\\
        & \os \intpull{\Sigma}{d-1}(\ldv{\chi} \iota_\chi L + \dd \iota_\chi \theta + \dbm \theta + \iota_{\dbm \chi} L) \\
        &= \intpull{\Sigma}{d-1} (\dbm \theta + \frac{1}{2}\dd (\iota_\chi \iota_\chi L) + \dd \iota_\chi \theta)\\
        &= \intpull{\Sigma}{d-1} \left(\dbm \theta + \dd (\iota_\chi \theta + \frac{1}{2} \iota_\chi \iota_\chi L)\right),
    \end{Align}
where we have used equations \eqref{equationa} and \eqref{equationb} to go from the fourth to the fifth line.
Using Stokes theorem, this can be rewritten as
\begin{equation}\label{eq:extendedsymplecticform}
    \Omega^{\mathrm{ext}}= \Omega + \intpull{\partial \Sigma}{d-2} (\iota_\chi \theta + \frac{1}{2} \iota_\chi \iota_\chi L),
\end{equation}
and the additional term reduces to a corner contribution. This corner contribution is exactly what is needed to remove the flux term in equation \eqref{eq:281}. To see this,
let us contract the extended symplectic form with a diffeomorphism
    \begin{Align}
        I_{\D_{\xi}} \Omega^{\mathrm{ext}} &= I_{\D_{\xi}} \intpull{\Sigma}{d-1} (\dbm (\theta + \iota_\chi L) + \ldv{\chi} (\theta + \iota_\chi L))\\
        &= \intpull{\Sigma}{d-1}(I_{\D_{\xi}} \dbm \theta + I_{\D_{\xi}} \ldv{\chi} \theta + I_{\D_{\xi}} \ldv{\chi} \iota_\chi L + I_{\D_{\xi}} \iota_{\dbm \chi} L - I_{\D_{\xi}} \iota_\chi \dbm L)\\
        &= \intpull{\Sigma}{d-1}(\ldv{\xi} \theta - \dbm I_{\D_{\xi}} \theta + I_{\D_{\xi}} \ldv{\chi} \theta + I_{\D_{\xi}}\ldv{\chi}\iota_\chi L + \iota_{\qty[\xi,\chi]}L - I_{\D_{\xi}} \iota_\chi \dbm L)\\
        &= \intpull{\Sigma}{d-1}(- \dbm I_{\D_{\xi}} \theta - \ldv{\chi} I_{\D_{\xi}} \theta + I_{\D_{\xi}} \ldv{\chi} \iota_\chi L + \iota_{\qty[\xi,\chi]}L - I_{\D_{\xi}} \iota_\chi \dbm L)\\
        &= \intpull{\Sigma}{d-1} (- \dbm I_{\D_{\xi}} \theta - \ldv{\chi} I_{\D_{\xi}} \theta - \ldv{\xi} \iota_\chi L + \ldv{\chi} \iota_{\xi} L + \iota_{\qty[\xi,\chi]}L + \iota_{\xi} \dbm L + \iota_\chi I_{\D_{\xi}} \dbm L)\\
        &= \intpull{\Sigma}{d-1}(-\dbm I_{\D_{\xi}} \theta - \ldv{\chi} I_{\D_{\xi}} \theta + \ldv{\chi} \iota_{\xi} L + \iota_{\xi} \dbm L)\\
        &= \intpull{\Sigma}{d-1}( - \dbm I_{\D_{\xi}} \theta - \ldv{\chi} I_{\D_{\xi}}\theta + \ldv{\chi} \iota_{\xi} L + \iota_{\xi} \dbm L)\\
        &= - \intpull{\Sigma}{d-1}(\dbm (I_{\D_{\xi}}\theta - \iota_{\xi} L) + \ldv{\chi} (I_{\D_{\xi}}\theta - \iota_{\xi} L)),
        \end{Align}
		\vspace{-2.5pt}\noindent
where in the second line we have used \eqref{equationc} and in the third line we used the fact that $\theta$ and $L$ are covariant. The first term in the last line is nothing but the 
Noether current \eqref{eq:diffeomorphismnoethercurrent} and the second term is exactly what arises from the variation of the embedding. In other words
\vspace{-2.5pt}
    \begin{Align}\label{eq:extendedhamiltonequation}
        I_{\D_{\xi}} \Omega^{\mathrm{ext}} &\os - \intpull{\Sigma}{d-1} (\dbm J_{\D_{\xi}} + \ldv{\chi} J_{\D_{\xi}})\\
        &= - \dbm (\intpull{\Sigma}{d-1} (J_{\D_{\xi}}))\\
        &= -\dbm H_{\xi}.
    \end{Align}
\vspace{-2.5pt}\noindent
This is a remarkable result and deserves a few comments. Introducing the embedding into the field space of the theory not only renders all diffeomorphisms integrable,
it does so without altering the form of the Noether charge. In practice, this means that one can compute the Noether current \eqref{eq:diffeomorphismnoethercurrent} and integrate it over
the hypersurface to find the Noether charge \eqref{eq:cornerdiffcharges}. The extended phase space formalism then ensures that this charge obeys Hamilton's equation for the extended
symplectic form \eqref{eq:extendedsymplecticform}. Moreover, it can be shown \cite{Ciambelli:2021vnn} that the Poisson bracket obtained from the extended symplectic form realizes
the algebra of diffeomorphisms faithfully
\vspace{-2.5pt}
\begin{Align}\label{eq:faitfulldiffrep}
	\qty{H_{\xi},H_{\zeta}}_{\mathrm{ext}} = - H_{\qty[\xi,\zeta]}.
\end{Align}
\vspace{-3pt}\noindent
As mentioned in the beginning of this subsection,
the flux appeared for transformations that moved the boundary. Within the extended phase space formalism, the displacement of the boundary is encoded in the variation of the embedding, and as such
the flux is removed from the equations.We emphasize that the system is still open and non-conservative. The fluxes are simply encoded in the phase space by the embedding map.
The extended phase space therefore provides us with Hamiltonian charges for every diffeomorphism, and a Poisson bracket arising from a symplectic structure that realizes
the diffeomorphisms algebra for transformations that admit a non-vanishing Noether charge. This last point is crucial for the quantization procedure.
As we will see later on, the Poisson bracket induced by the field space symplectic form is what allows us to relate quantum observables to their classical counterparts.
This is why the extended phase-space formalism is better suited to the corner proposal than, for example, the Barnich-Troessaert bracket \cite{Barnich:2009se,Barnich:2010eb,Barnich:2011mi}.
\paragraph{The example of electromagnetism}
As a simple example of the extended phase space formalism, we consider electromagnetism. In addition to the gauge symmetry, the Maxwell Lagrangian is also invariant under the global Poincar\'e
transformations\footnote{Since Yang--Mills theory is massless, it also possesses a conformal symmetry.
Although we will not discuss this symmetry here, it can be treated similarly to the Poincar\'e symmetry.}. These are an example of
local spacetime symmetries. Since they form only a subset of diffeomorphisms, this case can be worked out explicitly without undue complications and provides a concrete example of the extended phase-space formalism in action. Since the calculations are simple enough,
we will compute
the Noether charges for the global Poincaré symmetries directly from the symplectic form and prove that Hamilton's equation are satisfied in the extended phase space formalism.\par
 Maxwell theory is described by the action
\begin{equation}
	S_{\text{EM}}[A] = -\frac14 \int_M \dd^4 x \, F^{\mu\nu}F_{\mu\nu},
\end{equation}
where the field strength tensor is defined from the vector potential
\begin{equation}
	F_{\mu\nu} \defeq \partial_\mu A_\nu - \partial_\nu A_\mu.
\end{equation}
Taking the variation of the above gives
\begin{equation}\label{dS_YM}
	\dbm S_\text{YM} = \int_M \dd^4x\,\bigl(\partial_\mu F^{\mu\nu}\bigr)\dbm A_\nu - \int_M \dd^4x\,\partial_\mu\bigl(F^{\mu\nu}\dbm A_\nu\bigr).
\end{equation}
The first term produces Maxwell's equations and the second term is the symplectic potential current
\begin{align}
	\partial_\mu F^{\mu\nu}&= 0,\label{eq:EMEOM}\\
	\theta_{\mathrm{EM}}^\mu &= - F^{\mu\nu}\dbm A_\nu. 
\end{align}
The symplectic form is then given by
\begin{equation}\label{eq:EMsymplecticform}
	\Omega_{\text{EM}} = - \int_\Sigma \dd \Sigma_\mu \, \dbm F^{\mu\nu}\dbm A_\nu,
\end{equation}
where the field space wedge product is kept implicit in the notation.
Let us consider global translations first.
The Lagrangian density $\lem = -\tfrac{1}{4}F^{\mu\nu}F_{\mu\nu}$ is invariant under spacetime translations generated by a constant vector $\lambda^\mu$, $x^\mu \mapsto x'{}^\mu = x^\mu + \lambda^\mu$. These translations induce the following field variations:
\begin{align}
	\Delta_\lambda^{\text{T}} A^\mu &= \lambda^\sigma \partial_\sigma A^\mu, \label{Atranslation} \\
	\Delta_\lambda^{\text{T}} F^{\mu\nu} &= \lambda^\sigma \partial_\sigma F^{\mu\nu}. \label{Ftranslation} 
\end{align}
In order to compute the associated charge we need to contract the symplectic form \eqref{eq:EMsymplecticform} with the field space vector generated by the above transformation,
\begin{equation}
	\V_{\lambda}^{\text{T}} = \int_M \dd^4 x \,\Delta_\lambda^{\text{T}} A^{\mu}(x) \frac{\dbm}{\dbm A^{\mu}(x)}.
\end{equation}
We get
\begin{Align}
	\If_{\V_\lambda^{\text{T}}} \Omega_{\text{EM}} &= -\Sint{\mu} \lambda^{\sigma}\bigl(\partial_\sigma F^{\mu\nu} \dbm A_\nu - \dbm F^{\mu\nu} \partial_\sigma A_\nu \bigr) \\
	&= \Sint{\mu} \lambda^\sigma \dbm \left(F^{\mu\nu}\partial_\sigma A_\nu \right)- \Sint{\mu}\partial_\sigma \left(\lambda^\sigma F^{\mu\nu}\dbm A_\nu\right) \\
	&= \Sint{\mu} \lambda^\sigma \bigl(\dbm\left(F^{\mu\nu} \partial_\sigma A_\nu \right)-\partial_\sigma \left(F^{\mu\nu}\dbm A_\nu\right)\bigr) - \Sint{\mu}\partial_\sigma \bigl(\lambda^\sigma F^{\mu\nu}\dbm A_\nu- \lambda^\mu F^{\sigma \nu} \dbm A_\nu\bigr) \\
	&= \Sint{\mu}\lambda^\sigma \dbm \bigl(F^{\mu\nu}\partial_\sigma A_\nu + \delta_\sigma^\mu \lem \bigr) - 2\DSint{\mu}{\sigma} \lambda^\sigma F^{\mu\nu} \dbm A_\nu. \label{iTranslationOmega}
\end{Align}
where the equation of motion \eqref{eq:EMEOM} was used. We thus have
\begin{equation}\label{TranslationFlux}
	\If_{\V_{\lambda}^{\text{T}}}\Omega_{\text{EM}} = \dbm \tilde{H}_{\text{EM}}^{\rm T}[\lambda]  -2 \DSint{\mu}{\sigma} \lambda^\sigma F^{\mu\nu}\dbm A_\nu,
\end{equation}
where the charge $\tilde{H}_{\text{EM}}^{\rm T}[\lambda]$ is given by the integral of the canonical energy-momentum tensor $\tilde T \updown{\mu}{\sigma}$:
\begin{equation}\label{canonicalEMtensor}
	\tilde T \updown{\mu}{\sigma} := F^{\mu\nu}\partial_\sigma A_\nu + \delta^{\mu}_\sigma\lem
\end{equation}
We obtained the canonical translation charge with an additional corner term that can not be written as a total field space derivative. This additional term is an example
of a \textit{flux}, and in general it reflects the fact that the canonical energy-momentum tensor is not conserved, or that the system is dissipative
(there might be e.g., electromagnetic radiation going through the boundary). The reason it appears here is because translations move the location of the boundary.
In order to restore integrability of the charge, one needs to apply the extended phase space formalism.

Furthermore, the charge $\tilde{H}_{\text{EM}}^{\rm T}[\lambda]$ can be rewritten as
\begin{Align}
	\tilde{H}_{\text{EM}}^{\rm T}[\lambda] &= \Sint{\mu}\lambda^\sigma \bigl(F^{\mu\nu}\partial_\sigma A_\nu + \delta^\mu_\sigma \lem \bigr) \\
	&= \Sint{\mu}\lambda^\sigma \bigl(F^{\mu\nu}F_{\sigma\nu} + \partial_\nu(F^{\mu\nu}A_\sigma) +\delta^\mu_\sigma \lem \bigr) \\
	&= \Sint{\mu}\lambda^\sigma \bigl(F^{\mu\nu}F_{\sigma\nu} + \delta^\mu_\sigma \lem\bigr) + \DSint{\mu}{\nu} \lambda^\sigma F^{\mu\nu}A_\sigma \\
	&= H_{\text{EM}}^{\rm T}[\lambda] + H_{\text{EM}}[\lambda^\sigma A_\sigma]. \label{TranslationCharge}
\end{Align}
We thus obtain the translation charge $H_{\text{EM}}^{\rm T}[\lambda]$ defined\footnote{Note that the translation charge has support in the bulk. This is because global translations are not local transformations and thus, Noether's second theorem does not apply.} as the integral over $\Sigma$ of the symmetric energy-momentum tensor
\begin{equation}
	T \updown{\mu}{\sigma} := F^{\mu\nu}F_{\sigma\nu} + \delta^{\mu}_\sigma\lem
	\label{symmetricEMtensor}
\end{equation}
contracted with $\lambda^\sigma$.
The second term in \eqref{TranslationCharge} corresponds to a gauge charge, with the caveat that the corresponding gauge transformation is field dependent.
The presence of this particular boundary gauge charge reflects the standard issue about the gauge invariance of the canonical energy-momentum tensor
\eqref{canonicalEMtensor} in Maxwell theory, and can be understood as a consequence of the fact that the translations given by
\eqref{Atranslation}--\eqref{Ftranslation} do not commute with gauge transformations. One could get rid of the gauge charge contribution in
\eqref{TranslationCharge} if from the beginning one extends the definition of the transformation \eqref{Atranslation} to include a field dependent gauge
transformation in addition to a pure translation (see for instance \cite{Scheck:2012gam} or \cite{Mieling2017NoethersTA}, where this process is carried out
within the conventional framework of Noether's theorem.) The infinitesimal change in $A^\mu$ under such a generalized transformation is defined to be
\begin{equation}
	\tilde\Delta_\lambda^{\rm T}A^\mu = \lambda^\sigma\partial_\sigma A^\mu - \partial^\mu(\lambda^\sigma A_\sigma) = \lambda^\sigma F\downup{\sigma}{\mu},
	\label{Atranslation-generalized}
\end{equation}
while the field tensor still transforms as in \eqref{Ftranslation}. Note that this makes the field variation $\tilde\dbm_\lambda^{\rm T}A^\mu$ independent of the choice of gauge for $A^\mu$. For the extended transformation \eqref{Atranslation-generalized} we have
\begin{equation}
	\If_{\tilde V_\lambda^{\rm T}}\Omega_{\rm EM} = \dbm H_{\rm EM}^{\rm T}[\lambda] - 2\int_{\partial\Sigma} \dd\sigma_{\mu\sigma} \lambda^\sigma F^{\mu\nu}\dbm A_\nu,
\end{equation}
with the charge $H_{\rm EM}^{\rm T}[\lambda]$ corresponding to the symmetric energy-momentum tensor \eqref{symmetricEMtensor}.

Let us now turn our focus to Lorentz transformations. The potential and field strength transform as
\begin{align}
	\Delta_\lambda^{\text{L}} A_\mu &= \lambda\updown{\rho}{\sigma} x^\sigma \partial_\rho A_\mu + \lambda \updown{\nu}{\mu} A_\nu,\\
	\Delta_\lambda^{\text{L}} F^{\mu\nu} &= \lambda\updown{\rho}{\sigma} x^{\sigma} \partial_\rho F^{\mu\nu} - \lambda \updown{\mu}{\rho}F^{\rho\nu} - \lambda\updown{\nu}{\rho}F^{\mu\rho}, 
\end{align}
where $\lambda^{\mu\nu} = - \lambda^{\nu\mu}$ generates infinitesimal Lorentz transformations.
The contraction of the associated field space vector with the symplectic two-form gives
\begin{align}
	\If_{\V_{\lambda}^{\text{L}}} \Omega_{\text{EM}} &= - \Sint{\mu} \Bigr(\lambda \updown{\rho}{\sigma} x^\sigma \partial_\rho F^{\mu\nu} \dbm A_\nu - \lambda \updown{\mu}{\rho}F^{\rho\nu}\dbm A_\nu - \lambda \updown{\nu}{\rho}F^{\mu\rho} \dbm A_\nu  \\
	&\hspace{72pt}- \dbm F^{\mu\nu}\lambda \updown{\rho}{\sigma}x^{\sigma}\partial_\rho A_\nu-\dbm F^{\mu\nu} \lambda \updown{\rho}{\nu}A_{\rho}\Bigr) \\
	&= \Sint{\mu}\Bigl(\lambda\updown{\rho}{\sigma} \, \dbm \bigl(x^\sigma F^{\mu\nu}\partial_\rho A_\nu + F^{\mu\sigma}A_\rho\bigr) + \lambda \updown{\mu}{\rho}F^{\rho\nu}\dbm A_\nu -\partial_\rho \left(\lambda \updown{\rho}{\sigma} x^\sigma F^{\mu\nu}\dbm A_\nu \right)\Bigr) \\
	&= \Sint{\mu} \Bigl(\lambda\updown{\rho}{\sigma} \, \dbm \bigl(x^\sigma F^{\mu\nu}\partial_\rho A_\nu + F^{\mu\sigma}A_\rho\bigr) - \lambda \updown{\mu}{\sigma} x^\sigma F^{\rho\nu}\partial_\rho \dbm A_\nu  \\
	&\hspace{72pt}- \partial_\rho \left(\lambda \updown{\rho}{\sigma} x^\sigma F^{\mu\nu}\dbm A_\nu -\lambda \updown{\mu}{\sigma}x^\sigma F^{\rho\nu}\dbm A_\nu \right)\Bigr)
\end{align}
and we therefore have
\begin{equation}\label{LorentzFlux}
	\If_{\V_{\lambda}^{\text{L}}} \Omega_{\text{EM}} =\dbm \tilde{H}_{\text{EM}}^{\rm L}[\lambda] -2\DSint{\mu}{\rho} \lambda \updown{\rho}{\sigma}x^{\sigma}F^{\mu\nu}\dbm A_\nu ,
\end{equation}
where we again obtain a charge term
\begin{equation}
	\tilde H^{\rm L}_{\rm EM}[\lambda] := \Sint{\mu} \lambda \updown{\rho}{\sigma} \left(F^{\mu\nu}x^\sigma \partial_\rho A_\nu + A_\rho F^{\mu\sigma }+x^{\sigma}\delta^\mu_\rho \lem \right)
	\label{LorentzCharge}
\end{equation}
and a flux term which reflects the fact that the relativistic angular momentum is not conserved, because Lorentz transformations do not preserve the boundary.
The charge \eqref{LorentzCharge} can be written as
\begin{align}\label{LorentzCharge2}
	\tilde{H}^{L}_{\text{EM}}[\lambda] &= \Sint{\mu} \lambda \updown{\rho}{\sigma} \left(x^{\sigma}F^{\mu\nu}F_{\rho\nu} + x^{\sigma}F^{\mu\nu}\partial_\nu A_\rho + A_\rho F^{\mu\sigma} + \delta^\mu_\rho x^{\sigma}\lem \right)  \\
	&= \Sint{\mu} \lambda \updown{\rho}{\sigma} \bigl( x^{\sigma}F^{\mu\nu}F_{\rho\nu}+ \delta^\mu_\rho x^{\sigma}\lem\bigr) + \DSint{\mu}{\nu} \lambda\updown{\rho}{\sigma}x^{\sigma}F^{\mu\nu}A_\rho  \\
	&= H^{\rm L}_{\text{EM}}[\lambda] + H_{\text{EM}}[\lambda \updown{\rho}{\sigma}x^{\sigma} A_\rho].
\end{align}
After a short calculation, we find that the first term in \eqref{LorentzCharge2} can be expressed as
\begin{equation}
	H_{\text{EM}}^{L}[\lambda] = \frac12 \Sint{\mu} \lambda_{\rho\sigma}M^{\sigma\mu\rho},
\end{equation}
where
\begin{equation}
	M^{\sigma\rho\mu} := x^{\sigma}T^{\mu\rho}- x^{\rho}T^{\mu\sigma}.
\end{equation}
is the angular momentum tensor. 
Finally, similarly to the case of translations, the second term in \eqref{LorentzCharge2}
is a gauge charge corresponding to a field dependent gauge transformation. This can again be seen
as a consequence of the fact that Lorentz transformations do not commute with gauge transformations of the field $A_\mu$.\par
We now move to the extended phase space formalism. First let us introduce the notation of the extended phase space in the coordinate language.
The embedding maps a reference space $\mathcal{U}\subset \R^4$ to a subregion $U\subset M$
\begin{Align}
	\phi: \mathcal{U}&\longrightarrow U,\\
	 \X &\longmapsto \phi(\X) = x,
\end{Align} 
where we denote coordinates on the reference space by $\X,\mathtt{y},...$ and coordinates in the spacetime manifold by $x,y,...$. We further denote by $\mathcal{S}\subset \mathcal{U}$ the
subset of the reference space such that $\phi(\mathcal{S}) = \Sigma$ and $\phi(\partial \mathcal{S}) = \partial \Sigma$.
The Maxwell action for the subregion $U$ can then be written
\begin{align}\label{Actionphi}
	S_{\text{EM}}[A,\phi] = -\frac14\int_\mathcal{U} \dd^4\X\, J(\X)\, F^{\mu\nu}\bigl(\phi(\X)\bigr)F_{\mu\nu}\bigl(\phi(\X)\bigr),
\end{align}
where $J(\X)$ is the Jacobian
\begin{equation}
	J(\X) = \mathrm{det}\qty(\pdv{\phi(\X)^\mu}{\X^\nu}).
\end{equation}
The (on-shell) extended presymplectic form \eqref{eq:extendedsymplecticform} gives
\begin{Align}\label{extsymplformYM}
	\Omega^{\rm ext}_{\text{EM}}[A, \dbm A, \phi, \dbm \phi] = &-\int_{\cal S} \dd\Sigma_\mu\bigl(\phi(\X)\bigr)\,\dbm F^{\mu\nu} \dbm A_\nu  \\
	&+ \int_{\partial{\cal S}} \dd\sigma_{\mu\nu}\bigl(\phi(\X)\bigr)\,\bigl(2\chi^\mu F^{\nu\rho}\dbm A_\rho - \chi^\mu\chi^\nu L_{\rm EM} \bigr).
\end{Align}
(Here, as in the subsequent equations in this section, it is understood that each integrand is evaluated at $x = \phi(\X)$.)

Let us now compute the charges associated to the global symmetries that we just computed, in this extended formalism. We first compute the contraction of the extended presymplectic form with the field space vector $\V_\lambda^{\rm T}$ corresponding to global translations, with $\lambda^\mu$ being the constant vector generating the translation. 
Using the fact that $\If_{\V_{\lambda}^{\rm T}} \chi^\mu = - \lambda^\mu$, and recalling that the contraction of the non-extended presymplectic form $\Omega_{\rm EM}$ has already been calculated in \eqref{iTranslationOmega}, we get
\begin{Align}\label{translationcontraction}
	\If_{\V_\lambda^{\rm T}} \Omega^{\rm ext}_{\text{EM}} &= \If_{\V_\lambda^{\rm T}} \Omega_{\text{EM}} - 2\int_{\partial{\cal S}} \dd\sigma_{\mu\nu}\bigl(\phi(\X)\bigr)\, \bigl(\lambda^\mu F^{\nu\rho} \dbm A_\rho + \lambda^\sigma \chi^\mu F^{\nu\rho}\partial_\sigma A_\rho - \lambda^\mu \chi^\nu \lem\bigr)\\
	&= \int_{\cal S} \dd\Sigma_\mu\bigl(\phi(\X)\bigr)\, \dbm J_T^\mu + 2\int_{\partial{\cal S}} \dd\sigma_{\mu\nu}\bigl(\phi(\X)\bigr)\, \chi^\nu J_T^\mu ,
\end{Align}
where we have defined the translation charge current 
\begin{equation}
	J_T^\mu = \lambda^\sigma F^{\mu\nu}\partial_\sigma A_\nu + \lambda^\mu \lem.
\end{equation}
Note that the equations of motion imply that the translation current is conserved:
\begin{equation}
	\partial_\mu J^\mu_T = 0.
	\label{JT-cons}
\end{equation}
Using this fact and equation \eqref{embeddingvariation}, we get
\begin{equation}
	\dbm \left(\int_{\cal S} \dd\Sigma_\mu\bigl(\phi(\X)\bigr)\, J^\mu_T\right) = \int_{\cal S} \dd\Sigma_\mu\bigl(\phi(\X)\bigr)\, \dbm J_T^\mu + 2\int_{\partial{\cal S}} \dd\sigma_{\mu\nu}\bigl(\phi(\X)\bigr) \chi^\nu J_T^\mu.
	\label{dphi-JT}
\end{equation}
We thus conclude that
\begin{equation}
	\If_{\V^{\rm T}_\lambda} \Omega^{\rm ext}_{\text{EM}} = \dbm\left(\int_{\cal S} \dd\Sigma_\mu\bigl(\phi(\X)\bigr)\,\lambda^\sigma  \left(F^{\mu\nu}\partial_\sigma A_\nu + \dbm ^\mu_\sigma \lem \right)\right) = \dbm \tilde{H}_{\text{EM}}^{\rm T}[\lambda].
\end{equation}
Hence we see that, as expected, the addition of the embedding to the phase space and the use of the extended presymplectic form leads to the elimination of the flux term which appears in \eqref{TranslationFlux}.\par
We now move to the charges assocaited with Lorentz symmetries in the extended formalism. Now we have that $\If_{\V^{\text{L}}_\lambda}\chi^\mu = -\lambda\updown{\mu}{\nu}x^\nu$, where $\lambda\updown{\mu}{\nu}$ is the infinitesimal generator of the Lorentz transformation, and the contraction of the extended presymplectic form along a Lorentz transformation orbit gives
\begin{align}
	\If_{\V^{\text{L}}_\lambda} \Omega^{\rm ext}_{\text{EM}} &= \If_{\V^{\text{L}}_\lambda} \Omega_{\text{EM}}-2\int_{\partial{\cal S}} \dd\sigma_{\mu\nu}\bigl(\phi(\X)\bigr) \Bigl(\lambda \updown{\mu}{\rho}x^\rho F^{\nu\alpha} \dbm A_\alpha + \chi^\mu F^{\nu\alpha} \lambda \updown{\rho}{\sigma}x^{\sigma} \partial_\rho A_\alpha \notag \\
	&\hspace{150pt} + \chi^\mu F^{\nu\alpha}\lambda \updown{\rho}{\alpha} A_\rho - \chi^\nu \lambda \updown{\mu}{\rho}x^\rho \lem\Bigr) \notag \\
	&= \int_{\cal S} \dd\Sigma_\mu\bigl(\phi(\X)\bigr)\, \dbm J^\mu_L - 2 \int_{\partial{\cal S}} \dd\sigma_{\mu\nu}\bigl(\phi(\X)\bigr) \Bigl(\chi^\mu F^{\nu\alpha}\lambda \updown{\rho}{\sigma}x^\sigma\partial_\rho A_\alpha +  \chi^\mu F^{\nu\alpha}\lambda\updown{\rho}{\alpha}A_\rho \notag \\
	&\hspace{200pt} -\lambda \updown{\mu}{\rho}x^\rho \chi^\nu \lem \Bigr) \notag \\
	&= \int_{\cal S} \dd\Sigma_\mu\bigl(\phi(\X)\bigr)\, \dbm J^\mu_L + 2\int_{\partial{\cal S}} \dd\sigma_{\mu\nu}\bigl(\phi(\X)\bigr) \chi^\nu J^\mu_L,
\end{align}
where we defined the Lorentz charge current 
\begin{equation}
	J^\mu_L =\lambda\updown{\rho}{\sigma} x^\sigma F^{\mu\nu}\partial_\rho A_\nu +\lambda\updown{\rho}{\sigma} A_\rho F^{\mu\sigma} + \lambda\updown{\mu}{\sigma}x^\sigma \lem.
\end{equation}
A simple calculation shows that this current is also conserved: $\partial_\mu J^\mu_L = 0$. It follows that, as in \eqref{dphi-JT}, we have
\begin{equation}
	\dbm \left(\int_{\cal S} \dd\Sigma_\mu\bigl(\phi(\X)\bigr)\,J_L^\mu \right) = \int_{\cal S} \dd\Sigma_\mu\bigl(\phi(\X)\bigr)\, \dbm J^\mu_L + 2 \int_{\partial{\cal S}} \dd\sigma_{\mu\nu}\bigl(\phi(\X)\bigr) \chi^\alpha J^\mu_L,
\end{equation}
and we can finally write
\begin{equation}
	\If_{\V_\lambda^{\rm L}} \Omega^{\rm ext}_{\text{EM}} = \dbm \left(\int_{\cal S} \dd\Sigma_\mu\bigl(\phi(\X)\bigr)\, \lambda \updown{\rho}{\sigma} \left(x^\sigma F^{\mu\nu}\partial_\rho A_\nu  + \delta^\mu_\rho x^\sigma \lem+A_\rho F^{\mu\sigma}\right)\right) = \dbm \tilde{H}_{\text{EM}}^{\rm L}[\lambda].
\end{equation}
As in the translation case, the flux term is eliminated by the extended structure and we are left with the angular-momentum charge of electromagnetism.\par
This provides an interesting test case of the extended phase-space formalism which, to the best of our knowledge, has not been worked
out explicitly before \cite{Assanioussi:2023jyq}. While it is satisfying to see the flux cancel explicitly, the corresponding computations become significantly more
involved in the gravitational setting. In that case, we will therefore compute the charges directly
from the Noether current and rely on the general results of the extended phase-space formalism, rather than explicitly evaluating the contraction of the symplectic form.

\paragraph{Important relations}
We conclude this section with the proofs of the relations \eqref{equationa}, \eqref{equationb} and \eqref{equationc}.
The first relation concerns the variation of the field $\chi$. Consider a general integrated functional:
\begin{equation}
    \mathcal{I} = \intpull{U}{d} (\mathcal{A}),
\end{equation}
We have
\begin{equation}
\begin{aligned}
    \delta \mathcal{I} &= \int_U [\dbm \phi^*(\mathcal{A}) + \phi^*(\dbm\mathcal{A})]\\
    &= \intpull{U}{d} (\dbm \mathcal{A} + \ldv{\chi}\mathcal{A}).
\end{aligned}
\end{equation}
Taking the variation a second time yields
\begin{equation}
    \begin{aligned}
        0 = \dbm^2 \mathcal{I} &= \int_U [\dbm \phi^*(\dbm \mathcal{A} + \ldv{\chi} \mathcal{A}) + \phi^* (\dbm (\ldv{\chi}\mathcal{A}))]\\
        &= \intpull{U}{d} [\ldv{\chi} \dbm \mathcal{A} + \ldv{\chi}\ldv{\chi} \mathcal{A} + \ldv{\dbm \chi} \mathcal{A}- \ldv{\chi} \dbm \mathcal{A}]\\
        &= \intpull{U}{d} [(\ldv{\chi} \ldv{\chi} + \ldv{\dbm \chi}) \mathcal{A}] \\
        &= \intpull{U}{d} [(\ldv{\frac{1}{2}[\chi,\chi]}  + \ldv{\dbm \chi}) \mathcal{A}].
    \end{aligned}
\end{equation}
Note that the minus sign in front of the last term of the second line comes from the fact that $\chi$ is a field space one-form. We thus have
\begin{equation}\label{chivariation}
    \dbm \chi = -\frac{1}{2}[\chi,\chi].
\end{equation}
The second relation is the following
\begin{equation}
    \ldv{\chi} \iota_\chi L = \frac{1}{2} \iota_{\qty[\chi,\chi]}L + \frac{1}{2} \dd (\iota_\chi \iota _\chi L),
\end{equation}
which comes from the identity $[\ldv{\xi_1},\iota_{\xi_2}] = \iota_{[\xi_1,\xi_2]}$:
\begin{equation}
\begin{aligned}
    \ldv{\chi} \iota_\chi L &= [\ldv{\chi},\iota_\chi] L - \iota_\chi \ldv{\chi} L \\
    &= \iota_{\qty[\chi,\chi]} L - \iota_\chi \dd \iota_\chi L\\
    &= \iota_{\qty[\chi,\chi]} L - \ldv{\chi}(\iota_\chi L) + \dd (\iota_\chi \iota_\chi L).
\end{aligned}
\end{equation}
The third idenity is
\begin{equation}
\begin{aligned}
    \If_{\D_\xi} \iota_{\dbm \chi} L = -\frac{1}{2} \If_{\D_\xi} \iota_{\qty[\chi,\chi]} L = \iota_{\qty[\xi,\chi]} L,
\end{aligned}
\end{equation}
where \eqref{chicondition} and \eqref{chivariation} were used. \clearpage

\section{Corner Symmetries in Gravity}\label{sec:cornersymmetriesingravity}
We now set to apply the formalism developed in the previous section to the gravitational case. Gravitational corner charges first appeared in the Hamiltonian formalism in the seminal work
of Regge and Teitelboim
\cite{Regge:1974zd} (see also \cite{BenguriaCorderoTeitelboim1977}).
Later, Barnich and Brandt developed a framework, in the Lagrangian language, to relate asymptotically conserved corner quantities and gauge symmetries
\cite{BarnichBrandt2002}. Together with the advent of covariant phase-space methods, these results spurred
an abundance of work on corner charges and asymptotic symmetries in the 2000s, culminating in the modern framework of the corner proposal
\cite{Donnelly:2016auv,Speranza:2017gxd,Geiller:2017whh,Freidel:2020xyx,Freidel:2020svx,Freidel:2020ayo,Donnelly:2020xgu,Ciambelli:2021nmv,Donnelly:2022kfs,Ciambelli:2022vot,Ciambelli:2023bmn,CanepaCattaneo2024}.\par
The corner proposal can be stated in the following simple form \cite{Ciambelli:2022vot,Ciambelli:2023bmn}:
\begin{center}
\textit{``Gravity is described by a set of charges and their algebra at corners.''}
\end{center}
In other words, the claim is that corner charges encode all the information about the gravitational field. This statement is holographic in nature. Of particular
interest in this thesis is the notion
of \textit{local} holography, i.e. the idea that a local region of spacetime is described by the charges living on the boundary of that region. One natural question is then: which charges and which algebra?
The answer to which was given in the work of Ciambelli and Leigh \cite{Ciambelli:2021vnn,Ciambelli:2021nmv,Ciambelli:2022cfr},
where it was found that there exists a maximal subset of diffeomorphisms that can admit a non-vanishing Noether charge in any gravitational theory. These universal corner symmetries are thus the
natural starting point of the corner proposal.\par
This section is organized as follows. In Subsection~\ref{subsec:universalcornersymmetries}
we discuss the emergence of the universal corner symmetries from general considerations of diffeomorphisms and embeddings.
In Subsection~\ref{subsec:cornersymmetriessss}, we construct the Noether charges of spherically symmetric
Einstein--Hilbert gravity and discuss their algebra.

\subsection{Universal corner symmetries}\label{subsec:universalcornersymmetries}
We begin our discussion of corner symmetries in gravity by describing the emergence of the universal corner symmetry group and its associated subgroups. These results lie at the core of the corner proposal and will be used throughout this thesis.
\par We start by giving the general description of an embedded corner.
Let $M$ be a four-dimensional manifold\footnote{While we focus on the four-dimensional case here, the discussion extends straightforwardly to arbitrary spacetime dimension \cite{Ciambelli:2021vnn}.} and $S$ a codimension-2 manifold (the corner). The way $S$ is embedded into $M$ is defined by the injective map
\begin{equation}\label{embedding}
    \phi: S \longrightarrow M.
\end{equation}
The normal tangent space to the corner is the vertical sub-bundle
\begin{equation}
    V = \mathrm{ker}\qty(\dd \phi).
\end{equation}
Let $\sigma^\alpha: S \longrightarrow \mathbb{R}^{2},\, (\alpha = 1,2)$ be local coordinates on $S$. Let $y^M:M \longrightarrow \mathbb{R}^4$ be local coordinates on M. A choice of embedding \eqref{embedding}, is then given by $y^M(\sigma)$.
Without loss of generality---and without choosing a specific embedding---one can choose the $y^M$ coordinates such that
\begin{equation}\label{mcoordinate}
    y^M = (u^a,x^i),
\end{equation}
where $u^a, a = 1,2$ are the tangential coordinates on the corners and $x^i, i =0,2$ are the normal coordinates. More precisely, this split of coordinates is such
that the vertical sub-bundle is spanned by the normal coordinates
\begin{equation}
    V = \mathrm{span}\qty(\partial_i).
\end{equation}
One can then define the two normal one forms by
\begin{equation}
    n^i = \dd x^i -\omega^i_a(u,x) \dd u^a,
\end{equation}
where $\omega^i_a$ is an Ehresmann connection \cite{Ehresmann1951,Ehresmann1952,Marle2007}. This connection represents the ambiguity in the decomposition of the tangent bundle
\begin{equation}
    \mathrm{T}M = V \oplus H,
\end{equation}
where the horizontal bundle is defined as
\begin{equation}
    H = \mathrm{ker}(n^i) = \qty{\xi\in \mathrm
    T M : n^i(\xi) = 0}.
\end{equation}
The entire description given above does not necessitate the introduction of a metric. We note that, if a metric is introduced at this point, it can be written in the adapted
coordinates as
\begin{equation}
    ds^2 = h_{ij} n^i n^j + \gamma_{ab}\dd u^a \dd u^b.
\end{equation}
From this expression, it is clear that the Ehresmann connection encodes the mixing between the normal and tangential coordinates.\par
Noether's second theorem tells us that the charges associated with diffeomorphisms will localize on the codimension-2 surface $S$.
To analyze the algebra of these charges, it is therefore natural to expand the diffeomorphisms around the location of the corner. This is done most easily by introducing the
trivial embedding
\begin{equation}\label{trivial_embedding}
    y^M_0(\sigma) = (u_0^a(\sigma),x_0^i(\sigma)) = (\delta^a_\alpha \sigma^\alpha,0).
\end{equation}
In other words, the trivial embedding localizes the corner at the origin of the normal coordinate system and uses the corner coordinates $\sigma^\alpha$ as the tangential ones.
Expanding ``close to" this codimension-2 embedding thus corresponds to expanding around $x^i = 0$. Let us consider two vector fields $\xi, \zeta \in TM$. In the coordinate system \eqref{mcoordinate} they are expressed as
\begin{equation}
\begin{aligned}
    \xi &= \xi^a(u,x) \partial_a + \xi^i(u,x) \partial_i,\\
    \zeta &= \zeta^a(u,x) \partial_a + \zeta^i(u,x) \partial_i.
\end{aligned}
\end{equation}
 Expanding close to the surface $x^i = 0$, one gets
 \begin{equation}\label{vectorsexpansion}
 \begin{aligned}
    \xi^a(u,x) &= \xi^a_{(0)}(u) + x^j\xi^a_{(1) j}(u) + x^j x^k \xi^a_{(2)j k}(u) + \mathcal{O}(x^3),\\
    \xi^i(u,x) &= \xi^i_{(0)}(u) + x^j\xi^i_{(1) j}(u) + x^j x^k \xi^i_{(2)j k}(u) + \mathcal{O}(x^3),\\
    \zeta^a(u,x) &= \zeta^a_{(0)}(u) + x^j\zeta^a_{(1) j}(u) + x^j x^k \zeta^a_{(2)j k}(u) + \mathcal{O}(x^3),\\
    \zeta^i(u,x) &= \zeta^i_{(0)}(u) + x^j\zeta^i_{(1) j}(u) + x^j x^k \zeta^i_{(2)j k}(u) + \mathcal{O}(x^3).
 \end{aligned}
 \end{equation}
Now consider the vector field
\begin{equation}
   \eta= \qty[\xi,\zeta] = \qty(\qty[\xi,\zeta])^a \partial_a + \qty(\qty[\xi,\zeta])^i \partial_i.
\end{equation}
On the one hand, we may expand this bracket about the point $x^i=0$. On the other hand, we may first expand the vector fields in \eqref{vectorsexpansion}
and then compute their bracket term by term. If we truncate these series at a fixed order, the bracket can in principle produce contributions at higher orders
in the expansion of $\eta$. This leads to the natural question: does there exist a truncation order for which the bracket closes on the truncated space,
i.e.\ such that the bracket of two truncated vector field expansions contains no terms
beyond the order at which the expansions were cut off? This is far from obvious, since one might expect closure to hold only for the full expansion containing all orders.
The observation at the core of the corner proposal is that there actually exists a maximal subset of diffeomorphisms that closes an algebra at the zeroth order in
tangential directions and first order in the normal ones. More precisely, this algebra is generated by vectors of the type
\begin{equation}\label{ucagenerators}
\begin{aligned}
    \xi &= \xi^a_{(0)} \partial_a + \qty(\xi_{(0)}^i + x^j \xi_{(1)j}^i) \partial_i,\\
    \zeta &= \zeta^a_{(0)} \partial_a + \qty(\zeta_{(0)}^i + x^j \zeta_{(1) j}^i) \partial_i.
\end{aligned}
\end{equation}
Any additional term in the expansion \eqref{vectorsexpansion} results in a non-closing algebra, hence the term ``maximal". The Lie bracket of the vector fields
expressed in \eqref{ucagenerators} gives
\begin{equation}
\begin{aligned}\label{eq:liebracketucs}
    \qty[\xi,\zeta] &= \qty[\hat{\xi}_{(0)},\hat{\zeta}_{(0)}]^a \partial_a\\
    &\qquad + \qty(\hat{\xi}_{(0)} \qty(\zeta_{(0)}^i) - \hat{\zeta}_{(0)} \qty(\xi_{(0)}^i) +\zeta_{(1)j}^i \xi_{(0)}^j - \xi_{(1)j}^i \zeta_{(0)}^j) \partial_i\\
    &\qquad + x^j\qty(-\qty[\xi_{(1)},\zeta_{(1)}]^i{}_{\, j} + \hat{\xi}_{(0)} \qty(\zeta_{(1)j}^i) - \hat{\zeta}_{(0)} \qty(\xi_{(1)j}^i)) \partial_i, 
\end{aligned}
\end{equation}
where $\hat{\xi} = \xi^a \partial_a,\hat{\zeta} \equiv \zeta^a\partial_a$. We emphasize that, even though
it was suppressed in the notation for clarity, every object in the above formula depends on the tangential coordinates $u^a$.
In the first line we recognize the algebra of diffeomorphisms on the corner $\mathfrak{diff}\qty(S)$.
For each point of the corner, the objects $\xi^i_{(0)}$ and $\zeta^{i}_{(0)}$ generate constant shifts in the normal directions, and hence realize the abelian group of normal translations $\mathbb{R}^2$.
Likewise, at each point of the corner, the objects $\xi^i_{(1)j}$ and $\zeta^{i}_{(1)j}$
implement general linear transformations in the normal plane to the corner, and thus furnish the generators of $\mathrm{GL}\!\left(2,\mathbb{R}\right)$.
From the second line of \eqref{eq:liebracketucs}, we see that the normal translations are acted upon by the corner diffeomorphisms by tangential derivative and
by the general linear group by the standard matrix action. Finally, the first term on the last line reproduces the $\mathfrak{gl}(2,\mathbb{R})$ algebra,
while the next two terms show how the corner diffeomorphisms act on the $\mathfrak{gl}(2,\mathbb{R})$ generators.
The resulting complete structure is the algebra of the \textit{Universal Corner Symmetry} group
\begin{equation}\label{universalgroup}
    \mathrm{UCS} = \mathrm{Diff}(S) \ltimes \qty(\mathrm{GL}(2,\mathbb{R})\ltimes \qty(\mathbb{R}^2))^S.
\end{equation}
A few remarks on this result are in order:
\begin{itemize}
    \item The semidirect product emphasizes that, while the corner diffeomorphisms act on the linear transformations and the translations, the converse is not true. The same statement holds for the linear group acting on the translations. This is caused by
 the fact that, while every component in the series \eqref{vectorsexpansion} depend on the tangential coordinates $u^a$, they were stripped of their dependencies on the normal directions
 by the Taylor expansion.
    \item The $S$ in the exponent of the linear group and the translations means that there is a copy of the general linear group and the translation group at each
    point of the corner. These are related to supertranslations and superrotations
    \cite{BONDI1960,Bondi1962,Sachs1961,Sachs1962,Sachs1962b,NewmanPenrose1966BMS,AlessioEsposito2018BMS,Barnich:2009se,Barnich:2010eb,
    BarnichTroessaert2010Supertranslations,Barnich:2011mi,BarnichTroessaert2013Comments,Strominger2014BMS,CachazoStrominger2014SoftGraviton,
    KapecLysovPasterskiStrominger2014,Campiglia:2014yka,HeLysovMitraStrominger2015,CampigliaLaddha2015NewSymmetries,StromingerZhiboedov2016Memory,
    CompereLong2016Vacua,PasterskiStromingerZhiboedov2016Memories,MadlerWinicour2016BondiSachs,StromingerZhiboedov2017PairCreation,FlanaganNichols2017,HollandsIshibashiWald2017,
    CompereFiorucciRuzziconi2018,KolPorrati2019,Pasterski2019Superrotations,GodazgarGodazgarPope2019,Prabhu2019Supermomentum,FlanaganPrabhuShehzad2020,CampigliaPeraza2020,Freidel:2021fxf}
     and is directly related to them.
    \item The ``universal" in the name of the group, comes from the fact that this result is purely topological. In other words, no metric was introduced at any point in the derivation.
    This implies that the $\mathrm{UCS}$ group and its structure is not dependent on a particular theory of gravity, but is a universal characteristic of any gravitational theory defined on a manifold.
\end{itemize}
There are several important subgroups of the UCS that are known in the literature. First, one of the generator of the $\spl{2}$ subgroup was found by Carlip
 and Teitelboim in \cite{Carlip:1993sa} where it was called a boost symmetry. Next, the extended corner symmetry group
\begin{equation}\label{eq:ecsgroup}
    \mathrm{ECS} = \mathrm{Diff}(S) \ltimes \qty(\spl{2} \ltimes \R^2)^S,
\end{equation}  
was introduced by Donnelly and Freidel in \cite{Donnelly:2016auv}, where they recognized the boost symmetry of \cite{Carlip:1993sa} as a gauge fixing of the special linear group in \eqref{eq:ecsgroup}.
It was also discussed in \cite{Speranza:2017gxd}, and later reappeared in \cite{Ciambelli:2021vnn,Ciambelli:2021nmv} in the context of universal corner symmetries and the extended phase space.
It is the largest subgroup of the UCS for which non-vanishing Noether charges were found at finite distance corners.
It will be the main focus of this thesis. The next subgroup worth mentioning is the
Bondi--Metzner--Sachs--Weyl group
\begin{equation}
    \mathrm{BMSW} = (\mathrm{Diff}(S)\ltimes \R)\ltimes \R, 
\end{equation}
which is the group of symmetries for asymptotically flat Einstein--Hilbert gravity \cite{Freidel:2021fxf}. This group in turns contains the generalized BMS group \cite{Campiglia:2014yka}
\begin{equation}
    \mathrm{GBMS}= \mathrm{Diff}(S)\ltimes \R
\end{equation}
as well as the Extended BMS group, where only the conformal Killing vectors of $S$ are considered \cite{Barnich:2009se}, and the original BMS group
\cite{BONDI1960,Sachs1961,Bondi1962,Sachs1962,Sachs1962b}.
\par
As stated in the introduction, the UCS is the starting point of the corner proposal. We are however faced with our first technical
difficulty. There is no general prescription to classify all the unitary irreducible representations of infinite dimensional groups like the one of equation \eqref{universalgroup}.
The first angle of attack---and as a first consistency check of the proposal--- is to consider the two-dimensional version, where the corner becomes a point
and we are left with the finite dimensional groups
\begin{equation}
    \mathrm{UCS}_2 = \mathrm{GL}\qty(2,\R)\ltimes \R^2, \qquad \mathrm{ECS}_2 = \mathrm{SL}\qty(2,\R)\ltimes \R^2.
\end{equation}
The remainder of this thesis is devoted to applying the corner proposal to the two-dimensional extended corner symmetry group.
From now on, we will thus denote it by $\mathrm{ECS}$
 with the understanding that we are always talking about the two-dimensional version.
\subsection{Spherically symmetric gravity}\label{subsec:cornersymmetriessss}
In this section, we show that the Noether charges of spherically symmetric Einstein--Hilbert gravity generate the ECS algebra.
We start with the Einstein--Hilbert action in a four-dimensional spacetime $M$ (we work in units where $c=\hbar=1$)
\begin{equation}\label{eq:einsteinhilbert}
S_{\mathrm{EH}} = \frac{1}{16\pi G}\int_M \sqrt{-g} \,\dd^4 x \,R_4.       
\end{equation}
In the case of spherically symmetric manifold $M$, the metric can be written as
\begin{equation}\label{eq:sphericallysymmetricmetric}
    \dd s^2 = g_{ij}\dd x^i \dd x^j + \rho(x^i)^2 \dd \Omega_S^2,
\end{equation}
where $i,j = 0,1$, $\dd \Omega_S^2$ is the metric on the two sphere and $\rho$ is a scalar field that only depends on the coordinates $(x^0,x^1)$. In order to compute the Ricci scalar of the above metric, we rewrite it
\begin{equation}
    \dd s^2 = \frac{\rho^2}{L^2}\qty(\frac{L^2}{\rho^2} g_{ij} \dd x^i \dd x^j + L^2 \dd \Omega^2_S),
\end{equation}
where $L$ is a length scale introduced for dimensional reasons. Even though it is arbitrary, it will drop out of the final results.
The Manifold $M$ is therefore conformally equivalent to a product manifold $\tilde{M}_2\times S^2$
\begin{equation}
    \dd \tilde{s}^2 =  \tilde{g}_{ij}\dd x^i \dd x^j + L^2 \dd \Omega^2_S,
\end{equation}
with 
\begin{equation}\label{eq:tildetogconformaltransformation}
    \tilde{g}_{ij} = \frac{L^2}{\rho^2}g_{ij}.
\end{equation}
The Ricci scalar \( R_4 \) of the spherically symmetric geometry and the Ricci scalar \( \tilde{R}_4 \) of the conformally related geometry are connected through the standard conformal transformation formula:
\begin{equation}
    R_4 = L^2 \left( \rho^{-2} \tilde{R}_4 - 6 \rho^{-3} \tilde{\Box} \rho \right).
\end{equation}
Furthemore, the curvature of the conformal geometry is simply the sum of the curvature of the two-dimensional manifold $\tilde{M_2}$ and the one of the sphere
\begin{equation}
    \tilde{R}_4 = \tilde{R}_2 + \frac{2}{L^2}. 
\end{equation}
This allows us to write the action \eqref{eq:einsteinhilbert} as
\begin{Align}
    S_{\mathrm{EH}} &= \frac{1}{16\pi G}\frac{1}{L^2}\int_M \dd^4 x \sqrt{-\tilde{g}}\, \qty[\rho^2 \qty(\tilde{R}_2 + \frac{2}{L^2}) - 6 \rho \tilde{\Box} \rho]\\
    &= \frac{1}{4 G} \int_{\tilde{M}_2}\dd^2 x \sqrt{-\tilde{g}}\qty[\rho^2\qty(\tilde{R}_2+\frac{2}{L^2}) -6 \rho \tilde{\Box}\rho]\\
    &=\frac{1}{4 G} \int_{\tilde{M}_2}\dd^2 x \sqrt{-\tilde{g}}\qty[\rho^2\qty(\tilde{R}_2+\frac{2}{L^2})+6\partial_i \rho \partial^i \rho] -\frac{3}{4G}\stintbt{\partial \tilde{M}_2}{i}\sqrt{-\tilde{g}}\tilde{g}^{ji}\partial_j \qty(\rho^2).\\
\end{Align}

In order to connect the above action to two-dimensional dilaton theories, we define the dimensionless field
\begin{equation}\label{Phidef}
    \Phi = \frac{\rho^2}{4G}
\end{equation}
where $l_P=\sqrt{G\hbar}$ is the Planck length. Using this new field, the action reads
\begin{equation}\label{eq:dimreddilatontheory}
    S_{EH} = \hbar\int_{\tilde{M}_2}\dd^2x \sqrt{-\tilde{g}}\qty(\Phi \tilde{R}_2 +2 \frac{\Phi}{L^2}+\frac32 \frac{\partial_i \Phi \partial^i \Phi}{\Phi}) -3\hbar\stintbt{\partial \tilde{M}_2}{i}\sqrt{-\tilde{g}}\tilde{g}^{ji}\partial_b \Phi.
\end{equation}
We can now perform a final conformal transformation on the two-dimensional metric
\begin{equation}\label{eq:tildetohatconformaltransformation}
    \tilde{g} = \Phi^{-\frac32}\hat{g}.
\end{equation}
to get the final action 
\begin{equation}\label{eq:dimredaction}
        S_{\mathrm{EH}} = \int_{\tilde{M}_2} \dd^2 x \sqrt{-\hat{g}}\qty(\Phi \hat{R}_2 + \Phi^{-\frac12}\frac{2}{L^2}) - \frac32 \stintbt{\partial \tilde{M}_2}{i}\sqrt{-\hat{g}}\hat{g}^{ij}\partial_j\Phi.
\end{equation}
This proves the dynamical equivalence between spherically symmetric Einstein--Hilbert theory in four-dimensions and a two-dimensional dilaton gravity model with a $\Phi^{-\frac12}$ potential.
Finally we write the dictionary between the original four-dimensional metric \eqref{eq:sphericallysymmetricmetric} and the fields in the action \eqref{eq:dimreddilatontheory},
\vspace{-24pt}
\begin{center}
\begin{equation}\label{eq:dictionary}
\boxed{\begin{aligned}
    \Phi &= \frac{\rho^2}{4G}\\
    \hat g_{ij} &= \frac{\rho\, L^2}{(4G)^{3/2}}\, g_{ij} 
\end{aligned}%
}   
\end{equation}
\end{center}
In the context of the extended phase space, we consider the action related to a subregion embedded in the manifold $U \overset{\phi}{\rightarrow}M$. In this thesis,
we will only consider the cases where the boundary of the subregion is a null hypersurface defined by $\Phi = \mathrm{cste}$. The normal vector to this hypersurface is given by\footnote{Since the hypersurface
is null, the normal vector cannot be normalized. This means that there is no canonical choice of proportionality coefficient between the left-hand side and right hand side of equation \eqref{eq:nullnormal}.}
\begin{equation}\label{eq:nullnormal}
    n_i = \partial_i \Phi.
\end{equation}
For the subregion action, the boundary integral in \eqref{eq:dimredaction} becomes an integral on the boundary of $U$.
Since both the surface element and the integrand are proportional to the null normal vector, this term drops off. We will thus ignore it in the following.
\par

We are now ready to compute the charges of the dimensionally reduced theory.
 Varying the bulk action yields the following equations of motions
 \begin{align}
    \qty(\hat{\Box} \Phi - \Phi^{-\frac12}\frac{1}{L^2})\hat{g}_{ab} - \hat{\nabla}_{(a} \hat{\nabla}_{b)}\Phi &= 0,\label{eq:eomphi}\\
    \hat{R} &= \frac{\Phi^{-\frac32}}{L^2}\label{eq:eomg}.
 \end{align}
 and the following symplectic potential current
 \begin{equation}
\theta^{\mathrm{bulk}}_{i} = \Phi \hat{\nabla}^j \theta_{ij}- \theta_{ij}\hat{\nabla}^j \Phi,
 \end{equation}
 where 
 \begin{equation}
     \theta_{ij} = \delta \hat{g}_{ij} - \hat{g}_{ij}\hat{g}^{kl}\delta \hat{g}_{kl}.
 \end{equation}
The contraction of the symplectic potential with a field space vector gives
\begin{equation}
I_{\xi}\theta^i_{{\mathrm{bulk}}} \os \hat{\nabla}_b\qty(4 \xi^{[j}\hat{\nabla}^{i]}\Phi + \Phi \hat{\nabla}^{[j}\xi^{i]}) + \frac{3}{L^2} \Phi^{-\frac12}\xi^i. 
\end{equation}
The Noether current \eqref{eq:diffeomorphismnoethercurrent}
is given by
\begin{equation}
    J_\xi^{i} \os \hat{\nabla}_b Q^{ij}_\xi,
\end{equation}
where
\begin{equation}
    Q^{ij}_\xi = 4 \xi^{[j}\hat{\nabla}^{i]}\Phi + \Phi \hat{\nabla}^{[j}\xi^{i]}.
\end{equation}
Finally, the Noether charge is given by
\begin{Align}\label{eq:charges}
   H_{\xi}
   &= \phi^*\qty(\frac{\epsilon^{ij}}{2\sqrt{-\hg}}\qty[\xi^k \qty(4 \hg_{kj}\partial_i \Phi + \Phi \partial_j g_{ki}) + \Phi g_{ki}\partial_j\xi^k])
\end{Align}
where $\epsilon^{ab}$ is the Levi-Civita \textit{symbol}, $\phi$ is the corner embedding \eqref{embedding} and where we used the fact that the connection is torsion free. Note that this result does not depend on the arbitrary length scale $L$.
Using the results of Subsection \ref{sec:extendedphasespaceandambiguities}, we know that the charge \eqref{eq:charges} is Hamiltonian with respect to the extended symplectic form \eqref{eq:extendedsymplecticform}, and that
the corresponding symplectic Poisson bracket realizes the algebra of diffeomorphisms.\par

In order to connect to the universal corner symmetries discussed in Subsection \ref{subsec:universalcornersymmetries}, we use the
trivial embedding \eqref{trivial_embedding} which places the corner at the origin of the coordinate system $S =(x^0 =0,x^1=0)$. The Noether charge 
\eqref{eq:charges} can then be written
\begin{equation}\label{eq:2186}
    H_{\xi} =  \xi_{(0)}^i t_{i} + \xi_{(1)j}^i N_i^j,
\end{equation}
where 
\begin{align}
    t_i &= \frac{\epsilon^{kj}}{2 \sqrt{-\hg^{(0)}}}\qty(4 \hg^{(0)}_{ij}\Phi^{(1)}_k + \Phi_{(0)}g^{(1)}_{kij}),\\
    N_i^j &= \frac{\epsilon^{kj}}{2\sqrt{-\hg^{(0)}}}\Phi_{(0)} g^{(0)}_{ik},
\end{align}
where the subscript $(0)$ (respectively $(1)$) indicate the fields (respectively the fields derivatives) evaluated at the corner. We can write the Poisson bracket between two charges
\begin{Align}\label{eq:Hxizeta}
    \qty{H_{\xi},H_{\zeta}} &= \qty{\xi_{(0)}^i t_{i} + \xi_{(1)j}^k N_k^j,\zeta_{(0)}^l t_{l} + \zeta_{(1)m}^n N_n^m}\\
    &= \xi_{(0)}^i \zeta_{(0)}^l \qty{t_i,t_l} + \xi_{(1)j}^k\zeta_{(1)m}^n \qty{N_k^j,N_n^m} + \qty(\xi_{(1)j}^k\zeta_{(0)}^l - \xi_{(0)}^l\zeta_{(1)j}^k) \qty{N_k^j,t_{l}}.
\end{Align}
Denote $\eta = \qty[\xi,\zeta]$. From equation \eqref{eq:liebracketucs}, we know that 
\begin{align}
    \eta_{(0)}^i &= \zeta^{i}_{(1)j} \xi_{(0)}^j - \xi^{i}_{(1)j}\zeta^j_{(0)},\\
    \eta_{(0)j}^i &= - \qty[\xi_{(1)},\zeta_{(1)}]^i_j.
\end{align}
We can thus write
\begin{equation}\label{eq:Heta}
    -H_{\eta} =\qty(\xi^{i}_{(1)j}\zeta^j_{(0)}-\zeta^{i}_{(1)j} \xi_{(0)}^j ) t_i +\qty[\xi_{(1)},\zeta_{(1)}]^i_j N_i^j.
\end{equation}
Matching equation \eqref{eq:Hxizeta} with equation \eqref{eq:Heta}, we find
\begin{equation}
    \qty{N^i_j,N^k_l} = \delta^k_j N^i_l - \delta^i_l N^k_j,\quad \qty{N^i_j,t_k} = -\delta^i_k t_j + \frac12 \delta^i_j t_k,\quad \qty{t_i,t_j} = 0.
\end{equation}
Since $N^0_0 = -N_1^1$, the above algebra is exactly the one of the ECS.
Finally, we use the dictionary \eqref{eq:dictionary} to write the expression for the charges in terms of the original four-dimensional data
\begin{align}
    t_i &= \frac{\epsilon^{kj}}{G\sqrt{-g^{(0)}}}\qty[g^{(0)}_{ij}\rho_{(0)}\rho^{(1)}_k + \frac{\rho_{(0)}}{8}\qty(\partial_j \rho_{(0)} g^{(0)}_{ki} + \rho_{(0)}^2 g^{(1)}_{ki,j})],\label{eq:chargeti}\\
    N^j_i &= \frac{\epsilon^{kj}}{G\sqrt{-g^{(0)}}}\frac{\rho_{(0)}^2}{8}g^{(0)}_{ik}.\label{eq:chargenij}
\end{align}
We conclude by commenting on the dimensions of those charges. While the $\spl{2}$ charges are always dimensionless, the translation charges have dimensions that depend on the dimensions
of the coordinates. They are dimensionless if the coordinates are dimensionless and have dimension of energy if the coordinates have dimension of length.


\chapter{The Mathematics of Quantum Corners}\label{chapter3}
In the previous chapter, we gave an overview of the classical treatment of corner charges and corner algebras in gravitational theories.
In the present chapter, we turn to the quantum description. Classically, we have access to an underlying theory of gravitation, which allows us to 
derive the gravitational data associated with a corner in terms of Noether charges and their associated algebras.
In principle, one could do the same in the quantum setting: given a complete formulation of quantum gravity, together with a precise understanding of what
corners are in that context, the quantum data associated with a corner could be derived.
Although we do not have access to such a complete quantum description, there are still general inferences we can draw from basic properties of quantum systems.
In particular, the quantum states of a system admitting physical symmetries are organized into unitary irreducible representations of the corresponding symmetry group.
This is where the corner proposal becomes especially powerful. Since, in that framework, gravity is described by a set of charges and their
algebras at corners, we can infer---without detailed knowledge of the full theory---that quantum states of gravity must reside in representations of
the corner symmetry group. Even without adopting the strongest form of the corner proposal, one can make the more conservative statement that an interesting
subset of quantum-gravitational states should be organized in this way. Quantum corners are therefore simply states in the Hilbert spaces selected by the
representation theory of the corner symmetry group.
To connect quantum corners with the more familiar classical observables, we must reinterpret the latter in a slightly different way.
This requires introducing new mathematical structures, namely coadjoint orbits. Coadjoint orbits provide the classical analogue of irreducible unitary representations.
Making their precise connection to representations will, in turn, require some additional ingredients. A detailed account of these structures is the subject of the
present chapter.
The representation theory of the two-dimensional extended corner symmetry group was initiated in \cite{Ciambelli:2024qgi} and developed in full detail in
\cite{Varrin:2024sxe}. The coadjoint orbits of the quantum corner symmetry group, together with their geometric quantization, were obtained in \cite{Neri:2025fsh}. Finally, the relation between the classical and quantum corner observables was first pointed out in \cite{Varrin:2025okc}.
\par
Here is how this chapter is organized. In Section~\ref{sec:mathematicalbackground} we review the mathematical structures
needed to describe quantum corners and to connect them with the classical picture. We begin with Mackey's theory of induced representations, and provide the
representations of the ECS as an example of its application. We then introduce coadjoint orbits and briefly review their geometric quantization and Kirillov's
orbit method. Coadjoint orbits connect to classical observables via moment maps, and to quantum observables via generalized Perelomov coherent states. We therefore conclude the section
by presenting these objects and explaining how they fit together, thereby providing the notion of a semiclassical limit in the corner proposal.\par
Next, in Section~\ref{sec:quantumcornersymmetries}, we develop these structures explicitly for the quantum corner symmetry group.
We begin by explaining how this group arises from the two-dimensional ECS group and by identifying its Casimir operators.
We then use the theory of induced representations to derive its representation theory. Next, we analyze the coadjoint orbits and construct their geometric quantization,
recovering the same representations through Kirillov's orbit method. Finally, we build the coherent states and formulate the semiclassical limit,
together with the correspondence between quantum and classical observables.\par
Section~\ref{sec:mathematicalbackground} consists mostly of review material, drawn in part from \cite{Varrin:2024sxe,Varrin:2025okc}.
Section~\ref{sec:quantumcornersymmetries} is based on original results published in \cite{Varrin:2024sxe,Neri:2025fsh,Varrin:2025okc}.
Portions of \cite{Varrin:2024sxe,Varrin:2025okc} are directly reproduced here, and have been extended where needed.

\section{Mathematical background}\label{sec:mathematicalbackground}
In order to describe quantum corners and their physical implications, we first need to introduce the mathematical structures that characterizes them. At the core of the corner proposal
and quantum corners lies the theory of group representations. Given a group $G$, a map $U$ from the group to the linear operators on a vector space $\mathcal{H}$ is called a
representation if it is a group homomorphism with respect to the operator product
\begin{equation}
    U(g_1 g_2) = U(g_1)U(g_2).
\end{equation}
A representation is called irreducible if there exists no vector subspace inside of $\mathcal{H}$ that is invariant under the action of the group. If $\mathcal{H}$ is a Hilbert space,
a representation is called unitary if it preserves the scalar products, i.e. if the operators $U(g)$ are Hermitian with respect to that scalar product.\par
When the Hilbert space of the representation is taken
to describe a physical system, there is an important additional subtlety. In general, a quantum theory is defined by a Hilbert space $\mathcal{H}$ and the transition probabilities between different states
\begin{equation}
    \mathcal{P}_{\alpha\rightarrow \beta} = \abs{\braket{\alpha}{\beta}}^2,
\end{equation}
where $\ket{\alpha},\ket{\beta}\in\mathcal{H}$ are vectors in the Hilbert space, and the bracket denotes the scalar product. An immediate consequence is that the theory's predictions do not depend on the vectors phases. This fact is implemented by defining the states of the theory as the $\textit{ray operators}$ rather than the vectors themselves. The former are defined as equivalence classes of vectors, with the following equivalence relation 
\begin{equation}
    \ket{\alpha} \sim e^{i\theta}\ket{\alpha}
\end{equation}
for any $\theta \in \mathbb{R}$. That is, vectors that only differ by a phase are different representative elements of the same ray operator. The states are then elements of the projective Hilbert space
\begin{equation}
    P\mathcal{H} = \mathcal{H}/\sim.
\end{equation}
Such a Hilbert space can be obtained from a symmetry group by considering the projective representations of the latter. A representation $U:G\longrightarrow \mathrm{End}\qty(P\mathcal{H})$ is called projective, if the following equation is satisfied
\begin{equation}\label{eq:groupcentralextension}
    U_{g_1}U_{g_2} = \omega(g_1,g_2) U_{g_1\cdot g_2},
\end{equation}
where $\omega(g_1,g_2)$ are the 2-cocycles of the group. Due to a famous theorem by Bargmann and Mackey \cite{Bargmann1954,Mackey_1958,Mackey1976TheTO}, it is known that the projective representations of a path-connected Lie group are equivalent to
the ordinary unitary representations of its maximal central extension.
The ``maximal" encapsulates two different origins of the central extensions. The first one comes from the first homotopy group and is linked with the universal cover of the symmetry group. In the example of Poincaré symmetry,
this is the reason why we study representations of $\mathrm{SL}\left(2,\mathbb{C}\right)\ltimes \mathbb{R}^4$ instead of the original Poincaré
group with Lorentz symmetries. The second one is the addition of a central generator determined by the 2-cocycles of the algebra. Unlike Poincaré symmetries,
which do not allow non-trivial 2-cocycles, the Galilean group—the symmetry group of non-relativistic mechanics—does. The added central generator is then identified
with the mass of the non-relativistic particles \cite{LevyLeblond1963Galilei}. This last example highlights the strong physical significance of the projective nature of the representations.
The quantum physical systems associated with the corner symmetries 
should therefore be described by unitary irreducible representations of the maximally 
centrally extended version of that group. How do we find such representations?
The standard approach for semidirect-product groups, such as those underlying corner symmetries, is Mackey's theory of induced representations \cite{Mackey1951,Mackey1952,Mackey1953},
which is the subject of the next section.\par
\begin{figure}[h]
    \centering
    \includegraphics[width=0.7\linewidth]{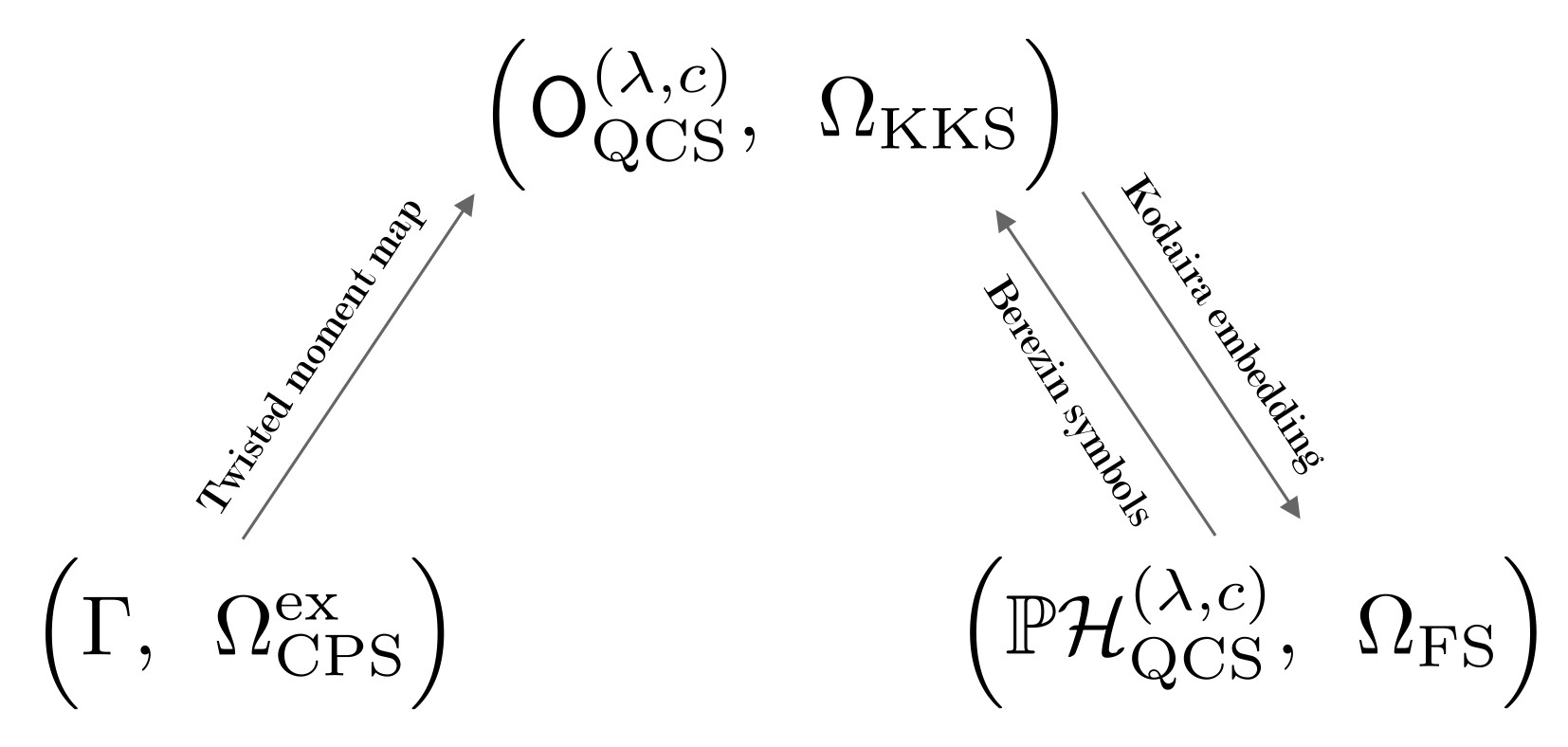}
    \caption{The three symplectic spaces appearing in the corner proposal: the coadjoint orbits equipped with the Kirillov--Kostant--Souriau form, the field space equipped with the extended symplectic form of the covariant phase-space formalism, and the projective Hilbert space equipped with the Fubini--Study form. The classical limit is obtained by identifying the point on the coadjoint orbit determined by the twisted moment map with the point determined by the Berezin symbols.
}
    \label{fig:mathematicaltriad}
\end{figure}

Most of the mathematical structure required by the corner proposal, however, arises when one attempts to connect
these representations to the classical side of the story, where geometry and fields live. In order to do so, it turns out to be very instructive to consider the representations
as arising through the quantization of the associated coadjoint orbits through Kirillov's method  \cite{Kirillov_1962,Kirillov1976,Kirillov1990,Kirillov1999,Kirillov2004}.
In addition to the purely mathematical interest of providing a successful example of the orbit method in a setting where it had not previously been realized,
coadjoint orbits are a crucial ingredient for understanding how representations relate to the classical phase space of the theory.
Indeed, coadjoint orbits connect to classical
fields via the moment maps of symplectic geometry, and to the projective Hilbert space via Perelomov generalized coherent states. The mathematical structure of the corner proposal which
we present here can be described as the relationship between three symplectic manifolds. The coadjoint orbits equipped with the Kostant--Kirillov--Souriau form,
the projective Hilbert space equipped with the Fubini-Study form, and the classical field space equipped with the CPS symplectic form. The relationships between these structures
are depicted in Figure \ref{fig:mathematicaltriad}.\par

This section is organized as follows. We begin in Subsection \ref{subsec:inducedrepresentation} with a review of Mackey's theory
of induced representations, with particular emphasis on the semidirect-product case. In  Subsection \ref{subsec:coadjointorbits} we introduce the theory of coadjoint orbits.
Finally, Subsection \ref{subsec:momentmaps} is devoted to relating coadjoint orbits to the projective Hilbert space via coherent states, and to the covariant phase space formalism via moment maps.

\subsection{Induced representations}\label{subsec:inducedrepresentation}

The theory of induced representations was developed by George Mackey in the fifties.
It is a generalization of the famous work of Wigner and Bargmann on the representation theory of the Poincaré group
\cite{Wigner:1939cj,Wigner1948,Wignerbargman1948}. Without delving into the mathematical details, we give here a self-contained presentation of the construction
of induced representations, with a focus on the case of semidirect groups. We start by giving general considerations on Mackey's theory and then move on to the cases of interest in the present work.\par

Given a a group $G$, a subgroup $H \subset G$ and a representation of the subgroup on a vector space $V$
\begin{equation}
    D:H \longrightarrow \mathrm{End}(V),\quad
    h \longmapsto D(h),
\end{equation}
one can construct a representation of the entire group, acting on vector-valued functions on the coset space $\psi :G/H\rightarrow V$\footnote{We assumed here the existence of a global section of the bundle $G\xrightarrow{\pi} G/H$ with canonical projection map $\pi$. While this is not a restriction of the general construction, it is true for semidirect groups and will thus suffice in our case.}.
 In the case of the semidirect group $G = H \ltimes N$, any group element can be decomposed as $g = h \cdot t$ for $h \in H$ and $t \in N$ and the induced representation is given by
\begin{equation}\label{inducedrepresentation}
    (U_g \psi)(p) = D(h)\psi(g^{-1} \rhd p),
\end{equation}
where $\rhd$ denotes the natural action of the group on its coset space.

 If $V$ is a Hilbert space and $D$ is unitary with respect to the inner product, one can always make $U$ into a unitary representation. However, even if the representation $D$ is irreducible, the representation $U$ is generally not. In order to obtain irreducible representations, one has to apply the \textit{little group} method, that we will now describe.\par
 The structure of semidirect product groups implies the existence of an automorphism on the normal subgroup given by
 \begin{equation}\label{automorphism}
     t \longmapsto \tilde{t}= h\cdot t\cdot h^{-1}.
 \end{equation}
 Given an irreducible unitary representation of the normal subgroup $\Xi: N \longrightarrow \mathrm{End}(\mathcal{H}^{(\Xi)})$, there exists two subgroup $\mathring{H} \subset H$ and $H^{(0)}\subset H$ and their associated representations
 \begin{Align}
     &\mathring{D}: \mathring{H}\longrightarrow \mathrm{End}\qty(\mathcal{H}^{\left(\mathring{D}\right)}),\\
     &D^{(0)}: H^{(0)}\ltimes N \longrightarrow \mathrm{End}\qty(\mathcal{H}^{(\Xi)}).
 \end{Align}
The representation $D^{(0)}$ is an extension of the representation $\Xi$
\begin{equation}
    D^{(0)}\mid_N = \Xi,
\end{equation}
that represents the automorphism \eqref{automorphism} on the Hilbert space $\mathcal{H}^{(\Xi)}$
\begin{equation}\label{automorphismrepresentation}
     \Xi(\tilde{t}) = \Xi(h\cdot t\cdot h^{-1}) = D^{(0)}(h)\, \Xi(t)\, D^{(0)}(h)^{-1}.
\end{equation} 
The unitary irreducible representations of the group $G$ can then be induced from representations of the form
\begin{equation}\label{productrepresentation}
     \mathring{U} = \mathring{D} \otimes D^{(0)}.
 \end{equation}
The two cases of interest in this thesis correspond to $H^{(0)}$ being trivial, which, as we will shortly see, arises when $N$ is Abelian, and the case where $H^{(0)} = \mathring{H} = H$.\par
To treat the Abelian case, we first note that the derivative of the representation $\Xi$ is a representation of the Lie algebra $\mathfrak{n}$. Therefore, given a basis
$\qty{T_a}$ of $n$, the vectors $\Xi'(T_a)$ span the Hilbert space $\mathcal{H}^{(\Xi)}$. Since the representation $D^{(0)}$ also acts on the Hilbert space of the
$\Xi$ representation, the representation $D'^{(0)}$ must be defined in terms of the enveloping algebra of $\mathfrak{n}$.
We denote the now-abelian normal subgroup $N=\mathbb{R}^n
$. Since the enveloping algebra is also abelian, the Lie algebra of $H$ cannot be represented in it and thus the extension $D^{0}$ is trivial
$D^{0}(h) = 1 $ and $H^0$ is the identity element. The representation of the automorphism \eqref{automorphismrepresentation} then reduces to 
\begin{equation}\label{automorphismabelian}
    \Xi(\tilde{t}) = \Xi(h\cdot t \cdot h^{-1}) =\Xi(t)\quad \forall t \in N.
\end{equation}
This equation then defines $\mathring{H}$ as the \textit{little group}. That is, the subgroup of $H$  whose automorphism on the normal subgroup $N$ preserves the representation $\Xi$ in the sense of \eqref{automorphismabelian}. In practice however, it will be easier to work with the equivalent stabilizer group that we will now define. Since the normal subgroup is a vector space, the semidirect group is defined by a matrix representation of the subgroup $H$ 
\begin{equation}\label{matrixrepresentation}
    \varphi: H \longrightarrow M_{n x n}\qty(\mathbb{R}),\quad
    h\longmapsto \varphi\qty[h],
\end{equation}
and the automorphism \eqref{automorphism} can be written
\begin{equation}
    \tilde{t} = \varphi[h] t,
\end{equation}
where $t$ is interpreted as a n-tuple in $\mathbb{R}^n$ and $\varphi[h]$ acts on it via the usual matrix multiplication. Furthermore, the irreducible unitary representations of the abelian group are 1-dimensional and $\Xi$ simply becomes the character of the representation. Introducing the dual $\mathbb{R}^{*n}$, and the pairing $\pairing{\cdot}{\cdot}: \mathbb{R}^{n} \times \mathbb{R}^{*n} \longrightarrow \mathbb{R}$, the elements $p\in\mathbb{R}^{*n}$ label the characters by
\begin{equation}\label{character}
    \Xi_p(t) = e^{i\pairing{t}{p}}.
\end{equation}
Introducing the dual action $\varphi^*$ on the dual space $\mathbb{R}^{*n}$
\begin{equation}
    \pairing{\varphi[h]t}{p} = \pairing{t}{\varphi^*[h^{-1}]p}, \quad \forall t \in \mathbb{R}^n, p\in\mathbb{R}^{*n},
\end{equation}
we see from equations \eqref{automorphismabelian} and \eqref{character} that the little group is isomorphic to the stabilizer
\begin{equation}
    H_p = \qty{h \in H\mid \varphi^*[h]
    p = p}.
\end{equation}
 We also define the orbit at a point $p\in \mathbb{R}^{*n}$ as
\begin{equation}
    \mathcal{O}_p = \qty{\varphi^*[h] p\mid h \in H}.
\end{equation}
The particularity of the Abelian normal subgroup case hinges on two isomorphisms. First, the orbit $\mathcal{O}_p$ is isomorphic to the coset space $H/H_p$. To identify, them we introduce a map\footnote{It can be shown that this map is well defined and unique up to right composition by any function that sends $\mathcal{O}_p$ to $H_p$.}
\begin{equation}
    \mathsf{h}:\mathcal{O}_p \longrightarrow H, \quad q\longmapsto \mathsf{h}_q,
\end{equation}
such that $\varphi^*[\mathsf{h}_q] p = q$ for all $q\in\mathcal{O}_p$. One can then identify a point $q \in \mathcal{O}_p$ with the coset $\mathsf{h}_q H_p \in H/H_p$. The second isomorphism is between the coset space $H/H_p$ and the coset space $G/G_p$, where $G_p = H_p \ltimes N$, which allows us to write the induced representation space as functions on the orbit $\mathcal{O}_p$. We are now finally ready to construct the induced representation. Given a unitary representation of the little group $\mathring{D}:H_p \longrightarrow \mathrm{End}(\mathring{\mathcal{H}})$ we construct the representation $D: G_p \longrightarrow \mathrm{End}(\mathring{\mathcal{H}})$ defined as
\begin{equation}
    D(\mathring{h},t) =  \Xi_p(t) \, \mathring{D}(\mathring{h}) = e^{i\pairing{t}{p}} \mathring{D}(\mathring{h}),
\end{equation}
for $\mathring{h}\in H_p$ and $t \in \mathbb{R}^n$. Using the isomorphisms described above, we can finally write the induced representation \eqref{inducedrepresentation} of $D$ to the entire group $G$ as
\begin{equation}\label{inducedrepresentationabelian}
    \qty(U_{(h,t)} \, \psi)(q) = e^{i\pairing{t}{q}} \mathring{D}\qty(\mathsf{h}_q^{-1} \cdot h \cdot \mathsf{h}_{\varphi^*[h^{-1}]})\psi\qty(\varphi^*[h^{-1}]q),
\end{equation}
where $q\in\mathcal{O}_p$. As a result, each distinct orbit will have its own set of representations.
In the Poincaré case, this is what leads to the different spin and mass states. Note that it can be shown that the representation \eqref{inducedrepresentationabelian}
does not depend on the choice of orbit representative. If the representation $\mathring{D}$ is irreducible,
Mackey theory ensures that the induced representations \eqref{inducedrepresentationabelian} form a complete set of irreducible unitary representations of $G$.
We also note that the intimate relationship between orbits and irreducible representations of a group go beyond the semi-product case.
This is the subject of the coadjoint orbit method of Kirillov that will be discussed in the next section.\par
If the normal subgroup is not Abelian, its representations are in general more complicated than the simple character \eqref{character}. However, the case of interest in this thesis is captured by the following specific instance. If the representation $D^{0}$ is defined globally, that is $G^{0} = G$, then by equation \eqref{automorphismrepresentation}, the little group is the entire group $H$. The representations \eqref{productrepresentation} are then already representations of the entire group and Mackey's theory ensure they cover all possible unitary irreducible representations. They are given explicitly by
\begin{equation}\label{eq217}
    U_{(h,t)} = D(h) \otimes D^{(0)}(h) \Xi(t),
\end{equation}
where $D:H\longrightarrow \mathrm{End}\qty(\mathcal{H}^{(D)})$ and $D^{(0)}:H\longrightarrow \mathrm{End}\qty(\mathcal{H}^{(\Xi)})$ are unitary irreducible representation of $H$.
 For more details, see for example \cite{Low:2019oon}.

\paragraph{The example of the ECS}
As an illustrative example of induced-representation theory,
and before applying it to the main subject of this thesis in Section \ref{sec:quantumcornersymmetries},
we apply the above formalism to the ECS group. While a somewhat lengthy example, the representation theory of the ECS group
is of obvious interest for the corner proposal,
even though it has not yet been used in any concrete application.\par
We start by inducing irreducible representations of $\spl{2}$ to the entire
$\mathrm{ECS}$ group. Although such representations are not generally unitary, they remain highly relevant in physics.
Indeed, this is the case of relativistic fields which generally are in non-unitary representations of the Poincaré group.
In fact, they can be obtained by inducing the finite dimensional irreducible representations of the Lorentz group, described using the local isomorphism to
$\mathrm{SU}(2)\times\mathrm{SU}(2)$, to the full inhomogeneous group by the method detailed in the previous section. Since the Lorentz group is non-compact,
those finite dimensional representations can not be unitary. In this section, we will apply the same analysis to the $\mathrm{ECS}$ group. The special linear group being
locally isomorphic to $\mathrm{SU}(2)$, we can construct the associated finite dimensional irreducible representations and induce them to the entire corner symmetry
group using Mackey's theory. This will serve as a straightforward application of the theory of induced representations and a good warm-up for the more complicated
cases ahead.
As previously mentioned, the induced representations will not generally be irreducible. In order to obtain irreducible representations, one needs to apply the little
group method, which we will tackle next.\par

Since $\mathfrak{sl}\qty(2,\mathbb{R}) \cong \mathfrak{su}\qty(2)$, the finite dimensional representation theory of $\spl{2}$ is equivalent to the representations of $\mathrm{SU}(2)$. To make this isomorphism more explicit, let us introduce the angular momentum basis of the algebra 
\begin{equation}
    J_0 = 2 L_0,\quad J_+ = L_+, \quad J_- = -L_-.
\end{equation}
For any $j\in \mathbb{N}$ or $j \in \frac{\mathbb{N}}{2}$ there exists a $2j+1$ dimensional representation spanned by the eigenvectors of $J_0$
\begin{equation}
    J^{(j)}_0 \psi^{(j)}_m = m \psi^{(j)}_m,
\end{equation}
where $m = -j,-j + 1,...,j$ and the $(j)$ subscript indicates that the generator is taken in the corresponding representation. The remaining generators of $\mathfrak{sl}(2,\mathbb{R})$ then act as raising and lowering operators 
\begin{align}
    J^{(j)}_\pm \psi^{(j)}_m &= \sqrt{j(j+1)-(m(m\pm1))} \psi^{(j)}_{m\pm1}.
\end{align}
such that $J^{(j)}_- \psi^{(j)}_{-j} = J^{(j)}_+ \psi^{(j)}_{j} = 0$.
 The Hilbert spaces carry a representation of the group by exponentiation of the generators: Any element $h \in \spl{2}$ can be written as
\begin{equation}
    h = \exp(h^i J_i),
\end{equation}
where $h^i,\, i=0,-,+$ are coordinates on the Lie group\footnote{Since $\mathrm{SL}\qty(2,\mathbb{R})$ is not simply connected, the exponentiation of an algebra element will be an element of the universal cover of the special linear group $\widetilde{\mathrm{SL}\qty(2,\mathbb{R})}$. However the fact that the maximal compact subgroup has integers values ($e^{2\pi i L_0}=\pm 1$) is exactly the condition needed for the representations to be of $\spl{2}$ rather than its universal cover.}.
The group representation can then be written 
\begin{equation}\label{jrepresentation}
    D^{(j)}(h) \psi_m = \sum_{k=0}^{\infty} \frac{(h^i J^{(j)}_i)^k}{k!} \psi_m, \quad \forall h \in \spl{2},
\end{equation}
 where the $j-$index of the vectors was dropped for simplicity.\par
We now construct the induced representations of the $\mathrm{ECS}$. In order to do so, we first observe that the coset space is isomorphic to $\mathbb{R}^2$:
\begin{equation}
    \mathrm{ECS}/\spl{2} \cong \mathbb{R}^2.
\end{equation}
The induced representation will therefore be constructed on vector-valued functions
\begin{equation}
    \psi: \mathbb{R}^2 \longrightarrow \mathcal{H}^{(j)},
\end{equation}
where $\mathcal{H}^{(j)}$ denotes the Hilbert space of the $j-$representation.
Any element $g \in \mathrm{ECS}$ can be written as $g = t\cdot h =\exp(t^a P_a)\exp(h^i J_i)$, where $a=-,+$. The induced representation \eqref{inducedrepresentation} can then be written
\begin{Align}\label{sl2rinducedrepresentation}
    (U^{(j)}_{(t\cdot h)} \psi_m)(x) &= D^{(j)}(h) \psi_m\Bigl(\varphi\qty[\exp(-h^i J_i)](x-t)\Bigr)\\
    &= \sum_{k=0}^{\infty}\frac{(h^i J^{(j)}_i)^k}{k!} \psi_m\Bigl(\varphi\qty[\exp(-h^i J_i)](x-t)\Bigr),
\end{Align}
where we have used that the action of the group on the coset space is simply given by
\begin{equation}
    (t\cdot h)\rhd x = \varphi[h]x + t.
\end{equation}
For any irreducible $j-$representation of $\spl{2}$, equation \eqref{sl2rinducedrepresentation} defines a representation of the $\mathrm{ECS}$ group. Let us now compute the representation of the algebra
\begin{equation}
    (U'_X \psi_m)(x) = \dv{\alpha}\qty(\qty(U^{(j)}_{\exp(\alpha X)}\psi_m)(x))\mid_{\alpha = 0}, 
\end{equation}
where $X\in\mathfrak{ecs}$.
The translations act as derivative operators
\begin{equation}\label{algreptranslation}
    \qty(U'_{P_\pm}\psi_m)(x) = -\pdv{\psi_m(x)}{x_\pm}.
\end{equation}
and the $\mathfrak{sl}\qty(2,\mathbb{R})$ generators act as
\begin{equation}\label{sl2ralgebraaction}
    (U'_{J_i} \psi_m)(x) = J^{(j)}_i \psi_m(x) + \qty(\varphi'\qty[J_i])\updown{a}{b} x^b \partial_a \psi_m(x),
\end{equation}
where $\varphi': \mathfrak{sl}(2,\mathbb{R}) \longrightarrow M_{2x2}\qty(\mathbb{R})$ is the matrix representation of the Lie algebra defined as
\begin{equation}
    \varphi'[X] = \dv{\alpha}\qty(\varphi\qty[-\alpha X]).
\end{equation}
To write this action more explicitly, let us introduce the explicit matrix representation
\begin{equation}\label{explicitmatrixrepresentation}
   \varphi'\qty[ J_0] = \mqty(1&&0\\0&&-1), \quad \varphi'\qty[J_+] = \mqty(0&&1\\0&&0),\quad \varphi'\qty[J_-] = \mqty(0&&0\\
   1&&0).
\end{equation}
We get
\begin{Align}\label{algrepsl2r}
    (U'_{J_0} \psi_m)(x) &= m \psi_m(x) + (x_- \partial_{x_-} - x_+ \partial_{x_+})\psi_m(x),\\
    (U'_{J_+}\psi_m)(x) &=\sqrt{j(j+1)-m(m+1)}\, \psi_{m+1}(x) + x_- \partial_{x_+} \psi_m(x)\\
    (U'_{J_-}\psi_m)(x) &= \sqrt{j(j+1)-m(m-1)}\, \psi_{m-1}(x) + x_+ \partial_{x_-} \psi_m(x).
\end{Align}
This action is reminiscent of the action of the Lorentz algebra on the fields of a relativistic field theory.
The first terms are the 'spin' parts, corresponding to a transformation of the fields in some internal space,
and the inhomogeneous terms are the 'orbital' parts corresponding to the actual spacetime transformations.
In particular, note that if we choose the trivial representation  $D^{(j=0)}$, only the inhomogeneous part remains.
The thus obtained representation is called the defining representation in \cite{Ciambelli:2022cfr}.
It has a vanishing Casimir value and is irreducible. In the analogy with the Poincaré group, these would correspond to scalar fields. 
However, as expected, the representation is not irreducible for a general $j$. To see this, one can use the induced representation of the algebra to compute
the action of the Casimir operator \eqref{eq:ecscubiccasimir} on an arbitrary $j$-representation and find that it is not a multiple of the identity.
Furthermore, since the representations of $\spl{2}$ from which we induced are not unitary, neither are the resulting representations of the $\mathrm{ECS}$.
One could instead induce the infinite dimensional unitary representations of the special linear group discussed in Section \ref{sec:quantumcornersymmetries}.
Although they would have a very similar form to \eqref{sl2rinducedrepresentation}, the unbounded nature of the $j$ index would allow for unitarity.
These representations are reminiscent of the 'expinor' theory of Harish-Chandra and Dirac \cite{dirac1945,harishchandra1947} (See also \cite{mukunda1993} for
a historical review), which does not have any known physical relevance.\par
We now move to the irreducible representations of the $\mathrm{ECS}$.
In order to apply the little group method to we need to consider the orbits of $\spl{2}$ in $\mathbb{R}^{*2} \cong \mathbb{R}^2$. To do so, let us introduce the $KAN$ matrix representation of the special linear group. Any element $g \in \spl{2}$ can be written uniquely as
\begin{equation}
    \varphi(g) = \mqty(\cos(\theta)&&-\sin(\theta)\\\sin(\theta)&&\cos(\theta))\,\mqty(\frac{1}{r}&&0\\0&&r)\,\mqty(1&&x\\0&&1).
\end{equation}
The parameters $\theta \in \qty[0,2\pi]$, $r\geq 0$ and $x\in\mathbb{R}$ are thus coordinates on the group and we will denote the group element defined by its $KAN$ decomposition by
\begin{equation}
    g_{\theta,r,x} = k_{\theta}\cdot a_r\cdot n_x.
\end{equation}
Let us now describe the orbits. Any point $(p_-,p_+)\in\mathbb{R}^{2}$, that is different from the origin, can be reached from the point $(1,0)$ by first scaling the element to the right size, and then rotating it to the right vector
\begin{equation}
    \mqty(p_-\\p_+) = \varphi\qty(k_{\tilde{\theta}})\, \varphi\qty(a_{\tilde{r}})\, \mqty(1\\0),
\end{equation}
for any $\qty(p_-,p_+) \neq (0,0)$ with
\begin{equation}\label{parameterofcanonicaltransformation}
    \tilde{r} = \sqrt{p_-^2+p_+^2},\quad \tilde{\theta} = \mathrm{sgn}\qty(p_+)\arccos(\frac{p_-}{\sqrt{p_-^2+p_+^2}}).
\end{equation}
The above construction shows that there exists only two distinct orbits of the action of $\spl{2}$ on $\mathbb{R}^2$: The non-trivial orbit $\mathcal{O}_{(1,0)}$ and the trivial one containing only the origin $\mathcal{O}_0$. Since the latter corresponds to a trivial action of the translations, we will only look at the former. Sticking with $(1,0)$ as the orbit representative, it is easy to see that the little group is simply the $N$ component of the $KAN$ decomposition
\begin{equation}
    H_{(1,0)} = \qty{n_x \mid x\in\mathbb{R}} \cong \mathbb{R}.
\end{equation}
Since the group is abelian, its irreducible unitary representations are once again fully described by its characters
\begin{equation}
    \mathring{D}^{(\nu)}:H_{(1,0)}\longrightarrow \mathbb{C}, \quad \mathring{D}^{(\nu)}(n_x) = e^{i \nu x},
\end{equation}
where the the distinct representations are labeled by the different values of $\nu \in \mathbb{R}$. We can now construct the induced representation which acts on functions on the orbit. To emphasize that we are now working in the state representation, we will denote these functions by $\psi(p) \equiv \ket{p}$
where $p = (p_-,p_+) \neq (0,0)$. For a $\mathrm{ECS}$ group element $g = t \cdot h$, the induced representation \eqref{inducedrepresentationabelian} is given by
\begin{equation}\label{ECSinducedrepresentation}
    U^{(\nu)}_{(t,h)} \ket{p} = e^{i \pairing{t}{p}} \mathring{D}^{(\nu)}\qty(\mathsf{h}^{-1}_p \cdot h \cdot \mathsf{h}_{\varphi^*[h^{-1}]p})\ket{\varphi^*\qty[h^{-1}]p},
\end{equation}
where we recall that $\mathsf{h}_p$ is the $\spl{2}$ group element that brings the representative to the point $p$. In this case, it is therefore simply given by 
\begin{equation}
    h_p = k_{\tilde{\theta}}\cdot a_{\tilde{r}},
\end{equation}
with $\tilde{\theta}$ and $\tilde{r}$ given by equation \eqref{parameterofcanonicaltransformation}. Since we chose irreducible unitary representations of the little group, Mackey's theory ensures that one can always equip the Hilbert space with a scalar product such that the induced representation is unitary. Equation \eqref{ECSinducedrepresentation} then describes the complete set of irreducible unitary representations of the $\mathrm{ECS}$ group. The representations are characterized by a single parameter $\nu \in \mathbb{R}$, reflecting the existence of a unique Casimir operator. In fact, one can show that $\nu$ is precisely the scalar value that $\mathcal{C}_{\mathrm{ECS}}$ takes on these representations. In order to do so, we can use the explicit matrix representation of the $\mathfrak{sl}(2,\mathbb{R})$ generators \eqref{explicitmatrixrepresentation}\footnote{Since we are now working with unitary representations, we choose the matrices of \eqref{explicitmatrixrepresentation} to define the generators multiplied by the imaginary unit. This ensures that the corresponding operators $U'^{(\nu)}\qty(L_i)$ are Hermitian}. The induced representation of the algebra can then be calculated using elementary matrix calculations. For the $\mathfrak{sl}(2,\mathbb{R})$ generators we get
\begin{Align}
    U'^{(\nu)}\qty(L_0) \ket{p} &= \qty[2\nu \frac{ p_- p_+}{\qty(p_-^2+p_+^2)^2} + \frac12 \qty(p_+ \partial_{p_+} - p_- \partial_{p_-})]\ket{p},\\
    U'^{(\nu)}\qty(L_-) \ket{p} &= \qty[\nu\frac{p_-^2-p_+^2}{(p_-^2 + p_+^2)^2} +  p_- \partial_{p_+}]\ket{p},\\
    U'^{(\nu)}\qty(L_+)\ket{p} &= \qty[\nu \frac{p_+^2-p_-^2}{\qty(p_-^2+p_+^2)^2}-  p_+ \partial_{p_-}]\ket{p}.
\end{Align}
The translations simply act by multiplication
\begin{equation}
    P_\pm \ket{p} = i p_\pm \ket{p}.
\end{equation}
Choosing the trivial representation of the little group ($\nu = 0$) and applying a Fourier transform, one obtains again the defining representation of the $\mathrm{ECS}$. 
Finally, it is straight forward to check that
\begin{equation}
U'^{(\nu)}\qty(\mathcal{C}_{\mathrm{ECS}}) \ket{p} = \nu \ket{p}.
\end{equation}
\par We conclude this section with a remark about the relationship between the field representation of the previous section and the state representation that was just constructed. In the context of quantum field theory, the field representation 
is related to the state representation by introducing field operators acting on the Hilbert space of the latter, $\hat{\psi}_m(x)$, that transform according to
\begin{equation}
    \qty(U^{(\nu)}_{(t,h)})^{-1} \hat{\psi}_m\bigl(\varphi[h]x + t\bigr)  U^{(\nu)}_{(t,h)} = D^{(j)}(h)\hat{\psi}_m(x),
\end{equation}
where $D^{(j)}$ is the representation defined in equation \eqref{jrepresentation}.
\subsection{Coadjoint orbits}\label{subsec:coadjointorbits}
We consider a Lie group $G$ with associated Lie algebra $\mathfrak{g}$. We denote by $X^a, a=1,...,\mathrm{dim}(G)$ a basis of the Lie algebra.
We can write the Lie bracket in this basis as
\begin{equation}
    \qty[X^a,X^b] = C^{ab}_c X^c,
\end{equation}
which defines the structure constants $C^{ab}_c$.
An important feature of Lie groups is that one can always construct a representation on its own algebra.
For each element $g\in G$ we define the inner automorphism map
\begin{Align}
    \Psi_g: G &\longrightarrow \mathrm{Aut}\qty(G),\\
       \Psi_g(h) &= g h g^{-1}.
\end{Align}
The adjoint action is then defined as the differential of $\Psi_g$ at the identity element $e\in G$ 
\begin{equation}
    \mathrm{Ad}_g \defeq \qty(\dd \Psi_g)_e: \mathfrak{g}^* \longrightarrow \mathfrak{g}^*.
\end{equation}
The map
\begin{equation}
    \mathrm{Ad}: G\longrightarrow \mathrm{Aut}(\mathfrak{g}), \quad g\mapsto \mathrm{Ad}_g,
\end{equation}
is called the adjoint representation of the group $G$. The fact that it is a representation follows directly from the definition of the inner automorphism.
$\Psi_{g g'}(h) = \Psi_{g}\circ \Psi_{g'}(h)$.
Given a Lie algebra element $X\in\mathfrak{g}$, the derivative of the adjoint representation $\dd\qty(\mathrm{Ad})_e \defeq \mathrm{ad}$ produces the adjoint representation of the algebra
\begin{equation}
    \ad{X} : \mathfrak{g}\longrightarrow \mathfrak{g} \quad Y \longmapsto \ad{X}\qty(Y) \defeq \qty[X,Y].
\end{equation}
With that definition, the structure constants can be seen as the matrix components of the adjoint action
\begin{equation}
    \qty(\ad{X^a})^b_c= C^{ab}_c. 
\end{equation}
For reasons that will soon become clear, it turns out to be more interesting to consider the action of the group on the dual algebra $\mathfrak{g}^*$, defined
as the set of linear functionals on the algebra
\begin{equation}
    \mathfrak{g}^* = \qty{m: \mathfrak{g} \longrightarrow \R}.
\end{equation}
The dual algebra is also called the \textit{coalgebra} and its elements \textit{covectors}. One interesting property of the coalgebra is that it is always a Poisson manifold.
This is seen  as follows.
Consider two functions on the coalgebra $f,g\in C^\infty(\mathfrak{g}^*)$. Their differential $\dd f,\dd g$ are elements of the cotangent bundle $\mathrm{T}^*\mathfrak{g}^*$.
Since the coalgebra is a linear space, we have the isomorphism
\begin{equation}
    \mathrm{T}^* \mathfrak{g}^* \equiv \mathfrak{g}^{**} \equiv \mathfrak{g}.
\end{equation}
We can then define the Poisson bracket
\begin{equation}\label{eq:poissonbracketcoalgebra}
    \qty{f,g}_{P}(m) = \pairing{m}{\qty[\dd f(m),\dd g(m)]}.    
\end{equation}
While every symplectic manifold is a Poisson manifold, it is a well known fact that the converse is not true. In general, the coalgebra is therefore not a symplectic manifold.
This can be seen easily by considering a odd-dimensional algebra. The standard procedure on Poisson manifolds, is to construct the symplectic foliation of the space, where the
individual leaves of the foliation host a symplectic structure. For coalgebras, this procedure is beautifully realized by the construction of the coadjoint orbits.
\par

We introduce
the dual basis $X^*_a$ of $\mathfrak{g}^*$ defined through the relation
\begin{equation}
    \pairing{X_b^*}{X^a} = \delta^a_b,
\end{equation}
where $\pairing{m}{X} \defeq  m(X)$ is the natural pairing between a vector space and its dual. The \textit{coadjoint action} of the group on $\mathfrak{g}^*$ is then defined as the dual map of the adjoint action 
\begin{equation}
    \pairing{\mathrm{Ad}^*_g m}{X} = \pairing{m}{\mathrm{Ad}_{g^{-1}} X}.
\end{equation}
Running through the entire group while acting on a point $m\in\mathfrak{g}^*$ defines a submanifold called the coadjoint orbit
\begin{equation}
    \mathsf{O}_m = \qty{\mathrm{Ad}^*_g m : g\in G}.
\end{equation}
There are two key results concerning coadjoint orbits. The first being that each orbit is completely characterized by the stabilizer subgroup
\begin{equation}\label{eq:stabilizersubgroupcoadjoint}
    H_m = \{\,g \in G : \mathrm{Ad}^*_g m = m \,\}.
\end{equation}
This is due to the famous isomorphism between the coset space obtained by the group quotient of $G$ by the stabilizer $H_m$ and the orbit of the point $m$
\begin{equation}\label{eq:coadjointisomorphism}
    \mathsf{O}_m \cong G / H_m.
\end{equation}
This directly follows from the bijectivity of the orbit map
\begin{Align}\label{eq:orbitmap}
    \tau: G/H_m &\longrightarrow \mathcal{O}_m,\\ gH_m &\mapsto \tau(gH_m) = \mathrm{Ad}^*_{g} m,
\end{Align}
where $gH_m \in G/H_m$ denotes a coset.
The second key point which is at the core of the theory of coadjoint orbit, is that any coadjoint orbit is naturally a symplectic manifold \cite{Souriau:1970,kostant1970,Kirillov2004}. 
This is seen as follows, for any element of the algebra $X$, define the fundamental vector field on the coadjoint orbits by its action on a functional $f\in C^\infty(\mathcal{O}_m)$
\begin{equation}\label{eq:fundamentalvectorfieldorbits}
    (X^\# f)(m) = \dv{t}\qty[f\qty(\mathrm{Ad}^*_{\exp(t X)}m)]\eval_{t=0}.
\end{equation}
The symplectic structure is then given by the well-known Kostant--Kirillov--Souriau (KKS) two-form, which is defined at the point $m\in\mathfrak{g}^*$ as
\begin{equation}
    \Omega^m_{\mathrm{KKS}}(X^\#,Y^\#) \defeq \pairing{m}{\qty[X,Y]}.
\end{equation}

The fundamental vector field is closely related to the linear coordinate functions on the coalgebra. Once a basis is given, they are defined as follows
\begin{equation}\label{eq:coordinatefunctiondef}
    \chi^a: \mathfrak{g}^* \longrightarrow \mathbb{R},\quad \chi^a(m) = \pairing{m}{X^a}.
\end{equation}
It then follows that these are Hamiltonian functions for the fundamental vector field
\begin{equation}\label{eq:coordinatehamiltonaian}
    \iota_{\qty(X^a)^\#}\Omega_{\mathrm{KKS}} = \dd \chi^a,
\end{equation}
where $\dd$ and $\iota$ respectively denote the deRham differential and the interior product on the coalgebra. The above equation also implies that the coordinate functions realize the algebra through the Poisson bracket induced by the KKS form
\begin{equation}\label{eq:kkscoordinatefunction}
    \qty{\chi^a,\chi^b}_{\mathrm{KKS}}\defeq \iota_{(X^b)^\#}\iota_{\qty(X^a)^\#} \Omega_{\mathrm{KKS}} = \qty(X^b)^\#(\chi^a) = C^{ab}_c \chi^c.
\end{equation}
On the coadjoint orbits, the Poisson bracket induced by the symplectic form coincides with the Poisson bracket defined in equation \eqref{eq:poissonbracketcoalgebra}.
Finally, we note that we can also introduce a local version of the coadjoint orbits through a point $m\in\mathfrak{g}^*$ using the coadjoint action of the algebra
\begin{equation}
    \mathcal{O}_m = \qty{\coad{X}m \mid X\in \mathfrak{g}}.
\end{equation}
The stabilizer subalgebra is then defined as 
\begin{equation}
    \mathfrak{h}_m = \qty{X \in \mathfrak{g} : \coad{X}m=0},
\end{equation}
and the local version of the isomorphism \eqref{eq:coadjointisomorphism} can be written
\begin{equation}
    \mathcal{O}_m  \equiv \mathfrak{g}/\mathfrak{h}_m.
\end{equation}
\par
Since coadjoint orbits always admit a symplectic structure, they are natural candidates for classical phase spaces of a theories admitting $G$ as its group of symmetries.
On the other hand, we know that the quantum counterparts of those classical phase spaces are the Hilbert spaces of the representations. It seems therefore natural to ask: Is there a link
between the coadjoint orbits of a group and its unitary irreducible representations? This is the subject of Kirillov's orbit method.
that we now briefly introduce.\par
The orbit method is a geometric quantization procedure. As such, the idea is to construct a quantization map, that brings classical observables---seen as functions on the coadjoint orbit---to quantum operator acting on the Hilbert space of the representation.
One necessary condition on that quantization map is that it is an algebra isomorphism, i.e. that the commutator of quantum operators reproduces the Poisson bracket on the coadjoint orbit. For
our purposes it will be sufficient to consider the quantization of the coordinate functions \eqref{eq:coordinatefunctiondef}. We therefore want a map $\chi^a \rightarrow \hat{\chi}^a$ such that
\begin{equation}\label{eq:quantizationvalid}
  [\hat{\chi}^a,\hat{\chi}^b]
  = i C^{ab}_c \hat{\chi}^c,
\end{equation}
where the bracket denotes the operator commutator and where we have used equation \eqref{eq:kkscoordinatefunction}.
We begin by introducing a complex line bundle over the coadjoint orbit
\begin{equation}
    \pi : L \rightarrow \mathsf{O}_m .
\end{equation}
Locally, this amounts to attaching a copy of $\mathbb{C}$ to each point of the orbit. Sections of this bundle, $\psi \in \Gamma(\mathsf{O}_m,L)$, will serve as wavefunctions on which the operators act.
The next ingredient is a hermitian structure on the bundle. A hermitian structure is, for each point $m \in \mathfrak{g}^*$, a map
\begin{equation}
    h: \Gamma(\mathsf{O}_m,L) \times \Gamma(\mathsf{O}_m,L) \rightarrow \C.
\end{equation}
that is linear in the second argument and antilinear in the first.
Finally, we introduce a connection on $L$, i.e.\ a family of linear maps assigning to each vector field $X^\# \in \mathrm{T}\mathsf{O}_m$ an operator
\begin{equation}
    \nabla_{X^\#} : \Gamma(\mathsf{O}_m,L) \longrightarrow \Gamma(\mathsf{O}_m,L),
\end{equation}
satisfying the Leibniz rule
\begin{equation}
    \nabla_{X^\#}(f\psi)=X^\#(f)\,\psi+f\,\nabla_{X^\#}\psi,
    \quad \forall\, f\in C^\infty(\mathsf{O}_m).
\end{equation}
The connection and the hermitian structure must be compatible
\begin{equation}\label{eq:hermitianconnectioncompatibility}
    X^\# \qty[h(\psi_1,\psi_2)] = h\qty(\nabla_{X^\#} \psi_1,\psi_2) + h(\psi_1,\nabla_{X^\#} \psi_2),
\end{equation}
where, in the left hand side, the hermitian structure evaluated on two wavefunctions is interpreted as a function on the coalgebra. Note that if the vector field is such that
the left hand side vanishes, equation \eqref{eq:hermitianconnectioncompatibility} is then a Hermicity condition for the operator $i\nabla_{X^\#}$.\par
Let us now consider the quantization map
\begin{equation}\label{eq:quantizationmap}
    \chi^a \longrightarrow \hat{\chi}^a = \chi^a  - i \nabla_{(X^a)^\#}, 
\end{equation}
where the first term is understood as the multiplication by the coordinate function. 
A direct computation shows that this map is valid in the sense of \eqref{eq:quantizationvalid}
if the curvature of the connection is the KKS form $R_\nabla = i \Omega_{\mathrm{KKS}}$. This can only be the case if the KKS form satisfies the following integrality condition in terms of its
cohomology class
\begin{equation}\label{eq:orbitquantizationcondition}
    [\Omega]\in H^2(\mathsf{O},2\pi \mathbb{Z}).
\end{equation}
This is the origin of spin quantization in the case of $\mathrm{SU}(2)$. In order to complete the quantization procedure one needs to deal with the fact that the wavefunctions depend on every coordinate of the coadjoint orbit. In other words, in terms of local
Darboux coordinate, they depend on both the position and the momentum variables. The last step in quantizing the orbits is therefore to choose an
integrable subbundle $\mathcal{P}\subset T\mathsf{O}_m$ such that, for each
$x\in \mathsf{O}_m$, the fiber $\mathcal{P}_x\subset T_x\mathsf{O}_m$ is a
Lagrangian subspace (with respect to the KKS symplectic form). This choice---called a \textit{polarization}---is not independent of the connection on the line bundle.
Indeed, after constructing operators via the quantization map~\eqref{eq:quantizationmap}, one retains only those differential operators that preserve the chosen polarization.
In practice, this is implemented by choosing a symplectic potential for the KKS form that is adapted to the polarization. This will become clearer when we consider explicit examples.
\par
Although the orbit method provides the correct geometric intuition in many important situations,
it should be regarded in general as a guiding principle rather than a universal theorem. For real reductive groups,
the expected correspondence between irreducible unitary representations and suitably quantized coadjoint orbits is only partial and remains conjectural in full
generality. Moreover, there are known mismatches in both directions, such as the complementary series of $\mathrm{SL}(2,\mathbb{R})$, which is not naturally
captured by a coadjoint orbit, and examples due to Torasso showing that admissible symplectic homogeneous spaces need not always produce unitary representations
\cite{Torasso1983,Vogan1998,masonbrown2022arthursconjecturesorbitmethod}. By contrast, the method is rigorously successful for connected, simply connected nilpotent Lie groups, where Kirillov's theorem
identifies irreducible unitary representations with coadjoint orbits \cite{Kirillov_1962,Kirillov2004}. For compact connected Lie groups, geometric quantization of
integral coadjoint orbits also gives a robust construction of irreducible representations, closely related to the Borel--Weil--Bott theorem; in particular, for
$\mathrm{SU}(2)$ one recovers the usual spin-$j$ representations from the quantization of the two-sphere \cite{peter2009orbitmethodcompactconnected,ALEKSEEV1988391}.
In theoretical physics, the orbit method has proved especially
fruitful in the description of spin and internal degrees of freedom \cite{Souriau:1970}, in the symplectic description of relativistic
particles via coadjoint orbits of the Poincar\'e group \cite{Souriau:1970,ahlouche2021}, in the study of BMS particles in the context of asymptotic symmetries \cite{Barnich:2014kra,Barnich:2015uva,Barnich:2021dta}, and in two-dimensional conformal
and gravitational theories, where coadjoint orbits of loop groups give rise to the geometric actions of Wess--Zumino--Witten models,
while coadjoint orbits of the Virasoro group govern the geometric actions associated with Polyakov--Virasoro gravity
\cite{Alekseev:1988ce,Witten:1987ty,Alekseev:2018pbv}.

\subsection{Coherent states and moment maps}\label{subsec:momentmaps}
Having introduced coadjoint orbits, we now explain how they relate to projective representations and to the classical phase space.
We begin with the representation-theoretic aspect, introducing generalized Gilmore--Perelomov coherent states,
and subsequently consider the symplectic viewpoint via moment maps. We conclude by relating the two objects and providing the correspondence between quantum and classical observables.
\paragraph{Coherent states}
Generalized Gilmore--Perelomov coherent states for a general Lie group can be specifically constructed to give a embedding of the coadjoint orbits into
the projective Hilbert space. Their construction is based on the fundamental isomorphism of coadjoint orbit theory \eqref{eq:coadjointisomorphism}. Their construction goes as
follows. Given a group $G$ with unitary representations $U$ on a Hilbert space $\mathcal{H}$,
pick a normalized reference state $\ket{\psi_0}\in \mathcal{H}$ and define its isotropy
subgroup 
\begin{equation}\label{eq:isotropysubgroup}
    H = \qty{g\in G : U(g) \ket{\psi_0} = e^{i \theta_g} \ket{\psi_0}}.
\end{equation} 
That is, the isotropy subgroup is given by the element that only act via a phase on the reference state.
Consider the principle $H$ bundle over the coset space $G/H$
\begin{equation}
    G \xlongrightarrow{\pi_G} G/H,\quad \pi_G(g) = \qty[g],
\end{equation}
where the braces denote the equivalence class.
Given a section $\sigma: G/H \longrightarrow G$, i.e. a choice of representative, we define the displacement operator as
\begin{equation}\label{eq:displacementoperator}
    S_\sigma([g]) = U(\sigma(\qty[g])).
\end{equation}
The generalized coherent states are then defined as the displaced reference state
\begin{equation}\label{eq:coherentstatedef}
    \ket{\sigma(g)} = S_\sigma(\qty[g]) \ket{\psi_0}.
\end{equation}
The system of coherent states is thus parametrized by elements of the coset $x \in G/H$. In general, one can parametrize
an element of the coset using the generators of the algebra $\mathfrak{g}/\mathfrak{h}$ as
\begin{equation}\label{eq:parametrizationcoset}
    g_{x} = e^{x_i X^i}, \quad i=1,...,\mathrm{dim}(G)- \mathrm{dim}(H).
\end{equation}
$x^i$ are then coordinates on the homogeneous manifold of the coset space and we denote $\ket{\sigma(g_x)}= \ket{x}$.
One of the nice properties of these states is that, up to a phase, the group representation acts on them via the natural group action on the cosets
\begin{equation}\label{eq:gropactiononcoherentstates}
    U(g) \ket{x} = e^{i\alpha_{g,x}} \ket{g \rhd x}. 
\end{equation}
where $\rhd$ is defined as
\begin{equation}\label{eq:leftgroupactioncosets}
    \rhd: G \times G/H \longrightarrow G/H, \quad (g_1,\qty[g_2]) \mapsto g_1 \rhd \qty[g_2]  = \qty[g_1 g_2]
\end{equation}
One might worry that the definition of the coherent state \eqref{eq:coherentstatedef} depends on the choice of representative $\sigma$. However, a different choice of section only changes the coherent state by a phase factor, which in turn corresponds to a different choice of coordinates on the coset space. Going back to our discussion about projective representations, this means that a different
choice of representative leaves the physical state, i.e. the ray, invariant. The isotropy subgroup \eqref{eq:isotropysubgroup} is then a stabilizer for the ray in which the reference
state sits. If one now chooses the reference state such that its isotropy subgroup is the same as the stabilizer subgroup of the coadjoint action \eqref{eq:stabilizersubgroupcoadjoint}\footnote{This is not always possible since there might be no normalizable eigenstates of the generators of the stabilizer subgroup.},
the coset space is isomorphic to the orbit by equation \eqref{eq:coadjointisomorphism}. To be precise, the coset will be isomorphic to the particular orbit related to the representation being used. Since the orbit is characterized by a point $m \in \mathfrak{g}^*$, we denote the associated representation by $(U^{(m)},\mathcal{H}^{(m)})$.
Thus, the system of coherent states is labeled by
points on the associated coadjoint orbit, yielding an embedding of the coadjoint orbits in the projective Hilbert space
\begin{Align}\label{eq:embeddingprojective}
    \Phi^{(m)} : G/H_m &\longrightarrow \mathbb{P}\qty(\mathcal{H}^{(m)}),\\
     [g]&\longmapsto \Phi^{(m)}(\qty[g]) = \qty[S_\sigma^{(m)}(g) \ket{m,\psi_0}],
\end{Align}
where the label of the representation was added to the Hilbert space element as well.
Note that the right hand side is nothing but the equivalence class of coherent states as defined in \eqref{eq:coherentstatedef}.
where the brackets denote the ray, i.e.\ the equivalence class of states differing by an overall phase. In particular,
for the parametrization \eqref{eq:parametrizationcoset} we get 
\begin{equation}
    \Phi^{(m)}(\qty[g_x]) =\qty[\ket{m,x}].
\end{equation}
This map is known as a Kodaira embedding and does not depend on the choice of representative $\sigma$. 
\par
For an irreducible representation, the family of coherent state is always complete. In other words, they form a basis of the Hilbert space. Another important property
is the resolution of the identity. We call a reference state $\ket{m,\psi_0}$, admissible if there exists a measure $\dd \mu(x)$ on the coset space such that
\begin{equation}
    \int_{G/H_m} \dd \mu(x) \ket{m,x}\bra{m,x} = \mathds{1}.
\end{equation}
\par
Thanks to the embedding \eqref{eq:embeddingprojective}, the coordinates $x^i$ can be understood as coordinates on the coadjoint orbit. The quantum states $\ket{m,x}$ are the key to relating the representation-theoretic structure to classical observables. To this end, we define the
\textit{Berezin symbol} as the expectation value of algebra operators in coherent states
\begin{equation}\label{eq:berezinsymbols}
    l^{(m)}_{x} : \mathfrak{g} \longrightarrow \mathbb{R}, 
    \qquad 
    l^{(m)}_{x}\qty(X) \defeq
    \mel{m,x}{\dd U(X)}{m,x}.
\end{equation}
Given an element of the algebra $X\in \mathfrak{g}$, we also define its associated \textit{Berezin function} as
\begin{equation}\label{eq:berezinfunctiondef}
    l_{(m)}^X : \mathsf{O}_m \longrightarrow \mathbb{R}, \quad l_{(m)}^X(x) \defeq l^{(m)}_{x}\qty(X).
\end{equation}
As functions on the orbit, these can naturally be identified with the classical observables of the theory. In order to see precisely how this is the case, we start by introducing
the Berezin Poisson bracket
\begin{equation}\label{eq:berezinbracket}
    \{ l_{(m)}^X, l_{(m)}^Y \}_B
    \coloneqq 
    \frac{1}{i\hbar}
    l_{(m)}^{\qty[X,Y]}.
\end{equation}
For the basis of the algebra, the Berezin bracket reads
\begin{equation}
    \qty{l_{(m)}^{X^a},l_{(m)}^{X^b}}_B = C^{ab}_c l_{(m)}^{X^c},
\end{equation}
which reproduces the Lie--Poisson structure on the coadjoint orbit.
The Berezin functions equipped with the bracket \eqref{eq:berezinbracket},
are reminiscent of the coordinate functions \eqref{eq:coordinatefunctiondef} equipped with the KKS bracket.
In fact, in the case where the orbits are Kähler, it can be shown explicitly that the two objects correspond. In order to do so, consider the
Maurer-Cartan form $\Theta \in \Omega(G,\mathfrak{g})$. As a form on the group G, we can define its operator-valued pull-back on the coset space by the section $\sigma$ 
\begin{equation}
    \Theta_\sigma^{(m)} \defeq \qty(S^{(m)}_{\sigma})^{-1}\dd S^{(m)}_{\sigma},
\end{equation} 
One can now define the Berry connection
\begin{equation}
    A^{(m)} = \mel{m,\psi_0}{\Theta^{(m)}_\sigma}{m,\psi_0}.
\end{equation}
This is a $U(1)$ connection on the coset bundle, and the phase ambiguity in the choice of representative acts as a gauge transformation.
The Fubini-Study two-form is defined as the curvature of that connection\footnote{More precisely, \eqref{eq:FubiniStudyform} is the pull-back,
via the Kodaira embedding, of the Fubini--Study two-form on the projective Hilbert space to the coset space.
We will, however, slightly abuse terminology and refer to this pull-back simply as the Fubini--Study two-form.
}
\begin{equation}\label{eq:FubiniStudyform}
   \Omega^{(m)}_{\mathrm{FS}} = \dd A^{(m)} =- \mel{m,\psi_0}{\Theta^{(m)}_{\sigma}\wedge \Theta^{(m)}_{\sigma}}{m,\psi_0},
\end{equation}
where $\dd$ denotes the de-Rham differential on the orbit/coset space and where we have used the Maurer-Cartan equation $\dd \Theta + \Theta \wedge \Theta = 0$.
The fact that the Kodaira embedding is Kähler implies that the Fubini-Study 2-form pulls back to the KKS form on the orbits \cite{Onofri_75,FABerezin_1975,coherentstateandkahlermanifolds,,coherentstateandkahlermanifolds,RAWNSLEY199045,Bordemann:1993zv,Schlichenmaier_2010},
that is
\begin{equation}\label{eq:KKSequalFS}
    \tau^*\Omega^{(m)}_{\mathrm{KKS}} =  \Omega^{(m)}_{\mathrm{FS}},
\end{equation}
where we have used the orbit map \eqref{eq:orbitmap} to pull-back the KKS form to the coset space and where the $m$ subscript denotes that this is the KKS form for the orbit $\mathsf{O}_m$.
Note that although the pullback of the Maurer–Cartan form depends on the choice of representative $\sigma$, the KKS form does not, since the stabilizer subgroup acts only by a phase on the reference state.
This reflects the familiar fact that, although the connection depends on the choice of representative, the curvature does not.\par

Let us now contract the above with the fundamental vector field on the cosets. The former is defined as the pushforward of the fundamental vector field on the orbits by the inverse of the orbit map \eqref{eq:orbitmap}
\begin{equation}
    \tilde{X} \defeq \qty(\tau^{-1})_* X^\#.
\end{equation}
Using the action on the cosets \eqref{eq:leftgroupactioncosets}, one can write the action of $\tilde{X}$ on a function $\tilde{f}\in C^{\infty}(G/H)$
\begin{equation}
    (\tilde{X}\tilde{f})(\qty[g]) \defeq \dv{t}\tilde{f}(\exp(t X)\triangleright \qty[g])\eval_{t=0}.
\end{equation}
We use the definition of the displacement operator \eqref{eq:displacementoperator} in order to evaluate the (pulled-back) Maurer-Cartan form on this vector field (we drop the $m$ subscript for clarity)
\begin{Align}
    \Theta_\sigma(\tilde{X})([g]) &= S_\sigma^{-1}\qty([g]) \dv{t}S_\sigma(\qty[e^{tX} g])\eval_{t=0}\\
    &= S_\sigma^{-1}\qty([g_0]) \dd U (X)  S_\sigma\qty([g_0]).\\
\end{Align}
We can now calculate the contraction of the Symplectic form with the fundamental vector field
\begin{Align}
    \iota_{\tilde{X}}\tau^*\Omega_{\mathrm{KKS}}
     &= -\mel{\psi_0}{\iota_{\tilde{X}}\qty(\Theta_\sigma \wedge \Theta_\sigma)}{\psi_0}\\
    &= -\mel{\psi_0}{\qty[\iota_{\tilde{X}}\Theta_\sigma,\Theta_\sigma]}{\psi_0}\\
    &= -\mel{\psi_0}{S_\sigma^{-1}\dd U\qty(X) \dd S_\sigma + \dd S_\sigma^{-1} \dd U \qty(X) S_\sigma}{\psi_0}\\
    &= -\dd \mel{\psi_0}{S_\sigma^{-1}\dd U \qty(X)S_\sigma}{\psi_0}.
\end{Align}
Before concluding, we note that, using the parametrization \eqref{eq:parametrizationcoset}, the ray $\qty[\ket{\psi_0}]$ is associated to the point in the coadjoint orbit with coordinates
$x=0$. We denote this point by $m_0\in\mathfrak{g}^*$ and characterize the orbits and representations by that point.

Evaluating the above at the point $\qty[g_x]\in G/H$ gives
\begin{Align}
    \qty(\iota_{\tilde{X}}\tau^*\Omega_{\mathrm{KKS}})_{\qty[g_x]} &= \qty(\iota_{X^\#}\Omega_{\mathrm{KKS}})_{\mathrm{Ad}^*_{g_x}m_0},\\
    &= - \dd \mel{\psi_0}{\qty(S^{m_0}_\sigma)^{-1}([g_x])\dd U \qty(X)S^{m_0}_\sigma([g_x])}{\psi_0} \ ,\\
     &= - \dd \mel{m_0,x}{\dd U \qty(X)}{m_0,x}\\
    &= - \dd l^X_{(m_0)}(x)
\end{Align}
That is, the Berezin function is the hamiltonian function of the fundamental vector field.
By equation \eqref{eq:coordinatehamiltonaian} we conclude
\begin{equation}\label{eq:bereziniscoordinate}
    l^{X^a}_{m} = -\chi^a\eval_{\mathsf{O}_{m}},
\end{equation}
 Where the sign difference arises only due to different conventions on the two symplectic manifolds. It immediatly follows that the Berezin bracket and the KKS one coincide for the Berezin functions.
\begin{equation}\label{eq:bracketcoincides}
\qty{l^X,l^Y}_{B} = \qty{l^X,l^Y}_{\mathrm{KKS}}.
\end{equation}
\par
We conclude this part by briefly commenting on this result. If one extends the definition of the Berezin symbol to quadratic polynomials in the enveloping algebra $U[\mathfrak{g}]$
\begin{equation}
    l_{x}(XY) = \mel{x}{\dd U(X)\dd U(Y)}{x},
\end{equation}
The associated Berezin function then admits the asymptotic expansion \cite{Berezin1975,FABerezin_1975,RAWNSLEY199045,Bordemann:1993zv,Karabegov1996,Schlichenmaier:2000lmh,Charles2003} 
\begin{equation}\label{eq:Berezinexpansion}
  l^{XY} = l^X l^Y + \hbar \mathsf{h}^{\bar{a}b}\partial_{\bar{a}} l^X \partial_b l^Y + \mathcal{O}\qty(\hbar^2),
\end{equation}
where $\mathsf{h}^{\bar{a}b}$ are the components of the inverse Kähler metric\footnote{In Berezin-Toeplitz quantization, the role of $\hbar$
is usually played by the inverse of the representation parameter.
For example in the case of $\mathrm{SU}(2)$, $\hbar$ is identified with $j^{-1}$ where $\mathcal{C}_{\mathrm{SU}(2)} = j(j+1) \mathds{1}$.}.
The full asymptotic expansion yields the Berezin–Toeplitz star product, which furnishes a deformation quantization of the underlying Kähler manifold.
Equation \eqref{eq:bracketcoincides} then reflects the fact that, for Kähler coadjoint orbits, the Berezin-Toeplitz quantization of the Kähler manifold coincides with the geometric quantization of the coadjoint orbit.
\paragraph{Moment maps}
The previous paragraph described how coadjoint orbits relate to the quantum phase space.
We now turn to their relationship with the classical phase space.
The classical phase space is described by a symplectic manifold $\Gamma,\Omega_\Gamma$ on which the group acts
\begin{Align}
    \triangleright_\Gamma: G\times \Gamma &\longrightarrow \Gamma,\\
    \qty(g,\phi) &\longmapsto g \triangleright_\Gamma \phi.
\end{Align}
The space $\Gamma$ can be thought of as the field space of the CPS formalism, and $\Omega_\Gamma$ the extended symplectic 2-form \eqref{eq:extendedsymplecticform}.
Using the group action, can also introduce a notion of fundamental vector field on the classical phase space.
We define the fundamental vector field associated with the algebra element $X\in\mathfrak{g}$ through its action on a function $F\in C^{\infty}(\Gamma)$
\begin{equation}
    (\mathcal{X}F)(\phi) \defeq \dv{t}\qty[F(\exp(t X)\triangleright_{\Gamma}\phi) ]\eval_{t=0}.
\end{equation}
Let us now introduce a map
\begin{equation}\label{eq:momentmap}
    \mu: \Gamma \longrightarrow \mathfrak{g}^*.
\end{equation}
We call $\mu$ a moment map if its pairing with a vector is the Hamiltonian function for the fundamental vector field
\begin{equation}\label{eq:fundamentalhamiltonian}
    \If_{\mathcal{X}} \Omega_\Gamma = -\dbm  \pairing{\mu}{X},
\end{equation}
and it satisfies the equivariance condition
\begin{equation}\label{eq:momentmapequivariance}
    \mu(g\triangleright_\Gamma \phi) = \mathrm{Ad}^*_{g} \mu(\phi).
\end{equation}
The pairing between a vector and the moment map is also called the comoment map
\begin{equation}\label{eq:comomentmap}
    \tilde{\mu}: \mathfrak{g}\longrightarrow C^{\infty}\qty(\Gamma), \quad \tilde{\mu}\qty(X) \defeq \pairing{\mu}{X}.
\end{equation}
One can show that the equivariance condition implies that the comoment map is an algebra anti-homomorphism.
\begin{Align}\label{eq:comomentmapequivariance}
    \{\tilde{\mu}\qty(X),\tilde{\mu}\qty(Y)\}_\Gamma &= \dbm \tilde{\mu}(Y)\qty(\mathcal{X})\\
    &= \pairing{\dbm \mu(\mathcal{X})}{Y}\\
    &= \pairing{\mathrm{ad}^*_X \mu}{Y}\\
    &= -\tilde{\mu}\qty([X,Y]),
\end{Align}
where we used the infinitesimal version of the equivariance condition \eqref{eq:momentmapequivariance}
\begin{equation}
    \dbm \mu (\mathcal{X}) = \mathrm{ad}_X^* \mu.
\end{equation}
Note that this equation is nothing else than the statement that the push forward of the fundamental vector field on the phase space by the moment map gives the fundamental
vector field on the coadjoint orbits
\begin{equation}
    \mu_* \mathcal{X} = X^\#.
\end{equation}
With all these definitions and results, it is clear that the moment map and its associated comoment map relates the classical field space to the abstract coadjoint orbits. The equivariance of the comoment map
\eqref{eq:comomentmapequivariance} is exactly the statement that the Noether charges realize the algebra of diffeomorphism in the extended CPS formalism \eqref{eq:faitfulldiffrep}. The moment map \eqref{eq:momentmap} associates a specific field
configuration to a particular point in a particular coadjoint orbit. Furthermore, for a basis of the algebra we can use the coordinate functions \eqref{eq:coordinatefunctiondef} to write
\begin{equation}
    \tilde{\mu}(X^a) = \chi^a \circ \mu.
\end{equation}
In other words, the comoment map of the basis vectors give the components of the moment map in the coalgebra. On a given orbit, the Berezin functions coincide with
the coordinate functions \eqref{eq:bereziniscoordinate}. It would therefore be natural to identify the comoment map with the Berezin function evaluated at the point selected by
the moment map. This would then be a concrete
instance of the familiar principle that expectation values of quantum operators in coherent states reproduce the classical values of the corresponding observables. There is however one
more subtlety that we need to cover before making this association. In the extended phase space formalism, the classical algebra of Noether charges always represents the diffeomorphism
algebra without any central extensions. On the quantum side however, we know that the physical representations are the unitary irreducible representations of the centrally extended version
of the group. The missing link, is the notion of twisted coadjoint orbits.
\par
The (local) twisted coadjoint orbits are defined as
\vspace{-2pt}\noindent
\begin{equation}\label{eq:twistedcoadjointorbit}
    \mathcal{O}^c_m = \qty{\coad{X}m + c \sigma_X \mid X\in\mathfrak{g}}
\end{equation}
\vspace{-2pt}\noindent
where $\sigma_X \in \mathfrak{g}^*$ is determined by the Lie algebra $2$-cocycle $B \in Z^2(\mathfrak{g},\mathbb{R})$, through
\vspace{-2pt}\noindent
\begin{equation}
    \pairing{\sigma_X}{Y} = -B(X,Y), \qquad X,Y \in \mathfrak{g},
\end{equation}
\vspace{-2pt}\noindent
and where $c$ is the twisting parameter. In the following, we restrict our attention to the case where a single non-trivial cocycle exists, that is, when the second Lie algebra cohomology group
\vspace{-2pt}\noindent
\begin{equation}
    H^2(\mathfrak{g},\mathbb{R}) \coloneqq Z^2(\mathfrak{g},\mathbb{R}) \big/ B^{2}\qty(\mathfrak{g},\mathbb{R}),
\end{equation}
\vspace{-2pt}\noindent
with $B^{2}\qty(\mathfrak{g},\mathbb{R})$ denoting the space of 2-coboundaries—i.e., trivial cocycles—
is one-dimensional. The twisted coadjoint orbit through $m\in\mathfrak{g}^*$
is then isomorphic to the ordinary coadjoint orbits of the maximally centrally extended algebra at the point $(m,c)\in \tilde{\mathfrak{g}}^* = \mathfrak{g}^* \oplus \mathbb{R}$
\vspace{-2pt}\noindent
\begin{equation}\label{eq:twistedcoadjointorbitisomorphism}
    \mathcal{O}^c_m \cong \tilde{\mathcal{O}}_{(m,c)}.
\end{equation}
\vspace{-2pt}\noindent
The twisting parameter is then nothing else than the value of the central element on the coadjoint orbits of $\tilde{\mathfrak{g}}$. Next, we define the twisted moment map
\vspace{-2pt}\noindent
\begin{equation}
    \mu^c : \Gamma \longrightarrow \tilde{\mathfrak{g}}^*, \quad \mu^c(\phi) = \qty(\mu(\phi),c). 
\end{equation}
\vspace{-2pt}\noindent
The associated twisted comoment map for a vector $(X,a)\in \tilde{\mathfrak{g}}$ is given by
\vspace{-2pt}\noindent
\begin{equation}\label{eq:twistedcomap}
     \tilde{\mu}^c(X,a) \defeq \pairing{\mu^c}{(X,a)} = \mu\qty(X) + c a,
\end{equation}
\vspace{-2pt}\noindent
and satisfies the twisted equivariance condition
\vspace{-2pt}\noindent
\begin{equation}
    \qty{\tilde{\mu}^c\qty(X,a),\tilde{\mu}^c\qty(Y,b)}_\Gamma = \tilde{\mu}\qty([X,Y]) + c \,B(X,Y),
\end{equation}
\vspace{-2pt}\noindent
which is the standard equivariance condition on the extended algebra $\tilde{\mathfrak{g}}$.
The correspondence between twisted coadjoint orbits of an algebra and the standard coadjoint orbits of its maximal central extension is the classical analogue of the relationship between projective representations of the algebra and ordinary representations of its maximal central extension in the quantum case.
\par
This suggests the following procedure for the identification of classical and quantum observables. On the classical side, the moment map~\eqref{eq:momentmap} assigns to each field configuration in the classical phase space a point in $\mathfrak{g}^*$.
On the quantum side, evaluating Berezin symbols \eqref{eq:berezinsymbols} yields a comoment map for the centrally extended algebra $\tilde{\mathfrak{g}}$,
and thus determines a point on the corresponding $\tilde{\mathcal{O}}_{(m,c)}$ coadjoint orbit through equation \eqref{eq:bereziniscoordinate}. In the untwisting limit $c\to 0$,
it is clear from the definition of twisted coadjoint orbits \eqref{eq:twistedcoadjointorbit} and the isomorphism \eqref{eq:twistedcoadjointorbitisomorphism}
that this point gets projected to a point in the untwisted orbit $\mathcal{O}_m$, which can then be identified with the point selected by the classical moment map.
We will construct this correspondence explicitly in the case of the QCS in the next section.\par
\section{Quantum Corner Symmetries}\label{sec:quantumcornersymmetries}
Up to this point, the discussion in this chapter has been fully general. We now move to the explicit construction of these mathematical structures in the
case of the two-dimensional corner symmetry group. We begin in Subsection~\ref{subsec:centralextensionsandcasimirs} by explaining how the quantum corner symmetry group
arises from the structure of the ECS algebra, and we discuss the Casimir invariant of this new algebra.
Next, in Subsection~\ref{subsec:representations}, we study the representations of the quantum corner symmetry group constructed via Mackey's theory
of induced representations. Subsection~\ref{subsec:coadjointorbits} is devoted to the coadjoint orbits of the quantum corner symmetry group and to their relation to the representations via geometric quantization.
Finally, in Subsection~\ref{subsec:quantumclassicalcorr}, we construct the coherent states of the quantum corner symmetry group, as well as the moment maps
and the classical Casimir function
on phase space. We conclude by explaining how these structures fit together, and by defining the semiclassical limit in the context of the corner proposal.
\subsection{Central extensions and Casimirs}\label{subsec:centralextensionsandcasimirs}
As we have already mentioned, the physical representations of a symmetry group are described
by the maximally centrally extended version of that group.
The maximal central extension of a group includes contributions from both the first homotopy group and the second cohomology group. We start by adressing the former while the latter will be discussed at the end of the section. The second cohomology group of an algebra is given by the set of 2-cocycles modulo trivial ones. Let us denote the generators of the $\mathfrak{ecs}$ by $X_i, \, i=1,...,5$. We have 
\begin{equation}
    \qty[X_i,X_j] = C^{k}_{ij} X_k,
\end{equation}
where $C^{k}_{ij}$ are the structure constants of the algebra. The possible central extensions are then described by 2-cocycles $B: \mathfrak{ecs}\otimes \mathfrak{ecs} \longrightarrow \mathfrak{u}(1)$ valued in the abelian Lie algebra of the unitary group. They enter the extended commutation relations as
\begin{equation}
    \qty[X_i,X_j]_B = C^{k}_{ij} X_k + B(X_i,X_j). 
\end{equation}
In order to describe an algebra extension, the cocycle must obey two properties. First, the cocycle is antisymmetric, which comes from the antisymmetricity of the Lie bracket. Second, the Jacobi identity of the algebra imposes the following conditions on the cocycle
\begin{equation}\label{jacobicocycle}
    C^p_{jk} B_{ip} + C^p_{ki} B_{jp} + C^{p}_{ij} B_{kp} = 0,
\end{equation}
where $B_{ij} = B(X_i,X_j)$. A 2-cocycle is called trivial if it can be reabsorbed into the definition of the structure constants. That is, if there exists coefficients $\lambda_k \in \mathbb{R}$ such that
\begin{equation}\label{eq:trivialcocycle}
    B_{ij} = C^k_{ij}\lambda_k.
\end{equation}
Determining the second cohomology group of the algebra is therefore equivalent to finding all antisymmetric matrices $B_{ij}$
satisfying relation \eqref{jacobicocycle}, while retaining only the non-trivial ones. 
\par
We denote the basis of the the $\mathfrak{ecs}$ algebra by
$\left(P_\pm,L_0,L_\pm,\right)$, where the $L$ operators form the $\mathfrak{sl}\left(2,\mathbb{R}\right)$ subalgebra
\begin{equation}
    [L_0,L_\pm] = \pm L_\pm, \qquad [L_-,L_+] = 2L_0,
\end{equation}
and the $P$ operators generate the normal translations 
\begin{equation}
    [P_-,P_+] = 0.
\end{equation}
 The semidirect product structure implies the following cross commutation relations
\begin{equation}
    [L_0,P_\pm] = \pm \frac12 P_\pm, \quad [L_\pm,P_\mp] = \mp P_\pm.
\end{equation}
The algebra admits one cubic Casimir \cite{Ciambelli:2022cfr}
\begin{equation}\label{eq:ecscubiccasimir}
    \mathcal{C}_{\mathrm{ECS}} = \frac12(L_- P_+^2 + L_+ P_-^2- 2 L_0 P_- P_+).
\end{equation}

Choosing the basis ordering $\qty{P_-,P_+,L_0,L_-,L_+}$ and using the explicit $\mathfrak{ecs}$ structure constants, we find that relation \eqref{jacobicocycle} holds for
matrices of the form
\begin{equation}\label{centralextensionmatrix}
    B = \mqty(0&&B_{12}&&B_{13}&&0&&-2 B_{23}\\
            -B_{12}&&0&&B_{23}&&-2B_{13}&&0\\
            -B_{13}&&-B_{23}&&0&&B_{34}&&B_{35}\\
            0&&2 B_{13}&&-B_{34}&&0&&B_{45}\\
            2 B_{23}&&0&&-B_{35}&&-B_{45}&&0).
\end{equation}
Upon removing trivial cocycles via \eqref{eq:trivialcocycle}, the only remaining central extension is between the translations,
\begin{equation}\label{centralextension}
    \qty[P_-,P_+] = Z,
\end{equation}
where we have dropped the $B$ subscript of the commutator for simplicity. This extension transforms the translational part of the $\mathfrak{ecs}$ into a three dimensional Heisenberg algebra. The resulting algebra was first introduced in \cite{Ciambelli:2024qgi} and is called the \textit{quantum corner symmetry} ($\mathrm{QCS}$)
algebra \footnote{In the mathematical literature, this structure is known as the Jacobi algebra \cite{jacobiforms}}.\par In order to obtain the maximally centrally extended group, one further needs to consider the universal covering of the special linear part. The physical representations of the $\mathrm{ECS}$ group are therefore equivalent to the irreducible unitary representations of
\begin{equation}\label{universalcoverqcs}
    \mathrm{QCS} = \widetilde{\spl{2}}\ltimes H_3,
\end{equation}
where the tilde denotes the universal cover, and $H_3$ is the three-dimensional Heisenberg group.\par Let us remark on these central extensions.
The extension arising from the cocycles is reminiscent of the Galilean mass operator, appearing as the central element extending the
commutator between translations and Galilean boosts. In that case, the extension is necessary to address massive non-relativistic
particles in both the quantum and classical frameworks. In the present case, however, the discussion in Subsection \ref{sec:extendedphasespaceandambiguities} shows that, within the extended phase-space formalism, the covariant phase space always realizes the corner symmetries without any non-trivial cocycles.
The appearance of the central element \eqref{centralextension} is therefore a purely quantum effect, akin to a quantum anomaly.
The universal cover is also an important feature. As mentioned earlier, passing to the universal cover $\widetilde{\mathrm{SL}}(2,\mathbb{R})$
is analogous to replacing the Lorentz group by $\mathrm{SL}(2,\mathbb{C})$ in quantum field theory, thereby allowing for half-integer spin representations,
i.e. fermions. Keeping with the corner proposal's analogy with
the Poincar\'e group in particle physics, representations of the corner symmetry group built from the universal cover of the special linear factor
(rather than from the standard $\spl{2}$)
may correspond to a class of spacetimes that appears only at the quantum gravity level. This possibility will be discussed further in Chapter~\ref{chapter4}.
\par
We now discuss the algebra $\mathfrak{qcs}$. One of the first questions one can ask about a Lie algebra is: what are its
Casimir operators? The central element is, of course, a trivial Casimir. The $\mathrm{ECS}$
Casimir however, no longer commutes with the translations. One can therefore ask if there exists a modification of the cubic Casimir \eqref{eq:ecscubiccasimir}.
that commutes with every generator of the $\mathrm{QCS}$. The easiest way to tackle that question is at the level of coadjoint orbits.
We start by computing the adjoint action of the group in the explicit basis $\qty{P_-,P_+,L_0,L_-,L_+,C}$. This can be done by expanding the
adjoint action of the group
\begin{equation}
    \mathrm{Ad}_{\exp(t X)} = \sum_{n=0}^\infty \frac{t^n}{n!}(\mathrm{ad}_X)^n.
\end{equation} 
For example, we can compute
\begin{Align}
    \mathrm{Ad}_{exp(t L_0)} L_+ &=  L_+ + t \qty[L_0,L_+] + \frac{t^2}{2}\qty[L_0,\qty[L_0,L_+]] + ...\\
    &= L_+\qty(1 + t + \frac{t^2}{2}+ ...)\\
    &= e^{t} L_+.
\end{Align}
Similar computations for every generator of the QCS produces Table \ref{table:adjointaction}. Next, we can consider the coadjoint action on the linear coordinate function \eqref{eq:coordinatefunctiondef}
\begin{equation}
    \mathrm{Ad}^*_{\exp (tX)} \, \chi^a (m) = \chi^a \qty(\mathrm{Ad}^*_{\exp(tX)}m), \quad \forall m\in \mathfrak{g}^*.
\end{equation}
Iterating over the full basis yields Table~\ref{table:coadjointaction}, where the lowercase symbols
denote the linear coordinate functions associated with the Lie-algebra elements labelled by the corresponding uppercase symbols.
The coadjoint orbits are characterized by functions $f\in C^\infty(\mathsf{O}_m)$ that are constants on the orbit. These are called the Casimir functions.
We can write a system of partial differential equations for these functions by noting that being constant on the orbit means they are annihilated by the fundamental
vector fields \eqref{eq:fundamentalvectorfieldorbits}
\begin{equation}
    (X^a)^\#(f) = A^{ab} \pdv{f}{x^b} = 0, \quad a = 1,...,5,
\end{equation}
where we omitted the fundamental vector field associated with the central element as its action is trivial.
Running through the entire basis and using the results in Table \ref{table:coadjointaction}, we find the following system
\begin{align}
    c \pdv{f}{p_+} + \frac{p_-}{2}\pdv{f}{l_0} + p_+ \pdv{f}{l_+} &= 0,\\
    c \pdv{f}{p_-} + \frac{p_+}{2}\pdv{f}{l_0} + p_- \pdv{f}{l_-} &= 0,\\
    \frac{p_-}{2} \pdv{f}{p_-} - \frac{p_+}{2}\pdv{f}{p_+} + l_- \pdv{f}{l_-} - l_+ \pdv{f}{l_+} &= 0,\\
    p_- \pdv{f}{p_+} + l_- \pdv{f}{l_0} + 2 l_0 \pdv{f}{l_+} &= 0,\\
    p_+ \pdv{f}{p_-} + l_+ \pdv{f}{l_0} + 2l_0 \pdv{f}{l_-} &= 0.
\end{align}
Which corresponds to the matrix
\begin{equation}
    \qty(A^{ab}) = \mqty(0&c&\frac{p_-}{2}&0&p_+\\c&0&\frac{p_+}{2}&p_-&0\\\frac{p_-}{2}&-\frac{p_+}{2}&0&l_-&-l_+\\0&p_-&l_-&0&2l_0\\p_+ &0&l_+&2l_0&0).
\end{equation}
This matrix is of rank $4$, which means that there exists one Casimir function in addition to the obvious central one. One can now reduce this matrix into its echelon form
and solve the system using the method of characteristics to find
\begin{equation}
    f = c(l_- l_+-l_0^2)- \frac12(p_+^2 l_- - p_-^2 l_+ + 2 p_- p_+ l_0).
\end{equation}
In the first term we recognize the Casimir function of the $\mathfrak{sl}(2,\R)^*$ coalgebra multiplied by the central element. The second term is the Casimir function of the
$\mathfrak{ecs}^*$ coalgebra. We therefore write
\begin{equation}\label{eq:QCSCasimirfunction}
    c_{\mathrm{QCS}} =  c\, c_{\mathrm{SL}\qty(2,\R)} + c_{\mathrm{ECS}}.
\end{equation}
This remarkable expression for the Casimir function provides a first hint of the rich structure of the QCS, whose consequences will reappear in the following sections.
Note that, in the limit where $c\to 0$, the Casimir function becomes the one of the non-centrally extended version of the group. This is closely related to the quantum-classical correspondence
discussed at the end of Section \ref{sec:mathematicalbackground} and will be discussed further in Subsection \ref{subsec:quantumclassicalcorr}.\par
To write the corresponding Casimir operator at the algebra level, one must account for the non-commuting nature of the generators.
This is done by writing the completely symmetrized version of \eqref{eq:QCSCasimirfunction}, and only then replacing the coordinates by their associated generators.
This procedure breaks the clean split of the function at the orbit level and we find
\begin{equation}\label{eq:qcscubiccasimir}
    \mathcal{C}_{\mathrm{QCS}} = Z\qty(L_- L_+-L_0\qty(L_0 + \frac32)) + \frac12(2 L_0 P_- P_+-L_+ P_-^2-L_- P_+^2).
\end{equation}

\begin{table}[h]
\centering
\caption{The adjoint action $\mathrm{Ad}_{\exp X}\,Y$}
\label{table:adjointaction}
\resizebox{\textwidth}{!}{
\renewcommand{\arraystretch}{1.4}
\begin{tabular}{c|cccccc}
\diagbox[width=3.2em,height=2\line]{$Y$}{$X$} & $tP_-$ & $tP_+$ & $tL_0$ & $tL_-$ & $tL_+$ & $tZ$ \\
\hline
$P_-$ & $P_-$ & $P_- - tC$ & $e^{-t/2}P_-$
      & $P_-$ & $P_- - tP_+$ & $P_-$ \\
$P_+$ & $P_+ + tC$ & $P_+$ & $e^{t/2}P_+$
      & $P_+ + tP_-$ & $P_+$ & $P_+$ \\
$L_0$ & $L_0 + \tfrac{t}{2}P_-$ & $L_0 - \tfrac{t}{2}P_+$ & $L_0$
      & $L_0 + tL_-$ & $L_0 - tL_+$ & $L_0$ \\
$L_-$ & $L_-$ & $L_- - tP_- + \tfrac{t^2}{2}C$ & $e^{-t}L_-$
      & $L_-$ & $L_- - 2tL_0 + t^2 L_+$ & $L_-$ \\
$L_+$ & $L_+ + tP_+ + \tfrac{t^2}{2}C$ & $L_+$ & $e^{t}L_+$
      & $L_+ + 2tL_0 + t^2 L_-$ & $L_+$ & $L_+$ \\
$C$   & $C$ & $C$ & $C$ & $C$ & $C$ & $C$
\end{tabular}
}
\end{table}

\begin{table}[h]
\centering
\caption{The coadjoint action $\mathrm{Ad}^*_{\exp (X)}\,y$}
\label{table:coadjointaction}
\resizebox{\textwidth}{!}{
\renewcommand{\arraystretch}{1.4}
\begin{tabular}{c|cccccc}
\diagbox[width=3.2em,height=2\line]{$y$}{$X$} & $tP_-$ & $tP_+$ & $tL_0$ & $tL_-$ & $tL_+$ & $tZ$ \\
\hline
$p_-$ & $p_-$ & $p_- + tc$ & $e^{t/2}p_-$
      & $p_-$ & $tp_+ + p_-$ & $p_-$ \\
$p_+$ & $p_+ - tc$ & $p_+$ & $e^{-t/2}p_+$
      & $p_+ - tp_-$ & $p_+$ & $p_+$ \\
$l_0$ & $l_0 - \tfrac{t}{2}p_-$ & $l_0 + \tfrac{t}{2}p_+$ & $l_0$
      & $l_0 - tl_-$ & $l_0 + tl_+$ & $l_0$ \\
$l_-$ & $l_-$ & $l_- + tp_- + \tfrac{t^2}{2}c$ & $e^{t}l_-$
      & $l_-$ & $2tl_0 + t^2 l_+ + l_-$ & $l_-$ \\
$l_+$ & $l_+ - tp_+ + \tfrac{t^2}{2}c$ & $l_+$ & $e^{-t}l_+$
      & $-2tl_0 + l_+ + t^2 l_-$ & $l_+$ & $l_+$ \\
$c$   & $c$ & $c$ & $c$ & $c$ & $c$ & $c$
\end{tabular}
}
\end{table}
\subsection{Representations}\label{subsec:representations}
In section \ref{sec:mathematicalbackground}, we argued that physical quantum systems are described by unitary irreducible projective representations of the symmetry group.
We further mentioned that these are equivalent ot the standard unitary irreducible representations of the maximally centrally extended version of that group. In the case of the
two-dimensional corner symmetry groups, the quantum systems are therefore described by the representations of the QCS group introduced earlier. This section is devoted to the construction of these representations, following
the formalism described in \S\ref{subsec:inducedrepresentation}.
We refer the mathematically inclined readers to \cite{jacobiforms,Berndt1998ElementsOT,Low:2012qyz,Low:2019oon}. For notational clarity, operators in a certain
representation will, from now on, be denoted by the operator itself, with the indices defining the representation in subscript.
\par
We start with the representations of the normal subgroup $H_3$. In an irreducible representation, Schur's lemma \cite{Schur1973NeueBD} imposes that the central element $C$ acts as a multiple of the identity.
Once the coefficient $c\in\mathbb{R}$ is fixed, Stone-Von Neumann's theorem \cite{vN1} ensures that there exists a unique unitary representation
\begin{equation}
    \Xi^{(c)} : H_3 \longrightarrow \mathrm{End}(\mathcal{F}),
\end{equation}
where $\mathcal{F}$ can be taken to be the Fock space of the quantum harmonic oscillator. The representation of the algebra can then be written
\begin{equation}\label{dXiP}
    P_-^{(c)} = \sqrt{c} \,a, \quad P_+^{(c)} = \sqrt{c}\,\da, \quad C^{(c)} = c,
\end{equation}
where $a.\da$ are the usual creation and annihilation operator
\begin{equation}
    \qty[a,\da] = 1.
\end{equation}
In \eqref{dXiP} (and in what follows) we assumed $c>0$. Although the negative case can be treated in a similar fashion, it is important to note that it gives rise
to a non-equivalent set of representations. The Weil representation of the special linear algebra
\begin{equation}
    \mathring{D}': \mathfrak{sl}\qty(2,\mathbb{R}) \longrightarrow \mathrm{End}(\mathcal{F}),
\end{equation}
expresses the generators of $\mathfrak{sl}\qty(2,\mathbb{R})$ in the universal algebra of the Heisenberg sub-algebra
\begin{equation}\label{metaplecticrepresentation}
    \mathring{L}_0 = \frac12\qty(\da a + \frac12), \quad \mathring{L}_- = \frac12 a a, \quad \mathring{L}_+ = \frac12 \da \da.
\end{equation}
By exponentiation, \eqref{metaplecticrepresentation} becomes a global
representation of $\widetilde{\spl{2}}$ that obeys condition \eqref{automorphismrepresentation}\footnote{More precisely, the Weil representation is a
representation of the metaplectic group $\mathrm{Mp}\qty(2,\mathbb{R})$, which is the double cover of $\spl{2}$. The universal cover of the metaplectic group is also
$\widetilde{\spl{2}}$ and the representation satisfying condition \eqref{automorphismrepresentation} is $\mathring{\tilde{D}} = \mathring{D}\circ \pi$, where 
$\pi: \widetilde{\spl{2}}\longrightarrow \mathrm{Mp}\qty(2,\mathbb{R})$ is the projection map. Since we will only write the explicit representation at the algebra level,
this detail does not matter here.}. Therefore, Mackey's theory ensures (c.f. \eqref{eq217}) that the complete set of irreducible unitary representations of the
$\widetilde{\mathrm{QCS}}$ are of the form
\begin{equation}\label{qcsrepresentations}
    U^{(\lambda,c)}: \widetilde{\mathrm{QCS}}\longrightarrow  \mathrm{End}(\mathcal{H}^{(\lambda)}\otimes \mathcal{F}),\quad U^{(\lambda,c)}_{(h,t)} = D^{(\lambda)}(h)\otimes \mathring{D}(h) \,\Xi^{(c)}(t),
\end{equation}
where $D^{(\lambda)}: \widetilde{\spl{2}}\longrightarrow \mathrm{End}(\mathcal{H}^{(\lambda)})$ are irreducible unitary representations of the universal cover of the special
linear group.
Following \cite{Kitaev:2017hnr}, the latter are completely defined through the value of the group element corresponding to the maximal compact subgroup
$e^{2\pi i L_0}$ which takes value $e^{2\pi i \epsilon}$ where $\epsilon \in \mathbb{R}$, and the parameter $\lambda$ that is related to the Casimir operator
through $\mathcal{C}_{\spl{2}}^{(\lambda)} \ket{\psi} = \lambda(1-\lambda)\ket{\psi}$ for any state $\ket{\psi}$.
The representation is then written on eigenstates of $L_0$ as
\begin{Align}
    L_0 \ket{\eta} &= \eta \ket{\eta},\\
    L_- \ket{\eta} &= \sqrt{(\eta-\lambda)(\eta+\lambda-1)}\ket{\eta-1},\\
    L_+ \ket{\eta} &= \sqrt{(\eta+\lambda)(\eta+1-\lambda)}\ket{\eta+1}.
\end{Align}
We will discuss the allowed values of $\eta$ and $\lambda$ in more detail when we turn to unitarity.\par

Using equation \eqref{qcsrepresentations}, the above representations, together with the Weil representation (\cref{dXiP,metaplecticrepresentation}),
describe the complete set of irreducible representations
of the $\widetilde{\mathrm{QCS}}$. They are labeled by three parameters $(\lambda,c,\epsilon)$ where $\lambda$ and $c$ are related to the two Casimirs, and $\epsilon$ is related to the parity of the representation.
We dropped the subscript $\epsilon$ from the representation labels for notational clarity.
To understand the role of the cubic Casimir \eqref{eq:qcscubiccasimir}, let us write the explicit algebra representation.
For a vector $X = L + P \in \mathfrak{qcs}$, it is simply given by
\begin{equation}
    U'^{(\lambda,c)}(X) = D'^{\qty(\lambda)}\qty(L) \otimes \mathring{D}'(L) \,\Xi'^{(c)}(P).
\end{equation}
Let us denote the elements of $\mathcal{H}^{(\lambda)}\otimes \mathcal{F}$ by $\ket{n} \otimes \ket{k}\equiv \ket{n,k}$.
We then have 
\begin{align}
  P_-^{(\lambda,c)} \ket{\eta,k} &= \sqrt{c}\sqrt{k} \ket{\eta,k-1},\\
    P_+^{(\lambda,c)} \ket{\eta,k} &= \sqrt{c}\sqrt{k+1}\ket{\eta,k+1},\\
    L_0^{(\lambda,c)} \ket{\eta,k} &= \qty(  \frac12 k + \frac14+\eta)\ket{\eta,k},\label{qcsrepl0}\\
    L_-^{(\lambda,c)} \ket{\eta,k} &=  \sqrt{(\eta-\lambda)(\eta+\lambda-1)}\ket{\eta-1,k} +\frac12 \sqrt{k(k-1)}\ket{\eta,k-2},\label{qcsreplm}\\
    L_+^{(\lambda,c)} \ket{\eta,k} &= \sqrt{(\eta+\lambda)(\eta+1-\lambda)} \ket{\eta+1,k} +\frac12 \sqrt{(k+1)(k+2)}\ket{\eta,k+2}\label{qcsreplp}.
\end{align}
It then simply follows that the action of the cubic Casimir gives
\begin{equation}\label{eq:casimiraction}
   \mathcal{C}^{(\lambda,c)}_{\mathrm{QCS}} \ket{\eta,k} =  \qty(c \lambda(1-\lambda) - \frac{3c}{16})\ket{\eta,k}.
\end{equation}
Remarkably, up to a additive constant that could be reabsorbed in the definition of the Casimir, the value of the cubic $\mathrm{QCS}$ Casimir in the irreducible representations is given by the value of the corresponding $\spl{2}$ Casimir, multiplied by the scalar value of the central element of the Heisenberg algebra. We note that, choosing the trivial representation of $\widetilde{\spl{2}}$, we recover the representation used in \cite{Ciambelli:2024qgi}. In the present conventions, this corresponds to a vanishing value of the cubic Casimir.\\
Let us now comment on unitarity. The complex structure compatible with the commutation relations (and the Hermiticity of the Casimir operators) is
\begin{equation}\label{complexstructure}
    L_0^\dag = L_0, \quad L_\pm^\dag=L_\mp,\quad P_\pm^\dag = P_\mp^\dag.
\end{equation}
Because of the tensor product structure of the representations \eqref{qcsrepresentations}, the unitarity conditions on the $\widetilde{\spl{2}}$ side and on the Weil
side are independent. Starting with the later, the unitarity of the representation for all $c\in \mathbb{R}\backslash \qty{0}$ is assured by Stone-von Neumann's theorem.
For the special linear part, the hermiticity of $L_0$ implies that $\eta\in\mathbb{R}$. The vectors $\ket{\eta}$ can always be taken to be orthonormal, as they
are eigenvectors of an Hermitian operator. Thus, the remaining conditions can be written
\begin{Align}
    \mel{\eta}{L_-L_+}{\eta} &= (\eta+\lambda)(\eta+1-\lambda) \geq 0,\\
    \mel{\eta}{L_+L_-}{\eta} &=  (\eta-\lambda)(\eta-1+\lambda)\geq 0,
\end{Align}
for all $\eta$ in a given representation.
These conditions are satisfied by the following classes of irreducible representations
\begin{itemize}
    \item \textbf{The trivial representation}
    \begin{equation*}
        \lambda = \epsilon = 0,\qquad \eta=0.
    \end{equation*}
    \item \textbf{The continuous series representations}
    \begin{equation*}
        \lambda > \abs{\epsilon}(1-\abs{\epsilon}), \,\text{for}\,\, \abs{\epsilon}\leq \frac12, \qquad \eta=\epsilon + \mathbb{Z}.
    \end{equation*}
    \item \textbf{The positive and negative discrete series representations}
    \begin{Align*}
        \lambda > 0, \, \epsilon=\lambda,\qquad \eta &= \lambda + \mathbb{N},\\
        \lambda >0,\, \epsilon=-\lambda, \qquad \eta &= -(\lambda + \mathbb{N}).
    \end{Align*}
\end{itemize}
The positive discrete series for $\lambda\in \frac{1}{n}\mathbb{Z}_{>0}$ are representations of $n$-fold covers of $\spl{2}$. In particular, $\lambda \in \frac12 \mathbb{Z}_{>0}$
are representations of the metaplectic group $Mp(2,\R)$ and $\lambda \in \mathbb{Z}_{>0}$ are genuine representations of $\spl{2}$.\par
We conclude this discussion of the representations by commenting on their rather distinctive structure. We stress that, for a semidirect-product group, it is highly non-generic for the representations to be realized on the tensor product of the constituent representations. For example, the irreducible unitary representations of the Poincar\'e group are not obtained
simply as tensor products of Lorentz representations with translation representations. This unique characteristic of the QCS is related to the particular form of its cubic Casimir function \eqref{eq:QCSCasimirfunction} discussed earlier.
While the reason behind this factorization of the Hilbert space is not made clear by the theory of induced representations---beyond the mathematical fact that there exists
a global little group---the geometric quantization of the coadjoint orbits offers a much more transparent perspective.
\subsection{Coadjoint orbits}\label{subsec:qcscoadjointorbits}
As explained in section \ref{sec:mathematicalbackground}, the coadjoint orbits are of crucial importance in the corner proposal, as they serve as the bridge between the quantum
and classical descriptions of corner symmetries. Beyond this physical motivation, the coadjoint orbits analysis offers an understanding of the peculiar form of the Casimir function \eqref{eq:QCSCasimirfunction} and of the
factorization of the Hilbert space of representations \eqref{qcsrepresentations}. Finally, since Kirillov's orbit method is not a general
theorem but rather a case-by-case prescription, the geometric quantization of these coadjoint orbits provides, a fully worked-out test of the method
for a new class of groups, whose consistency in this setting was, to the best of our knowledge, never established.\par
To describe the coadjoint orbits of the QCS, one would typically compute the KKS form explicitly on $\mathfrak{q}^*$ and then invert its restriction to the
coadjoint orbits---i.e.\ to the symplectic leaves defined by fixing the value of the Casimir function \eqref{eq:QCSCasimirfunction}---in order to obtain the Poisson brackets on those leaves. This was carried out for the ECS in \cite{Ciambelli:2022cfr}.
While instructive, the analysis is essentially local and therefore not very illuminating as regards the global topology of the orbits.
Keeping up with the surprisingly elegant structure of the QCS, there is, however, an easier way. To make this simplification manifest---and for later convenience---we
introduce the ``covariant''
QCS algebra basis $J\updown{a}{b},\,P^{a},\,Z$, with $a,b=0,1$, $J\updown{0}{0} = - J\updown{1}{1}$, and commutation relations
\begin{equation}\label{eq:qcsalgebra}
    [P^a, P^b] = \epsilon^{ab} Z, 
    \quad [P^c, J\updown{a}{b}] = \delta^c_b P^a - \tfrac{1}{2}\,\delta^a_b P^c, 
    \quad [J\updown{a}{b}, J\updown{c}{d}] = \delta^c_b J\updown{a}{d} - \delta^a_d J\updown{c}{b}.
\end{equation}
We further introduce the associated coordinate functions $\chi\updown{a}{b},\chi^a,\chi^z$ on $\mathfrak{qcs}^*$.
On a coadjoint orbit, their bracket induced by the KKS two form reproduces the algebra\footnote{Note that we dropped the $\mathrm{KKS}$ or $P$ subscript on the bracket since
they coincide on a coadjoint orbit.}
\begin{equation}
   \qty{\chi^a,\chi^b} = \epsilon^{ab} \chi^z,\quad \qty{\chi\updown{a}{b},\chi\updown{c}{d}} = \delta^c_b \chi\updown{a}{d}- \delta^a_d \chi\updown{c}{b},\quad \qty{\chi^c,\chi\updown{a}{b}} = \delta^c_b \chi^a -\frac12 \delta^a_b \chi^c.
\end{equation}
Note that in those coordinates, the Casimir function \eqref{eq:QCSCasimirfunction} takes the nice form
\begin{Align}
c_\mathrm{QCS} &= \frac{\chi^z}{2} \chi\updown{a}{b} \chi\updown{b}{a} + \frac12 \epsilon_{ac}\chi^c \chi\updown{a}{b} \chi^b,\\
&= \chi^z c_{\spl{2}} + c_{\mathrm{ECS}},
\end{Align}
where we defined the Casimir functions of the special linear group and the ECS
\begin{equation}\label{eq:casimirfunctionsl2recs}
   c_{\spl{2}} = \frac12 \chi\updown{a}{b}\chi\updown{b}{a},\quad c_{\mathrm{ECS}} = \frac12 \epsilon_{ac}\chi^c \chi\updown{a}{b}\chi^b.
\end{equation}
Now comes the crucial point: consider the following change of coordinates
\begin{equation}\label{eq:tildecoordinates}
    \tilde{\chi}\updown{a}{b}  = \chi\updown{a}{b} - \frac{1}{2\chi^z}\epsilon_{bc}\chi^c\chi^a.
\end{equation}
Note that, while the quadratic term would require some care to interpret at the algebra level,
we are of course free to perform arbitrary non-linear changes of coordinates on the coalgebra manifold. Writing the Poisson brackets in these coordinates is illuminating
\begin{equation}
    \qty{\tilde{\chi}\updown{a}{b},\tilde{\chi}\updown{c}{d}} = \delta^c_b \tilde{\chi}\updown{a}{d} - \delta^a_d \tilde{\chi}\updown{c}{b}, \quad \qty{\tilde{\chi}\updown{a}{b},\chi^c} = 0.
\end{equation}
We see that the new coordinates close an $\spl{2}$ algebra, but one that is completely decoupled from the Heisenberg sector. In these coordinates, the Casimir function reads
\begin{equation}
    c_{\mathrm{QCS}} = \frac{\chi^z}{2}\tilde{\chi}\updown{a}{b}\tilde{\chi}\updown{b}{a}.
\end{equation}
In other words, the QCS Casimir is precisely the Casimir of this decoupled $\spl{2}$ algebra, multiplied by the central element. This explains why, in equation \eqref{eq:casimiraction}, the QCS Casimir acts in this particular way.
Furthermore, this immediately implies that the coadjoint orbits factorize as the product of an $\spl{2}$ coadjoint orbit with a Heisenberg coadjoint orbit.
\begin{equation}\label{eq:qcsorbitfactorization}
    \mathsf{O}^{(\lambda,c)}_{\mathrm{QCS}} = \mathsf{O}^{(\lambda)}_{\spl{2}} \times \mathsf{O}^{(c)}_{H_3}.
\end{equation}
Thus, characterizing and quantizing the QCS coadjoint orbits can be done by performing the analysis independently for the special linear and Heisenberg orbits,
which greatly simplifies the issue. In order to give a complete example of the orbit method, we construct below the coadjoint orbits of the Heisenberg group and their
geometric quantization. Next, we give the coadjoint orbits of $\spl{2}$ and the geometric quantization of the hyperboloid orbits without any proof as to not overburden the presentation.
For further details on these aspects, we refer to the author's work \cite{Neri:2025fsh} (see also \cite{Witten:1987ty,hanh2003deformationquantizationquantumcoadjoint}).
\paragraph{Heisenberg coadjoint orbits}
The Heisenberg group $H_3$ can be realized as the set of $3\times3$ upper triangular matrices
\begin{equation}
\label{Eq.defrepHeisenberg}
    g(a,b,c) = \mqty(1&a&c\\0&1&b\\0&0&1),
\end{equation}
with the standard matrix multiplication as the group product.
The Heisenberg algebra $\mathfrak{h}_3$ is then isomorphic to the algebra of strictly upper triangular matrices with the matrix commutator as the Lie bracket. A convenient basis is
\begin{equation}
    X = \mqty(0&1&0\\0&0&0\\0&0&0), \quad P = \mqty(0&0&0\\0&0&1\\0&0&0), \quad Z=\mqty(0&0&1\\0&0&0\\0&0&0),
\end{equation}
with commutation relations
\begin{equation}
    [X,P]=Z.
\end{equation}
We now want to describe the elements of $\mathfrak{h}_3^*$ using the standard bilinear form on matrices as the pairing between vectors and covectors
\begin{equation}
    \pairing{M^*}{M} = \Tr[M^* M], \quad M\in\h,\,M^* \in \h^*.
\end{equation}
Using the above form, we observe that the orthogonal space to the algebra $\h_3^\perp$ is isomorphic to the upper triangular matrices. The dual algebra is therefore isomorphic to strictly lower triangular matrices
\begin{equation}\label{Eq.dualalgebraisomorphismwithmatrices}
    \h_3^* \cong \mathrm{M}_{3\times3}(\mathbb{R})/ \h_3^\perp,
\end{equation}
and is spanned by
\begin{equation}
   X^* = \mqty(0&0&0\\1&0&0\\0&0&0), \quad P^*=\mqty(0&0&0\\0&0&0\\0&1&0), \quad Z^* = \mqty(0&0&0\\0&0&0\\1&0&0).
\end{equation}
Equation \eqref{Eq.dualalgebraisomorphismwithmatrices} defines an equivalence relation where two matrices are equivalent
if they have the same strictly lower triangular part. Any matrix $M \in \mathrm{M}_{3\times3}\qty(\mathbb{R})$ can then be identified to an element
of the dual algebra by considering its equivalence class $[M]\in\h_3^*$. The coadjoint action of a group element is then simply given by the matrix product
\begin{equation}
   \mathrm{Ad}^*_{g} [M] =\qty[g M g^{-1}].
\end{equation}
A straightforward calculation gives
\begin{equation}
   \mathrm{Ad}^*_{g(a,b,c)}\qty[\mqty(0&0&0\\
   x&0&0\\z&p&0)] = \qty[\mqty(0&0&0\\x+b z&0&0\\z&p - a z&0)].
\end{equation}
The fact that $z$ is constant under the coadjoint action is in direct correspondence with the fact that $Z$ is a Casimir of the Heisenberg algebra.
At the level of the dual algebra, a coordinate function associated with a Casimir element is a Casimir function. Since the orbits are characterized by a fixed value of $z$,
points $m=(0,0,z)$ in the dual algebra with different values of $z$ belong to separate orbits. For each one, we can take such a point as the representative.
The above equation tells us that, given a value of $z\neq 0$\footnote{The $z=0$ orbits are simply points and will not be of interest to us.},
the orbits of the Heisenberg group are two-dimensional planes
\begin{equation}\label{Eq.heisenbergorbits}
     \mathsf{O}^{(z)}_{H_3}\equiv \mathsf{O}_{H_3}^{(0,0,z)} = \qty{(x,p,z)\mid x,p\in \mathbb{R}}.
\end{equation}
As expected, we have a foliation of the dual algebra in terms of coadjoint orbits
\begin{equation}
    \h_3^* \cong \mathbb{R}^3 = \bigcup_{z\in\mathbb{R}} \mathsf{O}_{H_3}^{(z)}.
\end{equation}

Using the coadjoint action, we can compute the fundamental vector fields \eqref{eq:fundamentalvectorfieldorbits} in the basis of $T\mathcal{O}_{(z)}^{H_3}$ associated with the $(x,p)$ coordinates on the orbit
\begin{equation}\label{eq:fundamentalvectorfieldsheisenberg}
    X^\# = -z \partial_{p}, \quad P^\# = z \partial_{x}, \quad Z^\# = 0.
\end{equation}
The only non-vanishing component of the KKS symplectic form at a point $m=(x,p,z)$ is given by
\begin{equation}
\Omega^{H_3}_{m}\qty(X^\#,P^\#) =\pairing{m}{[X,P]}=\pairing{(x,p,z)}{Z} = z,
\end{equation}
which implies
\begin{equation}
\label{Eq.H3KKS}
    \Omega^{H_3} = \frac{1}{z}\dd x\wedge \dd p.
\end{equation}
Since the orbits are contractible---they are isomorphic to $\R^2$---the KKS form is not only closed but exact. Indeed we can write $\Omega^{H_3} = - \dd \theta^{H_3}$ with $\theta^{H_3}=\frac{1}{z}p\dd x$.\par
Let us now apply the geometric quantization procedure we reviewed in Subsection \ref{subsec:coadjointorbits} to the Heisenberg orbits. We consider the trivial
line bundle over the coadjoint orbit with central element value $z$. Sections of this bundle are simply complex valued functions on the plane
\begin{equation}
    \phi: \mathcal{O}^{(z)}_{H_3}\cong \mathbb{R}^2 \longrightarrow \mathbb{C}.
\end{equation}
Since the bundle is trivial, we define the trivial point-wise hermitian structure
\begin{equation}\label{heisenberghermitian}
    \braket{\phi_1(x,p)}{\phi_2(x,p)}\defeq \overline{\phi_1(x,p)}\phi_2(x,p).
\end{equation}
We define the following connection on the trivial line bundle
\begin{equation}
    \nabla_{X^\#} \defeq  X^\# + \iota_{X^\#} \theta^{H_3}.
\end{equation} 
It is easy to check that the curvature of that connection is indeed the KKS form $\Omega^{H_3}$ up to a conventional sign and $i$ factor.
The quantization map \eqref{eq:quantizationmap} for the coordinate functions is then
\begin{equation}\label{Eq.quantizationmap}
    \chi^a \mapsto \hat{\chi}^a = \chi^a -i \qty(X^a)^{\#} - \iota_{\qty(X^a)^\#}\theta^{H_3}.
\end{equation}
Using the fundamental vector fields \eqref{eq:fundamentalvectorfieldsheisenberg} and the above equation, we can write the quantization of the coordinate functions explicitly
\begin{equation}
\label{Eq.H3quantization}
    \hat{\chi}^{X} = i z \,\partial_{p} + x,\quad
    \hat{\chi}^{P} = -i z \,\partial_{x},\quad
    \hat{\chi}^{Z} = z.
\end{equation}
We now need to choose a polarization that is preserved by the observables defined above. It is easy to see that for the simple choice $\partial_{p} \phi = 0$, we get
\begin{equation}
    \partial_{p} \qty(\hat{\chi}_{X}\phi) = \partial_{p} \qty(\hat{\chi}_{P}
    \phi) = 0,
\end{equation}
consistently with the fact that $\theta^{H_3}(\partial_p)=0$.
Polarized wavefunctions are therefore complex valued function of $x\in\R$, on which the observables act as 
\begin{equation}
\label{Eq.momentumoperatorheisenberg}
     \hat{\chi}^{X} \phi(x) = x \phi(x),\quad \hat{\chi}^{P}\phi(x) = -i z \partial_{x} \phi(x),\quad \hat{\chi}^{Z} \phi(x)= z \phi(x),
\end{equation}
which is the well known action of the Heisenberg group in quantum mechanics.

To construct the Hilbert space, we need to introduce a scalar product. Since we chose a real polarization, we need to use the \textit{half-form} construction \cite{bates_weinstein_quantization,Blau,carosso2018geometricquantization}.
On a given orbit, we denote by $Q\subset \mathcal{O}^{(z)}_{H_3}$ the submanifold corresponding to the chosen polarization
$\mathcal P = \mathrm{Span}\qty(\partial_{p})$ (an involutive distribution is integrable by Frobenius' theorem, $Q$ is the space of polarization leaves)
and construct its complexified cotangent bundle $\mathrm{T}^*Q^{\mathbb{C}}$. Sections of this bundle are complexified forms $\varpi$ on $\mathcal O_{H_3}^{(z)}$ that are
by construction orthogonal to the polarization, that is $\iota_\xi \varpi=0$, $\forall\xi\in\mathcal P$.
More concretely, in $(x,p)$ coordinates, the choice of $\mathcal P=\text{Span}(\partial_p)$ leads to $\mathrm{T}^*Q^{\mathbb{C}} = \qty{\lambda\, \dd x\mid \lambda \in C^{\infty}\qty(Q,\mathbb{C})}$.
We then introduce the square-root bundle $\delta_{\mathcal P} \longmapsto T^*Q$, whose sections are ``half-forms" in the sense that $\delta_{\mathcal P}^{\otimes 2}\cong T^*Q^{\mathbb{C}}$.
The wave functions are then defined as products of the form $\tilde{\phi} = \phi \nu$, where $\phi$ is a complex line bundle section and $\nu \in \Gamma(\delta_{\mathcal P})$.
In the simple case where $Q$ is one-dimensional, a polarized~\footnote{The covariant derivative on products $\tilde\phi=\phi\nu$ is given by the Leibniz rule.
Given the covariant derivative on $T^*Q^\C$, $\nabla_\xi\varpi=I_\xi\dd\varpi$, one defines the covariant derivative on $\delta_\mathcal P$ such that
\begin{equation*}
    2\nu\otimes\nabla_\xi\nu=\nabla_\xi(\nu\otimes\nu)=I_\xi \dd(\nu\otimes\nu).
\end{equation*}
} wave function $\tilde\phi$ can be written as $\phi_{1/2}(x)\sqrt{\dd x}$, with a $\phi_{1/2}$ that transforms as a density of weight $1/2$:
\begin{equation}
    \phi_{1/2}(x)\to\pqty{\dv{x'}{x}}^{1/2}\phi_{1/2}(x').
\end{equation}

The hermitian structure \eqref{heisenberghermitian} for polarized wave functions now becomes
\begin{equation}
    \braket{\tilde{\phi}_1(x)}{\tilde{\phi}_2(x)} = \overline{\phi_1(x)} \phi_2(x) \bar{\nu}_1 \nu_2,
\end{equation}
and is a volume form on the polarized submanifold $Q$. It can therefore naturally be integrated to obtain the scalar product. Modulo possible irrelevant functions that can be reabsorbed in the definition of the wave functions, we obtain the standard $L^2\qty(\R)$ scalar product
\begin{equation}
    \braket{\tilde{\phi}_1}{\tilde{\phi}_2} = \int_\R \dd x\, \overline{\phi_1(x)}\phi_2(x), 
\end{equation}
for which it is easy to check that the quantized momentum map are Hermitian operators.

Finally, we define the following change of coordinates
\begin{equation}
    \hat\chi^{A_\pm}=\frac{1}{\sqrt{2}}\pqty{\hat\chi^{X}\pm i\hat\chi^{P}}.
\end{equation}
Assuming $z>0$, we can use these to build creation and annihilation operators by~\footnote{For $z<0$, one just switches the roles of $\hat\chi^{A_-}$ and $\hat\chi^{A_+}$.}
\begin{equation}
\label{Eq.aidentification}
    \hat\chi^{A_-}=\sqrt{z}a,\qquad \hat\chi^{A_+}=\sqrt{z}a^\dagger,
\end{equation}
and check that $[a,a^\dagger]=\frac{1}{z}\hat\chi^{Z}=1$ indeed. We now introduce the Fock basis $\{\ket{k}\}$, constructed from the $\hat\chi^{X}$ (position) eigenstate basis using Hermite polynomials~\footnote{The $k$-th Hermite polynomial is
\begin{equation*}
    H_k(x) = (-1)^k e^{x^2} \frac{\dd^k}{\dd x^k} e^{-x^2}.
\end{equation*}}
\begin{equation}
\label{Eq.h3fockbasis}
    \ket{k} = \frac{1}{\pi^{1/4} \sqrt{2^k k!}}\int_{-\infty}^\infty  \dd x \, H_k(x E) \, e^{-\frac{x^2 E}{2}} \ket{x}, 
\end{equation}
where we denoted $\braket{x}{\phi} \equiv \phi(x)$. On these states, we obtain the expected action of $a$ and $a^\dagger$:
\begin{align}
    a\ket{k}=\sqrt{k}\ket{k-1},\qquad
    a^\dagger\ket{n}=\sqrt{k+1}\ket{k+1}.
\end{align}

\paragraph{$\spl{2}$ coadjoint orbits}
The coadjoint orbits of $\spl{2}$ are well known. To describe them, we work in the ``conformal'' basis $(D,H,K)$ of $\mathfrak{sl}(2,\mathbb{R})$, with commutation relations
\begin{equation}
    \qty[D,H] = H,\quad \qty[D,K] = -K,\quad \qty[K,H] = 2D\,.
\end{equation}
This basis will also play an important role later on. We introduce coordinates $(d,h,k)$ on $\mathfrak{sl}(2,\mathbb{R})^*$ by writing an arbitrary element of the coalgebra as
\begin{equation}
    m = d\,D^* + h\,H^* + k\,K^*,
\end{equation}
where $(D^*,H^*,K^*)$ denotes the dual basis. The Casimir function is then written as $c_{\spl{2}} = d^2- h k$ and
the coadjoint orbits are classified as follows
\begin{itemize}
    \item \textbf{Elliptic} $\mathsf{O}^{(\lambda)}_{\spl{2}}$ with $c_{\spl{2}} = \lambda^2 >0$
    \item \textbf{Null cone} $\mathsf{O}^{(0,\pm)}_{\spl{2}}$ with $c_{\spl{2}} =0$. These further split into the upper and lower cone and the origin
        \begin{enumerate}
        \item $\mathsf{O}^{0,+}_{\spl{2}}= \qty{(d,k,h) : d^2 = h k, \quad h+k > 0}$
        \item $\mathsf{O}^{0,-}_{\spl{2}}= \qty{(d,k,h) : d^2 = h k, \quad h+k < 0}$
        \item $\mathsf{O}^{0}_{\spl{2}}= (0,0,0)$
        \end{enumerate}
    \item \textbf{Hyperbolic} $\mathsf{O}^{\lambda,\pm}_{\spl{2}}$ with $c_{\spl{2}} = -\lambda^2 <0$. These further split into the upper and lower sheet hyperboloid.
    \begin{enumerate}
    \item $\mathsf{O}^{\lambda,+}_{\spl{2}} = \qty{(d,k,h): d^2 - h k = -\lambda^2, \,\, h+k>0}$
    \item $\mathsf{O}^{\lambda,-}_{\spl{2}} = \qty{(d,k,h): d^2 - h k = -\lambda^2,\,\, h+k <0}$
    \end{enumerate}
\end{itemize}
For the hyperbolic orbits,
the quantization map for the coordinate functions can be constructed in a similar way to what was done in the case of the Heisenberg orbits.
For the ladder operator, one then finds
 \begin{Align}\label{eq:ladderbasiscoordinatefunction}
    \hat{\chi}^{L_0} \ket{\eta} &= \eta \ket{\eta},\\
   \hat{\chi}^{L_-} \ket{\eta} &= \sqrt{(\eta-\lambda)(\eta+\lambda-1)}\ket{\eta-1},\\
   \hat{\chi}^{L_+} \ket{\eta} &= \sqrt{(\eta+\lambda)(\eta+1-\lambda)}\ket{\eta+1}.
\end{Align}
where $\eta \in \lambda + \mathbb{N}$ for $\mathsf{O}^{\lambda,+}_{\spl{2}}$ and $\eta \in -(\lambda + \mathbb{N})$ for $\mathsf{O}^{\lambda,-}_{\spl{2}}$. Note that the Casimir operator
acts as $\mathcal{C}_{\spl{2}} = \lambda(1-\lambda) \mathds{1}$. The classical Casimir function is therefore shifted by $\lambda$ in the quantum case. This is the standard Harish-Chandra
$\rho$-shift \cite{harishchandra}.
\paragraph{QCS coadjoint orbits}
We are now ready to describe the geometric quantization of the QCS coadjoint orbits. We will only describe the discrete series as they are the ones that will be used later on. Because of the orbit factorization \eqref{eq:qcsorbitfactorization}, the Hilbert space of the geometric
quantization will be the tensor product of the individual Hilbert spaces associated with each orbit. This corresponds to the description of the representations in Subsection \ref{subsec:representations}.
More precisely, if $\mathcal H^{(\lambda)}$ is the Hilbert space
arising from the geometric quantization of the orbit $\mathsf O_{\spl{2}}^{\lambda,\pm}$ (for the $\tilde \chi^a{}_b$ coordinates) and $\Fock^{(z)}$ is the Hilbert (Fock) space we constructed
out of the orbit $\mathsf O_{H_3}^{(z)}$, the total Hilbert space is simply the product
\begin{equation}\label{eq:qcshilbertspace}
    \mathcal H^{\lambda,z}=\mathcal H^{(\lambda)}\otimes \Fock^{(z)}.
\end{equation}
In the following, we are going to assume
$z>0$. A similar construction can be made for $z<0$.

In the factorized coordinates \eqref{eq:tildecoordinates}, the action of the QCS operators on the Hilbert space \eqref{eq:qcshilbertspace} can be trivially written
\begin{equation}
    \QCSop\updown{a}{b} = \hat{\tilde{\chi}}\updown{a}{b} \otimes \mathds{1}, \quad \QCSop^a = \mathds{1} \otimes \hat{\chi}^a, \quad \QCSop^z = \mathds{1}\otimes \hat{\chi}^z,
\end{equation}
where we denoted the QCS operators with a bold symbol to differentiate them from the ones acting on only one side of the tensor product. Let us now work in
a specific representation such that $\chi^z = c$. Inverting equation \eqref{eq:tildecoordinates},
the first operator can also be written in the original basis
\begin{equation}
    \QCSop\updown{a}{b} = \hat{\chi}\updown{a}{b} \otimes\mathds{1} + \mathds{1}\otimes \frac{1}{2 c} \epsilon_{bc}\chi^c \chi^a.
\end{equation}
Working in the ladder basis defined in equations \eqref{eq:ladderbasiscoordinatefunction} and \eqref{Eq.aidentification}, we find

\begin{Align}
    \QCSop^{L_0} &= \hat{\chi}^{L_0} \otimes \mathds{1} + \mathds{1}\otimes \frac12\qty(\da a + \frac12),\\
    \QCSop^{L_-} &= \hat{\chi}^{L_-}\otimes \mathds{1} + \mathds{1}\otimes \frac12 a a,\\
    \QCSop^{L_+} &= \hat{\chi}^{L_+}\otimes \mathds{1} + \mathds{1}\otimes \frac12 \da \da.
\end{Align}
In complete agreement with the representations obtained from Mackey's theory \eqref{qcsrepl0}, \eqref{qcsreplm} and \eqref{qcsreplp}.\par
This concludes our description of the coadjoint orbits of the QCS. The peculiar structure of the QCS, which we alluded to earlier, can now be understood as a consequence
of the existence of coordinates on the dual algebra in which the special linear and Heisenberg parts completely decouple. The resulting orbit structure therefore explains the form of the Hilbert space \eqref{eq:qcshilbertspace} and the action of the Casimir operator \eqref{eq:casimiraction}.
Moreover, we will use this decoupling again shortly to discuss the correspondence between quantum and classical operators.

\subsection{Coherent states, moment maps, and the classical limit}\label{subsec:quantumclassicalcorr}
We finally have all the tools necessary to describe quantum and classical QCS dynamical systems, as well as the correspondence between quantum and classical observables. In the spirit of
the corner proposal, we start with the quantum analysis of the coadjoint orbits by constructing the coherent states and the Berezin functions. We then discuss some facts about
a general QCS classical system and its associated moment maps. We conclude by the explicit construction of the quantum-classical correspondence mentioned at the end of Subsection \ref{subsec:momentmaps}.\par
\paragraph{QCS coherent states}
For concretness, we construct the coherent states on the
positive discrete series and comment on the negative series at the end. For readability, we work in a fixed representation $\mathcal{H}^{(\lambda)}_{\spl{2}}\otimes \Fock_c$
and denote the elements by $\ket{n,k} = \ket{n}\otimes \ket{k}, n,k\in\mathbb{N}$. The quantum operators will simply be denoted by the corresponding algebra element. With this notation, we have
\begin{Align}
     P_- \ket{n,k} &= \sqrt{ ck} \ket{n,k-1},\\
    P_+ \ket{n,k} &= \sqrt{c (k+1)} \ket{n,k+1},\\
  L_0 \ket{n,k} &= \qty(\lambda + n + \frac{k}{2} + \frac14)\ket{n,k},\\
    L_- \ket{n,k} &= \sqrt{n (n+2\lambda -1)}\ket{n-1,k} + \frac12\sqrt{k(k-1)}\ket{n,k-2},\\
    L_+ \ket{n,k} &= \sqrt{(n+1) (2 \lambda +n)}\ket{n+1,k}+\frac12\sqrt{(k+1)(k+2)}\ket{n,k+2}.
\end{Align}
The generalized Perelomov coherent states for the QCS discrete series were first introduced in \cite{Ciambelli:2025ztm}.
They are constructed on the reference state $\Omega = \ket{0,0}$ by first identifying the isotropy subgroup \eqref{eq:isotropysubgroup}. From the action of the generators on the
states, it is clear that the stabilizer algebra is generated by $(L_0,C)$. Since $L_0$ is the compact generator inside $\spl{2}$, we get
\begin{equation}
    H = U(1) \times Z(\mathrm{H}_3).
\end{equation}
where $Z(\mathrm{H}_3) \cong \mathbb{R}$ is the center of the Heisenberg group. Accordingly, the cosets are given by
\begin{equation}\label{eq:cosetspaceqcs}
    QCS/H \cong \spl{2}/U(1) \times \mathrm{H}_3/Z(\mathrm{H}_3) = \mathbb{D} \times \mathbb{R}^2.
\end{equation}
Let us parametrize the elements of the element of the homogeneous space by
\begin{equation}
    g_{\zeta,\alpha} =e^{\frac{1}{\sqrt{c}}(\alpha P_+ -\bar{\alpha}P_-)}  e^{c_\zeta L_+ - \bar{c}_\zeta L_-}\defeq \mathcal{D}(\alpha)\mathcal{S}(\zeta),
\end{equation}
where, denoting $\zeta = re^{i\theta}$
\begin{equation}
    c_\zeta =\tanh^{-1}(r)e^{-i\theta}.
\end{equation}
This particular parametrization is a choice of coordinates $\zeta \in \mathbb{D},\alpha \in \mathbb{C}$ on the cosets.
Any other parametrization would only alter the phase of the coherent state.
The QCS coherent state are then defined as
\begin{equation}\label{eq:qcscoherentstate}
    \ket{\zeta,\alpha} \defeq \mathcal{D}(\alpha)\mathcal{S}(\zeta)\ket{\Omega}.
\end{equation}
Recall that the $\spl{2}$ generators act on the Heisenberg side via the Weil representation.
It follows that one can write
\begin{equation}
    \ket{\zeta,\alpha} = \ket{\zeta}\otimes \ket{\alpha_\zeta},
\end{equation}
where $\ket{\zeta}$ is a $\spl{2}$ Perelomov coherent state, and $\ket{\alpha_\zeta}$ is a squeezed Glauber coherent state with
squeezing modulus $\tanh^{-1}(r)$ and squeezing angle $-(\theta + \pi)$. The disk coherent states $\ket{\zeta}$ were first introduced by Perelomov in \cite{Perelomov:1971bd}
(see also
\cite{APerelomov_1977} for a more general discussion). They can be written in the $\ket{n}$ basis as
\begin{equation}\label{eq:perelomovstateinnbasis}
    \ket{\zeta} = (1-\abs{\zeta}^2)^{\lambda} \sum_{n=0}^\infty \left(\frac{\Gamma(n+2\lambda)}{\Gamma(n+1)\Gamma(2\lambda)}\right)^\frac12 \zeta^n \ket{n}.
\end{equation}
These states are normalized but not othogonal
\begin{equation}\label{eq:zetaoverlap}
    \braket{\zeta}{\zeta'} = \frac{\qty(1- \abs{\zeta}^2)^\lambda \qty(1-\abs{\zeta'}^2)^\lambda }{(1-\bar{\zeta}\zeta')^{2\lambda}}.
\end{equation}
For any representation defined by $\lambda>\frac12$, they provide a resolution of the identity
\begin{equation}\label{eq:perelomovsl2rresolutionidentity}
   \frac{2\lambda-1}{\pi} \int_{\mathcal{D}} \dd\mu_(\zeta) \ket{\zeta}\bra{\zeta} = \mathds{1},
\end{equation}
where $\dd\mu(\zeta)$ is the $\mathrm{SU}(1,1)$ invariant measure on the unit disk
\begin{equation}
    \dd\mu(\zeta) = \frac{\dd^2 \zeta}{\qty(1-\zeta \bar{\zeta})^2}.
\end{equation}
The family of coherent states $\qty{\ket{\zeta} : \zeta \in \mathcal{D}}$ thus forms an overcomplete family for the Hilbert space $\mathcal{H}^{\lambda,+}_{\spl{2}}$. 
Note that, while the coherent states \eqref{eq:perelomovstateinnbasis} can be defined for any $\lambda>0$, the resolution of the identity \eqref{eq:perelomovsl2rresolutionidentity}
holds only for $\lambda>\tfrac12$. This leads to an interesting asymmetry in the quantum--classical correspondence. Since the representations remain perfectly well
defined for
$\lambda<\tfrac12$, the Berezin symbols \eqref{eq:berezinsymbols} are also well defined, and expectation values of quantum operators still produce classical functions.
However, in the absence of a resolution of the identity, the Berezin quantization map---which associates to a classical observable $A\in C^\infty(\mathbb{D})$ the operator
\begin{equation}
    \hat{A}^{(\lambda)} = \frac{2\lambda -1}{\pi}\int \dd \mu(\zeta)\, A(\zeta)\ket{\zeta}\bra{\zeta},
\end{equation}
with $\lambda$ playing the role of $\hbar^{-1}$---is not well defined. In particular, for $\lambda<\tfrac12$ one can no longer interpret
the QCS representations as arising from the Berezin quantization of the corresponding coadjoint orbits. This will have important physical consequences in Chapter \ref{chapter4}.\par
Glauber coherent states are the famous coherent states of the quantum oscillator. They can be expressed in the Fock basis as
\begin{equation}
    \ket{\alpha} = e^{-\frac{\abs{\alpha}^2}{2}}\sum_{k=0}^\infty \frac{\alpha^n}{\sqrt{n!}}\ket{n}, \quad \alpha \in \C.
\end{equation}
They are also normalized but not orthogonal
\begin{equation}
    \braket{\alpha}{\beta} = \exp(-\frac{\abs{\alpha-\beta}^2}{2} + i \Im(\bar{\beta}\alpha)),
\end{equation}
and also give a resolution of the identity in $\Fock_{c}$
\begin{equation}
    \frac{1}{\pi}\int \dd^2 \alpha \, \ket{\alpha}\bra{\alpha} = \mathds{1}.
\end{equation}
Because the squeezing operator $\mathcal{S}(\zeta)$ is unitary, this resolution of the identity stays valid for squeezed states
\begin{equation}
        \frac{1}{\pi}\int \dd^2 \alpha \, \ket{\alpha_\zeta}\bra{\alpha_\zeta} = \mathds{1}.
\end{equation}
\par

\par
To see the isomorphism between the coset space \eqref{eq:cosetspaceqcs} and the coadjoint orbit $\mathsf{O}^{(\lambda,+)}_{\spl{2}} \times \mathsf{O}^{(c)}_{H_3}$,
we introduce the isomorphism between the hyperboloid and the unit disk. In the conformal coordinates $(d,k,h)$ introduced earlier we can write
\footnote{The lower hyperboloïd orbit $\mathsf{O}^{\lambda,-}_{\spl{2}}$ is simply obtained by the transformation $(h,k)\mapsto (-h,-k)$.}
\begin{Align}
    \varphi: \mathbb{D}  &\longrightarrow \mathsf{O}^{(\lambda,+)}_{\spl{2}}\\
    \zeta &\longmapsto \varphi(\zeta) = \qty(-\lambda \frac{2 \Im(\zeta)}{1-\abs{\zeta}^2},\lambda \frac{\abs{1+\zeta}^2}{1-\abs{\zeta}^2},\lambda \frac{\abs{1-\zeta}^2}{1-\abs{\zeta}^2},).
\end{Align}
The isomorphism between the $\mathsf{O}_{H_3}^{(c)}$ and the complex plane is simply given by $\qty(\Re(\alpha),\Im(\alpha))\in \R^2$.\par
In which sense are the QCS coherent states ``classical"? To simplify the discussion, we will consider only the $\spl{2}$ part and relegate the complete treatment to chapter \ref{chapter4}.
In the context of the positive discrete series representation, the role of the classical phase space is played by the orbit $\mathsf{O}_{\spl{2}}^{(\lambda,+)}$. As mentioned earlier, in
the context of Berezin quantization, the ``semiclassical parameter" is the inverse of the representation parameter $\lambda^{-1}$. This can be seen from the overlap \eqref{eq:zetaoverlap} as follows.
The modulus of the overlap can be written as
\begin{equation}
    \abs{\braket{\zeta}{\zeta'}} = \cosh^{-2\lambda}\qty(\frac{d(\zeta,\zeta')}{2}),
\end{equation}
where 
\begin{equation}
    d(\zeta,\zeta') = 2 \tanh^{-1}\qty(\abs{\frac{\zeta-\zeta'}{1-\zeta \bar{\zeta}'}}),
\end{equation} 
is the geodesic distance on the Poincaré disk. Expanding the modulus for small values of this distance gives
\begin{equation}
    \abs{\braket{\zeta}{\zeta'}} = e^{-\frac{\lambda d^2}{4}\qty(1 + \frac{\lambda d^4}{96} + ....)}.
\end{equation}
We see that for distances $d^2\gg \frac{1}{\lambda}$, the overlap gets exponentially suppressed in the large $\lambda$ limit.
The dominant contribution comes from distances $d^2 \sim \frac{1}{\lambda}$ and we can write
\begin{equation}
    \braket{\zeta}{\zeta'} \approx e^{-\frac{\lambda}{4}d(\zeta,\zeta')^2}e^{-2i\lambda \mathrm{arg}\qty(1-\bar{\zeta}\zeta')}.
\end{equation}  
The first factor is a Gaussian heat kernel on the Poincaré disk with width $\sim\lambda^{-\frac12}$ \cite{McKean1970AnUB,davies1988}. The second factor is a phase that vanishes when $\zeta$ and $\zeta'$ are close enough that $1- \bar{\zeta}\zeta'$ is approximately real and positive.
Therefore, in the large $\lambda$ limit, the coherent state $\ket{\delta}$ approaches a delta function peaked on its classical value $\zeta \in \mathbb{D}$.
Another way to see this classical behavior in the large $\lambda$ limit which might be more intuitive to physicists is from the fluctuations.
By construction, the expectation values of the conformal basis 
in a $\spl{2}$ coherent states are given by the associated coordinates in the unit disk
\begin{Align}\label{eq:expectationvaluedkhinzeta}
    \expval{D}_\zeta &=-\lambda \frac{2 \Im(\zeta)}{1-\abs{\zeta}^2},\\
    \expval{K}_\zeta &= \lambda \frac{\abs{1+\zeta}^2}{1-\abs{\zeta}^2},\\
    \expval{H}_\zeta &=  \lambda \frac{\abs{1-\zeta}^2}{1-\abs{\zeta}^2}.
\end{Align}
Let us look at the operator $H$. We can compute its fluctuation
\begin{equation}
    \Delta H^2 = \expval{H^2}_\zeta - \expval{H}_\zeta^2 = \frac{\lambda}{2} \frac{\abs{1-\zeta}^4}{(1-\abs{\zeta}^2)^2}.
\end{equation}
The standard deviation is then defined as the square root of the fluctuation $\Delta H = \sqrt{\Delta H^2}$, and the relative deviation is given by
\begin{equation}
    \frac{\Delta H}{\expval{H}_\zeta} = \frac{1}{\sqrt{\lambda}}.
\end{equation}
Thus the relative deviation, which encodes how far the quantum value is from the classical one, vanishes in the large $\lambda$ limit. 
\paragraph{QCS classical Casimir}
The relationship between the coadjoint orbit of the QCS and the Noether charge of the extended covariant phase space formalism is encapsulated by twisted moment maps.
A priori, there may be several admissible choices of moment maps. We begin by showing that they are all equivalent in the following sense: the Casimir functional on field
space induced by any such choice is independent of that choice. In order to do so let us introduce the rescaled Casimir function on the orbits
\begin{equation}\label{eq:modifiedcasimir}
    c_{\mathrm{QCS}}' = \frac{1}{\chi^z}c_{\mathrm{QCS}}.
\end{equation}
This modified Casimir is nothing but the Casimir function \eqref{eq:QCSCasimirfunction} multiplied by a function that is constant on each orbits
\begin{equation}
    \chi^z\eval_{\mathsf{O}_{\mathrm{QCS}}^{(\lambda,c)}} = c.
\end{equation}
It is thus also a Casimir function. Let us now consider a field theory with phase space $\Gamma$ and symplectic form $\Omega_\Gamma$. We call the system a QCS dynamical
system if there exists comoment maps for the $\mathfrak{ecs}$
\begin{Align}\label{eq:qcsdynamicalsystemecsmomentmap}
    \tilde{\mu}(J^a_b) &= N^a_b,\\
    \tilde{\mu}(P^a)  &= t^a,\\
\end{Align}
where $N^a_b,t^a$ are functionals on the field space whose Poisson bracket defined by $\Omega_\Gamma$ reproduces the $\mathfrak{ecs}$ algebra. The twisted comoment maps on the $\mathfrak{qcs}$ algebra are then defined as
\begin{Align}\label{eq:momentmaplambda}
    \tilde{\mu}^{(c)}(J^a_b) &= N^a_b,\\
    \tilde{\mu}^{(c)}(P^a) &= t^a,\\
    \tilde{\mu}^{(c)}(Z) &= t^z = c.
\end{Align}
We then define the classical Casimir as the function on phase space given by
\begin{Align}\label{eq:classicalcasimir}
    C_{\mathrm{QCS}}^{\mathrm{cl}} &\defeq  \frac12 \,N^a_b t^b_a + \frac{1}{2\mu^z} \epsilon_{ac}t^c N^a_b t^b\\
    &=  C^{\mathrm{cl}}_{\spl{2}} + \frac{1}{c}C^\mathrm{cl}_{\mathrm{ECS}},
\end{Align}
Note that the classical Casimir and the Casimir functions are related by the pullback od the twisted moment map
\begin{equation}
  C^{\mathrm{cl}}_{\mathrm{QCS}} =  \qty(\mu^{(c)})^*(c_{\mathrm{QCS}}),\quad C^{\mathrm{cl}}_{\spl{2}} =  \qty(\mu^{(c)})^*(c_{\spl{2}}),\quad C^{\mathrm{cl}}_{\mathrm{ECS}} =  \qty(\mu^{(c)})^*(c_{\mathrm{ECS}}).
\end{equation}
\par
We now address the following question: if one selects a different set of moment maps \eqref{eq:momentmaplambda}, how does the classical Casimir function transform?  
Since any choice must satisfy the equivariance condition \eqref{eq:momentmapequivariance}, changing moment map corresponds to an automorphism on the
$\mathfrak{sl}\qty(2,\mathbb{R})$ algebra. They can be represented by $\spl{2}$ matrices $B^a_b$ with the moment map transforming as
\begin{equation}
    N'^a_b = B^a_c N^c_d \qty(B^{-1})^d_b.
\end{equation}
It is easy to see that the classical $\spl{2}$ Casimir is invariant under such a transformation
\begin{Align}
    C'^{\mathrm{cl}}_{\spl{2}} &= N'^a_b N'^b_a\\
    &= B^a_c (B^{-1})^d_b B^b_e (B^{-1})^n_a N^c_d N^e_n\\
    &= N^a_b N^b_a = C^{\mathrm{cl}}_{\spl{2}}.
\end{Align}
Let us now consider the Heisenberg part. It is well known that the automorphism group of the Heisenberg algebra---i.e. the group of transformations acting on $x$ and $p$ that preserve the canonical commutation relations---is $\spl{2}$. However one can also consider $\mathrm{GL}\qty(2,\mathbb{R})$ elements, as long as the central moment map is rescaled accordingly. More precisely, different moment map choices are related by the transformations
\begin{align}
    t'^a &= A^a_b t^b\\
    t'^z &= \det(A) t^z,
\end{align}
with $A \in \mathrm{GL}\qty(2,\mathbb{R})$. An additional constraint arises from the mixed commutation relations
\begin{align}
    \qty\big{t'^a,N'^c_d}_\Gamma &= \qty\big((AB^{-1})^a_d B^c_b - \tfrac{1}{2} \delta^c_d A^a_b) t^b\\
    &= \delta^a_d t'^c - \tfrac{1}{2} \delta^c_d t'^a\\
    &= \qty\big(\delta^a_d A^c_b - \tfrac{1}{2} \delta^c_d A^a_b)t^b.
\end{align}
By equating the coefficients inside the parentheses and examining the different index configurations, one readily finds that
\begin{equation}
    A = t\,B.
\end{equation}
Further taking the determinant on both sides of the above gives
\begin{equation}
    t = \pm \sqrt{\det A}.
\end{equation}
One can then compute how the classical ECS Casimir changes under a change of moment map
\begin{Align}
    C'^{cl}_{\mathrm{ECS}} &= \frac12 \epsilon_{ac}t'^c N'^a_b t'^b\\
    &= \frac12 \epsilon_{ac} A^c_d A^b_e B^a_m(B^{-1})_b^l N_l^m N^et^d\\
    &= \frac12 \qty(B^{-1}A)^l_e \epsilon_{mj} (B^{-1}A)^j_d N^m_l t^e t^d\\
   &= \frac12 \det(A) \epsilon_{ac}N^a_bt^b t^c\\
   &= \det(A)\, C^{cl}_{\mathrm{ECS}},
\end{Align}
where going from the second to the third line we used the fact that $B^T \epsilon = \epsilon B^{-1}$ for any matrix in $\spl{2}$.
Putting this result together with the invariance of the $\spl{2}$ Casimir and the transformation of the moment map of the central element, we find
\begin{equation}
    C'^{\mathrm{cl}}_{\mathrm{QCS}} = C^{\mathrm{cl}}_{\mathrm{QCS}}.
\end{equation}
That is, the classical Casimir is invariant under transformation of the moment maps.
While this results follows from Schur's lemma in the semi-simple case, no complete results are known in the general case.
This result furnishes a well defined---in the sense that it does not depend on the choice of moment maps--- classical counterpart to the quantum Casimir operator.
While the modified Casimir function \eqref{eq:modifiedcasimir} provides this identification, it is not adapted to the untwisting limit $c\to 0$.
In the following, we will therefore continue to use the original Casimir function, keeping in mind that its classical-phase-space counterpart
is only defined up to multiplication by an element of $\mathbb{R}^*$.\par
\paragraph{The classical limit}
Consider the coadjoint orbit $\mathsf{O}_{\mathrm{QCS}}^{\lambda,c}$.
Following the discussion on twisted coadjoint orbits of Subsection \ref{subsec:momentmaps} and the isomorphism \eqref{eq:twistedcoadjointorbitisomorphism}, this
can just as well be seen as a twisted coadjoint orbits of the $\mathrm{ECS}$. When the twisting parameter is taken to zero $c\to0$, the points in $\mathsf{O}_{\mathrm{QCS}}^{(\lambda,c)}$
get projected to an orbit of the $\mathrm{ECS}$. Which orbit? Consider a point $\tilde{p} = (p,c) \in \mathfrak{qcs}^*$ where $p \in \mathfrak{ecs}$. The QCS orbit
through the point $\tilde{p}$ gets deformed into the ECS coadjoint orbit through the point $p$ in the untwisting limit. This can be easily seen from the form of the Casimir function
\eqref{eq:QCSCasimirfunction} as well.\par
In order to define the classical limit,
we start by introducing the dual basis of the QCS algebra \eqref{eq:qcsalgebra} $(j\downup{a}{b},p_a,z)$ such that
\begin{equation}
    \pairing{j\downup{a}{b}}{J\updown{c}{d}} = \delta_a^c \delta_b^d ,\quad \pairing{p_a}{P^b} = \delta_a^b, \quad \pairing{z}{Z} = 1.
\end{equation}

By definition, a classical QCS dynamical system is a system where there exists $\mathfrak{ecs}$ comoment maps
\eqref{eq:qcsdynamicalsystemecsmomentmap}. For any field configuration $\varphi_0\in\Gamma$, this defines a point in $\mathfrak{ecs}^*$ through the moment map
\begin{equation}
  \mu(\varphi_0) = N^a_b[\varphi_0] j\downup{a}{b} + t^a[\varphi_0] p_a.
\end{equation}
One can associate a QCS coherent state to the field configuration $\varphi_0$ as follows. The Berezin symbols \eqref{eq:berezinsymbols} define a point in the coadjoint
orbit of the QCS---or twisted coadjoint orbit of the ECS---through
\begin{equation}
    \tilde{p}^{(\lambda,c)}(\zeta,\alpha) = l^{(\lambda,c)}_{\zeta,\alpha}\qty(J\updown{a}{b})j\downup{a}{b} + l^{(\lambda,c)}_{\zeta,\alpha}\qty(P^a)p_a + c z.
\end{equation} 
This point will project to a point in the $\mathfrak{ecs}^*$ in the untwisting limit $c\to 0$.
A coherent state $(\zeta_0,\alpha_0)$ is said to represent the classical field configuration $\phi_0$ if this point corresponds to the one provided by the classical moment map
\begin{equation}
    \lim_{c\to0}\tilde{p}^{\lambda,c}(\zeta_0,\alpha_0) = \mu(\varphi_0).
\end{equation}
Additionally, following the discussion about the classicality of coherent states earlier in this subsection, the state $\ket{\zeta_0,\alpha_0}$ will be sharply peaked
around its classical value $\tilde{p}^{\lambda,c}(\zeta_0,\alpha_0)$ in the large $\lambda$ limit. The classical limit in the corner proposal is therefore characterized
by the two limits $\lambda \to \infty$ and $c\to 0$. In order for these limits to be physically well defined, these two quantities should therefore be dimensionless.
Concretly, $\lambda$ is dimensionless if the $\spl{2}$ charges are dimensionless and $c$ is dimensionless if the translation charges are dimensionless. One can always
redefine the symmetry parameters $\xi_{(1)j}^i$ and $\xi_{(0)}^i$ in \eqref{eq:2186} such that this is the case.

\chapter{The Physics of Quantum Corners}\label{chapter4}
Now that we have described in details the mathematical structure associated with quantum corners, we move to the physical applications.
The modern framework around corner symmetries is motivated in part by the difficulty of defining local subsystems in gauge
theories and gravity.
A standard alternative is to characterize a spacetime subregion by its algebra of observables. This point of view is central to Algebraic Quantum Field Theory (AQFT)
\cite{Takesaki1970,Sewell1980,SusskindUglum1994,Bousso:1999xy,kay1991theorems,CasiniHuerta2007,Wall:2011hj,Casini:2011kv,Bousso:2015mna,
Bousso:2015wca,DongHarlowWall2016,Harlow:2016vwg,CardyTonni2016,FaulknerLewkowycz2017,LongoXu2018,Witten:2018zxz,KoellerLeichenauerLevineShahbaziMoghaddam2018,BalakrishnanFaulknerKhandkerWang2019,
CotlerEtAl2019Recovery,FaulknerLiWang2019,KamalPenington2019,CzechDeBoerGeLamprou2019,CeyhanFaulkner2020,
ParrikarRajgadiaSinghSorce2024,Gesteau2025LargeN}. On the other hand, the corner proposal allows for a state description of the subregions,
using the representation theory of the corner symmetries on the entangling surface.
Of course the description of subsystems in quantum theories is directly related to the notion of entanglement entropy.
In quantum gravity, spacetime itself is treated as a quantum system, and the entanglement entropy between subsystems should therefore be understood as the entanglement
entropy between distinct subregions of spacetime. In the present context, the upshot of this viewpoint is the emergence of the area law for entropy in the semiclassical limit.
In other words, we aim to interpret the Bekenstein--Hawking area term as arising from the entanglement entropy of the underlying quantum spacetime.
The application of the corner proposal to these questions began in \cite{Ciambelli:2024qgi}, where we proposed a way to glue two spacetime subregions
 using the representation theory of the QCS. This construction was later generalized to arbitrary QCS representations in \cite{Varrin:2024sxe}.
The computation of the entanglement entropy and the emergence of the area law for near-extremal Reissner--Nordstr\"om black holes were discussed in \cite{Ciambelli:2025ztm}.
These results were subsequently formalized and extended to general spherically symmetric spacetimes in \cite{Varrin:2025okc,Kowalski-Glikman:2025arealaw}.\par
Let us now describe the contents and main results of this chapter. Section \ref{sec:localsubsystems} starts with a general discussion about local subsystems and the entangling product.
We then turn to a concrete realization of these ideas within the framework of the corner proposal.
We first describe how to glue two corners together and subsequently analyze the reduced density operators and entanglement entropy for QCS coherent states.
The resulting entanglement entropy depends on the particular state $\ket{\zeta,\alpha}$ and the representation parameter $\lambda$. We then introduce a specific class of
coherent states, called \textit{classical states}, for which the entanglement entropy grows linearly in $\lambda$ at large $\lambda$. This linear behavior will later on
be related to the area law, justifying the name of these states as the ones that give the appropriate entropy in the semiclassical limit. Next, we discuss 
general quantum gravitational states and derive a generalized version of the entropy formula for non-abelian gauge theories. We conclude the section with comments on the relative
entropy between different QCS coherent states.\par\clearpage
We then move to connect these abstract results with physical quantities in Section \ref{sec:sssspacetime}. We start by giving the explicit values of the Noether charge computed
in Chapter \ref{chapter2}, for Static Spherically Symmetric (SSS) spacetimes. Next, we use the formalism developed in Chapter \ref{chapter3} to connect these classical charges to QCS coherent states.
This allows us to precisely connect the representation parameter $\lambda$ with the area of the entangling corner in Planck units. Finally, this is used to describe how the area law
emerges from the entanglement entropy of the classical states, which is the main result of this thesis.\par
Section~\ref{sec:goingfurther} is devoted to future directions and includes unpublished preliminary results. We start by discussing a broader class of classical states, that are obtained by
``dressing" the
$\spl{2}$ operators with the generator of boost. We then give a brief introduction to an alternative treatment of corner symmetries where the quantum states are taken in irreducible unitary 
representations of the UCS instead. Finally, we present the first steps in generalizing the work of this thesis to the higher dimensional case by discussing the three dimensional quantum corner symmetry group.
We give our final comments and outlooks in Section \ref{sec:conclusion}.\par

Parts of Subsection~\ref{subsec:gluing} are based on \cite{Ciambelli:2024qgi,Ciambelli:2025ztm}. Subsection~\ref{subsec:entanglemententropy}
draws in part on \cite{Ciambelli:2025ztm}, and also contains results that generalize those presented there. Section~\ref{sec:sssspacetime} is an extended version of the discussions
in \cite{Varrin:2025okc,Kowalski-Glikman:2025arealaw}. All original calculations underlying the new results in these sections were first performed by the author of this thesis.

\section{Local Subsystems}\label{sec:localsubsystems}
In quantum physics, the factorization of a system into its components is realized by the tensor product of Hilbert spaces.
In particular, one can define local subsystems in QFT as follows. Given two Cauchy slices $\Sigma,\bar{\Sigma}$ and the Hilbert
spaces
of the QFT associated with both slices
$\mathcal{H}_\Sigma,\mathcal{H}_{\bar{\Sigma}}$, the Hilbert space corresponding to the union of the
spatial regions is simply the tensor product of the individual Hilbert spaces
\begin{equation}\label{hilberttensorproduct}
    \mathcal{H}_{\Sigma\cup \bar{\Sigma}} = \mathcal{H}_{\Sigma}\otimes \mathcal{H}_{\bar{\Sigma}}.
\end{equation}
However, this simple prescription fails in the case of gauge theories.
To define a localized quantum subsystem in gauge theories, the initial data on the Cauchy slice has to obey gauge constraints
and the data on the interior and exterior slice cannot be specified independently. This results in the failure of the tensor
product factorization of the associated Hilbert spaces.
Instead, the glued Hilbert space $\mathcal{H}_{\Sigma\cup\bar{\Sigma}}$ is a proper subset of the naïve tensor product. This is called
an entangling product and its construction is at the core of the difficulty in defining local subsystems---or subregions--- in 
quantum gravity. This is of course related to the appearance of corner charges. Indeed, decomposing spacetime into two
subregions breaks diffeomorphism invariance,
which is reflected in the appearance of non-vanishing Noether charges on their common boundary. This shared boundary is a corner---called the entangling corner---
and as such should host a representation of the corner symmetry group.
The entangling product is therefore intimately related to the so-called \textit{edge modes} that carry the representations. A proposal of such a construction was
made in \cite{Donnelly:2016auv},
where the glued states are constructed as singlets
under the diagonal action of the corner symmetry group on both sides of the tensor product. This ensures that the contributions
from both subregions to the Noether charge cancel each other, restoring diffeomorphism invariance. This perspective is very close
to the one of quantum reference frames \cite{Aharonov1984Quantum,BartlettRudolphSpekkens2007,Giacomini:2017zju,Vanrietvelde:2018pgb,KrummHoehnMueller2021,HoehnKrummMueller2022,HoehnKotechaMele2023}.\par 
However, the perspective of the current work is quite
different. In our approach, any corner carries physical, non-vanishing Noether charges that could in principle be measured.
This is directly analogous to Gauss' law in electromagnetism: place a corner---a sphere---anywhere in spacetime, and there
is an associated electric charge given by the total charge enclosed by the sphere. In our construction, we therefore require that,
after gluing, the corner still supports non-trivial corner charges and their associated charge algebra.\par
In Subsection~\ref{subsec:gluing}, we begin by constructing the entangling product first proposed in \cite{Ciambelli:2024qgi}. We then apply the gluing to specific
quantum corner states and discuss the physics associated with tracing over one of the subsystems. Subsection~\ref{subsec:entanglemententropy}
is devoted to entanglement entropy and other information-theoretic quantities for QCS coherent states.

\subsection{Gluing}\label{subsec:gluing}
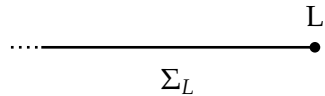
\begin{figure}[h]
    \centering
 \begin{tikzpicture}
  \draw[line width=1pt, dotted] (0,0) -- (.4,0); 
  \draw[line width=1pt] (.4,0) -- (4,0) node[midway, below,yshift = -4pt] {$\Sigma_L$}; 
  \fill (4,0) circle (2pt) node[above,yshift = 4pt]{L}; 
\end{tikzpicture}
    \caption{One spacelike segment connected to the boundary $L$. The Hilbert space describing this system is a representation of the corner symmetry group at $L$.}
    \label{fig:onesegment}
\end{figure}

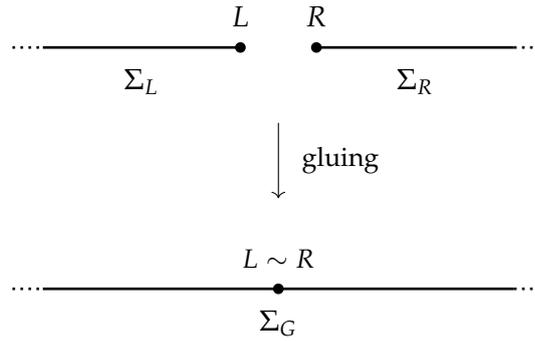
\begin{figure}
    \centering
    \begin{tikzpicture}
        \draw[line width=1pt, dotted] (0,0) -- (.4,0); 
        \draw[line width=1pt] (.4,0) -- (3,0) node[midway, below,yshift = -4pt] {$\Sigma_L$}; 
        \fill (3,0) circle (2pt) node[above,yshift = 4pt]{$L$}; 
        \draw[line width=1pt] (4,0) -- (6.6,0) node[midway, below,yshift = -4pt] {$\Sigma_R$};
        \draw[line width=1pt,dotted] (6.6,0) -- (7,0);
        \fill (4,0) circle (2pt)  node[above,yshift = 4pt]{$R$};

        \draw[->] (3.5,-1) -- (3.5,-2) node[midway,right]{\hspace{5pt}\small{gluing}};

        \draw[line width=1pt, dotted] (0,-3.2) -- (.4,-3.2);
        \draw[line width=1pt] (.4,-3.2) -- (6.6,-3.2) node[midway, below,yshift = -4pt] {$\Sigma_G$};
        \draw[line width=1pt,dotted] (6.6,-3.2) -- (7,-3.2);
        \fill (3.5,-3.2) circle (2pt) node[above,yshift = 4pt]{\small $L \sim R$};
        
    \end{tikzpicture}
    \caption{Two segments are glued together identifying the left and right corners. The Hilbert space associated with the Cauchy slice $\Sigma_G = \Sigma_L \cup \Sigma_R$ is given by a proper subspace of the tensor product of the left and right Hilbert space.}
    \label{fig:gluing}
\end{figure}

In the presence of spacetime boundaries, new degrees of freedom appear on the border: the edge modes.
These physical degrees of freedom carry a representation of the corner symmetry
group. In the case at hand, the representation theory of the corner symmetry group is taken to describes the quantum states of the edge modes. The gluing procedure
can thus be explicitly constructed as a condition on the tensor product of the states described in the previous sections.
Consider a spacelike segment $\Sigma_L$ connected to the corner $L$ as depicted in Figure \ref{fig:onesegment}.
To the corner of this Cauchy surface, we associate a Hilbert space $\mathcal{H}_L$ carrying a unitary irreducible representation of the corner symmetry group.
Let us now consider a second segment $\Sigma_R$ with corner $R$ and its own Hilbert space $\mathcal{H}_R$. Identifying the two corners and gluing the two segments
together, as depicted in Figure \ref{fig:gluing}, yields a Hilbert space $\mathcal{H}_G$ associated with a point of the glued segment. Note that the presence of
a Hilbert space associated with a bulk point is necessary, since any point can be taken as a fictitious boundary from which one can split into two subregions.
The glued Hilbert space is not the tensor product of the left and right ones, but a proper subspace
\begin{equation}\label{embeddinggluedhilbertspace}
    \mathcal{H}_G = \mathcal{H}_L \otimes_{\mathrm{QCS}} \mathcal{H}_R \subset \mathcal{H}_L \otimes \mathcal{H}_R = \tilde{\mathcal{H}}_G,
\end{equation}
where $\otimes_{\mathrm{QCS}}$ denotes the entangling product defined by the corner symmetry group $\mathrm{QCS}$, whose construction is given in what follows.

How do we implement this gluing procedure at the quantum level? We here propose that the quantum version of the gravitational constraints is the requirement
that a complete set of commuting observables (CSCO) of the two copies of the $\mathrm{QCS}$ should match. Eventually, we must obtain that the glued segment holds a representation of the corner
symmetry group. In order for the whole procedure to be consistent, the glued representation should be constructed from the available left and right operators. 
\par

\begin{figure}[h]
    \centering

    \begin{subfigure}[t]{0.48\textwidth}
        \centering
        \begin{tikzpicture}[scale=1.0, line cap=round, line join=round, baseline=(v)]
            \path[use as bounding box] (-2, -1) rectangle (2, 1);

            \coordinate (v) at (0,0);

            \draw[line width=1pt] (-2, 1) -- (v)
                node[midway, above left, xshift=2.5pt] {$p$};
            \draw[line width=1pt] (-2,-1) -- (v)
                node[midway, below left, xshift=2.5pt] {$q$};
            \draw[line width=1pt] (v) -- (2,0)
                node[midway, above, yshift=2pt] {$k$};

            \fill (v) circle (2pt);
        \end{tikzpicture}
        \caption{Three-point vertex.}
        \label{fig:feynmanthreelegs}
    \end{subfigure}
    \hfill
    \begin{subfigure}[t]{0.48\textwidth}
        \centering
        \begin{tikzpicture}[scale=1.0, line cap=round, line join=round, baseline=(v)]
            \path[use as bounding box] (-2, -1) rectangle (2, 1);

            \coordinate (v) at (0,0);

            \draw[line width=1pt] (-2,0) -- (v)
                node[midway, above, yshift=2pt] {$p$};
            \draw[line width=1pt] (v) -- (2,0)
                node[midway, above, yshift=2pt] {$k$};

            \fill (v) circle (2pt);
        \end{tikzpicture}
        \caption{Two-point diagram.}
        \label{fig:feynmantwolegs}
    \end{subfigure}

    \caption{A three legs Feynman diagram $(A)$ where momentum conservation is enforced by $\delta(p+q-k)$ and a two legs diagram $(B)$ where momentum conservation reads $\delta(p-q)$.}
    \label{fig:diagrams}
\end{figure}
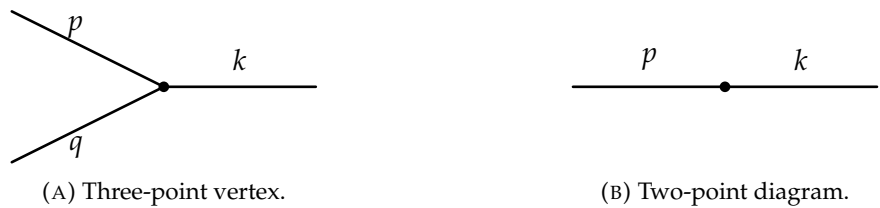

Before delving into the details, we discuss the interpretation of this gluing procedure. In the spirit of the analogy with the Poincar\'e group in quantum field theory, the first
angle comes from Feynman diagrams. Consider a diagram with a single vertex and three external legs, as in Figure~\ref{fig:feynmanthreelegs}.
Momentum conservation at the vertex is enforced by the delta function $\delta(p+q-k)$. One may likewise consider a bivalent diagram with one incoming
and one outgoing leg as in Figure \ref{fig:feynmantwolegs}, where momentum conservation is encoded in $\delta(p-q)$. These delta functions arise naturally because the fields are expanded in the
momentum basis.
Note that the momentum basis is simply a basis of eigenstates 
of the translation operators of the Poincar\'e algebra. In 
principle, one could instead expand the fields in a basis built 
from eigenstates of a different CSCO constructed from the generators of 
$\mathrm{ISO}(1,3)$---for instance $\qty(P^2, \mathbf{J}^2, J_z, 
|\mathbf{P}|)$, corresponding to a spherical wave 
decomposition. Since $\mathbf{J}$ does not commute with the 
individual components of $P^\mu$, such a basis would not 
consist of momentum eigenstates and the momentum-conserving delta 
functions at each vertex would be replaced by angular momentum 
coupling conditions (Clebsch--Gordan coefficients). The 
construction of Feynman diagrams and their associated rules 
therefore implicitly relies on a choice of CSCO. Formulating the full perturbative expansion 
natively in an angular momentum basis is significantly more 
involved. In practice, one computes the amplitudes in momentum 
space and subsequently projects them onto angular momentum 
eigenstates. This procedure is the subject of partial wave 
analysis~\cite{Jacob:1959at,Peters:2004qw}.

\par
Matching the values of a CSCO on the left and right copies of the QCS algebra is therefore akin to imposing a conservation, or continuity
condition across the entangling corner. To understand what this continuity condition constrains, let us look at the classical case.
In general relativity, gluing a subregion $U$ to its complement $\bar U$ across the entangling surface kinematically identifies the intrinsic metric on the two sides, $g_{ij}=\bar g_{ij}$.
Using this identification and imposing the bulk equations of motion, the stationarity of the action 
further implies that the conjugate momenta are matched $\pi^{ij}=\bar\pi^{ij}$ (or more generally a jump fixed by any distributional stress tensor on $N$) in accordance
with the junction conditions \cite{Barrabes:1991ng,Poisson} (see also Section $2.2$ of \cite{Chandrasekaran:2020wwn} for a similar discussion).
In the context of the corner proposal, the physical degrees of freedom are encoded in the corner charges: only certain components of the metric and of its
derivatives---namely those entering the charges---should be regarded as dynamical variables. Matching these components across the interface then enforces
equality of the charges computed from the two subregions. In the quantum theory, matching the charge eigenvalues can therefore be interpreted as a continuity
condition on the corresponding quantum metric components. This requires those components to be very sharply defined, which, by the Heisenberg uncertainty principle,
renders the non-commuting components correspondingly fuzzy.\par
We now construct the entangling product explicitly.
In order to perform the gluing, we need to identify the CSCO of the $\mathrm{QCS}$.
We start by defining the following self-adjoint operators
 \begin{Align}
     D &= \frac{i}{2}(L_+ - L_-),\\
         K &= \beta\left(L_0 + \frac12(L_++L_-)\right),\\
   H &= \frac{1}{\beta}\left(L_0 -\frac12 \qty( L_++L_-)\right),\\
    X &= \frac{1}{\sqrt{2 }}(P_++P_-),\\
    P &= \frac{i}{\sqrt{2 }}(P_+-P_-).
 \end{Align}
 The $H,K,D$ operators were introduced earlier and were called the conformal basis and the $X$ and $P$ operators correspond to the position and the momentum operators on the Heisenberg side\footnote{Note that the notation $X$ and $P$ is adopted solely to make contact with familiar structures, and should not be interpreted as referring to physical position or momentum operators.}.  
 The coefficient $\beta$ has the dimension of length and is introduced to assign canonical dimensions to the conformal algebra.
The conformal basis also appears in the context of conformal quantum mechanics \cite{deAlfaro:1976vlx}. There, $\beta$ is an arbitrary scale that, once fixed, plays a fundamental role in the physics of the system. The conformal mechanics states used here also appear in the study of the causal diamond in Minkowski spacetime \cite{Arzano:2020thh,Arzano:2021cjm,Arzano:2023pnf}. There, the parameter $\beta$ is related to the size of the causal diamond. It is therefore natural to think of this scale as defining the typical length of our subsystem.  As we will see later, it is related to the Unruh temperature associated with a corner.
\begin{equation}
    \qty[H,X] = -\frac{i}{\beta} P, \quad [K,P] = i\beta X,\quad \qty[H,P] = \qty[K,X] = 0.
\end{equation}
From the above equations, we can see that one choice CSCO is given by
$C,\mathcal{C}_\mathrm{QCS},H,P$.
Due to the tensor product structure of the representation, we can consider the $\spl{2}$ and the Weil group generated by $X$, $P$ independently.
The eigenstates of $P$ are simply given by\footnote{Here we again restrict ourselves to the $c>0$ case. The $c<0$ case can be treated similarly by switching the role of $X$ and $P$.}
\begin{equation}
    \ket{p}=  \qty(\frac{1}{\sqrt{\pi c}2^k k!})^{\frac12}\sum_{k=0}^\infty h_k(p) \ket{k} = e^{-\frac{p^2}{2c}}\sum_{k=0}^\infty H_k\qty(\frac{p}{\sqrt{c}}) \ket{k},
\end{equation}
where $H_k$ are the Hermite polynomials and the normalization coefficient of the wave function was chosen such that
\begin{equation}
   \braket{p}{p'} = \delta(p-p').
\end{equation}

For the $\spl{2}$ side, the eigenstates of $H$ are given by \cite{deAlfaro:1976vlx}
\begin{equation}\label{Ebasis}
    \ket{E} = \sum_{n=0}^\infty C_n(E) \ket{n},
\end{equation}
where
\begin{equation}
    C_n(E) = (2\beta)^{\lambda} \left(\frac{\Gamma(n+1)}{\Gamma(n+2\lambda)}\right)^\frac12 E^{\lambda-\frac12} e^{-\beta E} L_n^{2\lambda-1}(2 \beta E),
\end{equation}
where $L_n^{2\lambda-1}$ are the generalized Laguerre polynomials and where $E$ covers the positive real axis. The normalization coefficient was chosen such that
\begin{Align}
   \braket{E'}{E} &=  \sum_{n=0}^{\infty}C_n(E) C^*_n(E) = \delta(E'-E).
\end{Align}
The simultaneous eigenstates of $H$ and $P$ are therefore written $\ket{E,p} = \ket{E}\otimes \ket{p}$ and we have
\begin{align}
    H\ket{E,p} &= \qty(E + \frac{p^2}{2\beta})\ket{E,p},\label{eq:Heigenstate}\\
    P\ket{E,p} &= P \ket{E,p}.\label{eq:Peigenstate}
\end{align}
\par
In order to describe the gluing, we start from the pre-Hilbert space
\begin{equation}
    \tilde{\mathcal{H}}_G = \qty{\ket{E,p}_L\otimes \ket{E',p'}_R}.
\end{equation}
The glued Hilbert space is then obtained by equating the left and right action of CSOC $(\mathcal{C}_{\mathrm{QCS}},C,H,P)$. The Casimir operators simply forces the left and right states
to be in the same representation.
 Furthermore, equations \eqref{eq:Heigenstate} and \eqref{eq:Peigenstate} impose the glued states to be diagonal $E,p$ basis
\begin{equation}
    \mathcal{H}_G = \qty{\ket{E,p}_G = \ket{E,p}_L \otimes \ket{E,p}_R}.
\end{equation}
This gluing is a continuity condition for the variables describing the system.
The gluing procedure thus forces the left and right states to have the same values of $E$ and $p$.
A general glued state describing an entangling corner can therefore be written
\begin{equation}
    \ket{\psi_G} = \int \dd \Omega_{p,E}\,  \psi_G(E,p) \ket{E,p}_L \otimes \ket{E,p}_R,
\end{equation}
where we defined the measure on the gluing basis
\begin{equation}
    \int \dd \Omega_{p,E} = \int_{-\infty}^\infty \dd p \int_0^\infty \dd E.
\end{equation}
In general there is no a priori reason why the wave function $\psi_G(E,p)$ would be written as a product of independent wave functions $\psi_G(E,p) = \psi(E) \phi(p)$.
This is however the case for all of the states that we will consider. From now on we therefore write a general glued state
\begin{equation}\label{eq:generalstatefactorized}
        \ket{\psi,\phi}_G = \int \dd \Omega_{p,E}\,  \psi(E) \phi(p) \ket{E,p}_L \otimes \ket{E,p}_R.
\end{equation} 

Once the gluing is understood, the splitting of one corner into two follows the exact opposite path.
Start with a spatial segment and choose the corner at the point where you want to split the system. There is now one copy
of the Hilbert space associated with the chosen corner. The system is doubled by taking the diagonal tensor product and then relaxing the gauge constraints
\begin{equation}\label{splittingprocedure}
    \ket{E,p} \xlongrightarrow{\text{double}} \ket{E,p}_L \otimes \ket{E,p}_R \xlongrightarrow{\text{relax}} \ket{E,p}_L \otimes \ket{E',p'}_R.
\end{equation}
Thus, starting from a continuous Cauchy slice without boundaries we have a way to separate it into two independent localized subsystems.\footnote{In this section, we use the term ``Cauchy slice'' to make contact with the discussion of Hilbert-space non-factorization in gauge theories. The construction presented here is, however, completely general and applies to any codimension-1 surface on which the symplectic structure is defined.}
We call this inverse procedure the \textit{doubling map}. It can be written for a general state as
\begin{equation}
    \ket{\psi,\phi} = \int \dd\Omega_{p,E}\,\psi(E)\phi(p)\ket{E,p} \longrightarrow \int \dd \Omega_{p,E} \, \psi(E)\phi(p) \ket{E,p}_L\otimes \ket{E,p}_R.
\end{equation}
The doubling map allows one to factorize the Hilbert space associated with a corner into its local subsystems components. We note that this operation
resembles the factorization map in \cite{Jafferis:2019wkd}, but here, the use of symmetries uniquely fixes the degeneracy function appearing in the factorization.
While the representation theory of the $\mathrm{QCS}$ underpins the kinematics of the corner proposal, this map introduces additional information also derived from the symmetries. This extra structure connects to the dynamics of the original theory. Indeed, the procedure is built upon the value of the charges. Those are computed on-shell and know about the dynamics of the theory through Hamilton's equation. This interpretation is reminiscent of tensor networks, where, similarly, the factorization of the Hilbert space provides the additional structure missing from a simple quantum phase space without dynamics \cite{Cao:2016mst}.

We conclude this subsection by commenting on the glued algebra. As mentioned earlier, the glued corner should also host 
a representation of the QCS. In other words, there should be a way to construct glued operators that act on the glued state in the exact same way than left and right operators act on their respective copies of the Hilbert space.
A simple example of this can be done in the case where one picks the trivial representation on the $\spl{2}$ side. Then, the 
QCS representation is simply the Weil representation the Heisenberg Fock space. In that case, one can construct the glued $\mathrm{QCS}$ algebra on the glued Hilbert space explicitly
\begin{Align}\label{glued algebra}
    X_G &= \frac{1}{\sqrt{2}}\qty(X_L + X_R),\quad  P_G = \frac{1}{\sqrt{2}}(P_L + P_R),\\
    K_G &=  \frac12\qty(V_L + V_R + \frac{ X_L X_R}{C}), \quad H_G = \frac12\qty(K_L + K_R + \frac{P_L P_R}{C}),\\
    D_G & = \frac12\qty(D_L + D_R - i \frac{\qty(X_R P_L + X_L P_R)}{2}),
\end{Align} 
\noindent where we have used that the left and right central elements are equal and called $C$.
Although the irreducibility of this representation can be checked explicitly, it follows simply from the fact that the special linear operators
can be expressed as quadratic expressions of the translations, similar to the left and right cases.
In fact, the glued cubic Casimir acting on the glued Hilbert space gives the same constant value as the left and right ones. This confirms that the glued representation is indeed the same as the ones we started with.
\subsection{Entanglement entropy}\label{subsec:entanglemententropy}
Given the description of quantum states associated with a corner separating two subregions that we presented in the previous subsection
we can ask what the entanglement entropy between these subsystems is.
In this section, we study the entanglement entropy of two important classes of states, the vacuum state $\ket{\Omega} = \ket{n=0,k=0}$ and the coherent states described
in Section \eqref{sec:mathematicalbackground}.
We then analyze the behavior of the entropy for small and large values of the Casimir. We conclude the section with a discussion of a particular family of coherent states
that we call \textit{classical states}. These will be fundamental in our application of the formalism to static, spherically symmetric spacetimes in the next section.\par
We start with a general description of the computation of reduced density operators and Von-Neumann entanglement entropy. Given a corner in a state $\ket{\psi,\phi}$, 
the density operator associated with its local subsystems is given by
\begin{equation}
  \rho^{(\psi,\phi)} = \int \dd\Omega_{p,E} \dd \Omega_{p',E'}\, \psi(E)\psi^*(E')\phi(p)\phi^*(p')\bigl(\ket{E,p} \bra{E',p'}\bigr)_L\otimes\bigl(\ket{E,p} \bra{E',p'}\bigr)_R.
\end{equation}
In order to obtain the left reduced density matrix, we take the trace over the right degrees of freedom
\begin{equation}
    \rho^{(\psi,\phi)}_{L} \defeq \Tr_R\left(\rho^{(\psi,\phi)}\right) = \int\dd \Omega_{p,E} \,\abs{\psi(E)}^2\abs{\phi(p)}^2\bigl(\ket{E,p} \bra{E,p}\bigr)_L. 
\end{equation}
In order to compute the entanglement entropy, one first needs to perform a change of variables $E\to \tilde{E}, p\to \tilde{p}$ such that the new variables
are dimensionless. The entanglement entropy is then simply given by
\begin{Align}\label{entropyformula}
    S^{(\psi,\phi)}_L &= - \int\dd\Omega_{\tilde{p},\tilde{E}}\, \abs{\psi(\tilde{E})}^2\abs{\phi(\tilde{p})}^2 \ln\left(\abs{\psi(\tilde{E})}^2\abs{\phi(\tilde{p})}^2\right)\\
    &= - 2 \int_0^\infty \dd \tilde{E} \,\abs{\psi(E)}^2\ln(\abs{\psi(E)}) - 2\int_{-\infty}^{\infty} \dd \tilde{p} \, \abs{\phi(\tilde{p})}^2 \ln \left(\abs{\phi(\tilde{p})}\right),
\end{Align}
where we used the fact that the $\ket{\psi,\phi}$ state is normalized.
For the general states of the form \eqref{eq:generalstatefactorized}, the $\spl{2}$ and Heisenberg contribution therefore
factorize. In practice, this greatly simplifies the calculations as one can compute the two contributions separately.\par

Let us now consider the vacuum state.
We start by expressing this state in the gluing basis
\begin{Align}
    \ket{\Omega} &= \frac{\qty(2\beta)^{\lambda}}{\qty(c\pi)^\frac14} \qty(\Gamma(2\lambda))^{-\frac12}\int \dd \Omega_{p,E} \,E^{\lambda-\frac12} e^{-\beta E}\, e^{-\frac{p^2}{2c}} \ket{E,p}.
\end{Align}
Using the doubling map and tracing over the right degrees of freedom yields the reduced density matrix of the vacuum state
\begin{equation}\label{vacuumreduceddensityoperator}
    \rho^{\Omega}_L  = \frac{(2\beta)^{2\lambda}}{\sqrt{\pi c}\,\Gamma(2\lambda)}\int \dd \Omega_{p,E}\, E^{2\lambda-1}e^{-2\beta\left(E+\frac{p^2}{2\beta c}\right)} \ket{E,p}\bra{E,p}.   
\end{equation}
This density matrix represents a thermal state with temperature $T=2\beta$, whose 
density of states is controlled by the parameter $\lambda$. Its continuous energy levels are given by $\mathcal{E} = E + \frac{p^2}{2\beta c}$. 

\begin{figure}[H]
    \centering
    \includegraphics[width=0.6\linewidth]{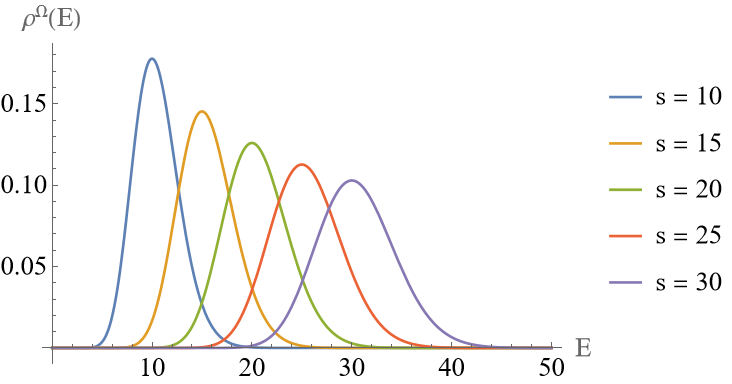}
    \caption{The distribution associated with the $\spl{2}$ part of the reduced density matrix of the vacuum state for unit temperature and different values of representation parameter.}
    \label{fig:rhoE}
\end{figure}

One observes that the operator factorizes into independent distributions
\begin{Align}
    \rho_L^\Omega(E)=\frac{(2\beta)^{2\lambda}}{\Gamma(2\lambda)}E^{2\lambda-1}e^{-2\beta E}\qquad\qquad 
    \rho_L^\Omega(p)=\frac{1}{\sqrt{\pi c}}e^{-\frac{p^2}{c}}.
\end{Align}
The first of the above corresponds exactly to the vacuum state density operator of conformal quantum mechanics \cite{deAlfaro:1976vlx}, and is indeed the part related to $\spl{2}$.
We plot this distribution for unit temperature and different values of $s = \lambda - \frac12$ in Figure \ref{fig:rhoE}. 

The average energy of the distribution is given by\footnote{The expectation value of the operator $H$ is the energy since $L_0$ is the Hamiltonian and $\beta \ev{H}_0 = \ev{L_0}_0$.}
\begin{equation}\label{groundstateenergy}
    \ev{H}_0 = \frac{\lambda}{\beta}. 
\end{equation}
In the large $\lambda$ limit, the above equation is reminiscent of the AdS/CFT correspondence, where the Casimir of the conformal algebra in the boundary theory is
related to the Beltrami-Laplace operator in the bulk
\begin{equation}\label{boundaryasimirbulkenergy}
    \lambda \approx  m L_{\mathrm{AdS}}, \quad m L_{\mathrm{AdS}} >>1,
\end{equation}
where $m$ is the mass of a bulk field and $L_\mathrm{AdS}$ is the length scale of the $\mathrm{AdS}$ spacetime \cite{Witten:1998qj}. This further supports the interpretation
of $\beta$ as a typical length scale of the system.
Moreover, as already observed, the partition 
function in \eqref{vacuumreduceddensityoperator} indicates a thermal behavior of the vacuum once restricted to the subregion. This suggests the following interpretation.
The vacuum state is a vacuum state of quantum gravity. It should therefore be interpreted as the absence of geometry rather than a flat geometry.\footnote{This is clearly stated in the theory of embadons explored in \cite{Ciambelli:2024swv}, in which the area of a corner is promoted 
to a quantum operator, and thus its spectrum becomes only semi-positive definite. Then, the vacuum state is associated with the zero eigenvalue of the area operator, 
interpreted as the absence of geometry, rather than as the geometry being flat. Indeed, the former interpretation pertains to quantum geometric states, while the 
latter would be the case for an operator associated to matter (or perturbative gravitational radiation) on a classical geometric background.}. Nonetheless, such a claim 
is made at the level of the global pure state, before partitioning it into subsystems. In fact, when an observer is restricted to a subsystem and does not have access 
to the information on the other side of the entangling corner, she traces out the degrees of freedom in the complementary inaccessible region, inevitably resulting in 
her experiencing thermal physics \cite{Gibbons:1977mu}. This phenomenon, originally discovered by Unruh in the framework of quantum field theory \cite{Unruh:1976db}, 
is a robust algebraic feature \cite{Bisognano:1976za} that we have discovered here applies also to our quantum geometric description. 

Before moving to the entanglement entropy, we note that the argument of the exponential in \eqref{vacuumreduceddensityoperator} is exactly the eigenvalue of the full 
operator $H$
\begin{equation}
    H\ket{E,p} = \left(E + \frac{p^2}{2\beta c }\right) \ket{E,p}.
\end{equation}
This allows us to write the reduced density operator in a basis-independent way
\begin{equation}
    \rho_L^\Omega = \frac{(2\beta)^{2\lambda}}{\Gamma(2\lambda)} \left(H-\frac{P^2}{2\beta c}\right)^{2\lambda-1} e^{-\beta H}.
\end{equation}
The presence of the $P$ operator differs from the vacuum density operator of \cite{deAlfaro:1976vlx}, illustrating the presence of the additional Heisenberg structure 
(that is, the translations) in our system. It is interesting to observe that we recover the de Alfaro-Fubini-Furlan theory  of \cite{deAlfaro:1976vlx} in the high 
temperature limit. We further remark that the modular Hamiltonian of the system defined by the entangling corner can be deduced from the equation above. In the $\lambda = \frac12$ 
case, for example, it takes the simple form $H_{\mathrm{mod}} = \beta H$.

We now turn our attention to the entanglement entropy.
Following the procedure described earlier, the vacuum entanglement entropy between the left and right subsystems gives
\begin{equation}\label{vacententr}
    S^\Omega = 2\lambda + \ln(\frac{\Gamma(2\lambda)}{2}) -(2\lambda-1) \psi^{(0)}(2\lambda) + \sigma_{\Omega},
\end{equation}
where $\psi^{(0)}$ is the digamma function and the constant $\sigma_\Omega$ comes from the $p$ integral
\begin{equation}
    \sigma_\Omega = \frac12(1 + \ln(\pi)).
\end{equation}
Note that, since the global state was pure,  we have dropped the $L$ index on the entropy since the right subregion yields the same result. 
The vacuum entropy is therefore given by a monotonously increasing function of the representation parameter $\lambda$. At small values of $\lambda$, 
the entropy is linear
\begin{equation}
    S^\Omega \sim 2\lambda-1, \quad \lambda-\frac12\ll1.     
\end{equation}
As $\lambda$ grows, the entanglement entropy transitions into a logarithmic behavior
\begin{equation}
    S^\Omega \sim \frac12 \ln(\lambda), \quad  \lambda\gg1.
\end{equation}
Since the effective dimension of the system associated with the vacuum state is given by $d_{\mathrm{eff}} \sim e^{S^\Omega}$, this result suggests that, for large 
\( \lambda \), the representation parameter can be interpreted as the effective number of degrees of freedom of the system. We will revisit this idea later when dealing 
with the general corner states.

Although the vacuum state is the natural starting point for the calculation of entanglement entropy, we know from the previous sections of this thesis that
the states that can be associated with classical notions such as geometry are the QCS coherent states. Because of the argument made at the beginning of this subsection,
we can treat
the $\spl{2}$ Perelomov states and the squeezed Glauber states separately. We start with the former.
For the gluing procedure, we  need to express the states \eqref{eq:perelomovstateinnbasis} in the $\ket{E}$ basis
\begin{Align}\label{perelomovinEbasis}
    \ket{\zeta} &= \frac{\qty(2\beta)^{\lambda} (1-\abs{\zeta}^2)^{\lambda}}{\sqrt{\Gamma(2\lambda)}} \int_0^\infty \dd E\,E^{\lambda-\frac12} e^{-\beta E}\sum_{n=0}^\infty \zeta^n L_n^{2\lambda-1}(2\beta E)\ket{E}\\
    &=  \frac{\qty(2\beta)^{\lambda}}{\sqrt{\Gamma(2\lambda)}}\qty(\frac{1-\abs{\zeta}^2}{(1-\zeta)^2})^{\lambda} \int_0^\infty \dd E\, E^{\lambda-\frac12} \, e^{-\beta E\left(\frac{1+\zeta }{1-\zeta }\right)}\ket{E},
\end{Align}
where we have used equation \eqref{Ebasis} and the generating function for the generalized Laguerre polynomials. The expression entering the entanglement entropy is the modulus square of the wave function
$\psi_\zeta(E) = \braket{E}{\zeta}$. Denoting $\zeta = r e^{i\theta}$, we compute
\begin{equation}
     \abs{\psi_\zeta(E)}^2 = \frac{\qty(2\beta)^{2\lambda}}{\Gamma(2\lambda)}\qty(\frac{1-r^2}{\Delta})^{2\lambda}E^{2\lambda-1}\exp(- 2\beta\frac{1-r^2}{\Delta} E).
\end{equation}
where $\Delta = 1-2r \cos\theta + r^2$.\par
The squeezed Glauber coherent states can be written in the momentum basis
\begin{equation}
    \ket{\alpha_\zeta} = \int_{-\infty}^\infty \dd p\, \phi_{\alpha,\zeta}(p) \ket{p}, 
\end{equation}
where the wave function $\phi_{\alpha,\zeta}(p) = \braket{p}{\alpha_\zeta}$ is given by \cite{Schumaker:1986tlu,Munguia-Gonzalez:2021qsy}
\begin{equation}\label{eq:squeezedwaavefunction}
    \phi_{\zeta,\alpha}(p) =\left(\frac{1-r^2}{\pi  c\,\Delta}\right)^{\!1/4} e^{ -\frac{i}{2}\arctan\!\left(\frac{r\sin\theta}{1-r\cos\theta}\right)}e^{ \frac{i\,x_0(p_0-2p)}{2c}}
\exp\!\left[
  - \frac{1}{2c}\frac{1+r\,e^{-i\theta}}{1-r\,e^{-i\theta}}(p-p_0)^2
\right],
\end{equation}
where $\alpha = \frac{1}{\sqrt{2c}}(x_0 + i p_0)$. The modulus square of the above wave function is
\begin{equation}\label{eq:modulusphipzetaalpha}
     \abs{\phi_{\alpha,\zeta}(p)}^2 = \qty(\frac{1-r^2}{\pi c \Delta})^{\frac12} \exp(-\frac{1-r^2}{c\Delta}(p-p_0)^2). 
\end{equation}
Following the same procedure for the entanglement entropy, we get
\begin{equation}
    S^{(\zeta,\alpha_\zeta)} = S^\Omega + \frac32 \ln(\frac{\Delta}{1-r^2}).
\end{equation} 

where we recall that $S^\Omega$ is the vacuum entanglement entropy given in equation \eqref{vacententr}.
We note that because of the Gaussian form of \eqref{eq:modulusphipzetaalpha}, the parameter $\alpha$ does not affect the entanglement entropy.
Interestingly, for Glauber coherent states, this result is well known in the context of quantum field theory and gravity in two dimensions, where it is related to the conformal flatness of two-dimensional geometries
\cite{Fiola:1994ir,Katsinis:2022fxu,Das:2005ah,Varadarajan:2016kei}. We also note that the entanglement entropy of the coherent states can be written
\begin{equation}
    S^{(\zeta,\alpha_\zeta)} = S^\Omega - \frac32 \ln(P_r(\theta)),
\end{equation}
where
\begin{equation}\label{eq:poissonkernel}
    P_r(\theta) = \sum_{n=-\infty}^\infty r^{\abs{n}} e^{in\theta} = \frac{1-r^2}{1-2r\cos(\theta)+r^2},
\end{equation}
is the Poisson kernel in the unit disk.
\par

Since $r$ takes values in $[0,1)$, the entanglement entropy of the coherent state can be considerably larger than the one of the vacuum state. In particular, it is 
interesting to consider the family of coherent states with $\theta = -\pi$ and $r= \tanh(\lambda)$. We will call such states \textit{classical states}, for reasons that will 
soon become clear. Looking at the definition \eqref{eq:qcscoherentstate}, the classical states can be written
\begin{equation}\label{eq:classicalstate}
    \ket{\psi_{\mathrm{cl}}(\lambda)} = \mathcal{D}(\alpha) e^{2 i \lambda D} \ket{\Omega}.
\end{equation}
 In the application of this 
formalism to static, spherically symmetric spacetimes in Section \ref{sec:sssspacetime}, these states can be identified with classical geometries, in the sense that they exhibit the
correct classical limit, which is why we refer to them as classical states. Their entanglement entropy is given by 
\begin{equation}
    S^{\psi_{cl}} = S^\Omega+ 3 \lambda.
\end{equation}
Conversely to the vacuum state, the state $\psi_{cl}$  has a linear dominant behavior in the large $\lambda$ limit
\begin{equation}\label{eq:asymptoticentropyclassicalstate}
    S^{\psi_{cl}} \sim 3\lambda + \frac12 \ln(\lambda), \quad \lambda \gg1.
\end{equation}
In particular the linear behavior for large values of the representation parameter $\lambda$ is what will produce the area law in the next section.
\par
The classical states are extremely relevant to the corner proposal, since they seem to be the only quantum states that produce the area law in the semiclassical limit.
We therefore take some time to comment on their properties.\par
We start by analyzing the special linear side of the classical states. The hyperbolic geodesic distance between the origin and the classical state is given by
\begin{equation}
    d(0,\zeta_{\mathrm{cl}}) = 2 \tanh^{-1}\qty(\abs{\zeta_{\mathrm{cl}}}) = 2\lambda.
\end{equation}
In the classical limit $\lambda\to \infty$, the classical states therefore get pushed to the boundary of the unit disk. The overlap $\abs{\braket{\zeta_{\mathrm{cl}}}{\zeta'}}$
is still a Gaussian of width $\lambda^{-\frac12}$, but its center moves to infinite hyperbolic distance from the bulk. Since the hyperboloid orbits of $\spl{2}$ are isomorphic to
euclidean AdS$_2$, these classical states could have an interesting interpretation in terms of the AdS$_2$/CFT$_1$ correspondence. We conclude the discussion about the 
special linear part of the classical state by noting that, the $D$ operator can be understood as generating boosts in the normal plane of the corner (see Section \ref{sec:sssspacetime} for a more detailed discussion).
Since the $\spl{2}$ part of the classical states can be written as
\begin{equation}
   \ket{ \psi_{\mathrm{cl}}(\lambda)} = e^{2i\lambda D} \ket{n=0},
\end{equation} 
this provides a physical interpretation of the coherent states \eqref{eq:classicalstate}, as those generated by the normal plane boost acting on the vacuum.
Fascinatingly, this links the classical interpretation with the quantum description above, and provides a rationale for its thermal behavior once restricted to a
subregion. Indeed, boosting a corner can be interpreted as moving along an accelerated trajectory in the normal plane, reminiscently of the Unruh effect \cite{Unruh:1976db}.
\par
There are also interesting observations to be made on the Heisenberg side of the classical states. We can compute the fluctuations and covariance of the operators
\begin{align}
    \Delta X^2 &= \frac{c}{2} \frac{\tilde{\Delta}}{1-r^2},\\
    \Delta P^2 &= \frac{c}{2} \frac{\Delta}{1-r^2},\\
\mathrm{Cov}\qty(X,P) &= - c \frac{r \sin(\theta)}{1-r^2},\\
\end{align}
where we remind the reader that $\mathrm{Cov}(X,P) = \frac12 \expval{XP + PX} - \expval{X}\expval{P}$ and where we defined $\tilde{\Delta} = 1+ 2r \cos\theta + r^2$. It is then easy to check that the squeezed coherent states saturate the Schrödinger uncertainty relation
\begin{equation}
    \Delta X^2 \Delta P^2 \geq \qty(\frac{c}{2})^2 + \mathrm{Cov}(X,P)^2.
\end{equation}
For the classical state with $\theta = -\pi$, the classical state further saturates the standard Heisenberg uncertainty
\begin{equation}
    \Delta X \Delta P \geq \frac{c}{2}.
\end{equation}
Furthermore, in the classical limit $\lambda\to\infty$, the coherent state with parameter $r=\tanh(\lambda)$ squeezes essentially all of the uncertainty
into $P$. The operator $X$, by contrast, becomes ultra-classical in the sense that its fluctuations vanish. This can also be seen at the level of relative fluctuations
\begin{Align}
    \lim_{\lambda\to\infty} \frac{\sqrt{\expval{X^2}_{\psi^{\mathrm{cl}}}}}{\expval{X}_{\psi^{\mathrm{cl}}}} &= 0,\\
\lim_{\lambda\to\infty} \frac{\sqrt{\expval{P^2}_{\psi^{\mathrm{cl}}}}}{\expval{P}_{\psi^{\mathrm{cl}}}} &= \infty.
\end{Align}
Of course for a general coherent state, the Heisenberg side is blind to the $\lambda \to \infty$ limit. The classical state is therefore a specific coherent state
that ensures that one of the Heisenberg operators becomes classical in the classical limit.
\par

\paragraph{General quantum corner states}
The most general quantum state associated with a corner is a superposition of states belonging to different representations. To construct such states, it is necessary to suitably sum over all representations. A natural mathematical framework for this summation is provided by the Plancherel measure, which arises in the direct integral decomposition of square-integrable functions on the $\mathrm{QCS}$ group with respect to the Haar measure
\begin{equation}
    L^2(\mathrm{QCS}) \cong\int_{\widehat{\mathrm{QCS}}} \dd\mu(\lambda,c)\, \mathrm{HS}(\mathcal{H}^{(\lambda,c)}) \cong \,\int_{\widehat{\mathrm{QCS}}} \dd\mu(\lambda,c) \qty(\mathcal{H}^{(\lambda,c)}\otimes \bar{\mathcal{H}}^{(\lambda,c)}),
\end{equation}
where $\widehat{\mathrm{QCS}}$ denotes the unitary dual of the group, $\dd\mu(\lambda,c)$ is the Plancherel measure, and $\mathrm{HS}(\mathcal{H}^{(\lambda,c)})$ denotes the Hilbert-Schmidt space over the representation space $\mathcal{H}^{(\lambda,c)}$.
Note that this formula can be seen as the non-compact generalization of the Peter--Weyl theorem. 
A general state can then be written as
\begin{Align}\label{eq:generalstate}
    \ket{\Phi} &= \int \dd \mu(\lambda,c)\, \Phi(\lambda,c) \intpE{p}{E} \psi(E)\phi(p)\ket{\lambda;E}\ket{c;p},
\end{Align}
where we now explicitly indicate the representation parameters $s$ and $c$ associated with the $\spl{2}$ and the Heisenberg states, respectively. The Plancherel measure, together with the associated Fourier inversion theorem, provides an orthogonal decomposition; in particular, operators belonging to distinct sectors are orthogonal. The completeness relation of the $\ket{s, c; E, p}$ states in a given Hilbert space now acts as a projector onto that Hilbert space
\begin{equation}
    \intpE{p}{E} \ket{\lambda,c;E,p}\bra{\lambda,c;E,p} = P_{(\lambda,c)},
\end{equation}
and the completeness relation on the total Hilbert space reads
\begin{equation}
    \int \dd \mu(\lambda,c)\, P_{(\lambda,c)} = \mathds{1},
\end{equation}
which is precisely the decomposition of a Hilbert space into distinct orthogonal spaces. The above completeness relation implies the normalization\footnote{If the representation is labeled by a discrete parameter, the delta function becomes a Kronecker delta.}
\begin{equation}
\braket{\lambda',c';E',p'}{\lambda,c;E,p} = \delta_\mu(\lambda-\lambda')\delta_\mu(c-c') \delta(E-E')\delta(p-p'),
\end{equation}
where $\delta_\mu$ is the adapted Dirac delta such that
\begin{equation}
    \int \dd \mu(\lambda,c) \,\Phi(\lambda,c) \delta_\mu(\lambda-\lambda') \delta_\mu(c-c') = \Phi(\lambda',c'),
\end{equation}
for any function $\Phi(\lambda,c)$. The general state is then normalized if the wave function is normalized with respect to the Plancherel measure
\begin{equation}
    \braket{\Phi} = \int \dd \mu(\lambda,c) \abs{\Phi(\lambda,c)}^2 = 1. 
\end{equation}
We can now construct a general corner state simply as
\begin{equation}
    \ket{\Phi} = \int \dd \mu(\lambda,c) \,\Phi(\lambda,c)\intpE{p}{E}  \psi(E)\phi(p) \ket{\lambda;E}\ket{c;p}.
\end{equation}
Next, we can split it into subsystems using the doubling map
\begin{equation}
    \ket{\Phi}\longmapsto \int \dd \mu(\lambda,c)\, \Phi(\lambda,c)\intpE{p}{E} \psi(E)\phi(p) \ket{\lambda;E}_L\ket{c;p}_L\otimes \ket{\lambda;E}_R\ket{c;p}_R.
\end{equation}
Note that the presence of the Casimirs in the maximally commuting subalgebra now becomes essential. It is reflected in the fact that the tensor product state
has the same value of $\lambda$ and $c$ on the left and on the right. Considering the density operator associated with the doubled state and tracing
over the right degrees of freedom yields 
\begin{equation}
    \rho_L = \int \dd \mu(\lambda,c)\, \dd E \abs{\Phi(\lambda)}^2 \abs{\psi(E)}^2 \abs{\phi(p)}^2  \ket{\lambda;E}\ket{c;p}\bra{\lambda;E}\bra{c;p}.
\end{equation}
The associated entanglement entropy is given by
\begin{equation}\label{eq:entanglemententropygeneralstate}
    S^{\psi,\phi} =\int \dd \mu(\lambda,c) \,  \abs{\Phi(\lambda,c)}^2 \left(-\ln(\abs{\Phi(\lambda,c)}^2) + S^{\psi,\phi}(\lambda,c)\right),
\end{equation}
where $S^{\psi,\phi}(\lambda,c)$ denotes the entanglement entropy of the representation $(\lambda,c)$ associated with the specific state determined by
the wave function $\psi(E)\phi(p)$, as computed in the previous sections. This quantity determines the contribution from each representation. 

In particular, using the Boltzmann interpretation of entropy, we can view it as a measure of the effective dimension of the representation in a particular state
\begin{equation}
    S^{\psi,\phi}(\lambda,c) = \ln( d_{\mathrm{eff}}(\lambda,c)).
\end{equation}
Plugging this back into the total entropy yields
\begin{equation}
    S^{\psi,\phi} = \int \dd \mu(\lambda,c)\, \abs{\Phi(\lambda,c)}^2\qty(-\ln(\abs{\Phi(\lambda,c)}^2)+ \ln(d_{\mathrm{eff}}(\lambda,c))).
\end{equation}
This result bears a striking resemblance to the entanglement entropy of non-Abelian gauge theories of Donnelly \cite{Donnelly:2014gva}
\begin{equation}\label{eq:Donnellyformula}
    S = \sum_R \abs{\psi(R)}^2\left(-\ln(\abs{\psi(R)}^2 + 2\ln(\mathrm{dim}R))\right),
\end{equation}
where, in our case, the role of the representation dimension of the compact gauge group is played by the exponential of the entanglement entropy of each representation
\begin{equation}\label{VARRINO}
    \mathrm{dim}R \sim \exp(S^{\psi,\phi}(\lambda,c)).
\end{equation}
Equation \eqref{eq:entanglemententropygeneralstate} is therefore a direct generalization of Donnelly's formula for the non-compact case. The fact that the entropy
only depends on the modulus $\abs{\Phi(\lambda,c)}$ and not on the relative phase is a consequence of the superselection rule, according to which states in different
representation cannot be in superposition. The second term in \eqref{eq:Donnellyformula} corresponds to the contribution from boundary degrees of freedom,
which introduce a maximally mixed state of dimension $\mathrm{dim}R$ at each corner. The factor of two accounts for the two endpoints of the segment.
This interpretation aligns with the understanding that the entanglement entropy $S^{\psi,\phi}(\lambda,c)$ originates from the presence of non-trivial representations
of the corner symmetries and is directly related to the gluing procedure, which produces mixed states. In the present work, we consider only a single corner,
and the factor of two is therefore absent.
\clearpage
We now focus our attention on a specific example that has illuminating physical consequences.
Since a complete description of the Plancherel measure for the QCS is still under investigation, we begin by restricting to the special linear component,
which corresponds to boundary-preserving diffeomorphisms. Although certain aspects of the QCS structure are absent in this setting, we expect the general computations
to exhibit similar features. Finally, before proceeding, we note that only the positive discrete series are considered in the present analysis.

The Plancherel measure for the discrete series of the universal cover of $\spl{2}$ is given by
\cite{PUKANSZKY1964}
\begin{equation}
    \dd \mu(\lambda) = \frac{2}{\pi^2}\lambda\,\dd \lambda,
\end{equation}
with $\lambda \in \R$. We consider a Gaussian wave function on the representation space, given by
\begin{equation}\label{eq:generalstategaussianwavefunction}
   \Phi(s) = \mathcal{N} \, e^{-\frac{(\lambda - \lambda_0)^2}{2}},
\end{equation}
and focus on the limit of large $\lambda_0$, where the wave function is sharply peaked at large values of the representation parameter $\lambda$, and thus contributes
significantly only in that regime. The normalization constant can be calculated using the Plancherel measure
\begin{equation}
    \mathcal{N} = \sqrt{\frac{\pi^{\frac32}}{2 \lambda_0}} + \mathcal{O}\qty(\lambda_0^{-\frac32}).
\end{equation}
For the state in each representation, we will consider the classical coherent state \eqref{eq:classicalstate}. Since the integral
only contributes at large values of $\lambda$, we can consider the asymptotic expansion \eqref{eq:asymptoticentropyclassicalstate}.
We then split equation \eqref{eq:entanglemententropygeneralstate} in two pieces\footnote{The ``bulk" and ``boundary" terminology is 
borrowed from \cite{Donnelly:2014gva}, and refers to degrees of freedom coming from the interior and at the edge, respectively. A similar split is performed in
\cite{Ball:2024hqe}. Note that the boundary term is accounted for by the logarithm of the effective dimension, by which we mean the exponential of the entropy,
see \eqref{VARRINO}.}
\begin{Align}
S^{\psi,\phi}&=S_{\text{bulk}}^{\psi,\phi}+S_{\text{boundary}}^{\psi,\phi}\\
S_{\text{bulk}}^{\psi,\phi}&=-\int \dd \mu(\lambda,c) \,  \abs{\Phi(\lambda,c)}^2 \ln(\abs{\Phi(\lambda,c)}^2)\\
S_{\text{boundary}}^{\psi,\phi}&=\int \dd \mu(\lambda,c) \,  \abs{\Phi(\lambda,c)}^2 S^{\psi,\phi}(\lambda,c)
\end{Align}
and compute their leading order terms
\begin{Align}\label{eq:firstandsecondterm}
    S_{\text{bulk}}^{\psi,\phi}= \ln(\lambda_0) + \mathcal{O}(1),\qquad
     S_{\text{boundary}}^{\psi,\phi}=\lambda_0 + \mathcal{O}\qty(\ln(\lambda_0)).
\end{Align}
As we will see in the next section, in the case of static spherically symmetric spacetimes
the parameter $\lambda$ corresponds to the area of the entangling corner. We can thus extract the following
physical interpretation from the result above. The general state defined by the wave function \eqref{eq:generalstategaussianwavefunction} can be seen as a
superposition of fuzzy area states sharply peaked around the classical value $\lambda_0$. Looking at equation \eqref{eq:firstandsecondterm}, we observe that the area
law emerges from the boundary contribution. This is an expected universal result which has also been realized in loop quantum gravity, string theory and matrix models
\cite{Donnelly:2008vx,Delcamp:2016eya,Bianchi:2024aim,Donnelly:2020teo,Frenkel:2023aft}. 
\clearpage
\paragraph{Relative entropy}
We conclude this section by discussing the relative entropy between coherent states. Relative entropy plays an important role in quantum information.
For example, it is a useful tool to describe the distinguishability of states: the probability of confounding two states after $N$ carefully prepared quantum measurements
is $p\sim e^{-S_{rel} N}$. Moreover, it satisfies a very useful monotonicity statement. Consider a completely positive trace preserving
operation ${\cal N}$. Then one can show \cite{Lindblad:1975kmh}
\begin{equation}
S\Bigl({\cal N}(\rho)\mid \mid {\cal N}(\sigma)\Bigr)\leq S\Bigl(\rho\mid \mid \sigma\Bigr). 
\end{equation}
This inequality is used in the proof of the generalized second law of thermodynamics \cite{Wall2010Rindler,Wall:2011hj}. While no physical applications of the relative entropy between
QCS coherent states are currently known, we expect these quantities to play an important role
in the corner proposal, and we therefore record their explicit expressions here.\par
Let us look at two coherent states in the same representation
\begin{Align}
    \ket{\zeta,\alpha_\zeta} &= \intpE{p}{E} \psi_\zeta(E)\phi_{\alpha,\zeta}(p)\ket{E,p},\\
    \ket{\zeta',\alpha'_{\tilde{\zeta}}} &= \intpE{p}{E} \psi_{\zeta'}(E)\phi_{\alpha',\zeta'}(p)\ket{E,p},
\end{Align}
where the wave functions are given in equations \eqref{perelomovinEbasis} and \eqref{eq:squeezedwaavefunction}.

The relative entanglement entropy between the reduced density operator associated with those two states is given by
\begin{Align}
S\Bigl((\zeta,\alpha_{\zeta})\mid \mid (\zeta',\alpha'_{\zeta'})\Bigr) &= \intpE{p}{E}  \abs{\psi_\zeta(E)}^2 \abs{\phi_{\alpha,\zeta} (p)}^2 \ln\qty(\frac{\abs{\psi_{\zeta}(E)}^2 \abs{\phi_{\alpha,\zeta}(p)}^2}{\abs{\psi_{\zeta'}(E)}^2 \abs{\phi_{\alpha',\zeta'}(p)}^2})\\
    &=\int \dd E \,\abs{\psi_{\zeta}(E)}^2\ln\qty(\frac{\abs{\psi_{\zeta}(E)}^2}{\abs{\psi_{\zeta'}(E)}^2}) \\& \quad +  \int \dd p \,\abs{\phi_{\alpha,\zeta}(p)}^2\ln\qty(\frac{\abs{\phi_{\alpha,\zeta}(p)}^2}{\abs{\phi_{\alpha',\zeta'}(p)}^2})
\end{Align}
Using the explicit wave functions, we get
\begin{Align}
    S\Bigl((\zeta,\alpha_{\zeta})\mid \mid (\zeta',\alpha'_{\zeta'})\Bigr) &= 2\qty(\Im\alpha - \Im\alpha')^2 \frac{1-r'^2}{\Delta'} \\
    &\qquad + \qty(2\lambda + \frac12)\qty[\frac{1-r'^2}{1-r^2}\frac{\Delta}{\Delta'}-\ln(\frac{1-r'^2}{1-r^2}\frac{\Delta}{\Delta'})-1].
\end{Align}
The first term is always positive and only vanishes for $\alpha = \alpha'$ and the classical inequality
$\ln(u)\leq u-1$ with equality if and only if $u=1$ implies that the second term is also always positive with equality only if $\zeta = \zeta'$.
Therefore the relative entropy between two QCS coherent states is always positive and vanishes only if the two states are equal, as is expected from relative
entropy. Interestingly, the unit-disk contribution to the relative entropy is completely determined by the Poisson kernel \eqref{eq:poissonkernel}
\begin{Align}
     S\Bigl((\zeta,\alpha_{\zeta})\mid \mid (\zeta',\alpha'_{\zeta'})\Bigr) &= 2(\Im\alpha-\Im\alpha')^2 P_{r'}(\theta') \\&\qquad+ \qty(2\lambda+\frac12)
     \qty[\frac{P_{r'}(\theta')}{P_{r}(\theta)}- \ln(\frac{P_{r'}(\theta')}{P_{r}(\theta)})-1].
\end{Align}
An interesting feature appears when one considers the relative entropy between a two infinitesimally close states $\rho$ and $\rho + \delta \rho$.
At first order in the perturbation, we find that this entropy vanishes which
is an expression of the first law of entanglement entropy \cite{BlancoCasiniHungMyers2013}. At second order we find
\begin{Align}
    S^{(2)}\Bigl({\rho+\delta \rho \mid\mid \rho }\Bigr) &= \frac{2(1-r^2)}{\Delta}(\delta \alpha_I)^2
+
\qty(4\lambda+1)
\left[
\frac{\left((1+r^2)\cos\theta-2r\right)^2}{\Delta^2(1-r^2)^2}\,\delta r^2 \right.
\\&\qquad -
\left. \frac{2r\sin\theta\left((1+r^2)\cos\theta-2r\right)}{\Delta^2(1-r^2)}\,\delta r\,\delta\theta
+
\frac{r^2\sin^2\theta}{\Delta^2}\,\delta\theta^2
\right],
\end{Align}
which is the Fisher information metric \cite{BraunsteinCaves1994StatisticalDistance}. To gather intuition we can look at the classical state where the
Fisher information metric becomes (we set $\alpha\in\R$ for simplicity)
\begin{equation}
    S\Bigl({\rho^{\mathrm{cl}}_\lambda+\delta \rho \mid\mid \rho^{\mathrm{cl}}_\lambda}\Bigr) = 4\qty(2\lambda + \frac12) \cosh^4(\lambda)\delta r^2 + \mathcal{O}(\delta r^3).
\end{equation}
This implies the following inequality for different representations
\begin{equation}
     S\Bigl({\rho^{\mathrm{cl}}_\lambda+\delta \rho \mid\mid \rho^{\mathrm{cl}}_\lambda}\Bigr) < 
    S\Bigl({\rho^{\mathrm{cl}}_{\tilde{\lambda}}+\delta \rho \mid\mid \rho^{\mathrm{cl}}_{\tilde{\lambda}}}\Bigr), \quad \lambda < \tilde{\lambda},
\end{equation}
This is consistent with interpreting the representation parameter as the effective size of the system, since the relative entropy always increases
with the size of the system that the reduced density operators describe~\cite{Wehrl:1978zz,Vedral:2002zz}. This is also consistent with $\lambda$ describing the area of
the entangling corner, as we will see in the next section.

\section{Static Spherically Symmetric Spacetimes}\label{sec:sssspacetime}
\vspace{-3pt}
We are now finally ready to dive in the main results of the formalism. In this section, we will study the static solutions to the theory studied in \ref{subsec:cornersymmetriessss}.
We will start by computing the charges \eqref{eq:charges} explicitly for that solution and construct the moment map. In order to do so, we introduce several standard coordinate systems adapted to the presence of an horizon.
 Next, we will relate those moment maps to the quantum structure by computing the Berezin symbols,
and explicitly construct the quantum to classical correspondence discussed at the end of Subsection \ref{subsec:quantumclassicalcorr}. This will allows us to show how the
area law emerges in the semiclassical limit for classical states. 
\vspace{-6pt}
\subsection{Classical charges}
\vspace{-2pt}
The static solution to spherically symmetric Einstein--Hilbert gravity can be written in Schwarzschild-like coordinates
\vspace{-3pt}
\begin{equation}\label{eq:sssspacetime}
    \dd s^2 = - f(\rho)dt^2 + \frac{d\rho^2}{f(\rho)} + \rho^2 \dd \Omega_S^2.
\end{equation}
\vspace{-3pt}\noindent
We will consider the case where a horizon is present---i.e. when $f(\rho)$ admits zeros. We define the horizon as the largest root $f(\rho_h) = 0$.
In order to identify the ECS charges (equations \eqref{eq:chargeti} and \eqref{eq:chargenij}), we need to expand the fields and diffeomorphisms near the corner
as we did in \ref{subsec:universalcornersymmetries}. The Schwarzschild coordinates are however not adapted to these expansions since the metric blows up on the horizon.
We thus start this subsection by constructing a system of coordinates adapted to the expansion.\par
Let us introduce the Tortoise coordinate
\begin{equation}
    \dv{\rho_*}{\rho} = \frac{1}{f(\rho)}, \quad \rho_* = \int^\rho \frac{d\rho'}{f(\rho')} + \mathrm{cste},
\end{equation}
where the constant is chosen such that near the horizon we have $r_*(r)\approx \frac{1}{2\kappa}\ln(\abs{r-r_h})$.
The metric can be written 
\begin{equation}
    ds^2 = f(\rho) \qty(-dt^2 + d\rho_*^2) + \rho^2 \dd\Omega_S^2.
\end{equation}
We further introduce null coordinates $u = t-\rho_*, v = t+\rho_*$ such that the metric becomes
\begin{equation}
    ds^2 = -f(\rho)\dd u \dd v + \rho^2\dd \Omega_S^2.
\end{equation}
Since the Tortoise coordinate blows up at $\rho=\rho_h$, the horizon in the null coordinates is located at $u\to + \infty$ or $v\to-\infty$.
Up to this point, we have not used the fact that the spacetime contains a horizon.
The horizon allow us to introduce dimensionless variables using the surface gravity $\kappa=\frac12 f'(\rho_h)$, by defining Kruskal-like coordinates
\begin{equation}
    U = - e^{-\kappa u}, \quad V = e^{\kappa v}.
\end{equation}
In these coordinates, the metric takes the form
\begin{equation}\label{eq:nullKSmetric}
    ds^2 = \frac{f(\rho)}{\kappa^2 UV}\dd U\dd V + \rho^2 \dd\Omega_S^2.
\end{equation}
The future horizon $(\rho=\rho_h,t\to\infty)$ corresponds to $U=0, V>0$ and the past horizon $(\rho=\rho_h, t\to -\infty)$ corresponds to $V=0,U<0$. The bifurcating point, where the future and past horizon meet 
is given by $U=V=0$.
Finally, we introduce Kruskal-Szekeres coordinates 
\begin{equation}
    U = T-X, \quad V = T + X,
\end{equation}
for which the metric becomes conformally flat 
\begin{equation}\label{eq:metricKScoordinates}
   ds^2 =  \frac{f(\rho)}{\kappa^2} e^{-2\kappa \rho_*}\qty(-\dd T^2 + \dd X^2) + \rho^2 \dd\Omega_S^2.
\end{equation}
These coordinates cover the maximally extended spacetime. The future and past horizons are simply given by the $T=\pm X$ lines and the bifurcating point sits at the origin $(T=0,X=0)$. \par
\begin{figure}[H]
    \centering
    \includegraphics[width=0.65\linewidth]{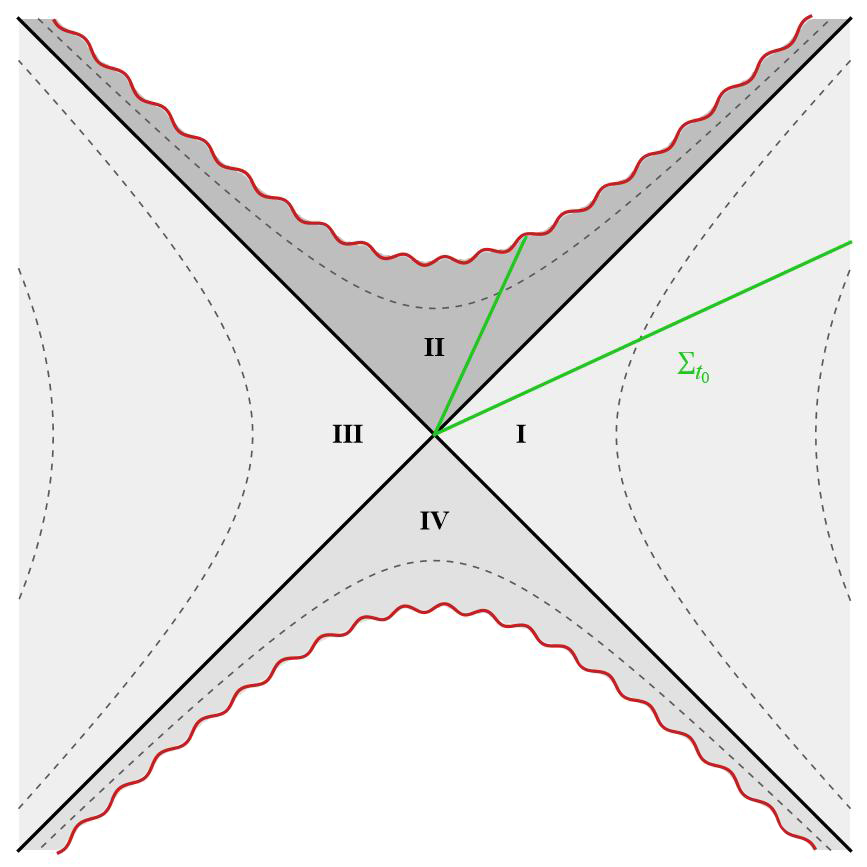}
    \caption{The maximally extended static, spherically symmetric spacetime shown in Kruskal-Szekeres coordinates $(X,T)$. The solid black lines represent the past
    and future event horizons, while the dashed grey lines are constant Schwarzschild radial coordinates surfaces $\rho = \mathrm{cste}$. The wavy red lines represent the past and future spacelike singularities. Region I is the right exterior region, described
    by the Schwarzschild coordinates $(t,\rho)$; Region II is the black-hole interior;
    Region III is the left exterior region, covered by a second copy of the static coordinates; and Region IV is the white-hole interior. The green line represents the constant Schwarzschild time slice
    on which the symplectic structure is defined.}
    \label{fig:ksdiagram}
\end{figure}

In order to see why they are particularly well adapted to the problem at hand, let us pick define our symplectic structure on a constant Schwarzschild time Cauchy slice $\Sigma_{t_0}$.
In the Kruskal-Szekeres coordinates, $\Sigma_{t_0}$ is a straight line through the origin
\begin{Align}
    T &= X \tanh(\kappa t_0), \quad U<0,V>0\\
    T &= X\coth(\kappa t_0), \quad U>0,V>0.
\end{Align}
It is timelike in the exterior region and spacelike inside the horizon, see Figure~\ref{fig:ksdiagram}.
Let us then focus on the horizon external region. The boundary of that subregion is the horizon, and the corner---the intersection between the boundary of the subregion and the Cauchy slice---is the bifurcating point.
We are therefore in the familiar situation of the trivial embedding where the corner sits at the origin of the coordinate system. Furthermore, the metric \eqref{eq:metricKScoordinates}, is completely regular at the corner.
We also note
that, using the expansion of the Tortoise coordinate just outside the horizon
\begin{equation}
 \rho_*(\rho) \approx \frac{1}{2\kappa}\ln(\rho-\rho_h),
\end{equation}
the near-horizon metric in Kruskal-Szekeres coordinates is Minkowski
\begin{Align}\label{eq:nearhorizonminkowski}
    ds^2 \approx 2\kappa \qty(-\dd T^2 + \dd X^2)+ \rho^2 \dd\Omega^2.
\end{Align}
\par
Expanding the diffeomorphisms around the corner as in \eqref{vectorsexpansion}, the vector field generators of the $\mathfrak{ecs}$ can be written as
\begin{Align}\label{eq:ecsvectorfieldsXT}
    \xi_{1} &= \frac12\qty(T \partial_X + X\partial_T),\\
    \xi_{2} &= \frac12\qty(X\partial_T - T \partial_X),\\
    \xi_{3} &= \frac12\qty(X\partial_X - T \partial_T),\\
    \xi_{4} &= \partial_T,\\
    \xi_{5} &= \partial_X,
\end{Align} 
where $\xi_{1},\xi_{2},\xi_{3}$ generate the $\spl{2}$ algebra and $\xi_4,\xi_5$ the normal translations. It is very interesting to note that, for the near-horizon metric \eqref{eq:nearhorizonminkowski},
the $\spl{2}$ triplet are the conformal Killing vectors. This setup is therefore very reminiscent of \cite{Arzano:2020thh,Arzano:2021cjm,Arzano:2023pnf},
and it would be interesting to study the connection further. In order to compute the charges \eqref{eq:charges}, it will be more convenient to work in the null coordinates where
the $\mathfrak{ecs}$ vector fields can be written
\begin{align}
    \xi_{1} &= \frac12\qty(V\partial_V - U \partial_U),\\
    \xi_{2} &= \frac12 \qty(V\partial_U- U\partial_V),\\
    \xi_{3} &= -\frac12(U\partial_V + V \partial_U),\\
    \xi_{4} &= \partial_V + \partial_U,\\
    \xi_{5} &=  \partial_V -\partial_U.
\end{align}
First, because the metric \eqref{eq:nullKSmetric} is purely off-diagonal, the only non-vanishing charge for $\spl{2}$ is\footnote{We choose $\epsilon^{UV}=1$.}
\begin{equation}\label{eq:chargexi1}
    H_{\xi_1} = \frac12\qty(N^V_V-N^U_U)\eval_{U=V=0} = \frac{\rho_h^2}{4 G},
\end{equation}
which is the area of the horizon in Planck units. Note that, since $\xi_1$ is the generator of boosts, this is nothing else than the standard
result that the Noether charge associated with the boost is the area \cite{Wald:1993nt}
The translation charges are slightly more involved but still straight forward.
Let us look at $t_U = H_{\frac12(\xi_4-\xi_5)}$. For a generic point that is not on the horizon, we find
\begin{equation}
    t_U = \frac{\rho f(\rho)}{16 G \kappa U}-\frac{\rho^2}{8 G U}\qty(\frac{f'(\rho)}{2\kappa}-1).
\end{equation}
To see that this vanishes at the bifurcating horizon, it suffices to first take the limit $V\mapsto 0$ for which
\begin{equation}
   \lim_{V\to0} f(\rho) = 0, \quad \lim_{V\to 0}f'(\rho) = 2\kappa.
\end{equation}
A similar argument applies to $t_V$. Hence, for a corner located at the bifurcation horizon,
the only non-vanishing $\mathrm{ECS}$ charge is \eqref{eq:chargexi1}.\par
We can now construct the moment maps. For diffeomorphism charges, specifying a moment map amounts to choosing a one-to-one correspondence between the vector fields
\eqref{eq:ecsvectorfieldsXT} and the abstract $\mathrm{ecs}$ generators $D,K,H,X,P$. At first sight,
there are of course several choices that satisfy the equivariance condition.
There is, however, a natural way to single out a preferred one. The moment maps are local objects and as such, they are blind to global properties of the group.
Looking at the $\spl{2}$ triplet in \eqref{eq:ecsvectorfieldsXT}, we observe that there is only one compact generator $\xi_2$.
It is therefore natural to associate this diffeomorphism with the compact generator in the abstract algebra---$L_0$. The equivariance condition
then forces\footnote{There is of course the usual sign freedom in the definition of $\tilde{\mu}(D)$ which corresponds to switching the role of $K$ and $H$.}
\begin{equation}\label{eq:classicalcomomentmapD}
    \tilde{\mu}(D) = H_{\xi_1} = \frac{\rho_h^2}{4 G}.
\end{equation}
All the other comomentum maps vanish.\par We can now write the moment map for the SSS spacetime with metric $g_{\mathrm{SSS}}$, defined by the line element \eqref{eq:sssspacetime}, and for the embedding $\phi_0$ placing the corner at the bifurcating horizon:
\begin{equation}
    \mu(g_{\mathrm{SSS}},\phi_0) = \frac{\rho_h^2}{4G}\, d^*,
\end{equation}
where $d^*$ is the covector dual to $D$. This point lies on the ECS coadjoint orbit characterized by
a vanishing value of the Casimir function \eqref{eq:casimirfunctionsl2recs}. It would be interesting to study how the coadjoint action moves this
point within the orbit, and to relate this motion to variations of the metric and embedding via the equivariance condition \eqref{eq:momentmapequivariance}.
For now, we turn to the quantum description.
\subsection{Quantum observables and the emergence of the area law}\label{subsec:quantumclassicalarealaw}
In order to connect the results of the previous subsection to the quantum theory, we need to compute the Berezin symbols. In order to do so, we note that
the QCS operators can be written as tensor products
\begin{align}
    D &= \ops{D}\otimes \mathds{1} + \mathds{1} \otimes \frac{XP + P X}{4 c},\\
        K &= \ops{K}\otimes \mathds{1} +\mathds{1} \otimes \frac{X^2}{2c},\\
    H &= \ops{H}\otimes \mathds{1} + \mathds{1} \otimes \frac{P^2}{2c},\\
    X &= \mathds{1}\otimes X,\\
    P &= \mathds{1}\otimes P.
\end{align}
The expectation value of an operator can then be computed independently for the $\spl{2}$ side and the Heisenberg side. The expectation values of $X,P$ and their quadratic
combination can be computed using the wavefunction \eqref{eq:squeezedwaavefunction} and the differential representation
\begin{equation}
    X \phi_{\zeta,\alpha}(p) = i \partial_p c \phi_{\zeta,\alpha}(p), \quad P \phi_{\zeta,\alpha}(p) = p\phi_{\zeta,\alpha}(p).
\end{equation}
The expectation values of the $D^{(S)},K^{(S)},H^{(S)}$ where already given in \eqref{eq:expectationvaluedkhinzeta}. Working in the negative discrete series corresponding to the orbit
$\mathsf{O}_{\mathrm{QCS}}^{(\lambda,-,c)}$, we get
\begin{align}
    \expval{D}_{\zeta,\alpha} &= -\frac{r\sin\theta}{2(1-r^2)}(4\lambda -1) + \Im\alpha \Re\alpha,\label{eq:berezinD}\\
    \expval{K}_{\zeta,\alpha}  &= -\frac{\tilde{\Delta}}{4(1-r^2)}\qty(4\lambda - 1) + \Re\alpha^2,\\
    \expval{H}_{\zeta,\alpha} &= -\frac{\Delta}{4(1-r^2)}\qty(4\lambda-1) + \Im\alpha^2,\\
    \expval{X}_{\zeta,\alpha} &= \sqrt{2c} \Re\alpha,\\
    \expval{P}_{\zeta,\alpha}&= \sqrt{2c}\Im\alpha.
\end{align}
Following the quantum-classical correspondence \ref{subsec:quantumclassicalcorr}, we need to choose QCS coherent state such that, in the $c\to 0$ limit, the above expectation values
become the classical comomentum maps. This can be done in the following way. For the translations, it suffices to pick a value of $\alpha$ that does not depend on $c$. The expectation values of
$X$ and $P$ then automatically vanish in the $c\to 0$ limit\footnote{One could also accommodate for non-vanishing translation charges by choosing $\alpha$ as an appropriate
function of $c$. Note that this can always be done since $c$ is dimensionless in our construction.}. Now, for any $\zeta = re^{i\theta}$, we can choose the following Glauber state
\begin{equation}
    \alpha_0 = \frac12\sqrt{\frac{4\lambda-1}{(1-r^2)}}\qty(\sqrt{\tilde{\Delta}} + i \sqrt{\Delta}),
\end{equation}
so that $\expval{K}_{\zeta,\alpha_0} = \expval{H}_{\zeta,\alpha_0} = 0$. Plugging this value of $\alpha$ back into equation \eqref{eq:berezinD} gives
\begin{equation}\label{eq:expvalD0}
    \expval{D}_{\zeta,\alpha_0} = \qty(4\lambda-1)\frac{\sqrt{1-2r^2 \cos(2\theta) + r^4}}{4(1-r^2)}. 
\end{equation} 
In particular, for any state with $\theta = -\pi$, we find the surprisingly simple result
\begin{equation}
    \expval{D}_{\zeta = -r,\alpha_0} = \lambda-\frac14.
\end{equation}
Reinstating the units of $\hbar$, we thus identify the representation parameter with the classical comoment map \eqref{eq:classicalcomomentmapD}
\begin{equation}\label{eq:beautiful}
    \lambda -\frac14 = \frac{\rho_h^2}{4G\hbar}.
\end{equation}\par
Let us take a moment to appreciate the beauty of this result. We started with QCS coherent states,
an abstract mathematical construction only based on the corner symmetry group, and argued
that the classical limit was defined in terms of the abstract representation parameter. Namely that the quantum state would concentrate on their classical values when $\lambda$
is very large. With the identification \eqref{eq:beautiful}, we observe that this classical limit is exactly the one expected of quantum gravity, where the Planck length is very small compared to the
macroscopic length scale of the problem
\begin{equation}
    \rho_h \gg l_p.
\end{equation}
Of course, the identification \eqref{eq:beautiful} holds for all values of $\lambda$. As mentioned earlier, however, Berezin quantization is only valid for
for $\lambda > \tfrac12$. Combining this bound with \eqref{eq:beautiful} then yields\footnote{While the inequality \eqref{eq:fundamentalinequality} is derived for $\theta = -\pi$ states, it is easy to see that this still holds in the general case, since the multiplicative factor on the right hand side
of \eqref{eq:expvalD0} is always greater or equal to one.}
\begin{equation}\label{eq:fundamentalinequality}
    \rho_h > l_p\,.
\end{equation}
In other words, within this construction the horizon scale $\rho_h$ cannot be taken below or at the Planck length.
This lower bound emerges from intrinsically quantum input---namely the existence of the coherent-state quantization of the QCS orbits---rather
than from semiclassical reasoning. In that sense, the corner proposal provides concrete support for the widely held expectation that $l_p$
sets a fundamental minimal length scale, beyond which the notion of sharply localized geometric data ceases to be operationally meaningful.
A similar results exists in string theory \cite{AmatiCiafaloniVeneziano1989}.
However, we emphasize that this bound arises as a quantization condition imposed on the underlying classical theory. In the spirit of the corner proposal,
one should keep in mind that there may
exist quantum states which do not result from quantizing any known classical theory of gravity, and which could therefore violate this bound.
\par
There is another interesting implication of the identification \eqref{eq:beautiful} related to the presence of the universal cover of $\spl{2}$ in the QCS.
As mentioned earlier, representations of $\widetilde{\mathrm{SL}}(2,\mathbb{R})$
that do not descend to $\spl{2}$---i.e. with $\lambda\notin\mathbb{Z}$---may not have a classical analogue, similarly to fermions in particle physics.
Restricting to representations of the special linear group,
the identification \eqref{eq:beautiful} implies that the corner area is quantized in discrete increments of the Planck area. Of course, this is a purely quantum effect.
Classically, the hyperboloids $\mathsf{O}_{\spl{2}}^{\lambda,\pm}$ admit a continuous ``radius'', and the discreteness arises only through the coadjoint orbit
quantization condition
\eqref{eq:orbitquantizationcondition}. This suggests an intriguing interpretation of the corner proposal: there may exist two classes of quantum spacetimes
---in analogy with the boson/fermion dichotomy in particle physics---such that only those admitting a classical geometric description exhibit area quantization
in the quantum theory.
\par
Let us now move to the entanglement entropy discussion. We consider the two subregions as the interior and exterior of the horizon, i.e. the regions I and II in Figure \ref{fig:ksdiagram}.
The boundary of the subregions is exactly the horizon defined by the $X=T$. The symplectic structure is defined on a constant $t$ slice and the corner is given
by the bifurcating horizon $X=T=0$. Now let us consider that the corner is in the classical state \eqref{eq:classicalstate}. In the semiclassical limit $\rho_h \gg l_p$, the entanglement entropy between
region I and II is then given by \eqref{eq:asymptoticentropyclassicalstate}
\begin{equation}\label{eq:arealaw}
    S^{\mathrm{cl}}_{\mathrm{I}\slash \mathrm{II}} \propto \frac{\rho_h^2}{4 l_p^2} + \frac12 \ln(\frac{\rho_h^2}{4 l_p^2}) + ...\,,\quad \rho_h \gg l_p,
\end{equation}
where the dots denote subleading terms in $\rho_h\slash l_p$. The first term is the Bekenstein--Hawking area law and the second term is a logarithmic quantum correction, which is
expected in most semiclassical gravity theories. Note that the overall proportionality coefficient comes from the asymptotic behavior of the entanglement entropy \eqref{eq:asymptoticentropyclassicalstate}.
Of course, one can always define a classical state $\psi^\mathrm{cl}(s \lambda)$, where $s \in \R$ will then modify the overall coefficient.
However,  the one-half coefficient in front of the logarithmic correction is universal, i.e. independent of $\lambda$. Although not straightforward,
it may be illuminating to compare this coefficient to the ones obtained in the Euclidean path integral formalism \cite{Harlow:2018tqv,Hernandez-Cuenca:2024icn} or to
interpret it as describing the field content of the corner proposal in the vein of \cite{Banerjee:2010qc,Sen:2012cj,Sen:2012kpz}.\par
Equation \eqref{eq:arealaw} is the central result of this thesis, and arguably the most significant outcome of the corner proposal to date.
It provides a quantum-gravitational perspective
on the area law, interpreting it as arising from the entanglement entropy between the spacetimes inside and outside the horizon.


\section{Going Further}\label{sec:goingfurther}
The fact that the corner proposal reproduces the area law from intrinsically quantum-gravitational
considerations provides strong support for the framework.
This result, and the formalism developed to obtain it, open up a wide range of interesting questions.
In this section, we offer a few preliminary considerations regarding some of these questions. We start by giving some additional classical states in Subsection \ref{subsec:moreclassicalstate}.
In Subsection \ref{subsec:ucs} we discuss the alternative perspective to start with the UCS and interpret the vanishing of the trace charge as a constrain on the quantum system. Finally,
in Subsection \ref{subsec:3D}, we present the starting point of the corner proposal in three dimensions.
\clearpage
\subsection{More classical states}\label{subsec:moreclassicalstate}
Since the emergence of the area law from the corner proposal relies on the classical states \eqref{eq:classicalstate}, one might ask whether there exist other states
with the same property in the QCS representations. In this subsection, we show that there is a large family of dressed states that also produce the crucial linear term in the entanglement entropy
for large $\lambda$. For simplicity, we only consider the special linear side.\par
We remind the reader that the classical state can be written as the operator $D$ acting on the vacuum by an amount $2\lambda$.
In a similar fashion, we can define the excited classical states
\begin{equation}
    \ket{\lambda_n} = e^{2is D}\ket{n}.
\end{equation}
Let us define the dressed rotation operator
\begin{equation}
    L_0^{(\lambda)} \defeq  U(\lambda) R U(\lambda)^\dag,
\end{equation}
Using the Campbell identity, we can compute
\begin{equation}
    L_0^{(\lambda)} = \frac12\qty(e^{-2\lambda}K +  e^{2\lambda} H).
\end{equation}
We find that $L_0^{(\lambda)} = c H + b D +a K$, with $\Delta = b^2-4ac < 0$, which means that $L_0^{(\lambda)}$
is an elliptic operator that admits a discrete spectrum. In fact, the eigenstates of $L_0^{(\lambda)}$ are exactly the excited classical states
\begin{Align}
    L_0^{(\lambda)} \ket{\lambda_n} &= r_n \ket{\lambda_n}.
\end{Align}
Moreover, we can define dressed ladder operators
\begin{align}
    L_\pm^{(\lambda)} \defeq U(\lambda) L_\pm U(\lambda)^\dag.
\end{align}
It is then easy to compute
\begin{Align}
    L_\pm^{(\lambda)} \ket{\lambda_n} &= U(\lambda) L_\pm \ket{n}\\
    &= \sqrt{r_n(r_n\pm1)-\lambda(\lambda-1)}U(\lambda)\ket{n\pm 1}\\
    &= \sqrt{r_n(r_n\pm 1)-\lambda(\lambda-1)} \ket{\lambda_{n\pm1}}.
\end{Align}
One can now define dressed coherent states
\begin{equation}
    \ket{\lambda_\zeta} = e^{c_\zeta L^s_+ -\bar{c_\zeta}L^s_-}\ket{\lambda_0}.
\end{equation}
Since the action of the dressed operators on the $\ket{\lambda_n}$ state coincides with the action of the original operators on the $\ket{n}$ states, we immediately deduce
\begin{equation}
    \ket{\lambda_\zeta} = \qty(1-\abs{\zeta}^2)^{\lambda}\sum_n \qty(\frac{\Gamma(n+2\lambda)}{\Gamma(2\lambda)\Gamma(n+1)})^\frac12\zeta^n \ket{\lambda_n}.
\end{equation}
How is this state written in the gluing basis $\ket{E}$? For a general state $\ket{\psi}$, define the wave function $\psi(E) = \braket{E}{\psi}$. The dilation operator then acts as \cite{deAlfaro:1976vlx}
\begin{equation}\label{DactiononE}
    D\psi(E) = i\qty(E\partial_E+\frac12)\psi(E).
\end{equation}
Let us then look at the wave function
\begin{equation}\label{defwavefunction}
    \psi_n(\lambda;E) \defeq \braket{E}{\lambda_n}.
\end{equation}
Using equation \eqref{DactiononE}, we find
\begin{equation}\label{pdvequationsE}
    \pdv{\lambda}\psi_n(\lambda;E) = -\qty(2E \partial_{E} + 1)\psi_n(\lambda;E).
\end{equation}
We can solve this equation using the method of characteristics: Look for solutions of this equation on curves $E(\lambda)$. We have
\begin{equation}
    \dv{\lambda} \psi_n(\lambda;E(\lambda)) = \pdv{\lambda}\psi_n(\lambda;E(\lambda)) + \dv{E(\lambda)}{\lambda}\pdv{E}\psi_n(\lambda;E(\lambda)).
\end{equation}
On the curve $E(\lambda)$ defined by
\begin{equation}\label{curveeq}
    \dv{E(\lambda)}{\lambda} = 2 E(\lambda),
\end{equation}
equation \eqref{pdvequationsE} becomes
\begin{equation}\label{odepsis}
    \dv{\lambda}\psi_n(\lambda;E(\lambda)) = -\psi_n(\lambda;E(\lambda))
\end{equation}
The original partial differential equation is then reduced to a system of ordinary differential equations. Equation \eqref{curveeq} is simply solved by
\begin{equation}\label{solutionEs}
    E(\lambda) = E_0 e^{2\lambda}
\end{equation}
Equation \eqref{odepsis} is solved by
\begin{equation}
    \psi_n(\lambda;E(\lambda)) = \psi_n(0,E_0) e^{-\lambda}
\end{equation}
Looking back at the definition of the wave function \eqref{defwavefunction}, we see that 
\begin{equation}
    \psi_n(\lambda=0;E) = C_n(E),
\end{equation}
where $C_n(E)$ is the wave function of the non-dressed state
\begin{equation}
    C_n(E) = 2^{\lambda}\qty(\frac{\Gamma(n+1)}{\Gamma(2\lambda+n)})^\frac12  E^{\lambda-\frac12}e^{- E}L_n^{2\lambda-1}(2 E).
\end{equation}
The solution to equation \eqref{pdvequationsE} is therefore
\begin{equation}
    \psi_n(\lambda,E) = e^{-\lambda} C_n(E e^{-2\lambda}),
\end{equation}
where we have used equation \eqref{solutionEs}.We can then write
\begin{align}
    \ket{\lambda_\zeta} &= \qty(1-\abs{\zeta}^2)^{\lambda}\sum_n \int \dd E\, \qty(\frac{\Gamma(n+2\lambda)}{\Gamma(2\lambda)\Gamma(n+1)})^{\frac12}\zeta^n e^{-\lambda} C_n(E e^{-2\lambda})\ket{E}\\
    &= \frac{\qty[t \qty(1-\abs{\zeta}^2)]^\lambda}{\sqrt{\Gamma(2\lambda)}}\int \dd E\, e^{-\frac{tE}{2}}E^{\lambda-\frac12}\sum_n\zeta^n L_n^{2\lambda-1}(tE)\ket{E},
\end{align}
where we denoted $t = 2 e^{-2\lambda}$. We can use the Laguerre generating function
\begin{equation}
    \sum_n \zeta^n L^{2\lambda-1}_n(t E) = \frac{1}{(1-\zeta)^{2\lambda}}e^{-\frac{\zeta t E}{1-\zeta}},
\end{equation}
to write
\begin{equation}
    \ket{\lambda_\zeta} = \qty(\frac{1-\abs{\zeta}^2}{(1-\zeta)^2})^{\lambda} \frac{t^{\lambda}}{\sqrt{\Gamma(2\lambda)}}\int \dd E\,E^{\lambda-\frac12}\, e^{\frac{E t}{2} \frac{\zeta+1}{\zeta-1}}\ket{E}.
\end{equation}
One can compute the norm of the wave function
\begin{equation}
    \abs{\psi_\zeta^\lambda(E)}^2 = \abs{\braket{\lambda_\zeta}{E}}^2 =P_{r}^{2\lambda}(\theta) \frac{t^{2\lambda}}{\Gamma(2\lambda)}\, E^{2\lambda-1}\, e^{-t E P_r(\theta)},
\end{equation}
where, for $\zeta = r e^{i\theta}$, $P_r(\theta)$ is the Poisson kernel \eqref{eq:poissonkernel}.
Comparing this norm to the norm of the non-dressed coherent states
\begin{align}
    \abs{\psi^\lambda_\zeta(E)}^2 
    &=\frac{\qty(2 P_{r}(\theta))^{2\lambda}}{\Gamma(2\lambda)}\,e^{-2\lambda}\,\qty(E e^{-2\lambda})^{2\lambda-1}\,e^{-2 E e^{-2\lambda}P_{r}(\theta)} = e^{-2\lambda}\abs{\psi_\zeta(Ee^{-2\lambda})}^2,
\end{align}
we can deduce the entanglement entropy of the dressed states
\begin{align}
    S^\zeta_\lambda &= -\int \dd E\, \abs{\psi^\lambda_\zeta(E)}^2 \ln\!\qty(\abs{\psi^\lambda_{\zeta}(E)}^2)\\
    &= - \int \dd E \, e^{-2\lambda} \abs{\psi_\zeta(Ee^{-2\lambda})}^2\qty[\ln\!\qty(\abs{\psi_\zeta(Ee^{-2\lambda})}^2)- 2\lambda]\\
    &= - \int \dd E\, \abs{\psi_\zeta(E)}^2 \qty[\ln\!\qty(\abs{\psi_\zeta(E)}^2) - 2\lambda]\\
    &= S^{\zeta} + 2 \lambda.
\end{align}
Therefore the family of dressed coherent states $\psi_\zeta^\lambda$ also have a linear term in $\lambda$ in their entanglement entropy and will therefore also produce an area term in the semi-classical limit.
Of course, these dressed coherent states are still coherent states. Indeed, the coordinates of the dressed coherent states in the $\ket{\lambda_n}$ basis are the same as the standard coherent states in the $\ket{n}$ one.
We can thus write
\begin{align}
    \ket{\lambda_\zeta} 
    &= \sum_n \braket{\zeta}{n} \ket{\lambda_n}\\
    &= U(\lambda)\sum_n \braket{\zeta}{n}\ket{n}\\
    &= U(\lambda) \ket{\zeta}.
\end{align}
Since the family of coherent states is invariant under the action of the group \eqref{eq:gropactiononcoherentstates}, we can write
\begin{equation}
    \ket{\lambda_\zeta} \sim \ket{e^{2i\lambda D} \rhd \zeta},
\end{equation}
where $\sim$ denotes equality up to a phase and $\rhd$ is the action on the coset which, in this specific case, is implemented by the Möbius transformation.
It is easy to see what this transformation does by computing the Berezin symbols of the conformal basis in these new dressed states. These follow simply 
from 
\begin{align}
    e^{-2i\lambda D}He^{2i\lambda D} &= e^{2\lambda}H,\\
    e^{-2i\lambda D}K e^{2i\lambda D} &= e^{-2\lambda}K.
\end{align}
So the $h$ (respectively $k$) coordinate gets multiplied by $e^{2\lambda}$ (respectively $e^{-2\lambda}$) compared to the non-dressed coherent states.
Of course the orbit is the same since the product of $h$ and $k$ is invariant under such transformation. In the classical limit $\lambda\to\infty$, the dressed states therefore suppress the
value of the $K$ charge and correspondingly scale the value of the $H$ charge while staying on the same hyperboloid.
\subsection{The corner proposal for the UCS}\label{subsec:ucs}
In this thesis we have used the ECS rather than the UCS because the classical theories we are aware of realize only the
traceless part of $\mathrm{GL}(2,\mathbb{R})$, i.e. $\spl{2}$. There is, however, an alternative perspective on the fate of
the trace charge in $\mathrm{GL}(2,\mathbb{R})$. One may regard the trace charge as present, but such that it happens to vanish in specific cases, for
instance in spherically symmetric Einstein--Hilbert gravity. The appropriate quantum description would then be formulated
in terms of $\mathrm{UCS}$ representations, while restricting to those points on the coadjoint orbits for which the trace charge vanishes.
These two approaches may be viewed as either imposing the vanishing-trace constraint at the classical level and then quantizing,
or quantizing the full system and imposing the constraint only afterwards.
\par
The $\mathrm{UCS}$ is different from its traceless version in a few different ways. First, it does not admit any non-trivial central extension.
Intuitively, this is because the central extension in the translation sector measures the area spanned by the two translation vectors in $\mathbb{R}^2$.
While $\spl{2}$ transformations preserve this area, the trace part of $\mathrm{GL}(2,\mathbb{R})$ acts by dilations and therefore rescales it.
Concretely, the Jacobi identity involving the trace generator $W$ and the two normal translations forbids a central extension between the latter.
A second key difference is that it has no Casimir invariants.
Consequently, its coadjoint orbits can be six-dimensional, as well as four- or two-dimensional \cite{Ciambelli:2022cfr}. On the quantum side, the representations can be
obtained via Mackey's induced representation theory. It turns out that there exists only one unitary irreducible representation \cite{Milad2023}
which can be expressed on functions $\psi(\vec{p},t)$, where $
\vec{p}\in \R^2\backslash \qty{\vec{0}}$, and $t\in \R^*$.
The scalar product is written with respect to the Haar measure $\dd\mu(t)$ on $\R^*$ seen as a group under
multiplication, and the Haar measure $\dd \mu(\vec{p})$ on the group
\begin{equation}
  K = \qty{\mqty(s&-t\\t&s): s,t \in \R, s^2+t^2>0},
\end{equation}
which is isomorphic to the orbits of $\gl{2}$ on $\R^2$ described by $\vec{p}$, trough the following identification
\begin{Align}
\gamma: \mathcal{O} &\rightarrow K,\\
\vec{p} &\mapsto \gamma(\vec{p}) = \frac{1}{\norm{\vec{p}}}\mqty(p_1 &-p_2\\p_2 &p_1).
\end{Align}
The invariant left (right) measure on $R^*$ is given by
\begin{equation}
  \dd\mu(t)= \frac{\dd t}{\abs{t}},
\end{equation}
and on $K$ by\footnote{Note that under a left group multiplication by the element
\begin{equation}
g= \mqty(s&-t\\t&s),
\end{equation}
we get $p_1' = s p_1 - tp_2, p_2' = tp_1 + s p_2$. The Jacobian gives
$\dd p_1'\dd p_2' = \det(g) \dd p_1 \dd p_2 = (s^2 + t^2)\dd p_1 \dd p_2$. Finally, one easily computes
$(p_1')^2 + (p_2')^2 = (s^2 + t^2)(p^2)$, hence the invariance of the measure. 
}
\begin{equation}
  \frac{\dd p_1 \dd p_2}{p^2}.
\end{equation}
The wave functions are $L^2$ with respect to this measure:
\begin{equation}\label{eq:innerproduct}
  \int_{\R^*} \frac{\dd t}{\abs{t}} \int_{\R^2\backslash \qty{\vec{0}}}\frac{\dd p_1 \dd p_2}{p^2} \, \abs{\psi(\vec{p},t)}^2 <\infty.
\end{equation}
Given an element $g$ of the UCS group written as $g=A \cdot x$ with $A\in \gl{2}$ and $\vec{x}\in \R^2$, the representation is written \cite{Milad2023}
\begin{equation}\label{eq:grouprepresentation}
 \qty( U_g \psi)(\vec{p},t) = \frac{\det(A)^\frac12 \norm{\vec p}}{\norm{\vec{p} A}}e^{2\pi i (\vec{p}\cdot \vec{x} + t^{-1}u_{p,A})}\psi(p A,v^{-1}_{p,A} t),
\end{equation} 
where $\vec{p}A$ is the matrix product between the 1x2 matrix $\vec{p}^T$ and the 2x2 matrix $A$
\begin{equation}
  \vec{p} A = \qty(p_1,p_2)\mqty(a&b\\c&d) = \qty(a p_1 + cp_2,bp_1 + d p_2),
\end{equation}
and where we defined
\begin{align}
 u_{p,A} &= \frac{(ac+bd)(p_1^2 - p_2^2)-(a^2 + b^2 -c^2 -d^2)p_1 p_2}{(ap_1 + cp_2)^2 + (bp_1 + d p_2)^2},\\
 v_{p,A} &= \frac{(ad-bc)(p_1^2 +p_2^2)}{(a p_1 + c p_2)^2 + (b p_1 + dp_2)^2}.
\end{align}
Using the matrix representation of the algebra generators, we can find their action on the wave function. Since the representation is defined for real matrix, we compute
\begin{equation}
\dv{\alpha}\qty(U_{e^{\alpha X}}\psi)\eval_{\alpha = 0}
\end{equation}
to obtain anti-hermitian generators, and then multiply the resulting operators by $i$ to obtain Hermitian operators.
For the $\gl{2}$ operators, we find
\begin{Align}\label{eq:ucsgeneratoraction}
  D\psi(\vec{p},t) &= \frac{i}{2}\qty(2t \frac{p_1^2 - p_2^2}{p^2}\partial_t + p_1 \partial_1 - p_2 \partial_2 + \frac{p_2^2 - p_1^2}{p^2}-i \frac{8\pi p_1 p_2}{tp^2})\psi(\vec{p},t) \\
  H \psi(\vec{p},t) &= -i\qty(2 \frac{tp_1 p_2}{p^2} \partial_t + p_1 \partial_2 - \frac{p_1 p_2}{p^2}+i 2\pi \frac{p_1^2 -p_2^2}{t p^2})\psi(\vec{p},t)\\
  K\psi(\vec{p},t) &= i\qty(2 \frac{t p_1 p_2}{p^2}\partial_t + p_2 \partial_1 - \frac{p_1 p_2}{p^2} + i 2 \pi \frac{p_1^2-p_2^2}{t p^2})\psi(\vec{p},t)\\
  W \psi(\vec{p},t) &=i (p_1 \partial_1 + p_2 \partial_2)\psi(\vec{p},t),\\
    P_a \psi(\vec{p},t) &= -2 \pi p_a \psi(\vec{p},t).
\end{Align}
One can verify that those operators are Hermitian with respect to the inner product \eqref{eq:innerproduct}. One can construct Perelomov coherent states by using the following
normalized fiducial state
\begin{equation}\label{eq:ucsfiducialstate}
    \psi_0(p,t) = \qty(\frac{p}{\sqrt{2\pi}\sigma_p})\qty(\frac{1}{(2\pi)^\frac14 \sqrt{\sigma_t}})\exp(-\frac{\norm{\vec{p}- \vec{p}_0}^2}{4\sigma_p^2})\exp(- \frac{\ln(\abs{t/t_0})^2}{4 \sigma_t^2}),
\end{equation}
where $p=\norm{\vec{p}}$. From the group representation \eqref{eq:grouprepresentation},
it is straightforward to see that the stabilizer is trivial. Indeed, there is no group element whose action can produce a phase that is independent of $\vec{p}$ and $t$.
Using the standard isomorphism of coadjoint orbits \eqref{eq:coadjointisomorphism}, it is therefore clear that the family of coherent states obtained by acting with the group on \eqref{eq:ucsfiducialstate}
describes the $6$-dimensional coadjoint orbit of the UCS.\par
Let us make two additional remarks about these representations and the choice of fiducial state. First, it is clear from \eqref{eq:ucsgeneratoraction}
that the expectation value of $W$ vanishes for any real-valued wave function, such as the fiducial state. Hence, any real-valued wave function automatically enforces
the vanishing of the trace charge at the classical level, and can therefore describe the known classical theories for which only the special linear part is active.
Second, the action of the $\mathrm{ECS}$ Casimir is particularly simple
\begin{equation}
    \mathcal{C}_{\mathrm{ECS}} \psi(\vec{p},t) = - 4\pi^3 \frac{p^2}{t}\,\psi(\vec{p},t).
\end{equation}
This makes it natural to take the translations together with $\mathcal{C}_{\mathrm{ECS}}$ as a CSCO, and the glued state will thus take the form
\begin{equation}
    \psi_G(\vec{p},t) = \psi_L(\vec{p},t) \otimes \psi_R(\vec{p},t).
\end{equation}
In principle, one could now compute the entanglement entropy and the Berezin symbols for a general coherent state.
Identifying the resulting expectation values with the classical charges would then make
it possible to determine which states correspond to the SSS solution, and which representation-theoretic parameters should be identified with the area.
Of course, the association is not as straightforward as in the QCS case,
since the orbits are not K\"ahler in general and the results of Section~\ref{sec:mathematicalbackground} therefore do not apply directly.

\subsection{The corner proposal in three dimensions}\label{subsec:3D}
One of the most natural directions for investigating the corner proposal is to reintroduce corner diffeomorphisms by going one dimension higher.
In a similar vein to our discussion of spherically symmetric spacetimes, the Hilbert space associated with the three-dimensional corner
symmetry group could then be related to cylindrically symmetric spacetimes. As we will see shortly, passing from two to three
dimensions introduces a number of new technical difficulties. We note, however, that the subsequent step from three to
four dimensions does not appear to introduce comparably many new complications. This stands in interesting contrast with other approaches to quantum gravity,
where three dimensions is essentially under control \cite{Witten:1988hc}, yet the genuine leap occurs in going to four dimensions.\par
In the three-dimensional case, the corner becomes a circle. The corresponding extended corner symmetry group is then
\begin{equation}
    \mathrm{ECS}_3 = \mathrm{Diff(S^1)}\ltimes\qty(\spl{2} \ltimes \R^2)^S.
\end{equation}
The first step in generalizing the corner proposal to that case is to compute the possible central extensions. As we will show in just a bit,
the maximally centrally extended version of the algebra can be written
\begin{equation}\label{eq:qcs3}
    \mathfrak{qcs}_3 = \mathrm{vir} \ltimes \qty(\widehat{L\mathfrak{sl}(2,\R)} \ltimes L \R^2),
\end{equation}
where the $L$ in front of an algebra denotes the corresponding loop algebra $L\mathfrak{a} = C^\infty(S^1,\mathfrak{a})$ and the hat denotes the centrally extended Kac-Moody version \cite{Moody1967,Moody1968,Kac1968,Kac1990}.
The absence of a central extension in the translational sector follows from the fact that the translation generators have conformal weight one.
This suggests that the central
extension in the two-dimensional case is accidental, much as the two-dimensional Poincar\'e group also admits a central extension. We note that the non-extended version of the corner symmetry group
---without the normal translations---can be represented by a $\spl{2}$ WZW model 
\cite{WessZumino1971,Witten1984WZW,KnizhnikZamolodchikov1984,Gawedzki1991NoncompactWZW,Teschner1999StructureConstants,Teschner2000OPE,MaldacenaOoguri2001I,MaldacenaOoguriSon2001II,MaldacenaOoguri2002III}
. This would also be the case for the extended version
if the $L\R^2$ algebra was centrally extended to  $L\mathfrak{h}_3$. This could be achieved by changing the conformal weights of the translation generators.
Although, at the time of writing, we are not aware of a compelling physical motivation for doing so, it is nevertheless interesting that the resulting group
would admit representations closely paralleling those of conformal field theories. It would in turn be fascinating to investigate the connections of the corner proposal with the 
Kerr/CFT correspondence \cite{Castro:2010fd,Bredberg:2011hp}.\par
Describing the full representation theory of the group associated with the three-dimensional quantum corner algebra \eqref{eq:qcs3} is a very challenging task.
There are, however, three possible lines of attack that may shed some light on the issue. First, as already mentioned, the structure of the algebra is reminiscent of a CFT.
The extensive literature on the subject may therefore contain useful insight into the construction of its representations. The other two are simply the theory of induced representations and the
orbit method. While the former does not transfer verbatim to the case of infinite dimensional group, some progress has been made in its generalizations \cite{KOSYAK20143395,wolf2012principalseriesrepresentationsinfinite}.
The orbit method for the higher-dimensional corner symmetry groups was started in \cite{Ciambelli:2022cfr} using the theory of Lie Algebroids. In that case, there is of course
the added complication of the quantization of the orbits.\par
We now prove that the maximally centrally extended version of the corner symmetry algebra in three dimensions is given by \eqref{eq:qcs3}. In this subsection
we denote the generator of $\spl{2}$ by $J_0,J_\pm$ and reserve the $L$ symbol for the generator of the Witt algebra $\mathfrak{diff}(S^1)$.
Since the $\spl{2}$ and $\R^2$ generators are function on the circle, we can decompose them into their Fourier modes.
\begin{equation}
    V_a(\theta) = \sum_{n=-\infty}^\infty V^n_a e^{in\theta},
\end{equation}
were $V^a,a= 1,...,5$ collectively denote the generators of the $\mathrm{ECS}_2$.
The Witt algebra generators can be written in the same basis as
\begin{equation}
L^m =  e^{in\theta}\pdv{\theta}.
\end{equation}
The commutation relations between the Witt algebra and the $L\mathfrak{ecs}_2$ algebra is then 
\begin{Align}
     \qty[L^m,L^n] &= i(n-m)L^{n+m}\\
    \qty[L^n,V_a^{m}] &= im V_a^{m+n}.
\end{Align}
Additionally the non-vanishing internal commutator of the $\mathfrak{ecs}_2^S$ are
\begin{Align}
    [J^n_0,J_\pm^m] &= \pm i J_\pm^{n+m}, \quad [J_-^n,J_+^m] = 2iJ_0^{n+m}\\
    [J_\pm^{n},P^m_\mp] &= 
 \mp i P^{n+m}_\pm,\quad [J_0^n,P_\pm^m] = \frac{i}{2}P_\pm^{n+m}.
\end{Align}
In order to find all non-trivial central extensions we write the most general central terms
\begin{Align}
    \qty[L^n,L^m] &= i(n-m)L^{n+m} + W^{n,m},\\
    \qty[L^n,V^a_m] &=i m V_a^{n+m} + M_a^{n,m},\\
    \qty[V_a^n,V_b^m] &= C\downup{ab}{c}V_c^{n+m} + T_{ab}^{n,m},
\end{Align}
where $C\downup{ab}{c}$ are the structure constants of the $\mathfrak{ecs}_2$. First of all, it is easy to see that the only Jacobi identities involving
the central terms $W^{n,m}$ are the of the $L-L-L$ type. Therefore these will have the standard Virasoro form 
\begin{equation}
    W^{n,m} = \frac{\omega}{12}m(m^2-1)\delta_{n+m,0}.
\end{equation}

Next we compute the Jacobi identity of the type $L-V-V$ and find
\begin{equation}\label{eq:equation6}
   n\left(M_a^{k,n+m} - M_a^{m,n+k}\right) + (k-m)M_a^{k+m,n} = 0
\end{equation}
From which we can immediately deduce 
\begin{equation}\label{M0eqs}
\boxed{M_a^{n,0} = 0,\,\, \forall n} \quad \boxed{M_a^{0,m+n} = M_a^{m,n}\left(\frac{n+m}{n}\right), \,\, n\neq 0}
\end{equation}

The $L-V-V$ Jacobi identity further gives
\begin{equation}\label{MandTrel}
   \boxed{ C\downup{ab}{c} M_c^{k,n+m} = m T_{ab}^{n,m+k} + n T_{ab}^{n+k,m}}
\end{equation}
Since we are only interested in non-trivial cocycles, we can always add coboundaries to the cocycles which transform them as
 \begin{equation}
     M_a^{n,m} \mapsto M_a^{n,m} + m \lambda_a(m+n).
 \end{equation}
From \eqref{M0eqs} and the coboundary freedom, we get
\begin{Align}
    (M')_a^{0,m+n} &= (M')_a^{m,n}\qty(\frac{n+m}{n})\\
    &= \left(M_a^{m,n} + n\lambda_a(m+n)\right)\qty(\frac{n+m}{n}).
\end{Align}
Choosing $\lambda_a(m+n) = \frac{(M')_a^{0,n+m}}{n+m}$, gives
\begin{equation}\label{eqplug}
    M_a^{m,n} = 0, \forall  m,n \, \text{such that}\, m+n\neq 0.
\end{equation}
First, plugging this result back into \eqref{MandTrel} gives
\begin{equation}
    T_{ab}^{n,m} = f_{ab}(n)\delta_{n+m,0},
\end{equation}
where 
\begin{equation}
    f_{ab}(n) = - f_{ba}(-n).
\end{equation}
Second, let us define $h_a(n) = M_a^{n,-n}$. Substituting $k=-n-m$ into equation \eqref{eq:equation6} gives a functional equation for $h_a(n)$
\begin{equation}
    n \qty(h^a(-n-m)-h^a(m))+ (-n-2m)h^a(-n) = 0.
\end{equation}
Putting $n=1$ in the above equation gives a recursion for $h^a$ which is solved by
\begin{equation}
    h^a(n) = c^a n,
\end{equation}
where $c^a =-h^a(-1)$. Finally, we can use the coboundary freedom $M^{'a}_{n,-n} = M^a_{n,-n} - n c^a$ to remove this term. This shows
that there are no non-trivial central extensions between the Witt sector and the $\spl{2}$ one.
\par
Let us denote the $\spl{2}$ sector of the $\mathfrak{ecs}_2^S$ by greek letters $\alpha,\beta,...= 1,2,3$. In other words, we can write
\begin{equation}
    \qty[V_{\alpha}^n,V_{\beta}^m] =  C\downup{\alpha\beta}{\gamma} V_{\gamma}^{m+n},
\end{equation}
where $C\downup{\alpha\beta}{\gamma}$ are the structure constant of the $\mathfrak{sl}\qty(2,\R)$ algebra.
The central terms in that sector are written
\begin{equation}
    T_{\alpha\beta}^{n,m} = f_{\alpha\beta}(n) \delta_{m,-n},
\end{equation}
Furthermore, the $V-V-V$ Jacobi identity gives
\begin{equation}\label{sl2rJacobi}
    C\downup{\alpha\beta}{\delta} T_{\delta\gamma}^{m+n,k} + C\downup{\beta\gamma}{\delta}T_{\delta\alpha}^{n+k,m} + C\downup{\gamma\alpha}{\delta} T_{\delta\beta}^{k+m,n} = 0.
\end{equation}
Since the algebra is semi-simple, we can raise and lower the indices using the Cartan-Killing metric and we have the relation
\begin{equation}
    C\updown{\alpha}{\gamma\delta}C\downup{\beta}{\delta \gamma} = \delta_\beta^\alpha.
\end{equation}
Putting $m=k=0$ in equation \eqref{sl2rJacobi} and multiplying by $C\updown{\beta\alpha}{\sigma}$ we get
\begin{Align}
 0 &= T_{\alpha\beta}^{n,0} + \left(C\updown{\gamma\sigma}{\alpha}C\downup{\gamma\beta}{\delta}- C\updown{\delta\gamma}{\alpha}C\downup{\beta\gamma}{\sigma}\right) T_{\delta\sigma}^{n,0} \\
  &= T^{n,0}_{\alpha\beta}-C\downup{\alpha\beta}{\gamma}C\downup{\gamma}{\delta\sigma} T_{\delta \sigma}^{n,0},
\end{Align}
where we used Jacobi's identity on the structure constants. Next, we use the coboundary freedom 
$T^{n,0}_{\alpha\beta} \mapsto T_{\alpha\beta}^{n,0} - C\downup{\alpha\beta}{\gamma} \mu_\gamma(n)$, and choose
\begin{equation}
    \mu_\gamma(n) = C\downup{\gamma}{\delta\sigma}T_{\delta\sigma}^{n,0},
\end{equation}
which puts $T'^{n,0}_{\alpha\beta}$ to zero. Putting $k= 0$ in \eqref{sl2rJacobi} we now get
\begin{equation}
    C\downup{\beta\gamma}{\delta} T_{\delta\alpha}^{n,m} + C\downup{\gamma\alpha}{\delta} T_{\delta\beta}^{m,n} = 0.
\end{equation}
This means that $T_{\alpha\beta}^{n,m}$ is an invariant tensor under the adjoint representation of $\spl{2}$. Since $\spl{2}$ is simple, the only candidate is the Killing form
\begin{equation}
    T_{\alpha\beta}^{n,m} = K\left(V_{\alpha},V_{\beta}\right) f(n) \delta_{m,-n},
\end{equation}
where $f(n) = -f(-n)$.
Note that it is always possible to find a basis where
\begin{equation}
    K(V_{\alpha},V_{\beta}) = \eta_{\alpha\beta},
\end{equation}
where $\eta$ is the Minkowski metric of signature $(+,-,-)$. Using this form of the killing metric into \eqref{sl2rJacobi} ,we get the following equation
\begin{equation}
    \left(f(m+n) + f(n+k)+ f(k+m)\right)\delta_{n+k+m,0} = 0.
\end{equation}
Solving for $k$ using the delta function and choosing $m=1$ gives
\begin{equation}
    f(n+1) - f(n) = f(1).
\end{equation}
Using the fact that $f(0) = 0$ we finally get
\begin{equation}
    f(n) = \kappa n,
\end{equation}
Thus, the most general form of the central extension in the $\spl{2}$ sector is
\begin{equation}
    [V^n_{\alpha},V^m_{\beta} ]= C\downup{\alpha\beta}{\gamma}V_{\gamma}^{n+m} + n \, \kappa K(V_{\alpha},V_{\beta}) \delta_{m+n,0}. 
\end{equation}
Which is simply the affine Kac-Moody algebra $\hat{\mathfrak{sl}}\qty(2,\R)$.\par
Next, we look at the extensions between $\spl{2}$ and the translations. We use the $J_0,J_\pm$ basis of the special linear algebra and denote
\begin{Align}
    [J^n_0,P^m_+] &= \frac12 P_+^{m+n} + T_{0+}^{n,m},\,\,  [J^n_0,P^m_-] = -\frac12 P_-^{m+n} + T_{0-}^{n,m},\\
    [J^n_-, P_+^m] &= P_-^{n+m} + T_{-+}^{n,m},\,\, [J^n_+,P_-^m] = - P_+^{n+m} + T_{+-}^{n,m},\\
    [J^n_+, P_+^m] &= T_{++}^{n,m},\qquad [J^n_-,P_-^m] = T_{--}^{n,m}
\end{Align}
The $P_+,J_0,J_+$ Jacobi gives
\begin{equation}
    \left(\frac12 f_{++}(n)+ f_{++}(k+n)\right)\delta_{m+n+k,0} = 0\, \Rightarrow f_{++} = 0. 
\end{equation}
Similarly, the $P_-,J_0,J_-$ Jacobi gives
\begin{equation}
    f_{--} = 0.
\end{equation}
The $J_+,P_+,J_-$ Jacobi identity gives
\begin{equation}
    \left(2 f_{0+}(k+m) + f_{+-}(k)\right) \delta_{m+n+k} = 0,
\end{equation}
This implies that both $f_{0+}$ and $f_{+-}$ are constant. Since their value at $0$ must vanish, the constant is zero. The $J_-,P_-,J_+$ Jacobi identity gives the similar result
\begin{equation}
    f_{0-} = f_{-+} = 0.
\end{equation}
There are therefore no non-trivial central extension between the $\spl{2}$ sector and the translation sector.\par
We are left with translations. 
Using the Jacobi identity between two modes of $P_\pm$ and the zero mode of $J_0$ gives a vanishing central extension for two similar translations. Furthermore, using the Jacobi identity for $J_0,P_-,P_+$ and denoting
\begin{equation}
    \qty[P_-^n,P_+^m] = C(n) \delta_{n,-m},
\end{equation}
we get
\begin{equation}
    C(n+m)- C(m) = 0.
\end{equation}
Thus $C = cste$. The only constant that satisfies equation \eqref{MandTrel} is zero and the translations are not centrally extended.


\section{Conclusion}\label{sec:conclusion}
In this thesis, we have developed a symmetry-based approach to quantum gravity centered on the physics of spacetime corners. The guiding idea has been that, in the
presence of boundaries, diffeomorphisms can acquire non-vanishing Noether charges and therefore become physical symmetries. This shifts the focus from the traditional
bulk description of gravity to the algebraic and representation-theoretic structures carried by codimension-2 boundaries. In that sense, the corner proposal provides
a bottom-up framework for quantum gravity, in which the corner symmetry group plays an analogous role to the Poincaré group in relativistic physics.

The first main result of this thesis was to establish the classical symmetry framework underlying this proposal in Chapter \ref{chapter2}. In
Section~\ref{sec:covariantphasespace}, we used the covariant phase-space formalism, together with its extended version, in which the boundary embedding is treated
dynamically, to show how integrable gravitational charges arise at corners and how their algebra faithfully realizes the algebra of spacetime diffeomorphisms.
In particular, the analysis of spherically symmetric Einstein--Hilbert gravity in Section~\ref{sec:cornersymmetriesingravity} showed that the relevant physical symmetry
group reduces to the two-dimensional extended corner symmetry group introduced there. This provides the classical structure that we aim to recover in the semiclassical
limit of the quantum theory developed below.

The second main achievement was the complete representation-theoretic analysis of the corresponding two-dimensional corner symmetry group in Chapter \ref{chapter3}.
In Section~\ref{sec:quantumcornersymmetries}, we constructed its central extensions, identified the Casimir operators of the maximally centrally extended version,
and derived its irreducible unitary representations. This yields an explicit candidate family of Hilbert spaces for quantum gravity within the corner proposal.
Beyond the classification itself, 
this result gives concrete content to the idea that quantum gravitational states should be organized by corner symmetries in direct analogy with Wigner's classification
in relativistic quantum theory.

A third important result was the construction of the semiclassical bridge between these quantum representations and the classical corner charges.
To this end, we analyzed the coadjoint orbits of the quantum corner symmetry group. We related them to classical observables via twisted moment maps
and to quantum observables via generalized Perelomov coherent states and Berezin symbols, as reviewed in Section~\ref{sec:mathematicalbackground}
and constructed explicitly for corner symmetries in Section~\ref{sec:quantumcornersymmetries}.
This provides a precise correspondence between quantum and classical observables, relating the objects of Chapter~\ref{chapter2} to the representation-theoretic data.

The fourth central result concerns the description of local subsystems in quantum gravity. In Chapter~\ref{chapter4}, we showed that the
corner framework naturally accommodates a gluing prescription for spacetime subregions. Rather than a naive tensor-product factorization, the Hilbert spaces
associated with adjacent regions are combined via an entangling product, defined by requiring that the charges computed from the two subregions agree on their
identified boundary, as described in Section~\ref{sec:localsubsystems}. This construction makes explicit the role of corner degrees of freedom in the definition of
quantum gravitational subsystems,
and provides a concrete realization of the idea that the obstruction to factorization in gauge theories and gravity is encoded in boundary symmetry data.

Building on this gluing construction, we then computed the entanglement entropy associated with quantum corner states. A distinguished class of coherent states,
referred to as classical states in Section~\ref{sec:localsubsystems}, was shown to possess an entropy that grows linearly with the representation parameter
$\lambda$ in the large-$\lambda$ regime. This behavior is precisely the one needed for a semiclassical interpretation in terms of geometric area.
The main physical result of the thesis is obtained when this quantum information-theoretic structure is matched to the classical geometry of static, spherically
symmetric spacetimes. Using the explicit Noether charges derived in Section~\ref{sec:cornersymmetriesingravity} and the coherent-state technology
of Chapter~\ref{chapter3}, we established in Section~\ref{sec:sssspacetime} a precise relation between the representation parameter
$\lambda$ and the area of the entangling corner in Planck units. For the classical coherent states singled out previously, the entanglement entropy then reproduces,
in the semiclassical limit, the Bekenstein--Hawking area law,
\begin{equation*}
S \sim \frac{A}{4 \ell_p^2} + \frac{1}{2}\ln\!\left(\frac{A}{4\ell_p^2}\right) + \cdots.
\label{eq:conclusion-area-law}
\end{equation*}
This is, in our view, the central result of the thesis. Within the
corner proposal, the black-hole entropy formula emerges as an entanglement entropy of quantum spacetime subregions,
derived only from symmetries and representation theory.

Taken together, these results provide substantial evidence that corner symmetries capture non-trivial and physically meaningful degrees of freedom of quantum gravity.
The framework developed here supplies, in a unified manner, a classical charge algebra, its quantum representations, a semiclassical sector, a prescription for local
subsystems, and a derivation of the entropy area law. While many open problems remain, these achievements show that the corner proposal is not merely a formal
reorganization of gravitational symmetries, but a productive and predictive framework in which structural questions of quantum gravity can be addressed concretely.

Beyond the future directions mentioned in Section~\ref{sec:goingfurther}, many avenues remain open already within the two-dimensional framework.
One natural direction would be to extend the analysis of Chapter~\ref{chapter4} beyond static spacetimes.
This would, for instance, make it possible to study Friedmann--Lemaître--Robertson--Walker spacetimes and to develop the quantum-cosmological aspect of the corner
proposal. Another important extension would be to incorporate matter fields. One could then investigate gravitational collapse and black-hole formation in a quantum
gravitational setting, and perhaps gain new insight into the generalized second law of thermodynamics.

A second direction we would like to emphasize concerns the emergence of spacetime itself. One of the strengths of the corner proposal is that it provides a Hilbert
space for quantum gravity without relying on the quantization of a pre-existing classical theory. At the same time, this is also one of its conceptual
challenges, since the notions of spacetime and classical geometry are no longer fundamental in the construction.
A natural question is therefore the following: how does classical spacetime emerge from the representation-theoretic data of the corner proposal?
One possible direction would be to construct \textit{corner networks}, describing several local subregions of spacetime glued together, in a way inspired
by tensor-network ideas and their relation to holography. Tensor networks have long suggested that patterns of entanglement may admit a geometric
interpretation \cite{Swingle2012EntanglementRenormalization,VanRaamsdonk2010BuildingSpacetime}, and more recent developments have shown that they can provide useful toy models of bulk-boundary
duality and quantum error correction \cite{PastawskiYoshidaHarlowPreskill2015,HaydenNezamiQiThomasWalterYang2016}. In the present context, such a construction would provide a possible framework
in which spacetime geometries could emerge from the entanglement and gluing of corner degrees of freedom.
For example, one could try to define an effective notion of distance or metric on the space of corner networks from relational data, in the spirit of \cite{Cao:2016mst}
and study whether a suitable coarse-graining of the network gives rise to an emergent classical geometry.\par
Overall, the corner proposal for quantum gravity constitutes a promising avenue of research.
It offers a novel perspective on the problem of quantum gravity, which has challenged physicists for more than a century.
The results presented in this thesis suggest that this framework is both viable and fruitful, and they open the way to a broad research program with the potential to
shed new light on the quantum structure of spacetime.


\printbibliography[heading=bibintoc]



\appendix 



\end{document}